\documentclass[12pt]{article}
\pdfoutput=1
\usepackage[english]{babel}
\usepackage{color}
\usepackage{graphicx}
\usepackage{latexsym}
\usepackage{amssymb, amsmath, amsfonts}
\usepackage{indentfirst}
\usepackage{float,times}
\usepackage{bbm}
\usepackage{subfigure}
\usepackage{slashed}
\usepackage{ifthen}
\usepackage{mathrsfs}
\usepackage{array}
\usepackage{feynmf}
\usepackage{pdfpages}
\usepackage{pxfonts}

\usepackage{blindtext}
\usepackage[
bookmarks]{hyperref}

\hypersetup{colorlinks, bookmarksopen, bookmarksnumbered,
citecolor=verdeCP3, linkcolor=bluCP3, pdfstartview=FitH, urlcolor=rossoCP3
}

\newcommand{\pnt}{\rule[-2mm]{0mm}{6mm}}

\newcommand{\virg}[1]{`#1'}

\def\s#1{\setbox0=\hbox{$#1$}%
\rlap{\ifdim\wd0>.7em\kern.22\wd0\else\kern.1\wd0\fi /}#1}
\def\circa#1{\,\raise.3ex\hbox{$#1$\kern-.75em\lower1ex\hbox{$\sim$}}\,}

\newskip\humongous \humongous=0pt plus 1000pt minus 1000pt

\newif\ifdtup

\def\oldreffmt#1{\rlap{[#1]} \hbox to 2\parindent{}}

\def\figfmt#1{\rlap{Figure {#1}} \hbox to 1in{}}

\def\beq{\begin{equation}}
\def\eeq{\end{equation}}
\def\bea{\begin{eqnarray}}
\def\eea{\end{eqnarray}}

\def\bq{\begin{quote}}
\def\eq{\end{quote}}

 \newcommand{\ba}{\begin{array}}
\newcommand{\ea}{\end{array}}
\newcommand{\bi}{\begin{itemize}}
\newcommand{\ei}{\end{itemize}}
\newcommand{\bn}{\begin{enumerate}}
\newcommand{\en}{\end{enumerate}}
\newcommand{\bc}{\begin{center}}
\newcommand{\ec}{\end{center}}

\newcommand{\no}{\nonumber}

 \newcommand{\gsim}{\lower.7ex\hbox{$\;\stackrel{\textstyle>}{\sim}\;$}}
\newcommand{\lsim}{\lower.7ex\hbox{$\;\stackrel{\textstyle<}{\sim}\;$}}

\def\met{{\slashed{E}}_T}
\def\mpt{{\slashed{p}}_T}

\def\be{\begin{equation}} 
\def\ee{\end{equation}} 
\def\bea{\begin{eqnarray}}
\def\eea{\end{eqnarray}}
\def\ba{\begin{array}}
\def\ea{\end{array}}
\def\kslash{\raise.15ex\hbox{/}\kern-.57em k}
\newcommand{\bear}{\begin{eqnarray}}
\newcommand{\eear}{\end{eqnarray}}

\newcommand{\nn}{~\nonumber\\}

\def\beq{\begin{equation}}
\def\eeq{\end{equation}}
\def\bal{\begin{align}}
\def\eal{\end{align}}

\def\drawbox#1#2{\hrule height#2pt
        \hbox{\vrule width#2pt height#1pt \kern#1pt
              \vrule width#2pt}
              \hrule height#2pt}

\def\Asym#1#2{\vcenter{\vbox{\drawbox{#1}{#2}
              \kern-#2pt 
              \drawbox{#1}{#2}}}}

\definecolor{rossoCP3}{cmyk}{0,.88,.77,.40}
\definecolor{verdeCP3}{rgb}{0.09765625, 0.57421875, 0.1015625}
\definecolor{bluCP3}{rgb}{0, 0.23, 0.67}

\begin{document}

\setlength{\unitlength}{1mm}
\begin{fmffile}{diagrammi}

\includepdf[noautoscale=true, scale=1.06, offset=35 -11]{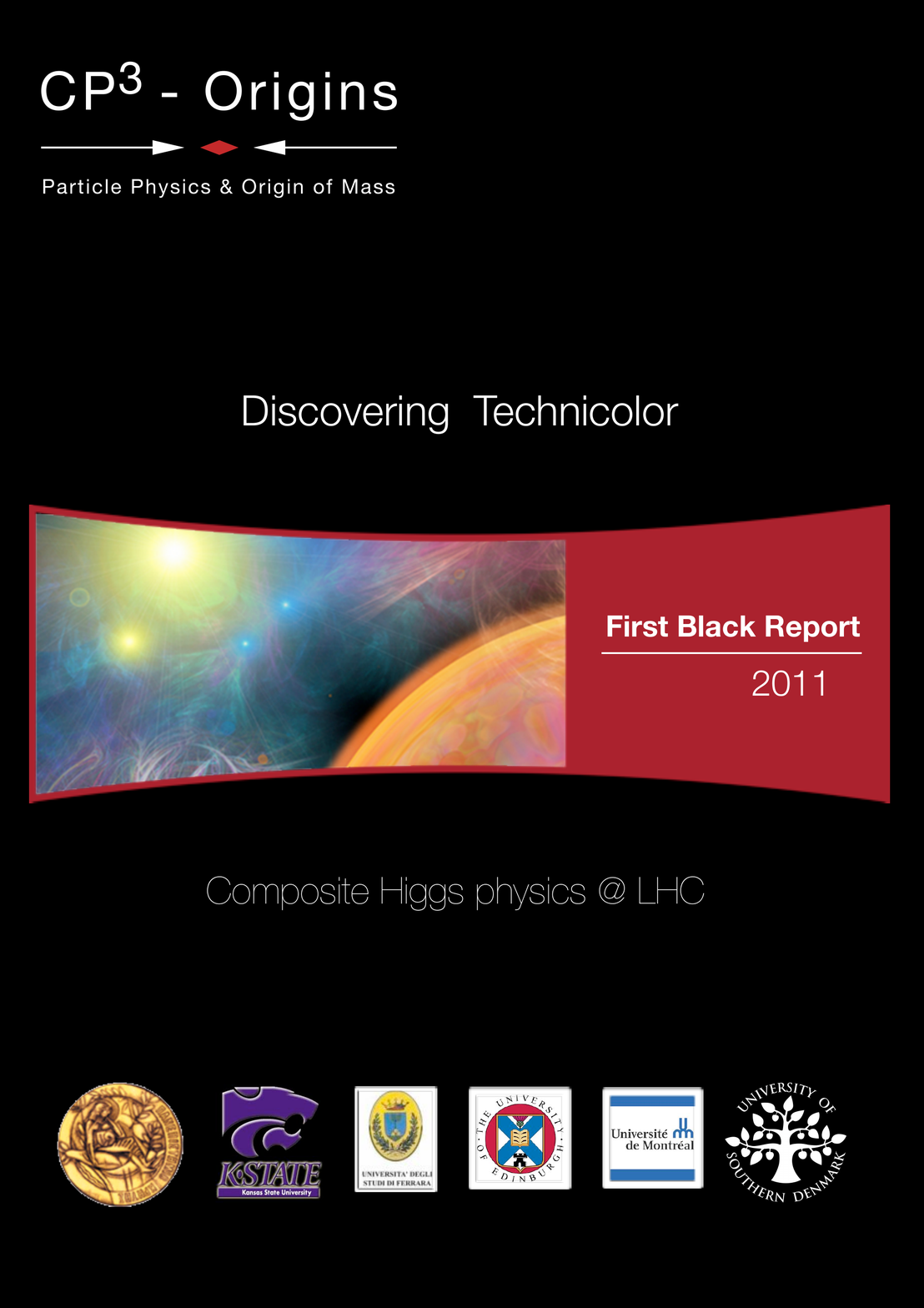}

\begin{titlepage}
\begin{flushright}
\end{flushright}
\begin{center}
{\Huge {\color{rossoCP3}
\bf Discovering Technicolor }} 
\end{center}
\par \vskip .2in \noindent
{  \large {\color{black}J.R. Andersen$^{\color{rossoCP3} {\varheartsuit}}$ O. Antipin$^{\color{rossoCP3} {\varheartsuit}}$ G. Azuelos$^{\color{rossoCP3} {\blacktriangle}}$  L. Del Debbio$^{\color{rossoCP3} {\clubsuit}}$ 
E. Del Nobile$^{\color{rossoCP3} {\varheartsuit}}$  \\~\\ S. Di Chiara$^{\color{rossoCP3} {\varheartsuit}}$  T. Hapola$^{\color{rossoCP3} {\varheartsuit}}$ 
M. J\"arvinen$^{\color{rossoCP3} {\blacklozenge}}$  P.J. Lowdon$^{\color{rossoCP3} {\clubsuit}}$ Y. Maravin$^{\color{rossoCP3} {\spadesuit}}$  \\~\\ I. Masina$^{\color{rossoCP3} {\blacktriangledown} \color{rossoCP3} {\varheartsuit}}$ 
M. Nardecchia$^{\color{rossoCP3} {\varheartsuit}}$ C. Pica$^{\color{rossoCP3} {\varheartsuit}}$  F. Sannino$^{\color{rossoCP3} {\varheartsuit}}$ 
}}
\\~\\
 \noindent
  {\it 
$^{\color{rossoCP3} {\varheartsuit}}${CP}$^{ \bf 3}${-Origins}, University of Southern Denmark, Odense, Denmark \\ 
$^{\color{rossoCP3} {\blacktriangle}}$Universite de Montr\' eal, Montr\' eal, Canada and TRIUMF, Vancouver, Canada \\ 
$^{\color{rossoCP3} {\clubsuit}}$Tait Institute, University of Edinburgh, Edinburgh, Scotland, UK \\ 
$^{\color{rossoCP3} {\blacklozenge}}$Crete Center for Theoretical Physics, University of Crete, Heraklion, Greece \\
$^{\color{rossoCP3} {\spadesuit}}$Kansas State University, Manhattan, KS, USA \\ 
$^{\color{rossoCP3} {\blacktriangledown}}$Universit\`a degli Studi di Ferrara and INFN Sez. di Ferrara, Italy 
}
 
  \par \vskip .1in \noindent
 \begin{abstract}
We provide a pedagogical introduction to extensions of the Standard Model in which the Higgs is composite. These extensions are known as models of dynamical electroweak symmetry breaking or, in brief, Technicolor. Material covered includes: motivations for Technicolor, the construction of underlying gauge theories leading to minimal models of Technicolor, the comparison with electroweak precision data, the  low energy effective theory, the spectrum of the states common to most of the Technicolor models, the decays of the composite particles and the experimental signals at the Large Hadron Collider. The level of the presentation is aimed at readers familiar with the Standard Model but who have little or no prior exposure to Technicolor. Several extensions of the Standard Model featuring a composite Higgs can be reduced to the effective Lagrangian introduced in the text.

We establish the relevant experimental benchmarks for Vanilla, Running, Walking, and Custodial Technicolor, and a natural fourth family of leptons, by laying out the framework to discover these models at the Large Hadron Collider.

\end{abstract} 
\vfill
\begin{flushright}
{\small \it  CP$^3$-Origins-2011-13 \\  CCTP-2011-11 }
\end{flushright}
\end{titlepage}

         \newpage

\def\baselinestretch{1.0}
\tiny
\normalsize

\tableofcontents

\newpage

\section{The need to go beyond}

The energy scale at which the Large Hadron Collider experiment (LHC) operates is determined by the need to complete the Standard Model (SM) of particle interactions and, in particular, to understand the origin of mass of the elementary particles. Together with classical general relativity the SM constitutes one of the most successful models of nature.  We shall, however, argue that experimental results and theoretical arguments call for a more fundamental description of nature.

\begin{figure}[hb]
\label{figureSMplot1} 
\center{
\includegraphics[width=12cm]{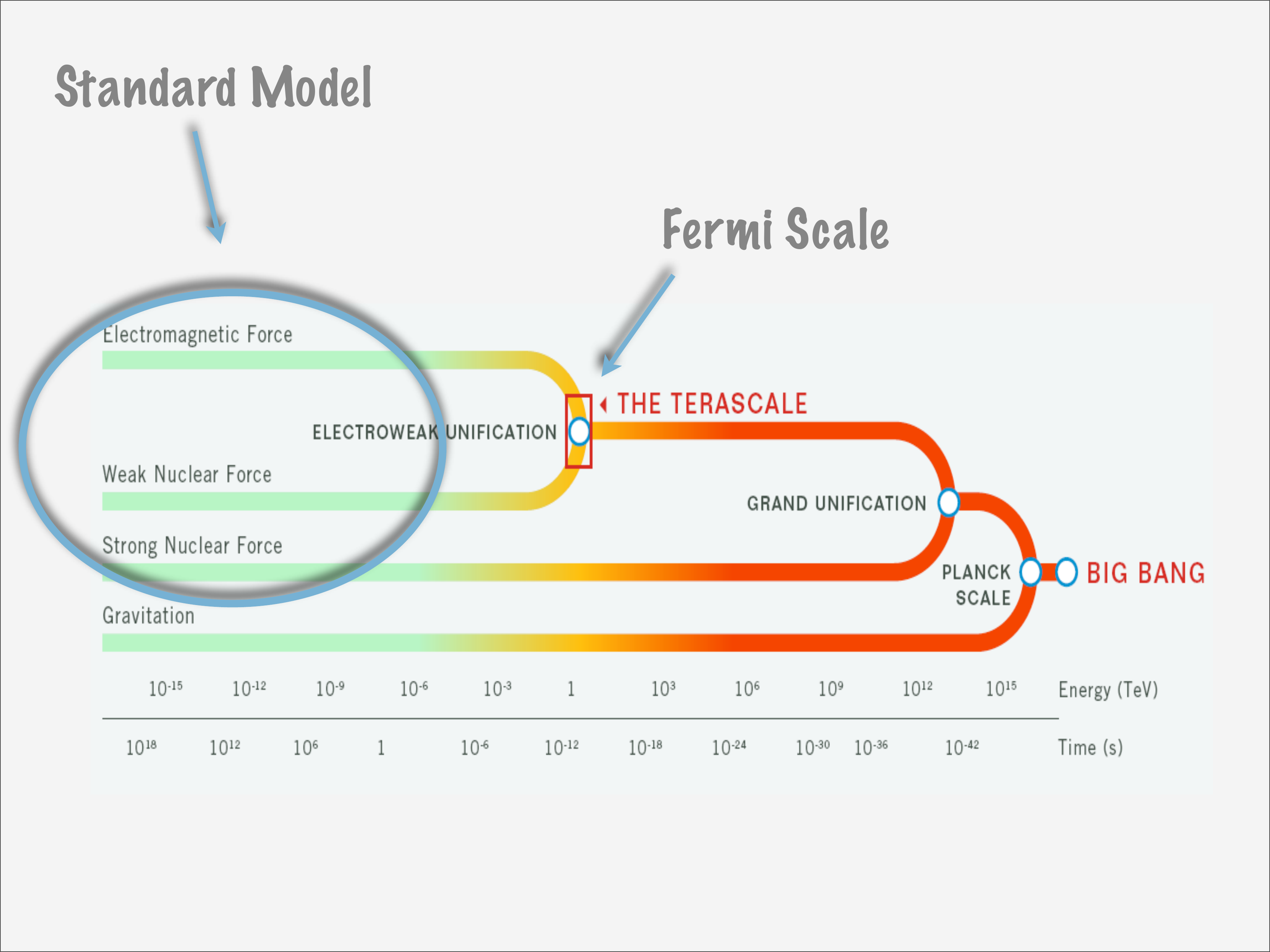}
\caption{Cartoon representing the various forces of  nature. At very high energies one may imagine that all the low-energy forces unify in a single force.}}
\end{figure}

In Fig.~\ref{figureSMplot1}, we schematically represent, in green, the known forces of  nature. The SM of particle physics describes the strong, weak and electromagnetic forces. The yellow region represents the energy scale around the TeV scale and is being explored  at the LHC, while the red  part of the diagram is speculative.  

\begin{figure}[ht]
\begin{minipage}{16pc}
\vskip1.5cm\hskip -.5cm\includegraphics[width=18pc]{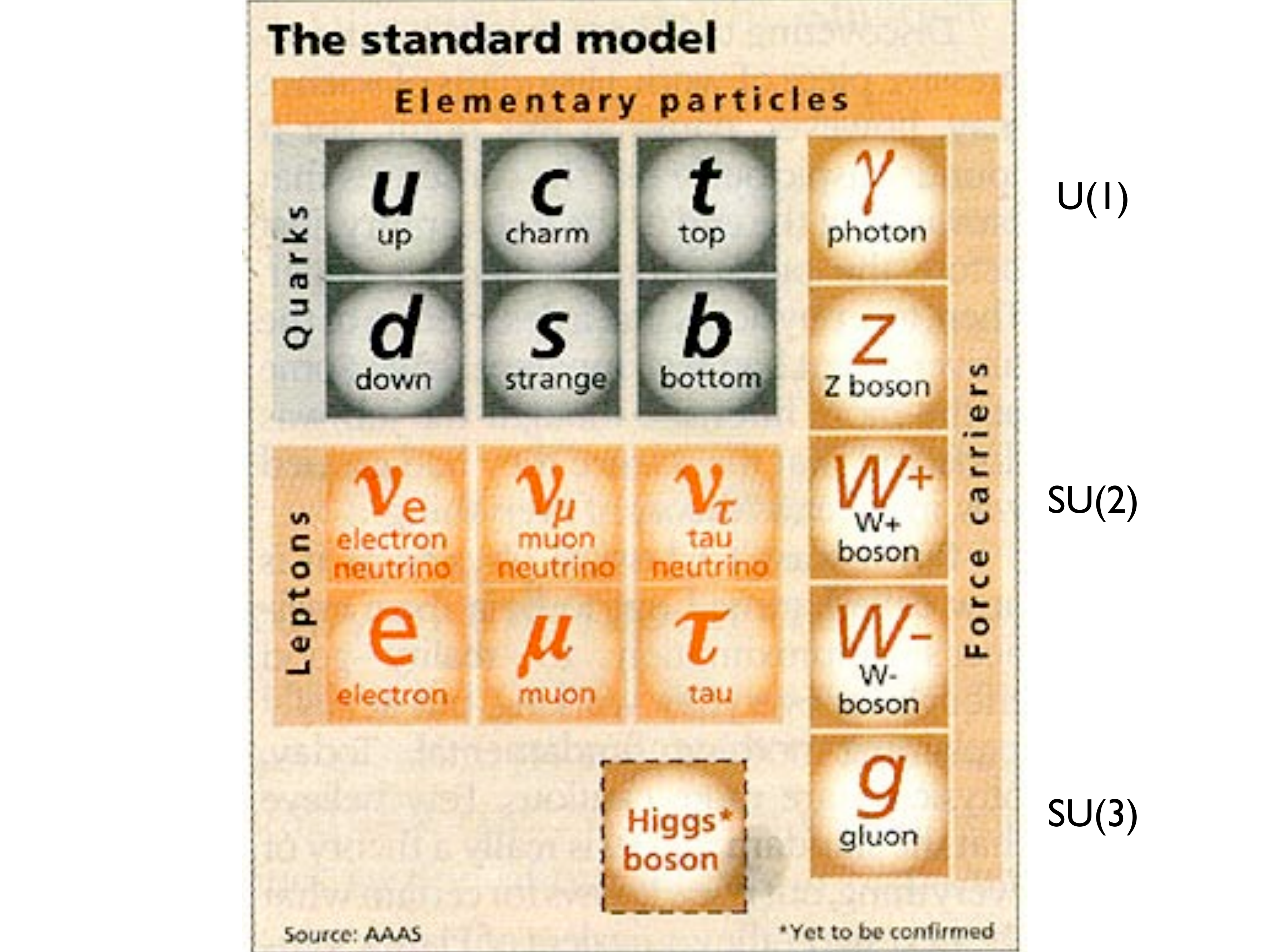}
\caption{Postage stamp representing all of the elementary particles which constitute the SM. The forces are mandated with the  $SU(3)\times SU(2) \times U(1)$ gauge group.}
\label{SMstamp}
\end{minipage}\hspace{3pc}%
\begin{minipage}{20pc}
\vskip -.7cm\hskip -1.3cm\includegraphics[width=28pc]{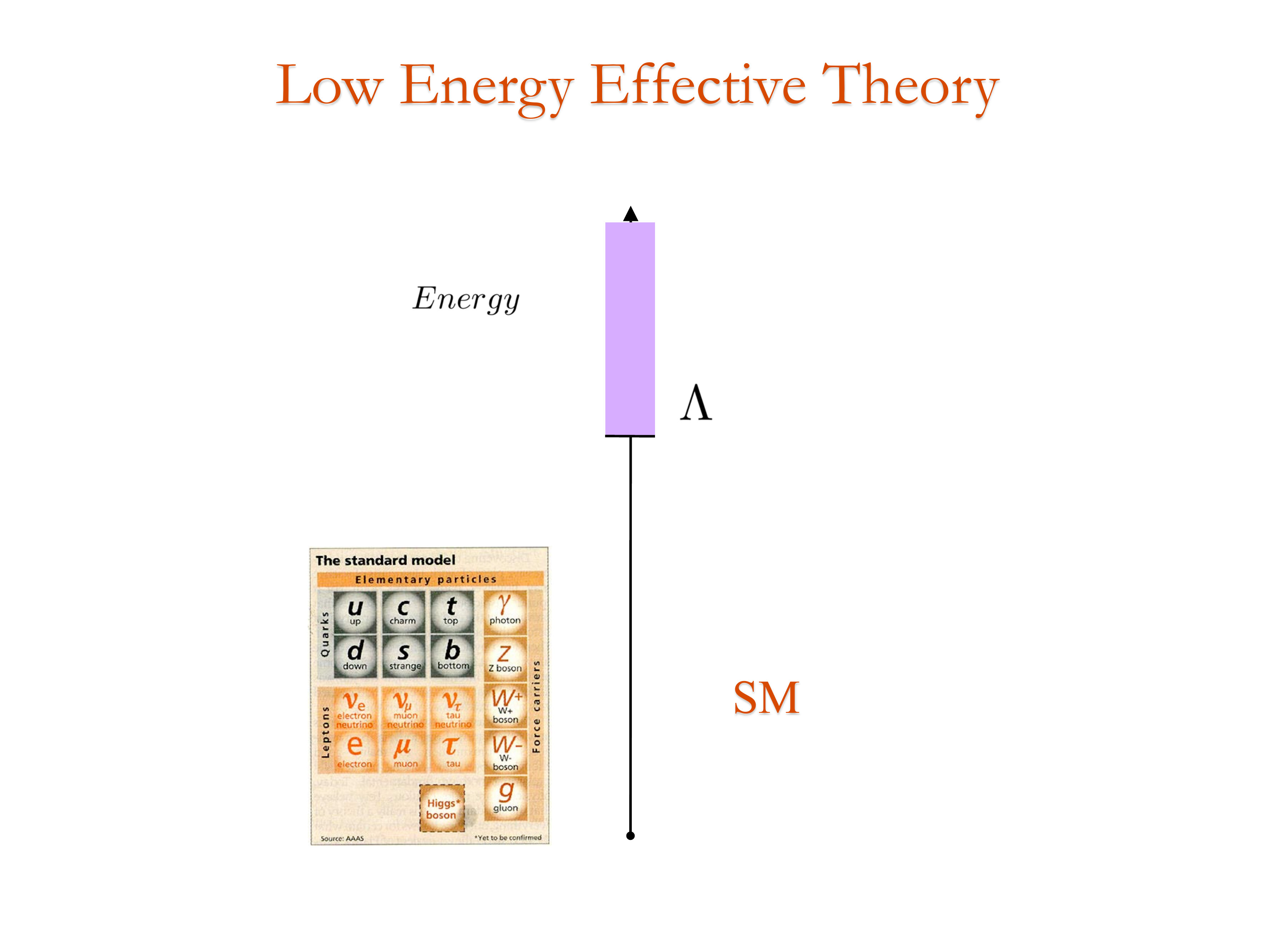}
\vskip -1cm\caption{The SM can be viewed as a low-energy theory valid up to a high energy scale $\Lambda$.}
\label{lowenergy}
\end{minipage} 
\end{figure}

All of the known elementary particles constituting the SM fit on the postage stamp shown in Fig.~\ref{SMstamp}. Interactions among quarks and leptons are carried by gauge bosons. Massless gluons mediate the strong force among quarks while the massive  gauge bosons, i.e. the $Z$ and $W$, mediate the weak force and interact with both quarks and leptons. Finally, the massless photon, the quantum of light, interacts with all of the electrically charged particles. The SM Higgs  is introduced to provide mass to the elementary particles and in its minimal version does not feel strong interactions. The interactions emerge naturally by invoking a gauge principle which is intimately linked to the underlying symmetries relating the various particles of the SM. 
The asterisk on the Higgs boson in the postage stamp indicates that it has not yet been observed. Intriguingly the Higgs is the only fundamental scalar of the SM.


The SM can be viewed as a low-energy effective theory valid up to an energy scale $\Lambda$, as schematically represented in Fig.~\ref{lowenergy}. Above this scale new interactions, symmetries, extra dimensional worlds or any other extension could emerge. At sufficiently low energies with respect to this scale one expresses the existence of new physics via effective operators. The success of the SM is due to the fact that most of the corrections to its physical observables depend only logarithmically on this scale $\Lambda$. In fact, in the SM there exists only one operator which acquires corrections quadratic in $\Lambda$. This is the squared mass operator of the Higgs boson.  Since $\Lambda$ is expected to be the highest possible scale, in four dimensions the Planck scale (assuming that we have only the SM and gravity), it is hard to explain {\it naturally}  why the mass of the Higgs is of the order of the Electroweak (EW) scale. This is the hierarchy problem. Due to the occurrence of quadratic corrections in the cutoff this SM sector is most sensitive to the existence of new physics.

\subsection{The Higgs}
It is a fact that the Higgs allows for a direct and economical way of spontaneously breaking the electroweak symmetry.  It generates simultaneously the masses of the quarks and leptons without introducing Flavor Changing Neutral Currents (FCNC)s at the tree level.  
 The Higgs sector of the SM
possesses, when the gauge couplings are switched off, an
$SU(2)_\text{L} \times SU(2)_\text{R}$ symmetry. The full symmetry group can be made explicit when re-writing the Higgs
doublet field \begin{eqnarray} H=\frac{1}{\sqrt{2}}\left(%
\begin{array}{c}
  \pi_2 + i\, \pi_1 \\
  \sigma - i\, \pi_3 \\
\end{array}%
\right)\end{eqnarray} as the right column of the following two by two matrix:
\begin{eqnarray}
\frac{1}{\sqrt{2}}\left(\sigma + i\,
\vec{\tau}\cdot\vec{\pi} \right) \equiv M
 \ .
\end{eqnarray}
The first column can be identified with the column vector $i\tau_2H^{\ast}$ while the second with $H$. 
$\tau^2$ is the second Pauli matrix. 
The $SU(2)_\text{L}\times SU(2)_\text{R}$ group acts linearly on $M$ according
to:
\begin{eqnarray}
M\rightarrow g_L M g_R^{\dagger} \qquad {\rm and} \qquad g_{L/R} \in SU(2)_\text{L/R}\ .
\end{eqnarray}
One can verify that:
\begin{eqnarray}
M\frac{\left(1-\tau^3\right)}{2} = \left(0\ , \, H\right) \ . \qquad
M\frac{\left(1+\tau^3\right)}{2} = \left(i\,\tau_2H^{\ast} \ , \, 0\right) \ .
\end{eqnarray}
The $SU(2)_\text{L}$ symmetry is gauged by introducing the weak gauge
bosons $W^a$ with $a=1,2,3$. The hypercharge generator is taken to
be the third generator of $SU(2)_\text{R}$. The ordinary covariant
derivative acting on the Higgs, in the present notation, is:
\begin{eqnarray}
D_{\mu}M=\partial_{\mu}M -i\,g\,W_{\mu}M + i\,g^{\prime}M\,B_{\mu} \ , \qquad {\rm
with}\qquad W_{\mu}=W_{\mu}^{a}\frac{\tau^{a}}{2} \ ,\quad
B_{\mu}=B_{\mu}\frac{\tau^{3}}{2} \ .
\end{eqnarray}
The Higgs Lagrangian is 
\begin{eqnarray}
{\cal L}&=&\frac{1}{2}{\rm Tr} \left[D_{\mu}M^{\dagger}
D^{\mu}M\right]-\frac{m_M^2}{2} {\rm
Tr}\left[M^{\dagger}M\right] - \frac{\lambda}{4}\,{\rm
Tr}\left[M^{\dagger}M\right]^2 \ .
\end{eqnarray}
At this point one {\it assumes} that the mass squared of
the Higgs field is negative and this leads to the electroweak
symmetry breaking. Except for the Higgs mass term the other SM operators have dimensionless couplings meaning that the natural scale for the SM is encoded in the Higgs mass\footnote{The mass of the proton is due mainly to strong interactions, however its value cannot be determined within QCD since the associated renormalization group invariant scale must be fixed to an hadronic observable.}. We recall that the Higgs Lagrangian has a familiar form since it is identical to the linear $\sigma$ Lagrangian which was introduced long ago to describe chiral symmetry breaking in QCD with two light flavors. 

At the tree level, when taking $m_M^2$ negative and the self-coupling $\lambda$ positive, one determines:
\begin{equation}
\langle \sigma \rangle^2 \equiv v^2=\frac{|m_M^2|}{\lambda} \qquad {\rm and} \qquad \sigma = v + h \ ,
\end{equation} 
where $h$ is the Higgs field. 
The global symmetry breaks to its diagonal subgroup:
\begin{eqnarray}
SU(2)_\text{L}\times SU(2)_\text{R} \rightarrow SU(2)_\text{V} \ .
\end{eqnarray}
To be more precise the $SU(2)_\text{R}$ symmetry is already broken explicitly by our choice of gauging only an $U(1)_Y$ subgroup of it and hence the actual symmetry breaking pattern is:
\begin{eqnarray}
SU(2)_\text{L}\times U(1)_Y \rightarrow U(1)_Q \ ,
\end{eqnarray}
with $U(1)_Q$ the electromagnetic abelian gauge symmetry. According to the Nambu-Goldstone's theorem three massless degrees of freedom appear, i.e. $\pi^{\pm}$ and $\pi^0$. In the unitary gauge these Goldstones become the longitudinal degree of freedom of the massive elecetroweak gauge bosons. Substituting the vacuum value for $\sigma$ in the Higgs Lagrangian the gauge bosons quadratic terms read:
\begin{equation}
\frac{v^2}{8}\, \left[g^2\,\left(W_{\mu}^1
W^{\mu,1} +W_{\mu}^2 W^{\mu,2}\right)+ \left(g\,W_{\mu}^3 -
g^{\prime}\,B_{\mu}\right)^2\right]  \ . \end{equation}
 The $Z_{\mu}$ and the photon $A_{\mu}$ gauge bosons are:
\begin{eqnarray}
Z_{\mu}&=&\cos\theta_W\, W_{\mu}^3 - \sin\theta_{W}B_{\mu} \ ,\nonumber \\
A_{\mu}&=&\cos\theta_W\, B_{\mu} + \sin\theta_{W}W_{\mu}^3 \ ,
\end{eqnarray}
with $\tan\theta_{W}=g^{\prime}/g$ while the charged massive vector bosons are
$W^{\pm}_{\mu}=(W^1\pm i\,W^2_{\mu})/\sqrt{2}$. 
The bosons masses $M^2_W=g^2\,v^2/4$ due to
the custodial symmetry satisfy the tree level relation $M^2_Z=M^2_W/\cos^2\theta_{W}$. 
Holding fixed the EW scale $v$ the mass squared of the Higgs boson is $2\lambda v^2_\text{EW}$ and hence it increases with $\lambda$.

Besides breaking the electroweak symmetry dynamically the ordinary Higgs serves also the purpose to provide mass to all of the SM particles via the Yukawa terms of the type:
\beq -Y_d^{ij}\bar{Q}_L^i H d_R^j  - Y_u^{ij}\bar{Q}_L^i (i\tau_2 H^{\ast}) u_R^j  + {\rm h.c.}\ , \eeq
where $Y_{q}$ is the Yukawa coupling constant, $Q_L$ is the
left-handed Dirac spinor of quarks, $H$ the Higgs doublet and
 $q$ the right-handed Weyl spinor for the quark and $i,j$ the flavor indices. The $SU(2)_\text{L}$ weak and spinor indices are suppressed. 

When considering quantum corrections the Higgs mass acquires large quantum corrections  proportional to the scale of the cutoff squared. 
\begin{equation}
{M_H}^2_{\rm ren} - M_H^2 = \frac{k g^2\Lambda^2}{16 \pi^2} \ .
\end{equation}
Here $g$ is and electroweak constant and k a numerical factor depending on the specific model, expected to be ${\cal O}(1)$.  $\Lambda$ is the highest energy above which the SM is no longer a valid description of Nature and a large fine tuning of the parameters of the Lagrangian is needed to offset the effects of the cutoff. This large fine tuning is needed because there are no symmetries protecting the Higgs mass operator from large corrections which would hence destabilize the Fermi scale (i.e. the electroweak scale). This problem is the one we referred above as the hierarchy problem of the SM.

The constant value of the Higgs field evaluated on the ground state is determined by the measured mass of the $W$ boson. On the other hand, the value of the SM Higgs mass  ($M_H$) is constrained only indirectly by the electroweak precision data. The preferred value of the Higgs mass (obtained by the standard fit which excludes direct Higgs searches at LEP and Tevatron) is $M_H=95.7^{+30.6}_{-24.2}$~GeV at 68\% confidence level (CL) with a 95\% CL upper limit $M_H < 171.5$~GeV, as given by the generic fitting package Gfitter \cite{Flacher:2008zq}. The corresponding results obtained by a fit including the direct Higgs searches produces $M_H=120.6^{+17.9}_{-5.2}$~GeV at 68\% confidence level (CL) with a 95\% CL upper limit $M_H < 155.3$~GeV, as reported on \href{http://gfitter.desy.de/GSM/}{http://gfitter.desy.de/GSM/} by the Gfitter Group \footnote{All the plots and numerical results we use in this section are reported by the Gfitter Group and can be found at the web-address: \href{http://gfitter.desy.de/GSM/}{\rm http://gfitter.desy.de/GSM/}.}.

\begin{figure}[t]
\begin{minipage}{16pc}
\includegraphics[width=16pc]{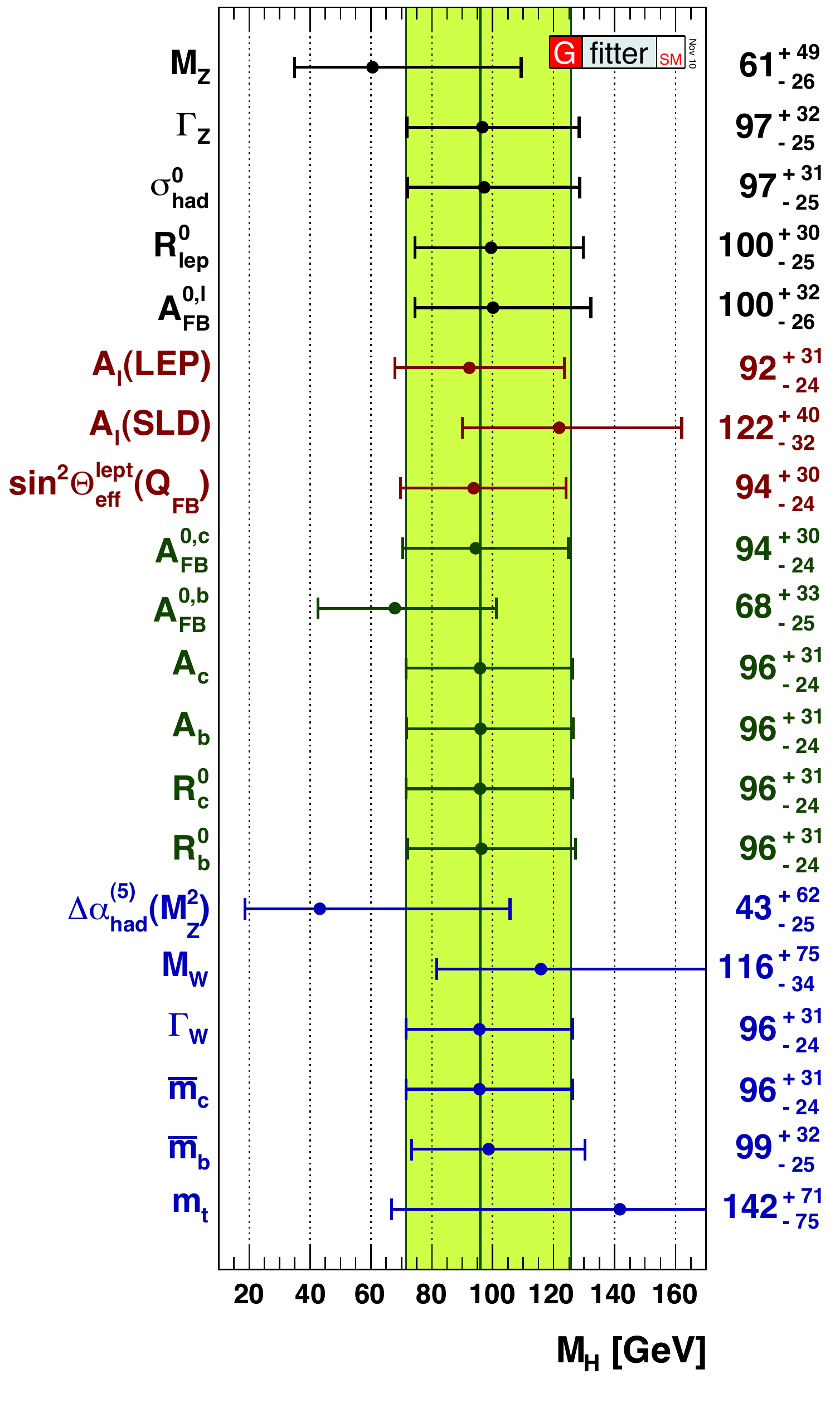}
\caption{Values of the Higgs mass from the standard fit (which does not take into account direct Higgs searches) obtained by excluding different electroweak observables. The green band represent the $1\sigma$ error range around the best fit value of $M_H$.}
\label{fig1}
\end{minipage}\hspace{1pc}
\begin{minipage}{20pc}
\includegraphics[width=20pc]{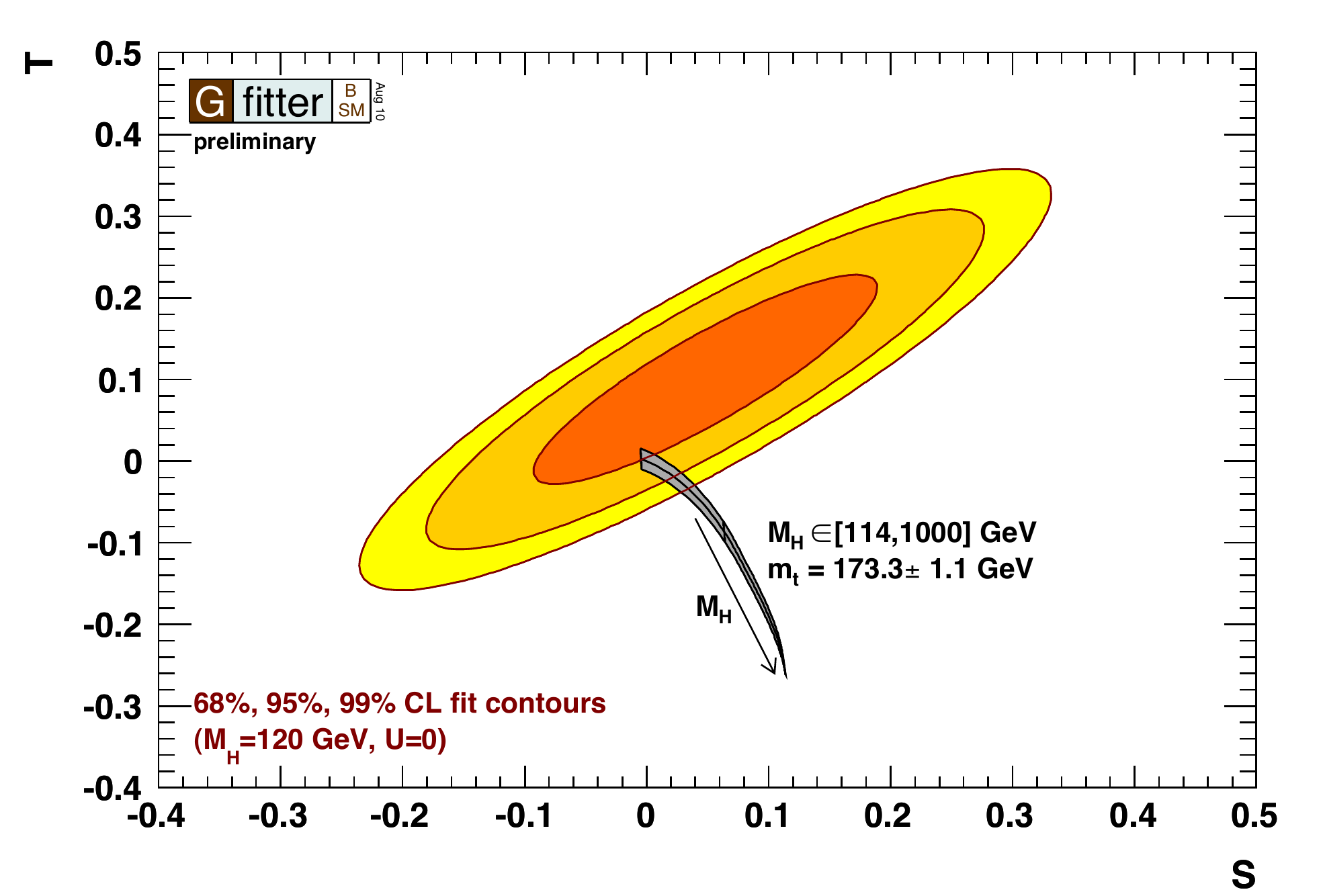}
\caption{The 68\%, 95\%, and 99\% CL countours of the electroweak parameters $S$ and $T$ determined from different observables derived from a fit to the electroweak precision data. The gray area gives the SM prediction with $m_t$ and $M_H$ varied as shown. $M_H = 120$ GeV and $m_t = 173.1$ GeV defines the reference point at which all oblique parameters vanish.}
\label{s06_stu_contours}
\end{minipage} 
\end{figure}
The final result of the average of all of the measures, however, has a Pearson's chi-square ($\chi^2$) test of 17.5 for 14 degrees
of freedom.
A Higgs heavier than $155.3$~GeV is compatible with precision tests if we allow simultaneously new physics to compensate for the effects of the heavier value of the mass. The precision measurements of direct interest for the Higgs sector are often reported using the $S$ and $T$ parameters as shown in Fig.~\ref{s06_stu_contours}. From this graph one deduces that a heavy Higgs is compatible with data at the expense of a large value of the $T$ parameter.  Actually, even the lower direct experimental limit on the Higgs mass can be evaded with suitable extensions of the SM Higgs sector.

Many more questions need an answer if the Higgs is found at the LHC: Is it composite? How many Higgs fields are there in nature? Are there hidden sectors?

\subsection{Riddles}

Why do we expect that there is new physics awaiting to be discovered? Of course, we still have to observe the Higgs, but this cannot be everything. Even with the Higgs discovered, the SM has both conceptual problems and phenomenological shortcomings. In fact, theoretical arguments indicate that the SM is not the ultimate description of nature: 
 \begin{itemize}
\item{{\bf Hierarchy Problem:} The Higgs sector is highly fine-tuned. We have no natural explanation of the large hierarchy between the Planck and the electroweak scales.}
\item{{\bf Strong CP problem:} There is no natural explanation for the smallness of the electric dipole moment of the neutron within the SM. This problem is also known as the strong CP problem.}
\item{{\bf Origin of patterns:} The SM can fit, but cannot explain the number of matter generations and their mass spectrum. }
\item{{\bf Unification of the forces:} Why do we have so many different interactions? It is appealing to imagine that the SM forces could unify into a single Grand Unified Theory (GUT). We could imagine that at very high energy scales gravity also becomes part of a unified description of nature.}
\end{itemize}
There is no doubt that the SM is incomplete since we cannot even account for a number of basic observations:  
\begin{itemize}

\item{{\bf Neutrino physics:}
Only recently it has been possible to have some definite answers about properties of neutrinos. We now know that they have a tiny mass, which can be naturally accommodated in extensions of the SM, featuring for example a  see-saw  mechanism. We do not yet know if the neutrinos have a Dirac or a Majorana nature.}

\item{{\bf Origin of bright and dark mass:}
Leptons, quarks and the gauge bosons mediating the weak interactions possess a rest mass. Within the SM this mass can be accounted for by the Higgs mechanism, 
which constitutes the electroweak symmetry breaking sector of the SM. However, the associated 
Higgs particle has not yet been discovered. Besides, the SM cannot account for the observed large fraction of dark mass of the universe. What is interesting is that in the universe the dark matter is about five times more abundant than the known baryonic matter, i.e. bright matter. We do not know why the ratio of dark to bright matter is of order unity.}

\item{{\bf Matter-antimatter asymmetry:} 
 From our everyday experience we know that there is very little bright antimatter in the universe. The SM fails to predict the observed excess of matter. }
 
 \end{itemize}
 
These arguments do not imply that the SM is necessarily incorrect, but it must  be extended to answer any of the questions raised above. The truth is that we do not have an answer to the basic question: What lies beneath the SM?

A number of possible generalizations have been
conceived (see \cite{Ellis:2009pz,Altarelli:2009bz,Giudice:2007qj,Mangano:2008ag,Altarelli:2008yi,Barbieri:2008zz} for reviews). Such extensions are introduced on the base of one or more guiding principles or prejudices. Two  technical reviews are \cite{DeRoeck:2009id,Accomando:2006ga}. 

In the models we will consider here the electroweak symmetry breaks via a
fermion bilinear condensate. The Higgs sector of the SM
becomes an effective description of a more fundamental fermionic
theory. This is similar to the Ginzburg-Landau theory of
superconductivity. If the force underlying the fermion condensate driving electroweak symmetry
breaking is due to a strongly interacting gauge theory these models are termed
Technicolor (TC). 

TC, in brief, is an additional non-abelian and strongly interacting gauge theory augmented
with (techni)fermions transforming under a given representation of the gauge group.
The Higgs Lagrangian is replaced by a suitable
new fermion sector interacting strongly via a new gauge interaction (technicolor). Schematically:
\begin{eqnarray}
\mathcal{L}_{Higgs} \rightarrow -\frac{1}{4}F_{\mu\nu}F^{\mu\nu} + i \bar{Q} \gamma_{\mu}D^{\mu} Q  + \ldots\ ,
\end{eqnarray}
where, to be as general as possible, we have left unspecified the underlying nonabelian gauge group and the associated technifermion ($Q$) representation. The dots represent new sectors which may even be needed to avoid, for example, anomalies introduced by the technifermions. 
The intrinsic scale of the new theory is expected to be less or of the order of a few TeV.
The chiral-flavor symmetries of this theory, as for ordinary QCD,
break spontaneously when the technifermion condensate $\bar{Q} Q $ forms. It is
possible to choose the fermion charges in such a way that there is,
at least, a weak left-handed doublet of technifermions and the
associated right-handed one which is a weak singlet. The covariant derivative contains the new gauge field as well as the electroweak ones. The condensate
spontaneously breaks the electroweak symmetry down to the
electromagnetic and weak interactions.
The Higgs is now interpreted as the lightest scalar field with the same quantum numbers of the fermion-antifermion composite field. The Lagrangian part responsible for the mass-generation of the ordinary fermions will also be modified since the Higgs particle is no longer an elementary object. 

Models of electroweak symmetry breaking via new strongly interacting
theories of TC type \cite{Weinberg:1979bn,Susskind:1978ms}
are a mature subject. The two uptodate reviews on which this work is based are \cite{Sannino:2009za,Sannino:2008ha}.  For older nice reviews, updated till 2002, see \cite{Hill:2002ap,Lane:2002wv}. 

One of the main difficulties in
constructing such extensions
of the SM is the very
limited knowledge about generic strongly interacting theories. This
has led theorists to consider specific models of TC which
resemble ordinary QCD and for which the large
body of experimental data at low energies can be directly exported
to make predictions at high energies. To reduce the tension with experimental constraints new strongly coupled theories with dynamics different from the one featured by a scaled up version of QCD are needed \cite{Sannino:2004qp}. 

We will review models of dynamical electroweak symmetry breaking making use of new type of four dimensional gauge theories \cite{Sannino:2004qp,Dietrich:2005jn,Dietrich:2005wk} and their low energy effective description 
\cite{Foadi:2007ue} useful for collider phenomenology. The phase structure of a large number of  strongly interacting nonsupersymmetric theories, as function of number of underlying colors has been uncovered via traditional nonperturbative methods \cite{Dietrich:2006cm} as well as novel ones \cite{Pica:2010mt,Ryttov:2007cx}. 

The theoretical part of this report should be integrated with earlier reviews  \cite{Sannino:2008ha,Hill:2002ap,Lane:2002wv,Shrock:2007km,Sarkar:1995dd,Chanowitz:1988ae,Farhi:1980xs,Kaul:1981uk,Chivukula:2000mb} on the various subjects treated here.

\newpage
\section{Dynamical Electroweak Symmetry Breaking}

It is a fact that the SM does not fail, when experimentally tested, to describe all of the known forces to a very high degree of experimental accuracy. This is true even if we include gravity. Why is it so successful?

The SM is a low energy effective theory valid up to a scale $\Lambda$ above which new interactions, symmetries, extra dimensional worlds or any possible extension can emerge. At sufficiently low energies with respect to the cutoff scale $\Lambda$ one expresses the existence of new physics via effective operators. The success of the SM is due to the fact that most of the corrections to its physical observable depend only logarithmically on the cutoff scale $\Lambda$. 

Superrenormalizable operators are very sensitive to the cut off scale. In the SM there exists only one operator with naive mass dimension two which acquires corrections quadratic in $\Lambda$. This is the squared mass operator of the Higgs boson.  Since $\Lambda$ is expected to be the highest possible scale, in four dimensions the Planck scale, it is hard to explain {\it naturally}  why the mass of the Higgs is of the order of the electroweak scale. The Higgs is also the only particle predicted in the SM yet to be directly produced in experiments. Due to the occurrence of quadratic corrections in the cutoff this is the SM sector highly sensitve to the existence of new physics. 

In Nature we have already observed Higgs-type mechanisms. Ordinary superconductivity and chiral symmetry breaking in QCD are two time-honored examples. In both cases the mechanism has an underlying dynamical origin with the Higgs-like particle being a composite object of fermionic fields.

\subsection{Superconductivity versus electroweak symmetry breaking }
The breaking of the electroweak theory is a relativistic screening effect. It is useful to parallel it to ordinary superconductivity which is also a screening phenomenon albeit non-relativistic. The two phenomena happen at a temperature lower than a critical one. In the case of superconductivity one defines a density of superconductive electrons $n_s$ and to it one associates a macroscopic wave function $\psi$ such that  its modulus squared
\begin{eqnarray}
|\psi|^2 = n_C = \frac{n_s}{2} \ ,
\end{eqnarray}
is the density of Cooper's pairs. That we are describing a nonrelativistic system is manifest in the fact that the macroscopic wave function squared, in natural units, has mass dimension three while the modulus squared of the Higgs wave function evaluated at the minimum is equal to $\langle |H|^2 \rangle = v^2/2$ and has mass dimension two, i.e. is a relativistic wave function. One can adjust the units by considering, instead of the wave functions, the Meissner-Mass of the photon in the superconductor which is
\begin{equation}
 M^2=q^2n_s/(4m_e) \ ,
 \end{equation}
  with $q=-2e$ and $2m_e$ the charge and the mass of a Cooper pair which is constituted by two electrons. In the electroweak theory the Meissner-Mass of the photon  is compared  with the relativistic mass of the $W$ gauge boson
  \begin{equation}
  M^2_W=g^2{v^2}/{4}\ ,
  \end{equation}
  with $g$ the weak coupling constant and $v$ the electroweak scale.  In a superconductor the relevant scale is given by the density of superconductive electrons typically of the order of $n_s\sim 4\times 10^{28}~\text{m}^{-3}$ yielding a screening length of the order of $
\xi = 1/M\sim 10^{-6}{\rm cm}$. In the weak interaction case we measure directly the mass of the weak gauge boson which is of the order of $80$~GeV yielding a weak screening length $\xi_W=1/M_W\sim 10^{-15}{\rm cm}$.  

For a superconductive system it is clear from the outset that the wave function $\psi$ is not a fundamental degree of freedom, however for the Higgs we are not yet sure about its origin. The Ginzburg-Landau effective theory in terms of $\psi$ and the photon degree of freedom describes the spontaneous breaking of the $U(1)_Q$ electric symmetry and it is the equivalent of the Higgs Lagrangian.  

If the Higgs is due to a macroscopic relativistic screening phenomenon we expect it to be an effective description of a more fundamental system with possibly an underlying new strong gauge dynamics replacing the role of the phonons in the superconductive case. A dynamically generated Higgs system solves the problem of the quadratic divergences by replacing the cutoff $\Lambda$ with the weak energy scale itself, i.e. the scale of  compositness.  An underlying strongly coupled asymptotically free gauge theory, a la QCD,  is an example. 
 
 \subsection{From color to Technicolor}
In fact even in complete absence of the Higgs sector in the SM the electroweak symmetry breaks \cite{Farhi:1980xs} due to the condensation of the following quark bilinear in QCD: 
\beq \langle\bar u_Lu_R + \bar d_Ld_R\rangle \neq 0 \ . \label{qcd-condensate}\eeq
  This mechanism, however, cannot account for the whole contribution to the weak gauge bosons masses. If QCD was the only source contributing to the spontaneous breaking of the electroweak symmetry one would have
\beq M_W = \frac{gF_\pi}{2} \sim 29~{\rm MeV}\ , \eeq 
with $F_{\pi}\simeq 93$~MeV the pion decay constant. This contribution is very small with respect to the actual value of the $W$ mass that one typically neglects it. 

According to the original idea of TC \cite{Weinberg:1979bn,Susskind:1978ms} one augments the SM with another gauge interaction similar to QCD but with a new dynamical scale of the order of the electroweak one. It is sufficient that the new gauge theory is asymptotically free and has global symmetry able to contain the SM $SU(2)_\text{L}\times U(1)_Y$ symmetries. It is also required that the new global symmetries break dynamically in such a way that the embedded $SU(2)_\text{L}\times U(1)_Y$  breaks to the electromagnetic abelian charge $U(1)_Q$ . The dynamically generated scale will then be fit to the electroweak one. 

Note that, except in certain cases, dynamical behaviors are typically nonuniversal which means that different gauge groups and/or matter representations will, in general, possesses very different dynamics. 

The simplest example of TC theory is the scaled up version of QCD, i.e. an $SU(N_\text{TC})$  nonabelian gauge theory with two Dirac Fermions transforming according to the fundamental representation or the gauge group. We need at least two Dirac flavors  to realize the $SU(2)_\text{L} \times SU(2)_\text{R}$ symmetry of the SM discussed in the SM Higgs section. One simply chooses the scale of the theory to be such that the new pion decaying constant is: \beq F_\pi^\text{TC} = v \simeq 246~ {\rm GeV} \ . \eeq 
The flavor symmetries, for any $N_\text{TC}$ larger than 2 are $SU(2)_\text{L} \times SU(2)_\text{R} \times U(1)_\text{V}$ which spontaneously break to $SU(2)_\text{V} \times U(1)_\text{V}$. It is natural to embed the electroweak symmetries within the present TC model in a way that the hypercharge corresponds to the third generator of $SU(2)_\text{R}$. This simple dynamical model correctly accounts for the electroweak symmetry breaking. The new technibaryon number $U(1)_\text{V}$ can break due to not yet specified new interactions. 
In order to get some indication on the dynamics and spectrum of this theory one can use the 't Hooft large N limit  \cite{'tHooft:1973jz,Witten:1979kh,'tHooft:1980xb}.  For example the intrinsic scale of the theory is related to the QCD one via:
\beq \Lambda_{\rm TC} \sim
\sqrt{\frac{3}{N_\text{TC}}}\frac{F_\pi^\text{TC}}{F_\pi}\Lambda_{\rm QCD} \ . \eeq
At this point it is straightforward to use the QCD phenomenology for describing the experimental signatures and dynamics of a composite Higgs.  

\subsection{Constraints from electroweak precision data}
\label{5}

The relevant corrections due to the presence of new physics trying to modify the electroweak breaking sector of the SM appear in the vacuum polarizations of the electroweak gauge bosons. These can be parameterized in terms of the three 
quantities $S$, $T$, and $U$ (the oblique parameters) 
\cite{Peskin:1990zt,Peskin:1991sw,Kennedy:1990ib,Altarelli:1990zd}, and confronted with the electroweak precision data. Recently, due to the increase precision of the measurements reported by LEP II, the list of interesting parameters to compute has been extended \cite{hep-ph/9306267,Barbieri:2004qk}.  We show below also the relation with the traditional one \cite{Peskin:1990zt}.  Defining with  $Q^2\equiv -q^2$ the Euclidean transferred momentum entering in a generic two point function vacuum polarization associated to the electroweak gauge bosons, and denoting derivatives with respect to $-Q^2$ with a prime we have \cite{Barbieri:2004qk}: 
\begin{eqnarray}
\hat{S} &\equiv & g^2 \ \Pi_{W^3B}^\prime (0) \ , \\
\hat{T} &\equiv & \frac{g^2}{M_W^2}\left[ \Pi_{W^3W^3}(0) -
\Pi_{W^+W^-}(0) \right] \ , \\
W &\equiv & \frac{g^2M_W^2}{2} \left[\Pi^{\prime\prime}_{W^3W^3}(0)\right] \ , \\
Y &\equiv & \frac{g'^2M_W^2}{2} \left[\Pi^{\prime\prime}_{BB}(0)\right] \ , \\
\hat{U} &\equiv & -g^2 \left[\Pi^\prime_{W^3W^3}(0)-
\Pi^\prime_{W^+W^-}(0)\right]\ , \\
V &\equiv & \frac{g^2 \, M^2_W}{2}\left[\Pi^{\prime\prime}_{W^3W^3}(0)-
\Pi^{\prime\prime}_{W^+W^-}(0)\right] \ , \\
X &\equiv & \frac{g g'\,M_W^2}{2} \ \Pi_{W^3B}^{\prime\prime}(0) \ .
\end{eqnarray}
Here $\Pi_V(Q^2)$ with $V=\{W^3B,\, W^3W^3,\, W^+W^-,\, BB\}$ represents the
self-energy of the vector bosons. The
electroweak couplings are the ones associated to the physical electroweak gauge bosons:
\begin{eqnarray}
\frac{1}{g^2} \equiv  \Pi^\prime_{W^+W^-}(0)
 \ , \qquad \frac{1}{g'^2}
\equiv  \Pi^\prime_{BB}(0) \ ,
\end{eqnarray}
while $G_F$ is
\begin{eqnarray}
\frac{1}{\sqrt{2}G_F}=-4\Pi_{W^+W^-}(0) \ ,
\end{eqnarray}
as in \cite{Chivukula:2004af}. $\hat{S}$ and $\hat{T}$ lend their name
from the well known Peskin-Takeuchi parameters $S$ and $T$ which are related to the new ones via
\cite{Barbieri:2004qk,Chivukula:2004af}:
\begin{eqnarray}
\frac{\alpha S}{4s_W^2} =  \hat{S} - Y - W  \ , \qquad 
\alpha T = \hat{T}- \frac{s_W^2}{1-s_W^2}Y \ .
\end{eqnarray}
Here $\alpha$ is the electromagnetic structure constant and $s_W=\sin \theta_W $
is the weak mixing angle. Therefore in the case where $W=Y=0$ we
have the simple relation
\begin{eqnarray}
\hat{S} &=& \frac{\alpha S}{4s_W^2} \ , \qquad 
\hat{T}= \alpha T \ .
\end{eqnarray}
The result of the the fit is shown in
Fig.~\ref{s06_stu_contours}. If the value of the Higgs mass increases the central value of the $S$ parameter moves to the left towards negative values. 

\begin{figure}[t]
\begin{center}
\includegraphics[width=8truecm]{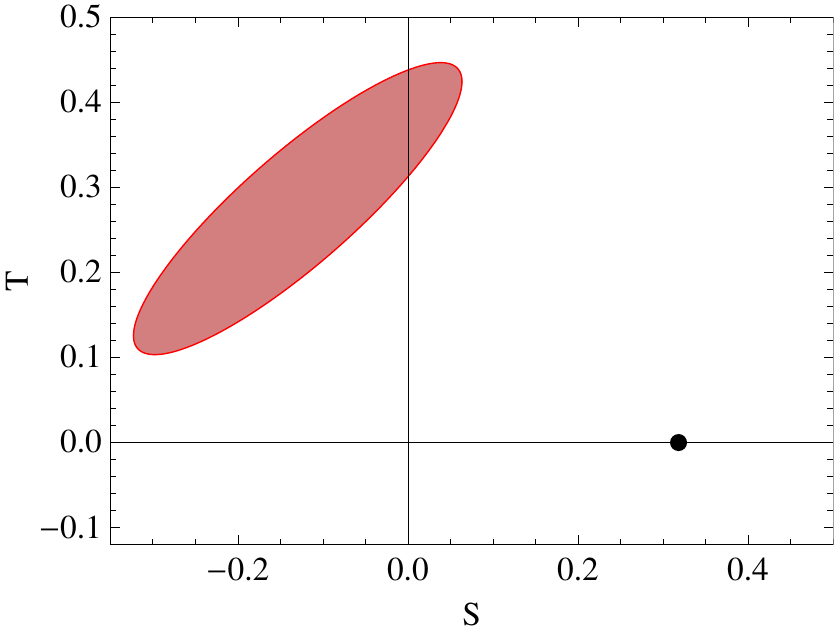}
\caption{$T$ versus $S$  for $SU(3)$ TC with one technifermion doublet (the black point) versus precision data for a one TeV composite Higgs mass (the shaded area).} \label{TCSproblem}
\end{center}
\end{figure}

\noindent
In TC it is easy to have a vanishing $T$ parameter while typically $S$ is positive. Besides, the composite Higgs is typically heavy with respect to the Fermi scale, at least for technifermions in the fundamental representation of the gauge group and for a small number of techniflavors. The oldest TC models featuring QCD dynamics with three technicolors and a doublet of electroweak gauged techniflavors deviate a few sigma from the current precision tests as summarized in Fig.~\ref{TCSproblem}.
Clearly it is desirable to reduce the tension between the precision data and a possible dynamical mechanism underlying the electroweak symmetry breaking. It is possible to imagine different ways to achieve this goal and some of the earlier attempts have been summarized in \cite{Peskin:2001rw}. 

The computation of the $S$ parameter in TC theories requires the knowledge of nonperturbative dynamics making difficult the precise knowledge of the contribution to  $S$. For example, it is not clear what is the exact value of the composite Higgs mass relative to the Fermi scale and, to be on the safe side, one typically takes it to be quite large, of the order at least of the TeV. However in certain models it may be substantially lighter due to the intrinsic dynamics. We will discuss the electroweak parameters later in this chapter. 

 It is, however, instructive to provide a simple estimate of the contribution to $S$ which allows to guide model builders. Consider a one-loop exchange of $N_D$ doublets of techniquarks transforming according to the representation $R_\text{TC}$ of the underlying TC gauge theory and with dynamically generated mass $\Sigma{(0)}$ assumed to be larger than the weak intermediate gauge bosons masses. Indicating with $d(R_{\rm TC})$ the dimension of the techniquark representation, and to leading order in $M_{W}/\Sigma(0)$ one finds:
 \begin{eqnarray}
S_{\rm naive} = N_D \frac{d(R_{\rm TC})}{6\pi} \ .
\end{eqnarray} 
This naive value provides, in general, only a rough estimate of the exact value of $S$. 
However, it is clear from the formula above that, the more TC matter is gauged under the electroweak theory the larger is the $S$ parameter and that the final $S$ parameter is expected to be positive. 

Attention must be paid to the fact that the specific model-estimate of the whole $S$ parameter, to compare with the experimental value, receives contributions also from other sectors. Such a contribution can be taken sufficiently large and negative to compensate for the positive value from the composite Higgs dynamics. To be concrete: Consider an extension of the SM in which the Higgs is composite but we also have new heavy (with a mass of the order of the electroweak) fourth family of Dirac leptons. In this case a sufficiently large splitting of the new lepton masses can strongly reduce and even offset the positive value of $S$. We will discuss this case in detail when presenting the Minimal Walking Technicolor (MWT) model. The contribution of the new sector ($S_{\rm NS}$) above, and also in  many other cases, is perturbatively under control and the total $S$ can be written as:
\begin{eqnarray}
S = S_{\rm TC} + S_{\rm NS} \ .
\end{eqnarray}
 The parameter $T$ will be, in general, modified and one has to make sure that the corrections do not spoil the agreement with this parameter.  From the discussion above it is clear that TC models can be constrained, via precision measurements, only model by model and the effects of possible new sectors must be properly included. We presented the constraints coming from $S$ using the underlying gauge theory information. However, in practice, these constraints apply directly to the physical spectrum. 
 To be concrete we will present in Section~\ref{pass} a model of walking TC passing the precision tests.

\subsection{Standard Model fermion masses}

Since in a purely TC model  the Higgs is a composite particle the Yukawa terms, when written in terms of the underlying TC fields, amount to four-fermion operators. The latter can be naturally interpreted as a low energy operator induced by a new strongly coupled gauge interaction emerging at energies higher than the electroweak theory. These type of theories have been termed Extended Technicolor (ETC) interactions \cite{Eichten:1979ah,Dimopoulos:1979es}. 

\begin{figure}[tp]
\begin{center}
\mbox{
\subfigure{\resizebox{!}{0.23\linewidth}{\includegraphics[clip=true]{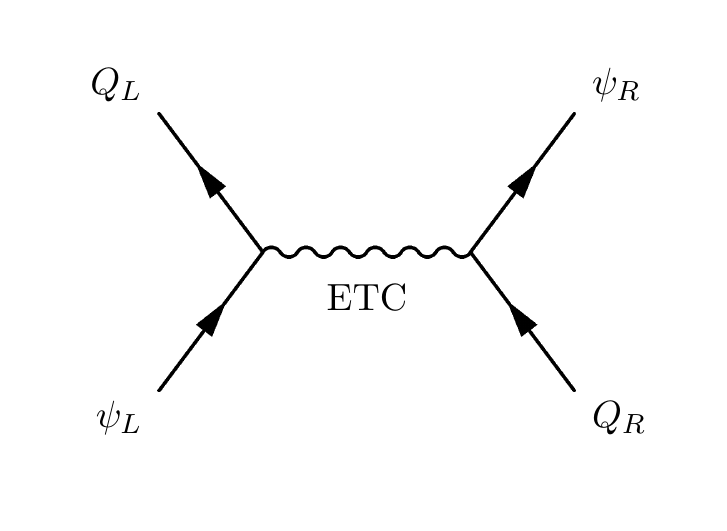}}}\qquad \qquad
\subfigure{\resizebox{!}{0.23\linewidth}{\includegraphics[clip=true]{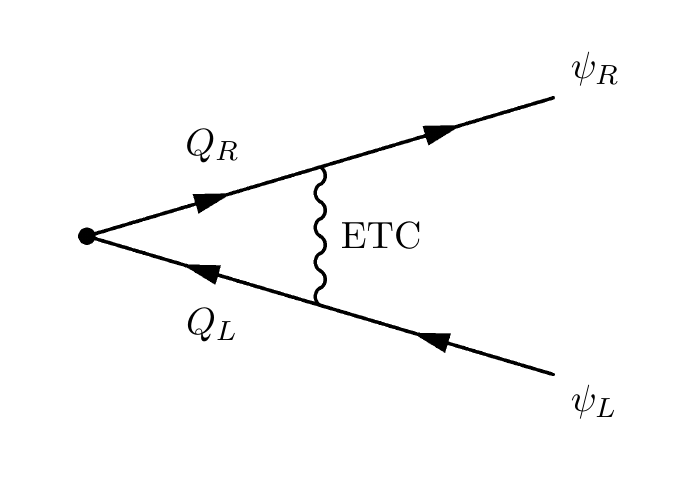}}}
}
\caption{Left panel: ETC  gauge boson interaction involving
techniquarks and SM fermions. Right panel: Diagram contribution to the mass to the SM fermions.}
\label{etcint}
\end{center}
\end{figure}

In the literature various extensions have been considered and we will mention them later in the text.  Here we will describe the simplest ETC model in which the ETC interactions connect the chiral symmetries of the techniquarks to those of the SM fermions (see left panel of Fig.~\ref{etcint}).

When TC chiral symmetry breaking occurs it leads to the diagram in the right panel of
Fig.~\ref{etcint}. Let's start with the case in which the ETC dynamics is represented by a $SU(N_\text{ETC})$ gauge group with: 
\beq N_\text{ETC} = N_\text{TC} + N_g \ , \eeq
and $N_g$ is the number of SM
generations. In order to give masses to all of the SM fermions, in this scheme, one needs a condensate for each SM fermion. This can be achieved by using as technifermion matter a complete generation of quarks and leptons (including a neutrino right) but now gauged with respect to the TC interactions.  

The ETC gauge group is assumed to spontaneously break $N_g$ times down to
$SU(N_\text{TC})$ permitting
three different mass scales, one  for each SM family. This type of TC with associated ETC is
termed the \emph{one family model} \cite{Farhi:1979zx}.
The heavy masses are provided by the
breaking at low energy and the light masses are provided by breaking
at higher energy scales. 
This model does not, per se, explain how the
gauge group is broken several times, neither is the breaking of weak isospin
symmetry accounted for. For example we cannot explain why the neutrino have masses much smaller than the associated electrons. See, however, \cite{Appelquist:2004ai} for progress on these issues. Schematically one has $SU(N_\text{TC} + 3)$ which breaks to  $SU(N_\text{TC} + 2)$ at the scale 
$\Lambda_1$ providing the first generation of fermions with a typical mass $m_1 \sim {4\pi
  (F_\pi^\text{TC})^3}/{\Lambda_1^2}$. At this point the gauge group breaks to $SU(N_\text{TC} + 1)$ with dynamical scale $\Lambda_2 $ leading to a second generation mass of the order of $m_2 \sim{4\pi
  (F_\pi^\text{TC})^3}/{\Lambda_2^2}$. Finally the last breaking
$SU(N_\text{TC} )$ at scale 
$\Lambda_3$ leading to the last generation mass $m_3 \sim {4\pi
  (F_\pi^\text{TC})^3}/{\Lambda_3^2}$. 
 
Without specifying an ETC one can write down the most general type of four-fermion operators involving TC particles $Q$ and ordinary fermionic fields $\psi$.  Following the notation of Hill and Simmons \cite{Hill:2002ap} we write:
\beq \alpha_{ab}\frac{\bar Q\gamma_\mu T^aQ\bar\psi \gamma^\mu
  T^b\psi}{\Lambda_\text{ETC}^2} +
\beta_{ab}\frac{\bar Q\gamma_\mu T^aQ\bar Q\gamma^\mu
  T^bQ}{\Lambda_\text{ETC}^2} + 
\gamma_{ab}\frac{\bar\psi\gamma_\mu T^a\psi\bar\psi\gamma^\mu
  T^b\psi}{\Lambda_\text{ETC}^2} \ , \eeq
where the $T$s are unspecified ETC generators. After performing a Fierz rearrangement one has:
\beq \alpha_{ab}\frac{\bar QT^aQ\bar\psi T^b\psi}{\Lambda_\text{ETC}^2} +
\beta_{ab}\frac{\bar QT^aQ\bar QT^bQ}{\Lambda_\text{ETC}^2} +
\gamma_{ab}\frac{\bar\psi T^a\psi\bar\psi T^b\psi}{\Lambda_\text{ETC}^2}
+ \ldots \ , \label{etc} \eeq
The coefficients parametrize the ignorance on the specific ETC physics. To be more specific, the $\alpha$-terms, after the TC particles have condensed, lead to mass terms for the SM fermions
\beq m_q \approx \frac{g_\text{ETC}^2}{M_\text{ETC}^2}\langle \bar
QQ\rangle_\text{ETC} \ , \eeq
where $m_q$ is the mass of {\em e.g.}~a SM quark, $g_\text{ETC}$ is the ETC gauge 
coupling constant evaluated at the ETC scale, $M_\text{ETC}$ is the mass of
an ETC gauge boson and $\langle \bar QQ\rangle_\text{ETC}$ is the
TC condensate where the operator is evaluated at the ETC
scale. Note that we have not explicitly considered the different scales for the different generations of ordinary fermions but this should be taken into account for any realistic model. 

\begin{figure}[tbh]
\begin{center}
\includegraphics[width=7truecm,height=4truecm]{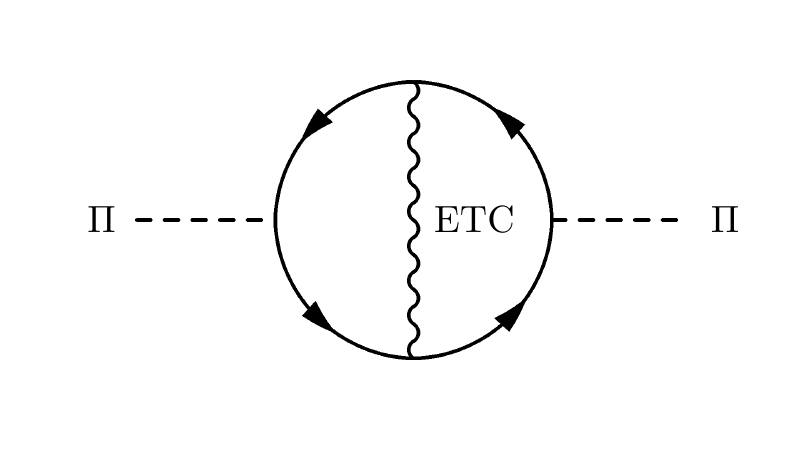}
\caption{Leading contribution to the mass of the TC pseudo
  Goldstone bosons via an exchange of an ETC gauge boson.} \label{masspgb}
\end{center}
\end{figure}

The $\beta$-terms of Eq.~(\ref{etc}) provide masses for
pseudo Goldstone bosons and also provide masses for techniaxions
\cite{Hill:2002ap}, see Fig.~\ref{masspgb}. 
The last class of terms, namely the $\gamma$-terms of
Eq.~(\ref{etc}) induce FCNCs. For example it may generate the following terms:
\beq \frac{1}{\Lambda_\text{ETC}^2}(\bar s\gamma^5d)(\bar s\gamma^5d) +
\frac{1}{\Lambda_\text{ETC}^2}(\bar \mu\gamma^5e)(\bar e\gamma^5e) + 
\ldots \ , \label{FCNC} \eeq
where $s,d,\mu,e$ denote the strange and down quark, the muon
and the electron, respectively. The first term is a $\Delta S=2$
flavor-changing neutral current interaction affecting the
$K_L-K_S$ mass difference which is measured accurately. The experimental bounds on these type of operators, together with the very {\it naive} assumption that ETC will generate $\gamma$-terms with coefficients of order one, leads to a constraint on the ETC scale to be of the order of or larger than $10^3$
TeV \cite{Eichten:1979ah}. This should be the lightest ETC scale which in turn puts an upper limit on how large the ordinary fermionic masses can be. The naive estimate is that  one can account up to around 100 MeV mass for a QCD-like TC theory, implying that the top quark mass value cannot be achieved.

The second term of Eq.~(\ref{FCNC}) induces flavor
changing processes in the leptonic sector such as $\mu\rightarrow e\bar ee,
e\gamma$ which are not observed.
It is clear that, both for the precision measurements  and the fermion masses, a better theory of the flavor is needed. For the ETC dynamics interesting  developments recently appeared in the literature  \cite{Ryttov:2010fu,Ryttov:2010hs,Ryttov:2010jt,Ryttov:2010kc}. We note that nonperturbative chiral gauge theories dynamics is expected to play a relevant role in models of ETC since it allows, at least in principle, the self breaking of the gauge symmetry. Recent progress on the phase diagrams of these theories has appeared in \cite{Sannino:2009za}.  

\begin{figure}[h]
\hskip -1cm
\includegraphics[width=16truecm]{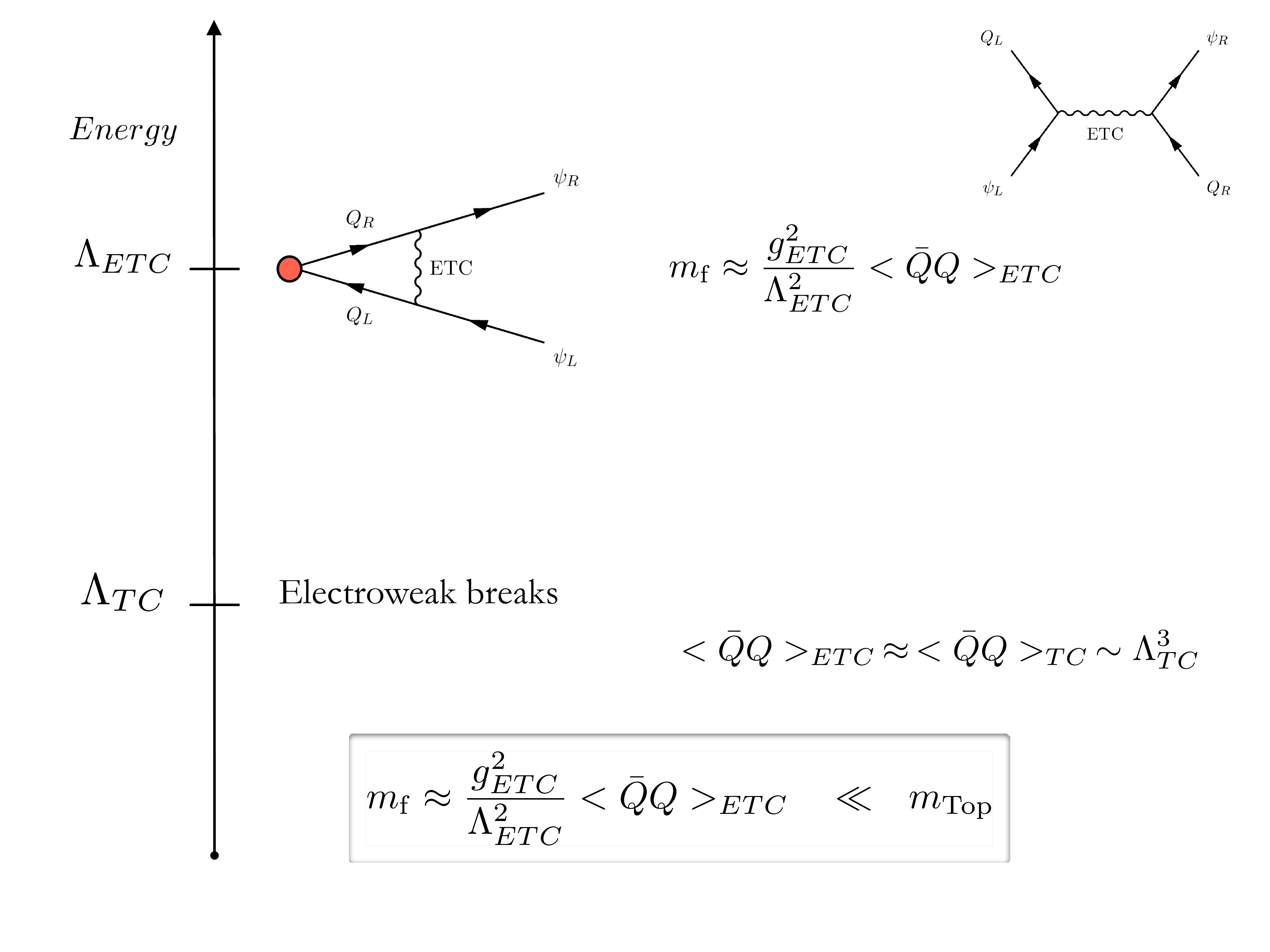}
\caption{Cartoon of the expected ETC dynamics starting at high energies with a more fundamental gauge interaction and the generation of the fermion masses assuming QCD-like dynamics.}
\label{ETC-dyn}
\end{figure}

In Fig.~\ref{ETC-dyn} we show the ordering of the relevant scales involved in the generation of the ordinary fermion masses via ETC dynamics, and the generation of the fermion masses (for a single generation and focussing on the top quark)  assuming QCD-like dynamics for TC.

\subsection{Walking}
To better understand in which direction one should go to modify the QCD dynamics, we analyze the TC condensate. 
The value of the TC condensate used when giving mass to the ordinary fermions should be evaluated not at the TC scale but at the ETC one. Via the renormalization group one can relate the condensate at the two scales via:
\beq \langle\bar QQ\rangle_\text{ETC} =
\exp\left(\int_{\Lambda_\text{TC}}^{\Lambda_\text{ETC}}
\text{d}(\ln\mu)\gamma_m(\alpha(\mu))\right)\langle\bar QQ\rangle_\text{TC} \ ,
\label{rad-cor-tc-cond}
\eeq 
where $\gamma_m$ is the anomalous dimension of the techniquark mass-operator. The boundaries of the integral 
 are at the ETC scale and the TC one.
For TC theories with a running of the coupling constant similar to the one in QCD, i.e.
\beq \alpha(\mu) \propto \frac{1}{\ln\mu} \ , \quad {\rm for}\ \mu >
\Lambda_\text{TC} \ , \eeq
this implies that the anomalous dimension of the techniquark masses $\gamma_m \propto
\alpha(\mu)$. When computing the integral one gets
\beq \langle\bar QQ\rangle_\text{ETC} \sim
\ln\left(\frac{\Lambda_\text{ETC}}{\Lambda_\text{TC}}\right)^{\gamma_m}
\langle\bar QQ\rangle_\text{TC} \ , \label{QCD-like-enh} \eeq
which is a logarithmic enhancement of the operator. We can hence neglect this correction and use directly the value of the condensate at the TC scale when estimating the generated fermionic mass:
\beq m_q \approx \frac{g_\text{ETC}^2}{M_\text{ETC}^2}\Lambda_\text{TC}^3 \ , \qquad 
 \langle \bar QQ\rangle_\text{TC} \sim \Lambda_\text{TC}^3 \ . \eeq

\begin{figure}
\centering
\begin{tabular}{cc}
\resizebox{6.0cm}{!}{\includegraphics{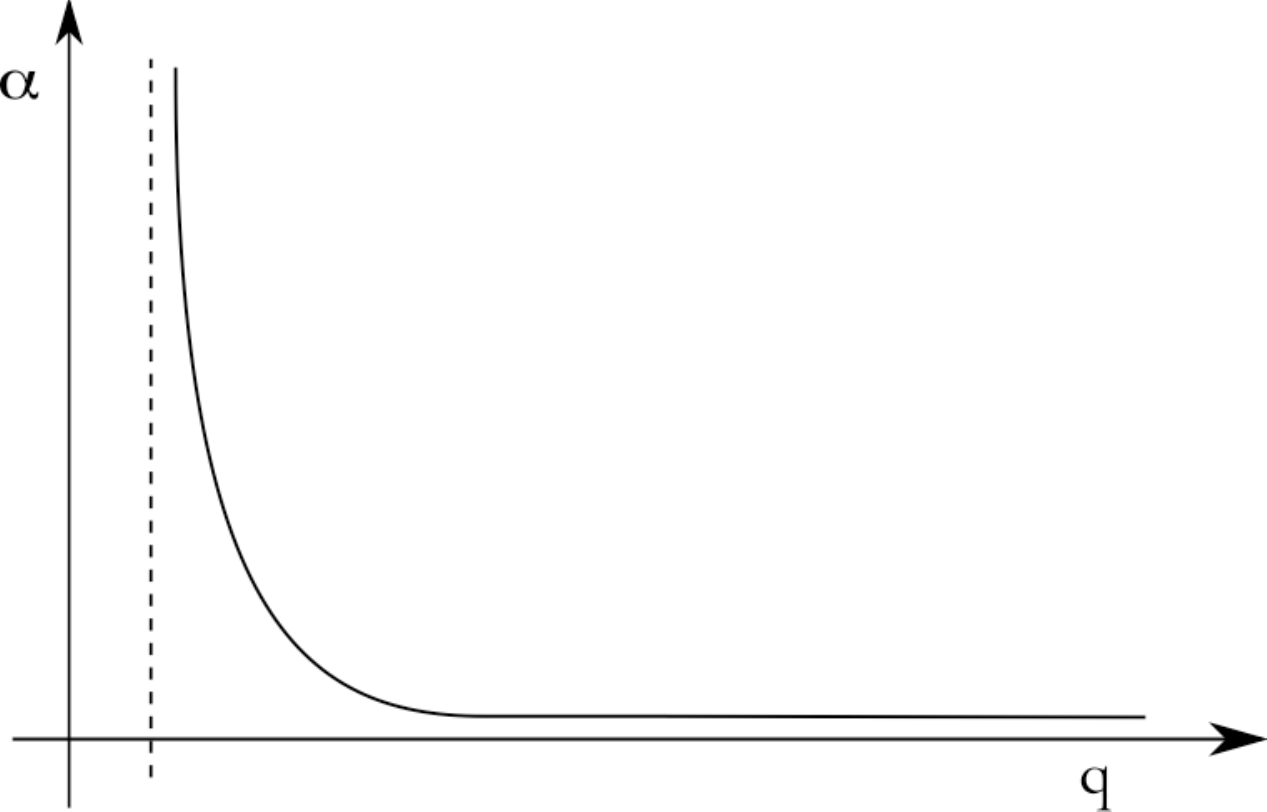}} ~~~&~~ \resizebox{6.0cm}{!}{\includegraphics{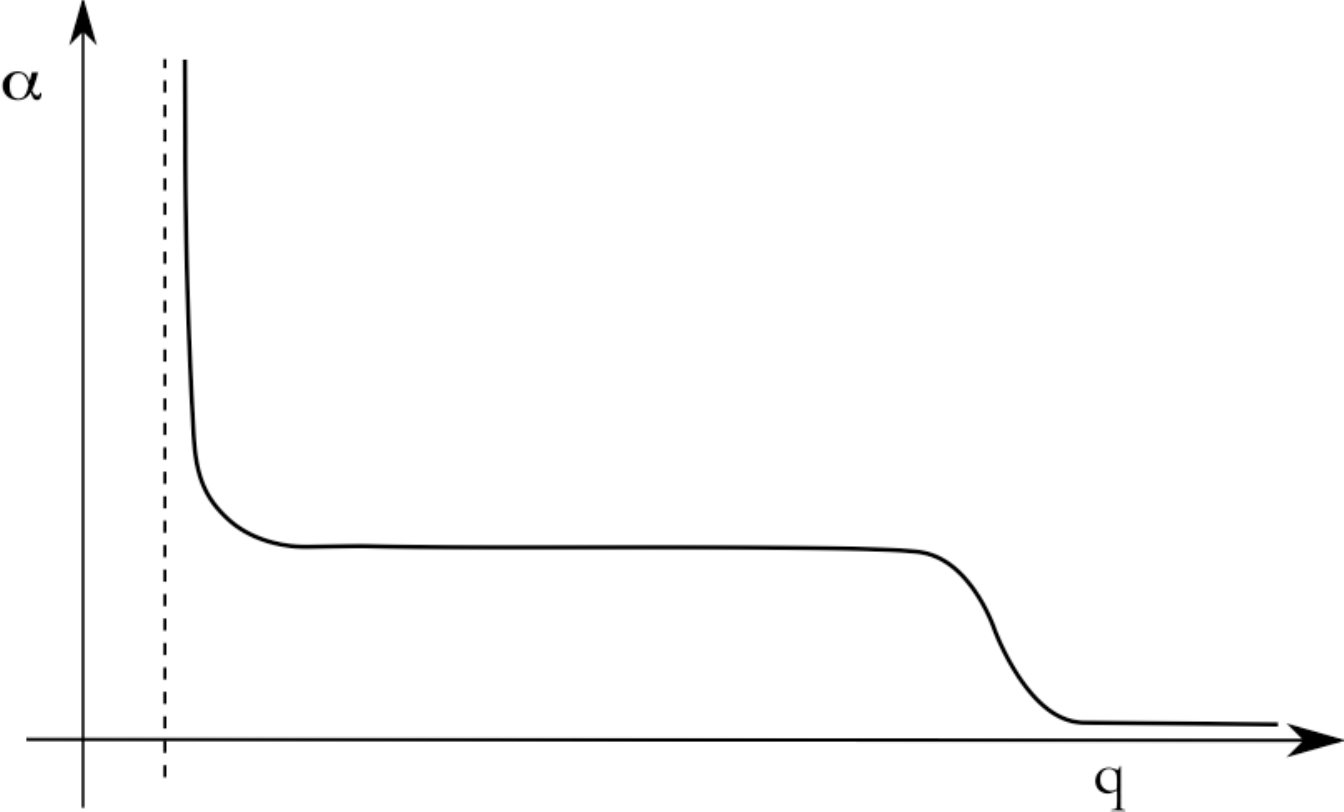}} \\&\\
&~~~~\resizebox{6.0cm}{!}{\includegraphics{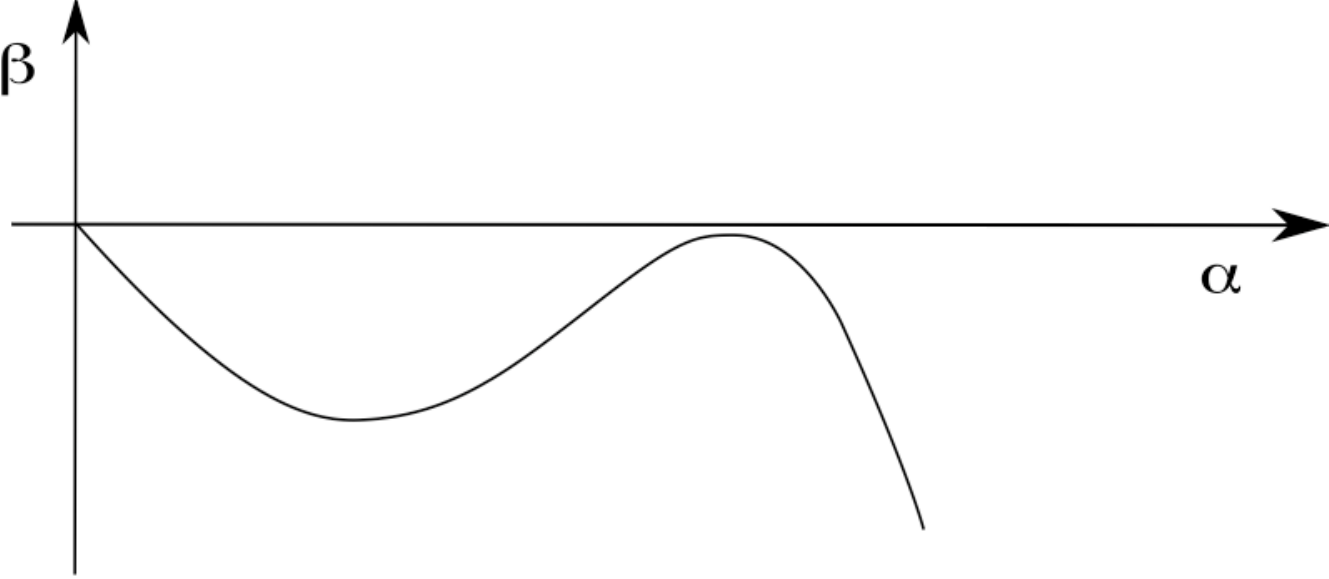}} 
\end{tabular}
\caption{Top left panel: QCD-like behavior of the coupling constant as function of the momentum (Running). Top right panel: walking-like behavior of the coupling constant as function of the momentum (Walking). Bottom right panel: cartoon of the beta function associated to a generic walking theory.}
\label{walkbeta}
\end{figure}

The tension between having to reduce the FCNCs and at the same time provide a sufficiently large mass for the heavy fermions in the SM as well as the pseudo-Goldstones can be reduced if the dynamics of the underlying TC theory is different from the one of QCD. The computation of the TC condensate at different scales shows that  if the dynamics is such that the TC coupling does not {\it run} to the UV fixed point but rather slowly reduces to zero one achieves a net enhancement of the condensate itself with respect to the value estimated earlier.  This can be achieved if the theory has a near conformal fixed point. This kind of dynamics has been denoted as of {\it walking} type. In Fig.~\ref{walkbeta} the comparison between a running and walking behavior of the coupling is qualitatively represented. 

 In the walking regime:
\beq \langle\bar QQ\rangle_\text{ETC} \sim
\left(\frac{\Lambda_\text{ETC}}{\Lambda_\text{TC}}\right)^{\gamma_m(\alpha^*)}
\langle\bar QQ\rangle_\text{TC} \ , \label{walking-enh} \eeq
which is a much larger contribution than in QCD dynamics  \cite{Yamawaki:1985zg,Holdom:1984sk,Holdom:1981rm,Appelquist:1986an}. Here $\gamma_m$ is evaluated at the would be fixed point value $\alpha^*$.  Walking can help resolving the problem of FCNCs in
TC models since with a large enhancement of the $\langle\bar
QQ\rangle$ condensate the four-Fermi operators involving SM fermions and
technifermions and the ones involving technifermions are enhanced by a factor of
$\Lambda_\text{ETC}/\Lambda_\text{TC}$ to the $\gamma_m$  power while the one involving only SM fermions is not enhanced.

We note that {\it walking} is not a fundamental property for a successful model of the origin of mass of the elementary fermions featuring TC. In fact several alternative ideas already exist in the literature (see \cite{Antola:2009wq,Antola:2010nt} and references therein).  However, a near conformal theory would still be useful to reduce the contributions to the precision data and, possibly, provide a light composite Higgs of much interest to LHC physics \cite{Dietrich:2005jn}.

\subsection{Ideal walking}

There are several issues associated with the original idea of walking: 
\begin{itemize}
\item{Since the number of flavors cannot be changed continuously it is not possible to  get arbitrarily close to the lower end of the conformal window. This applies to the TC theory {\it in isolation} i.e. before coupling it to the SM and without taking into account the ETC interactions. }
\item{It is hard to achieve large anomalous dimensions of the fermion mass operator even near the lower end of the conformal window for ordinary gauge theories.}
\item{It is not always possible to neglect the interplay of the four fermion interactions on the TC dynamics. }
\end{itemize}
In \cite{Fukano:2010yv} it has been argued that it is possible to {\it solve} simultaneously all the problems above by consistently taking into account the effects of the four-fermion interactions on the phase diagram of strongly interacting theories for any matter representation as function of the number of colors and flavors.  A positive effect is that the anomalous dimension of the mass increases beyond the unity value at the lower boundary of the new conformal window and can get sufficiently large to yield the correct mass for the top quark. It has also been shown that the conformal window, for any representation, shrinks with respect to the case in which the four-fermion interactions are neglected. This analysis derives from the study of the gauged Nambu-Jona-Lasinio phase diagram \cite{Kondo:1993jq}. 

It has been made the further {\it unexpected} discovery that when the extended TC sector, responsible for giving masses to the SM fermions, is sufficiently strongly coupled, the TC theory, in isolation, must feature an infrared fixed point in order for the full model to be phenomenologically viable and correctly break the electroweak symmetry \cite{Fukano:2009zm}.

\newpage
\section{Phenomenology of Minimal Technicolor}
 The existence of a new weak doublet of technifermions amounting to, at least, a global $SU(2)_\text{L}\times SU(2)_\text{R}$ symmetry later opportunely gauged under the electroweak interactions is the bedrock on which models of TC are built on. 
 
It is therefore natural to construct first minimal models of TC passing precision tests while also reducing the FCNC problem by featuring near conformal dynamics. By minimal we mean with the smallest fermionic matter content. These models were put forward recently in\cite{Sannino:2004qp,Dietrich:2005jn}.  To be concrete we describe here the (N)MWT \cite{Sannino:2008ha} extension of the SM. 

 The extended SM gauge group is now $SU(2)_\text{TC}\times SU(3)_\text{C}\times SU(2)_\text{L}\times U(1)_Y$ and the field content of the TC sector is constituted by four techni-fermions and one techni-gluon all in the adjoint representation of $SU(2)_\text{TC}$.  The model features also a pair of Dirac leptons, whose left-handed components are assembled in a weak doublet, necessary to cancel the Witten anomaly \cite{Witten:1982fp} arising when gauging the new technifermions with respect to the weak interactions. Summarizing, the fermionic particle content of the MWT is given explicitly by
 \beq Q_L^a=\left(\begin{array}{c} U^{a} \\D^{a} \end{array}\right)_L , \qquad U_R^a \
, \quad D_R^a \ ,  \qquad a=1,2,3 \ ,\eeq with $a$ being the adjoint color index of $SU(2)$. The left handed fields are arranged in three
doublets of the $SU(2)_\text{L}$ weak interactions in the standard fashion. The condensate is $\langle \bar{U}U + \bar{D}D \rangle$ which
correctly breaks the electroweak symmetry as already argued for ordinary QCD in Eq.~(\ref{qcd-condensate}).

The model as described so far suffers from the Witten topological anomaly \cite{Witten:1982fp}. However, this can easily be solved by
adding a new weakly charged fermionic doublet which is a TC singlet \cite{Dietrich:2005jn}. Schematically: 
\beq L_L =
\left(
\begin{array}{c} N \\ E \end{array} \right)_L , \qquad N_R \ ,~E_R \
. \eeq In general, the gauge anomalies cancel using the following
generic hypercharge assignment
\begin{align}
Y(Q_L)=&\frac{y}{2} \ ,&\qquad Y(U_R,D_R)&=\left(\frac{y+1}{2},\frac{y-1}{2}\right) \ , \label{assign1} \\
Y(L_L)=& -3\frac{y}{2} \ ,&\qquad
Y(N_R,E_R)&=\left(\frac{-3y+1}{2},\frac{-3y-1}{2}\right) \ \label{assign2} ,
\end{align}
where the parameter $y$ can take any real value \cite{Dietrich:2005jn}. In our notation
the electric charge is $Q=T^3 + Y$, where $T^3$ is the weak
isospin generator. One recovers the SM hypercharge
assignment for $y=1/3$.
\begin{figure}
\centerline{\includegraphics[width=9cm]{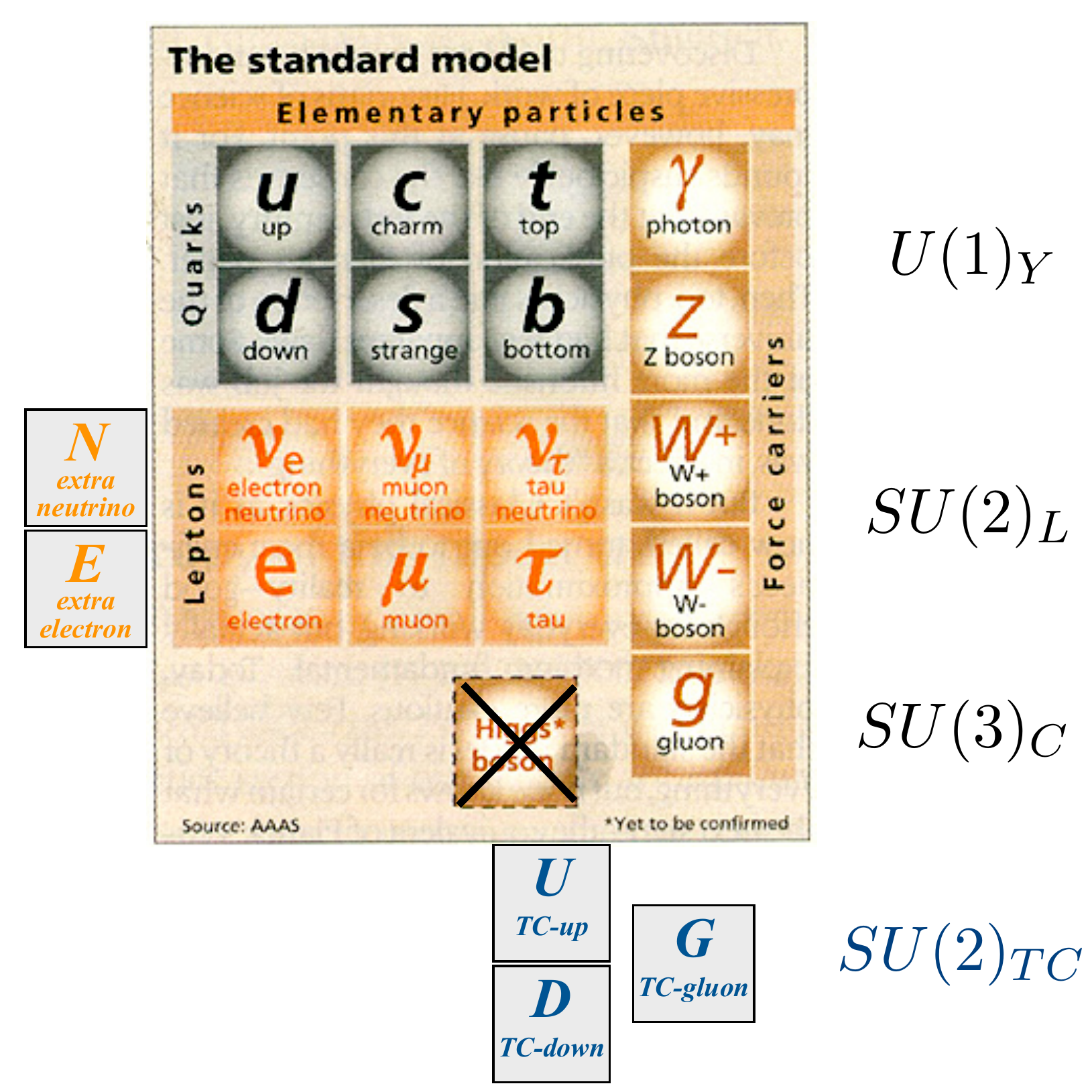}}
\label{MWTposter}
\caption{Cartoon of the Minimal Walking Technicolor Model extension of the SM.}
\end{figure}
To discuss the symmetry properties of the
theory it is
convenient to use the Weyl basis for the fermions and arrange them in the following vector transforming according to the
fundamental representation of $SU(4)$
\beq Q= \begin{pmatrix}
U_L \\
D_L \\
-i\sigma^2 U_R^* \\
-i\sigma^2 D_R^*
\end{pmatrix},
\label{SU(4)multiplet} \eeq where $U_L$ and $D_L$ are the left
handed techniup and technidown, respectively and $U_R$ and $D_R$ are
the corresponding right handed particles. Assuming the standard
breaking to the maximal diagonal subgroup, the $SU(4)$ symmetry
spontaneously breaks to $SO(4)$. Such a breaking is driven by the
following condensate \beq \langle Q_i^\alpha Q_j^\beta
\epsilon_{\alpha \beta} E^{ij} \rangle = - 2 \, \langle \overline{U}_R U_L
+ \overline{D}_R D_L\rangle \ , \label{conde}
 \eeq
where the indices $i,j=1,\ldots,4$ denote the components
of the tetraplet of $Q$, and the Greek indices indicate the ordinary
spin. The matrix $E$ is a $4\times 4$ matrix defined in terms
of the 2-dimensional unit matrix as
 \beq E=\left(
\begin{array}{cc}
0 & \mathbbm{1} \\
\mathbbm{1} & 0
\end{array}
\right) \ . \eeq

Here 
$\epsilon_{\alpha \beta}=-i\sigma_{\alpha\beta}^2$ and $\langle
 U_L^{\alpha} {{U_R}^{\ast}}^{\beta} \epsilon_{\alpha\beta} \rangle=
 -\langle  \overline{U}_R U_L
 \rangle$. A similar expression holds for the $D$ techniquark.
The above condensate is invariant under an $SO(4)$ symmetry. This leaves us with nine broken  generators with associated Goldstone bosons, of which three become the longitudinal degrees of  freedom of the weak gauge bosons.

Replacing the Higgs sector of the SM with the MWT the Lagrangian now
reads:
\begin{eqnarray}
\mathcal{L}_H &\rightarrow &  -\frac{1}{4}{\cal F}_{\mu\nu}^a {\cal F}^{a\mu\nu} + i\bar{Q}_L
\gamma^{\mu}D_{\mu}Q_L + i\bar{U}_R \gamma^{\mu}D_{\mu}U_R +
i\bar{D}_R \gamma^{\mu}D_{\mu}D_R \nonumber \\
&& +i \bar{L}_L \gamma^{\mu} D_{\mu} {L}_L 
+ i\bar{E}_R \gamma^{\mu}D_{\mu}E_R
+ i\bar{N}_R \gamma^{\mu}D_{\mu}N_R 
\end{eqnarray}
with the TC field strength ${\cal F}_{\mu\nu}^a =
\partial_{\mu}{\cal A}_{\nu}^a - \partial_{\nu}{\cal A}_{\mu}^a + g_\text{TC} \epsilon^{abc} {\cal A}_{\mu}^b
{\cal A}_{\nu}^c,\ a,b,c=1,\ldots,3$.
For the left handed techniquarks the covariant derivative is:

\begin{eqnarray}
D_{\mu} Q^a_L &=& \left(\delta^{ac}\partial_{\mu} + g_\text{TC}{\cal
A}_{\mu}^b \epsilon^{abc} - i\frac{g}{2} \vec{W}_{\mu}\cdot
\vec{\tau}\delta^{ac} -i g'\frac{y}{2} B_{\mu} \delta^{ac}\right)
Q_L^c \ .
\end{eqnarray}
${\cal A}_{\mu}$ are the techni gauge bosons, $W_{\mu}$ are the
gauge bosons associated to $SU(2)_\text{L}$ and $B_{\mu}$ is the gauge
boson associated to the hypercharge. $\tau^a$ are the Pauli matrices
and $\epsilon^{abc}$ is the fully antisymmetric symbol. In the case
of right handed techniquarks the third term containing the weak
interactions disappears and the hypercharge $y/2$ has to be replaced
accordingly to Eq.~(\ref{assign1}). For the
left-handed leptons the second term containing the TC
interactions disappears and $y/2$ changes to $-3y/2$. Only the last
term is present for the right handed leptons with an appropriate
hypercharge assignment.

\subsection{Low energy theory for MWT}

We construct the effective theory for MWT including composite scalars and vector bosons, their self interactions, and their interactions with the electroweak gauge fields and the SM fermions.

\subsubsection{Scalar sector} \label{sec:scalar}
The relevant effective theory for the Higgs sector at the electroweak scale consists, in our model, of a composite Higgs $\sigma$ and its pseudoscalar partner $\Theta$, as well as nine pseudoscalar Goldstone bosons and their scalar partners. These
can be assembled in the matrix
\begin{eqnarray}
M = \left[\frac{\sigma+i{\Theta}}{2} + \sqrt{2}(i\Pi^a+\widetilde{\Pi}^a)\,X^a\right]E \ ,
\label{M}
\end{eqnarray}
which transforms under the full $SU(4)$ group according to
\begin{eqnarray}
M\rightarrow uMu^T \ , \qquad {\rm with} \qquad u\in SU(4) \ .
\end{eqnarray}
The $X^a$'s, $a=1,\ldots,9$ are the generators of the $SU(4)$ group which do not leave  the Vacuum Expectation Value (VEV) of $M$ invariant
\begin{eqnarray}
\langle M \rangle = \frac{v}{2}E
 \ .
\end{eqnarray}
Note that the notation used is such that $\sigma$ is a \emph{scalar}
while the $\Pi^a$'s are \emph{pseudoscalars}. It is convenient to
separate the fifteen generators of $SU(4)$ into the six that leave the
vacuum invariant, $S^a$, and the remaining nine that do not, $X^a$.
Then the $S^a$ generators of the $SO(4)$ subgroup satisfy the relation
\begin{eqnarray}
S^a\,E + E\,{S^a}^{T} = 0 \ ,\qquad {\rm with}\qquad  a=1,\ldots  ,  6 \ ,
\end{eqnarray}
so that $uEu^T=E$, for $u\in SO(4)$. The explicit realization of the generators and the embedding of the electroweak generators in the $SU(4)$ algebra are shown in Appendix \ref{appgen}.
With the tilde fields included, the matrix $M$ is invariant in form under $U(4) \equiv SU(4) \times U(1)_{\rm
A}$, rather than just $SU(4)$. However the $U(1)_{\rm A}$ axial symmetry is anomalous, and is therefore broken at the quantum level.

The connection between the composite scalars and the underlying techniquarks can be derived from the transformation properties under $SU(4)$, by observing that the elements of the matrix $M$ transform like techniquark bilinears:
\begin{eqnarray}
M_{ij} \sim Q_i^\alpha Q_j^\beta \varepsilon_{\alpha\beta} \quad\quad\quad {\rm with}\ i,j=1\dots 4.
\label{M-composite}
\end{eqnarray}
Using this expression, and the basis matrices given in Appendix \ref{appgen}, the scalar fields can be related to the wavefunctions of the techniquark bound states. This gives the following charge eigenstates:
\begin{eqnarray}
\begin{array}{rclcrcl}
v+H & \equiv & \sigma \sim  \overline{U}U+\overline{D}D  &,~~~~ &
\Theta  &\sim& i \left(\overline{U} \gamma^5 U+\overline{D} \gamma^5 D\right) \ ,  \\
A^0 & \equiv & \widetilde{\Pi}^3  \sim  \overline{U}U-\overline{D}D &,~~~~ &
\Pi^0 & \equiv & \Pi^3 \sim i \left(\overline{U} \gamma^5 U-\overline{D} \gamma^5 D\right) \ , \\
A^+ & \equiv & {\displaystyle \frac{\widetilde{\Pi}^1 - i \widetilde{\Pi}^2}{\sqrt{2}}} \sim \overline{D}U &,~~~~&
\Pi^+ & \equiv & {\displaystyle \frac{\Pi^1 - i \Pi^2}{\sqrt{2}}} \sim i \overline{D} \gamma^5 U \ , \\
A^- & \equiv & {\displaystyle \frac{\widetilde{\Pi}^1 + i \widetilde{\Pi}^2}{\sqrt{2}}} \sim \overline{U}D &,~~~~&
\Pi^- & \equiv & {\displaystyle \frac{\Pi^1 + i \Pi^2}{\sqrt{2}}} \sim i \overline{U} \gamma^5 D \ ,
\end{array}
\label{TM-eigenstates}
\end{eqnarray}
for the technimesons, and
\begin{eqnarray}
\begin{array}{rcl}
\Pi_{UU} & \equiv & {\displaystyle \frac{\Pi^4 + i \Pi^5 + \Pi^6 + i \Pi^7}{2}} \sim U^T C U \ , \\
\Pi_{DD} & \equiv & {\displaystyle \frac{\Pi^4 + i \Pi^5 - \Pi^6 - i \Pi^7}{2}} \sim D^T C D \ , \\
\Pi_{UD} & \equiv & {\displaystyle \frac{\Pi^8 + i \Pi^9}{\sqrt{2}}} \sim U^T C D \ , \\
\widetilde{\Pi}_{UU} & \equiv &
{\displaystyle \frac{\widetilde{\Pi}^4 + i \widetilde{\Pi}^5 + \widetilde{\Pi}^6 + i \widetilde{\Pi}^7}{2}} \sim i U^T C \gamma^5 U \ , \\
\widetilde{\Pi}_{DD} & \equiv &
{\displaystyle \frac{\widetilde{\Pi}^4 + i \widetilde{\Pi}^5 - \widetilde{\Pi}^6 - i \widetilde{\Pi}^7}{2}} \sim i D^T C \gamma^5 D  \ , \\
\widetilde{\Pi}_{UD} & \equiv & {\displaystyle \frac{\widetilde{\Pi}^8 + i \widetilde{\Pi}^9}{\sqrt{2}}} \sim i U^T C \gamma^5 D \ ,
\end{array}
\label{TB-eigenstates}
\end{eqnarray}
for the technibaryons, where $U\equiv (U_L,U_R)^T$ and $D\equiv (D_L,D_R)^T$ are Dirac technifermions, and $C$ is the charge conjugation matrix, needed to form Lorentz-invariant objects. To these technibaryon charge eigenstates we must add the corresponding charge conjugate states ({\em e.g.}~$\Pi_{UU}\rightarrow \Pi_{\overline{U}\overline{U}}$).

Three of the nine Goldstone bosons ($\Pi^{\pm}, \Pi^0$) associated with the relative broken generators become the longitudinal degrees of freedom of
the massive weak gauge bosons, while the extra six Goldstone bosons will acquire a mass due to ETC interactions as well as the
electroweak interactions per se. Using a bottom up approach we will not commit to a specific ETC theory but limit ourself to introduce the minimal low energy operators  needed to construct a phenomenologically viable theory. The new Higgs Lagrangian is
\begin{eqnarray}
{\cal L}_{\rm Higgs} &=& \frac{1}{2}{\rm Tr}\left[D_{\mu}M D^{\mu}M^{\dagger}\right] - {\cal V}(M) + {\cal L}_{\rm ETC} \ ,
\label{Letc}
\end{eqnarray}
where the potential reads
\begin{eqnarray}
{\cal V}(M) & = & - \frac{m_M^2}{2}{\rm Tr}[MM^{\dagger}] +\frac{\lambda}{4} {\rm Tr}\left[MM^{\dagger} \right]^2 
+ \lambda^\prime {\rm Tr}\left[M M^{\dagger} M M^{\dagger}\right] \nonumber \\
& - & 2\lambda^{\prime\prime} \left[{\rm Det}(M) + {\rm Det}(M^\dagger)\right] \ ,
\label{Vdef}
\end{eqnarray}
and ${\cal L}_{\rm ETC}$ contains all terms which are generated by the ETC interactions, and not by the chiral symmetry breaking sector. Notice that the determinant terms (which are renormalizable) explicitly break the U(1)$_{\rm A}$ symmetry, and give mass to $\Theta$, which would otherwise be a massless Goldstone boson. 


In order to give masses to the remaining uneaten Goldstone boson we add  this term which is generated in the ETC sector: 
\begin{eqnarray}
{\cal L}_{\rm ETC} \supset \frac{m_{\rm ETC}^2}{4}\ {\rm Tr}\left[M B M^\dagger B + M M^\dagger \right]  \ , \label{VETCdef}
\end{eqnarray}
and $B\equiv 2\sqrt{2}S^4$ is a specific generator in the $SU(4)$ algebra. 

The potential ${\cal V}(M)$  produces a VEV
which parameterizes the techniquark condensate, and spontaneously
breaks $SU(4)$ to $SO(4)$. In terms of the model parameters the VEV is
\begin{eqnarray}
v^2=\langle \sigma \rangle^2 = \frac{m_M^2}{\lambda + \lambda^\prime - \lambda^{\prime\prime} } \ ,
\label{VEV}
\end{eqnarray}
while the Higgs mass is
\begin{eqnarray}
M_H^2 = 2\ m_M^2 \ .
\end{eqnarray}
The linear combination $\lambda + \lambda^{\prime} -
\lambda^{\prime\prime}$ corresponds to the Higgs self coupling in
the SM. The three pseudoscalar mesons $\Pi^\pm$, $\Pi^0$ correspond
to the three massless Goldstone bosons which are absorbed by the
longitudinal degrees of freedom of the $W^\pm$ and $Z$ boson. The
remaining six uneaten Goldstone bosons are technibaryons, and all
acquire tree-level degenerate mass through the ETC interaction in (\ref{VETCdef}):
\begin{eqnarray}
M_{\Pi_{UU}}^2 = M_{\Pi_{UD}}^2 = M_{\Pi_{DD}}^2 = m_{\rm ETC}^2  \ .
\end{eqnarray}
The remaining scalar and pseudoscalar masses are
\begin{eqnarray}
M_{\Theta}^2 & = & 4 v^2 \lambda^{\prime\prime} \nonumber \\
M_{A^\pm}^2 = M_{A^0}^2 & = & 2 v^2 \left(\lambda^{\prime}+\lambda^{\prime\prime}\right)
\end{eqnarray}
for the technimesons, and
\begin{eqnarray}
M_{\widetilde{\Pi}_{UU}}^2 = M_{\widetilde{\Pi}_{UD}}^2 = M_{\widetilde{\Pi}_{DD}}^2 =
m_{\rm ETC}^2 + 2 v^2 \left(\lambda^{\prime} + \lambda^{\prime\prime }\right) \ ,
\end{eqnarray}
for the technibaryons. 
To gain insight on some of the mass relations one can use \cite{Hong:2004td}.

\subsubsection{Vector bosons}
The composite vector bosons of a theory with a global $SU(4)$ symmetry are conveniently described by the four-dimensional traceless Hermitian matrix
\begin{eqnarray}
A^\mu = A^{a\mu} \ T^a \ ,
\end{eqnarray}
where $T^a$ are the $SU(4)$ generators: $T^a=S^a$, for $a=1, \dots ,6$, and $T^{a+6}=X^a$, for $a=1, \dots ,9$. Under an arbitrary $SU(4)$ transformation, $A^\mu$ transforms like
\begin{equation}
A^\mu \ \rightarrow \ u\ A^\mu \ u^\dagger \ ,\ \ \ {\rm where} \ u\in SU(4) \ .
\label{vector-transform}
\end{equation}
Eq.~(\ref{vector-transform}), together with the tracelessness of the matrix $A_\mu$, gives the connection with the techniquark bilinears:
\begin{equation}
A^{\mu,j}_{i}  \sim \ Q^{\alpha}_i  \sigma^{\mu}_{\alpha \dot{\beta}}  \bar{Q}^{\dot{\beta},j}
- \frac{1}{4} \delta_{i}^j Q^{\alpha}_k  \sigma^{\mu}_{\alpha \dot{\beta}} \bar{Q}^{\dot{\beta},k} \ .
\end{equation}
Then we find the following relations between the charge eigenstates and the wavefunctions of the composite objects:
\begin{eqnarray}
\begin{array}{rclcrcl}
v^{0\mu} & \equiv & A^{3\mu} \sim \bar{U} \gamma^\mu U - \bar{D} \gamma^\mu D & , & 
a^{0\mu} & \equiv & A^{9\mu} \sim \bar{U} \gamma^\mu \gamma^5 U - \bar{D} \gamma^\mu \gamma^5 D \\
v^{+\mu} & \equiv & {\displaystyle \frac{A^{1\mu}-i A^{2\mu}}{\sqrt{2}}} \sim \bar{D} \gamma^\mu U & , &
a^{+\mu} & \equiv & {\displaystyle \frac{A^{7\mu}-i A^{8\mu}}{\sqrt{2}}} \sim  \bar{D} \gamma^\mu  \gamma^5 U \\
v^{-\mu} & \equiv & {\displaystyle \frac{A^{1\mu}+i A^{2\mu}}{\sqrt{2}}} \sim  \bar{U} \gamma^\mu D & , &  
a^{-\mu} & \equiv & {\displaystyle \frac{A^{7\mu}+i A^{8\mu}}{\sqrt{2}}} \sim  \bar{U} \gamma^\mu  \gamma^5 D \\
v^{4\mu} & \equiv & A^{4\mu} \sim \bar{U} \gamma^\mu U + \bar{D} \gamma^\mu D  & , & & &
\end{array}
\label{TMV-eigenstates}
\end{eqnarray}
for the vector mesons, and
\begin{eqnarray}
\begin{array}{rcl}
x_{UU}^\mu & \equiv & {\displaystyle \frac{A^{10\mu}+i A^{11\mu}+A^{12\mu}+ i A^{13\mu}}{2}} \sim   U^T C \gamma^\mu \gamma^5 U \ , \\
x_{DD}^\mu & \equiv & {\displaystyle \frac{A^{10\mu}+i A^{11\mu}-A^{12\mu}- i A^{13\mu}}{2}} \sim   D^T C \gamma^\mu \gamma^5 D \ , \\
x_{UD}^\mu & \equiv & {\displaystyle \frac{A^{14\mu}+i A^{15\mu}}{\sqrt{2}}} \sim  D^T C \gamma^\mu \gamma^5 U \ , \\
s_{UD}^\mu & \equiv & {\displaystyle \frac{A^{6\mu}-i A^{5\mu}}{\sqrt{2}}} \sim U^T C \gamma^\mu  D \ ,
\end{array}
\label{TBV-eigenstates}
\end{eqnarray}
for the vector baryons.

There are different approaches on how to introduce vector mesons at the effective Lagrangian level. At the tree level they are all equivalent.

Based on this premise, the minimal kinetic Lagrangian is:
\begin{eqnarray}
{\cal L}_{\rm kinetic} = -\frac{1}{2}{\rm Tr}\Big[\widetilde{W}_{\mu\nu}\widetilde{W}^{\mu\nu}\Big] - \frac{1}{4}B_{\mu\nu}B^{\mu\nu}
-\frac{1}{2}{\rm Tr}\Big[F_{\mu\nu}F^{\mu\nu}\Big] + m^2 \ {\rm Tr}\Big[C_\mu C^\mu\Big] \ ,
\label{massterm}
\end{eqnarray}
where $\widetilde{W}_{\mu\nu}$ and $B_{\mu\nu}$ are the ordinary field strength tensors for the electroweak gauge fields. Strictly speaking the terms above are not only kinetic ones since the Lagrangian contains a mass term as well as self interactions. The tilde on $W^a$ indicates  that the associated states are not yet the SM weak triplets: in fact these states mix with the composite vectors to form mass eigenstates corresponding to the ordinary $W$ and $Z$ bosons. $F_{\mu\nu}$ is the field strength tensor for the new $SU(4)$ vector bosons,
\begin{eqnarray}
F_{\mu\nu} & = & \partial_\mu A_\nu - \partial_\nu A_\mu - i\tilde{g}\left[A_\mu,A_\nu\right]\ ,
\label{strength}
\end{eqnarray}
and the vector field $C_\mu$ is defined by
\begin{eqnarray}
C_\mu \ \equiv \ A_\mu \ - \ \frac{g}{\tilde{g}}\ G_\mu \ .
\end{eqnarray}
and $G_{\mu}$ is given by
\beq
g\ G_{\mu} = g\ W^a_\mu \ L^a + g^{\prime}\ B_\mu Y\ ,
\eeq
where $L^a$ and $Y$ are the generators of the left-handed and hypercharge transformations, as defined in Appendix  \ref{appgen}, with $Y$. The parameter $\tilde{g}$ represents the coupling among the vectors and the ratio $\frac{g}{\tilde{g}}$ is phenomenologically very important because it sets the mixing among gauge eigenstates and composite vectors eigenstates.
The mass term in Eq.~(\ref{massterm}) is gauge invariant, and gives a degenerate mass to all composite vector bosons, while leaving the actual gauge bosons massless. (The latter acquires mass as usual from the covariant derivative term of the scalar matrix $M$, after spontaneous symmetry breaking.)

The $C_\mu$ fields couple with $M$ via gauge invariant operators. Up
to dimension four operators the Lagrangian is
\begin{eqnarray}
{\cal L}_{\rm M-C} & = & \tilde{g}^2\ r_1 \ {\rm Tr}\left[C_\mu C^\mu M M^\dagger\right]
+ \tilde{g}^2\ r_2 \ {\rm Tr}\left[C_\mu M {C^\mu}^T M^\dagger \right] \nonumber \\
& + & i \ \tilde{g} \frac{\ r_3}{2} \ {\rm Tr}\left[C_\mu \left(M (D^\mu M)^\dagger - (D^\mu M) M^\dagger \right) \right]
+ \tilde{g}^2\ s \ {\rm Tr}\left[C_\mu C^\mu \right] {\rm Tr}\left[M M^\dagger \right] \ . \nonumber \\
\end{eqnarray}
The dimensionless parameters $r_1$, $r_2$, $r_3$, $s$ parameterize
the strength of the interactions between the composite scalars and
vectors in units of $\tilde{g}$, and are therefore naturally
expected to be of order one. However, notice that for
$r_1=r_2=r_3=0$ the overall Lagrangian possesses two independent
$SU(2)_\text{L} \times U(1)_R \times U(1)_\text{V}$ global
symmetries. One for the terms involving $M$ and one for the terms
involving $C_\mu$~\footnote{The gauge fields explicitly break the
original $SU(4)$ global symmetry to $SU(2)_\text{L} \times U(1)_R \times U(1)_\text{V}$, where $U(1)_R$ is the $T^3$ part of
$SU(2)_\text{R}$, in the $SU(2)_\text{L} \times SU(2)_\text{R} \times U(1)_\text{V}$ subgroup of $SU(4)$.}. The Higgs potential
only breaks the symmetry associated with $M$, while leaving the
symmetry in the vector sector unbroken. This {\em enhanced symmetry}
guarantees that all $r$-terms are still zero after loop corrections.
Moreover if one chooses $r_1$, $r_2$, $r_3$ to be small the near enhanced symmetry will protect these values against large corrections \cite{Casalbuoni:1995qt,Appelquist:1999dq}.


\subsubsection{Fermions and Yukawa interactions}
The fermionic content of the effective theory consists of the SM quarks and leptons, the new lepton doublet $L=(N,E)$ introduced to cure the Witten anomaly, and a composite techniquark-technigluon doublet. 

We now consider the limit according to which the $SU(4)$ symmetry is, at first, extended to ordinary quarks and leptons. Of course, we will need to break this symmetry to accommodate the SM phenomenology. We start by arranging the $SU(2)$ doublets in $SU(4)$ multiplets as we did for the techniquarks in Eq.~(\ref{SU(4)multiplet}). We therefore introduce the four component vectors $q^i$ and $l^i$,
\begin{eqnarray} 
q^i= \begin{pmatrix}
u^i_L \\
d^i_L \\
-i\sigma^2 {u^i_R}^* \\
-i\sigma^2 {d^i_R}^*
\end{pmatrix}\ , \quad
l^i= \begin{pmatrix}
\nu^i_L \\
e^i_L \\
-i\sigma^2 {\nu^i_R}^* \\
-i\sigma^2 {e^i_R}^*
\end{pmatrix}\ ,
\end{eqnarray}
where $i$ is the generation index. Note that such an extended $SU(4)$ symmetry automatically predicts the presence of a right handed neutrino for each generation. In addition to the SM fields there is an $SU(4)$ multiplet for the new leptons,
\begin{eqnarray}
L = \begin{pmatrix}
N_L \\
E_L \\
-i\sigma^2 {N_R}^* \\
-i\sigma^2 {E_R}^*
\end{pmatrix}\ ,
\end{eqnarray}
and a multiplet for the techniquark-technigluon bound state,
\begin{eqnarray} 
\widetilde{Q}= \begin{pmatrix}
\widetilde{U}_L \\
\widetilde{D}_L \\
-i\sigma^2 {\widetilde{U}_R}^* \\
-i\sigma^2 {\widetilde{D}_R}^*
\end{pmatrix}\ .
\end{eqnarray}
The techniquark-technigluon states, $\tilde{Q}$, being bound states of the underlying MWT model,  have a dynamical mass.

With this arrangement, the electroweak covariant derivative for the fermion fields can be written
\begin{eqnarray}
D_\mu \ = \  \partial_\mu \  - \  i \ g \ G_\mu (Y_{\rm V})  \ , 
\end{eqnarray}
where $Y_{\rm V}=1/3$ for the quarks, $Y_{\rm V}=-1$ for the leptons, $Y_{\rm V}=-3y$ for the new lepton doublet, and $Y_{\rm V}=y$ for the techniquark-technigluon bound state. 
Based on this matter content, we write the following gauge part of the fermion Lagrangian:
\begin{eqnarray}
{\cal L}_{\rm fermion} & = & i\ \overline{q}^i_{\dot{\alpha}}  \overline{\sigma}^{\mu,\dot{\alpha} \beta} D_\mu  q^i_\beta 
+ i\ \overline{l}^i_{\dot{\alpha}}  \overline{\sigma}^{\mu,\dot{\alpha} \beta} D_\mu  l^i_\beta 
+ i\ \overline{L}_{\dot{\alpha}}  \overline{\sigma}^{\mu,\dot{\alpha} \beta} D_\mu  L_\beta 
+ i\ \overline{\widetilde{Q}}_{\dot{\alpha}}  \overline{\sigma}^{\mu,\dot{\alpha} \beta} D_\mu  \widetilde{Q}_\beta  \nonumber \\
& + & x\ \overline{\widetilde{Q}}_{\dot{\alpha}}  \overline{\sigma}^{\mu,\dot{\alpha} \beta} C_\mu  \widetilde{Q}_\beta \ .
\label{fermion-kinetic}
\end{eqnarray} 
We now turn to the issue of providing masses to the SM fermions. In the first chapter the simplest ETC model has been briefly reviewed. Many extensions of TC have been suggested in the literature to address this problem. Some of the extensions use another strongly coupled gauge dynamics,  others introduce fundamental scalars. Many variants of the schemes presented above exist and a review of the major models is the one by Hill and Simmons \cite{Hill:2002ap}. At the moment there is not yet a consensus on which is the correct ETC. In our phenomenological approach we will parameterize our ignorance about a complete ETC theory by simply coupling the fermions to our low energy effective Higgs through the ordinary effective SM Yukawa interactions and we assume that any dangerous FCNC operator is strongly suppressed and therefore negligible. 
 
A discussion regarding the implications of having a natural fourth family of leptons is presented in detail in Section~\ref{4thleptons} of this report.

\subsubsection{Weinberg Sum Rules}
In order to
make contact with the underlying gauge theory, and discriminate between different classes of models, we make
use of the Weinberg Sum Rules (WSR)s. In \cite{Appelquist:1998xf} it was argued that the zeroth WSR -- which is nothing but
the definition of the $S$ parameter --
\begin{equation}
S=4\pi\left[\frac{F_V^2}{M_V^2}-\frac{F_A^2}{M_A^2}\right] \ , \label{eq:WSR0} 
\end{equation}
 and the first WSR,  
 \begin{equation}
 F_V^2 - F_A^2 = F_\pi^2 \ , \label{eq:WSR1} 
\end{equation}
 do not receive significant contributions from the
near conformal region, and are therefore unaffected. 
In these equations $M_V$ ($M_A$) and $F_V$ ($F_A$) are mass and decay constant of the vector-vector (axial-vector) meson, respectively, in the limit of zero electroweak gauge couplings. $F_\pi$ is the decay constant of the pions: since this is a model of dynamical electroweak symmetry breaking, $F_\pi=246$ GeV. The heavy vector boson masses are:
\begin{eqnarray}
M_V^2 &=& m^2 + \frac{\tilde{g}^2\ (s-r_2)\ v^2}{4} \nonumber \ , \\
M_A^2 &=& m^2 + \frac{\tilde{g}^2\ (s+r_2)\ v^2}{4} \ ,
\label{eq:masses}
\end{eqnarray}
and
\begin{eqnarray}
F_V & = & \frac{\sqrt{2}M_V}{\tilde{g}} \ ,  \nonumber \\
F_A & = & \frac{\sqrt{2}M_A}{\tilde{g}}\chi \ , \nonumber \\
F_\pi^2 & = & \left(1+2\omega\right)F_V^2-F_A^2 \ ,
\label{eq:FVFAFP}
\end{eqnarray}
where
\begin{eqnarray}
\omega \equiv \frac{v^2 \tilde{g}^2}{4 M_V^2}(1+r_2-r_3) \ , \quad \quad 
\chi \equiv 1-\frac{v^2\ \tilde{g}^2\ r_3}{4 M_A^2} \ . \label{eq:chi}
\end{eqnarray}
Then Eqs.~(\ref{eq:WSR0}) and (\ref{eq:WSR1}) give
\begin{eqnarray}
\label{eq:s_of_chi}
& & S=\frac{8\pi}{\tilde{g}^2}\left(1-\chi^2\right) \ , \label{eq:S} \\
& & r_2 = r_3-1 \ .
\end{eqnarray}
The second WSR, corresponding to a zero on the right hand side of the following equation, does receive important contributions from the near conformal region, and is modified to
\begin{eqnarray}
F_V^2 M_V^2 - F_A^2 M_A^2 = a \frac{8\pi^2}{d(R)} F_\pi^4 \ , \label{eq:WSR2}
\end{eqnarray}
where $a$ is expected to be positive and ${\cal O}(1)$, and $d(R)$ is the dimension of the representation of
the underlying fermions~\cite{Appelquist:1998xf}. For each of these sum rules a more general spectrum would involve a sum over all the vector
and axial states. 

In the effective Lagrangian we codify the walking behavior in $a$ being positive and ${\cal O}(1)$,
and the minimality of the theory in $S$ being small. A small $S$ is both due to the small number of flavors in
the underlying theory and to the near conformal dynamics, which reduces the contribution to $S$ relative to a
running theory~\cite{Appelquist:1998xf,Sundrum:1991rf,Kurachi:2006mu}.
 
 \subsubsection{Passing the electroweak precision tests}
\label{pass}
\begin{figure}[!t]
\centering
\begin{tabular}{cc}
\resizebox{7.1cm}{!}{\includegraphics{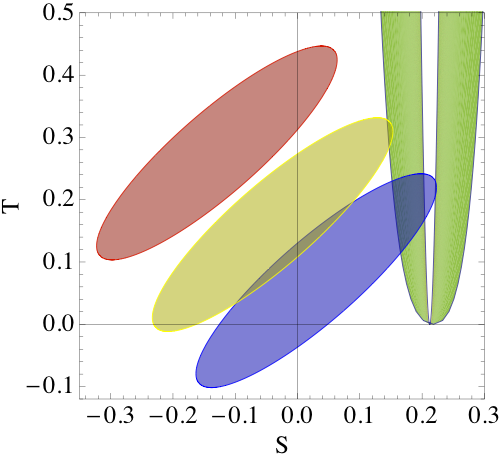}}
~~&~~~~
\resizebox{7.1cm}{!}{\includegraphics{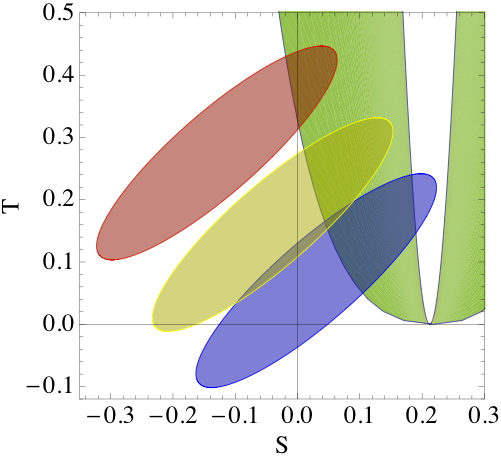}}
\end{tabular}
\caption{The ellipses represent the 90\% confidence region for the $S$ and $T$ parameters. The ellipses, from lower to higher, are obtained for a reference Higgs mass of 117 GeV, 300 GeV, and 1 TeV, respectively. The contribution from the TC sector of the MWT theory per se and from the new leptons is expressed by the green region. The left panel has been obtained using a SM type hypercharge assignment while the right one is for $y=1$. }
\label{ST}
\end{figure}

We have studied the effects of the lepton family on the
electroweak parameters in~\cite{Dietrich:2005jn}, we summarize here the main results
in Fig.~\ref{ST}, where we have used the updated experimental values for S and T given in \cite{Nakamura:2010zzi}.  The ellipses represent the 90\%
confidence region for the $S$ and $T$ parameters. The ellipses, from lower to higher, are obtained for a reference Higgs mass of 117 GeV, 300 GeV, and 1 TeV, respectively. The
contribution from the MWT theory per se and of the new leptons  \cite{He:2001tp} is expressed by the green region. The left panel has been obtained using a SM type hypercharge assignment while the right one is for $y=1$. In both pictures the regions of overlap between the theory and the precision contours are achieved when the upper component of the weak isospin doublet is lighter than the lower component. The opposite case leads to a total $S$ which is larger than the one predicted within the new strongly coupled dynamics per se.  This is due to the sign of the hypercharge for the new leptons. The mass range used in the plots is $M_Z \leqslant m_{E,N} \leqslant 10~M_Z$. The plots have been obtained assuming a Dirac mass for the new neutral lepton (in the case of a SM hypercharge assignment).

The analysis for the Majorana mass case has been performed in \cite{Kainulainen:2006wq} where one can again show that it is possible to be within the 90\% contours. 

\subsubsection{The Next to Minimal Walking Technicolor Theory (NMWT)}
\label{4}

The theory with three technicolors contains an even number of electroweak doublets, and hence
it is not subject to a Witten anomaly.  
The doublet of technifermions, is then represented again as:
\bear
Q_L^{\{C_1,C_2 \}}
=
\left(\begin{array}{l}U^{\{C_1,C_2 \}}\\ D^{\{C_1,C_2 \}}\end{array}\right)_L \ ,
\qquad 
Q_R^{\{C_1,C_2\}}&=&\left(U_R^{\{C_1,C_2\}},~ D_R^{\{C_1,C_2\}}\right) \ .
\nn
\eear
Here $C_i=1,2,3$ is the technicolor index and $Q_{L(R)}$ is a doublet (singlet) with respect 
to the weak interactions. 
Since the two-index symmetric representation of $SU(3)$ is complex the flavor symmetry is $SU(2)_\text{L}\times SU(2)_\text{R}\times U(1)$. 
Only three Goldstones emerge and
are absorbed in the longitudinal components of the weak vector bosons.

Gauge anomalies are absent with the choice $Y=0$ for the hypercharge of the left-handed technifermions:
\bear
Q_L^{(Q)}
=
\left(\begin{array}{l} U^{(+1/2)}\\ D^{(-1/2)}\end{array}\right)_L \ .
\eear
Consistency requires for the right-handed technifermions (isospin singlets):
\bear
Q_R^{(Q)}&=&\left(U_R^{(+1/2)},~D_R^{-1/2}\right) \ ,
\nn
Y&=&~~+1/2,~-1/2 \ .
\eear 
All of these states will be bound into hadrons. There is no need for an associated fourth family of leptons, and hence it is not expected to be observed in the experiments.

Here the low-lying technibaryons are fermions constructed with three techniquarks in the following way:
\begin{eqnarray}
B_{f_1,f_2,f_3;\alpha} = Q^{\{C_1,C_2 \}}_{L;\alpha,
f_1}Q^{\{ C_3,C_4 \}}_{L;\beta,f_2} Q^{\{ C_5,C_6 \}}_{L;\gamma,f_3}\epsilon^{\beta \gamma}
\epsilon_{C_1 C_3 C_5}\epsilon_{C_2 C_4 C_6} \ .
\end{eqnarray}
where $f_i=1,2$ corresponds to $U$ and $D$ flavors, and we are not specifying the flavor symmetrization which in any 
event will have to be such that the full technibaryon wave function is fully antisymmetrized in technicolor, flavor and spin. 
$\alpha$, $\beta$, and $\gamma$ assume the values of one or 
two and represent the ordinary spin. Similarly we can construct different technibaryons using only right-handed fields or a mixture of left- and right-handed ones.

\subsection{Beyond MWT}

When going beyond MWT one finds new and interesting theories able to break the electroweak symmetry while featuring a walking dynamics and yet not at odds with precision measurements, at least when comparing with the naive $S$ parameter.  A compendium of these theories can be found in \cite{Dietrich:2006cm}. Here we will review only the principal type of models one can construct.


\subsubsection{Partially Gauged Technicolor\label{pgt}}

A small modification of the traditional TC approach, which neither
involves additional particle species nor more complicated gauge groups, 
allows constructing several other viable candidates. It consists in letting 
only one doublet of techniquarks transform non-trivially under the electroweak
symmetries with the rest being electroweak singlets, as first suggested in 
\cite{Dietrich:2005jn} and later also used in \cite{Christensen:2005cb}.
Still, all techniquarks transform under the TC gauge group. Thereby only one techniquark doublet contributes directly\footnote{Via TC interactions all of the matter content of the theory will affect physical observables associated to the sector coupled to the electroweak symmetry.} to the oblique 
parameter which is thus kept to a minimum for theories which need more
than one family of techniquarks to be quasi-conformal. It is the condensation 
of that first electroweakly charged family that breaks the electroweak 
symmetry. The techniquarks which are uncharged under the electroweak gauge group are
natural building blocks for components of dark matter.

\subsubsection{Split Technicolor}

We summarize here also another
possibility \cite{Dietrich:2005jn} according to which we keep the technifermions gauged under the electroweak theory in the fundamental 
representation of the $SU(N)$ TC group while still reducing the number of techniflavors needed to be
near the conformal window. Like for the partially gauged case described above 
this can be achieved by adding matter uncharged under the weak interactions. 
The difference with Section~\ref{pgt} is that this part of matter transforms 
under a different representation of the TC gauge group than the 
part coupled directly to the electroweak sector. For example, for definiteness let's choose it to be a 
massless Weyl fermion in the adjoint representation of the TC gauge 
group. The resulting theory has the same matter content as $N_f$-flavor super QCD but without the scalars;
hence the name {\it Split Technicolor}.  The matter content of {\it Split Technicolor} lies between that of super QCD and QCD-like theories with 
matter in the fundamental representation.  We note that a split
TC-like theory has been used in \cite{Hsu:2004mf}, to
investigate the strong CP problem. 

Split Technicolor shares some features with theories of split
supersymmetry advocated and studied in
\cite{ArkaniHamed:2004fb,Giudice:2004tc} as possible extensions of the
SM.
Clearly, we have introduced Split Technicolor---differently
from split supersymmetry---to address the hierarchy problem. This is why we
do not expect new scalars to appear at energy scales higher than the one of
the electroweak theory unless one tries to supersymmetrize the model at higher energies.

In \cite{Ryttov:2008xe} one can find  an explicit example of (near) conformal TC with two types of technifermions, i.e. transforming according to two different representations of the underlying TC gauge group \cite{Dietrich:2006cm,Lane:1989ej}. The model possesses a number of interesting properties to recommend it over the earlier models of dynamical electroweak symmetry breaking:
\begin{itemize}
\item
 Features the lowest possible value of the naive $S$ parameter \cite{Peskin:1990zt,{Peskin:1991sw}} while possessing a dynamics which is near conformal.

 \item Contains, overall, the lowest possible number of fermions.

\item Yields natural DM candidates.
\end{itemize}
Due to the above properties we term this model {\it Ultra Minimal near conformal Technicolor} (UMT). It is constituted by an $SU(2)$ TC gauge group with two Dirac flavors in the fundamental representation also carrying electroweak charges, as well as, two additional Weyl fermions in the adjoint representation but singlets under the SM gauge groups.

By arranging the additional fermions in higher dimensional representations, it is possible to construct models which have a particle content smaller than the one of partially gauged TC theories. In fact instead of considering additional fundamental flavors we shall consider adjoint flavors. Note that for two colors there exists only one distinct two-indexed representation.

\subsection{Vanilla Technicolor}

Despite the different envisioned underlying gauge dynamics it is a fact that the SM structure alone requires the extensions to contain, at least, the following chiral symmetry breaking pattern (insisting on keeping the custodial symmetry of the SM): 
\beq
\label{basepattern}
SU(2)_\text{L} \times SU(2)_\text{R} \to SU(2)_\text{V} \ .
\eeq

We will call this common sector of any TC extension of the SM, the {\it vanilla} sector. The reason for such a name is that the vanilla sector is common to old models of TC featuring running and walking dynamics. It is worth mentioning that the {\it vanilla} sector is common not only to TC extensions but to several extensions, even of extra-dimensional type, in which the Higgs sector can be viewed as composite. In fact, the effective Lagrangian we are about to introduce can  be used for modeling several extensions with a common {vanilla} sector respecting the same constraints spelled out in \cite{Foadi:2007ue}. The natural candidate for a walking TC model featuring exactly this global symmetry is NMWT \cite{Sannino:2004qp}.

Based on the {\it vanilla symmetry} breaking pattern we describe the low energy spectrum in terms of the lightest spin one vector and axial-vector iso-triplets $V^{\pm,0}, A^{\pm,0}$ as well as the lightest iso-singlet scalar resonance $H$. In QCD the equivalent states are the $\rho^{\pm,0}$, $a_1^{\pm,0}$ and the $f_0(600)$ \cite{Belyaev:2008yj}.  It has been argued in \cite{Sannino:2008ha,Hong:2004td}, using Large N arguments, and in \cite{Dietrich:2005jn,Dietrich:2006cm}, using the saturation of the trace of the energy momentum tensor,  that  models of dynamical electroweak symmetry breaking featuring (near) conformal dynamics  contain a composite Higgs state which is light with respect to the new strongly coupled scale ($4\,\pi \, v$ with $v \simeq 246$~GeV). These indications have led to the construction of models of TC with a naturally {\it light composite} Higgs. Recent investigations using Schwinger-Dyson \cite{Doff:2008xx} and gauge-gravity dualities \cite{Fabbrichesi:2008ga} also arrived to the conclusion that the composite Higgs can be light \footnote{The Higgs boson here is identified with the lightest $0^{++}$ state of the theory. Calling it a dilaton or a meson makes no physical difference since these two states mix  at the 100\%  level and both couple to the trace of the stress energy momentum tensor of the theory. In the construction of the low energy effective theory saturating the trace anomaly there is no way to distinguish these states.}.  The 3 technipions $\Pi^{\pm ,0}$ produced in the symmetry breaking become the longitudinal components of the $W$ and $Z$ bosons. 

The composite spin one and spin zero states and their interaction with the SM fields are described via the following effective Lagrangian which we developed, first for minimal models of walking TC  \cite{Foadi:2007ue,Appelquist:1999dq}: 

\begin{eqnarray}
{\cal L}_{\rm boson}&=&-\frac{1}{2}{\rm Tr}\left[\widetilde{W}_{\mu\nu}\widetilde{W}^{\mu\nu}\right]
-\frac{1}{4}\widetilde{B}_{\mu\nu}\widetilde{B}^{\mu\nu}
-\frac{1}{2}{\rm Tr}\left[F_{{\rm L}\mu\nu} F_{\rm L}^{\mu\nu}+F_{{\rm R}\mu\nu} F_{\rm R}^{\mu\nu}\right] \nonumber \\
&+& m^2\ {\rm Tr}\left[C_{{\rm L}\mu}^2+C_{{\rm R}\mu}^2\right]
+\frac{1}{2}{\rm Tr}\left[D_\mu M D^\mu M^\dagger\right]
-\tilde{g^2}\ r_2\ {\rm Tr}\left[C_{{\rm L}\mu} M C_{\rm R}^\mu M^\dagger\right] \nonumber \\
&-&\frac{i\ \tilde{g}\ r_3}{4}{\rm Tr}\left[C_{{\rm L}\mu}\left(M D^\mu M^\dagger-D^\mu M M^\dagger\right)
+ C_{{\rm R}\mu}\left(M^\dagger D^\mu M-D^\mu M^\dagger M\right) \right] \nonumber \\
&+&\frac{\tilde{g}^2 s}{4} {\rm Tr}\left[C_{{\rm L}\mu}^2+C_{{\rm R}\mu}^2\right] {\rm Tr}\left[M M^\dagger\right]
+\frac{\mu^2}{2} {\rm Tr}\left[M M^\dagger\right]-\frac{\lambda}{4}{\rm Tr}\left[M M^\dagger\right]^2 \ ,
\label{eq:boson}
\end{eqnarray}
where $\widetilde{W}_{\mu\nu}$ and $\widetilde{B}_{\mu\nu}$ are the ordinary electroweak field strength tensors, $F_{{\rm L/R}\mu\nu}$ are the field strength tensors associated to the vector meson fields $A_{\rm L/R\mu}$~\footnote{In \cite{Foadi:2007ue}, where the chiral symmetry is $SU(4)$, there is an additional term whose coefficient is labeled $r_1$. With an $SU(N) \times SU(N)$ chiral symmetry this term is just identical to the $s$ term.}, and the $C_{{\rm L}\mu}$ and $C_{{\rm R}\mu}$ fields are
\begin{eqnarray}
C_{{\rm L}\mu}\equiv A_{{\rm L}\mu}-\frac{g}{\tilde{g}}\widetilde{W_\mu}\ , \quad
C_{{\rm R}\mu}\equiv A_{{\rm R}\mu}-\frac{g^\prime}{\tilde{g}}\widetilde{B_\mu}\ .
\end{eqnarray}
The 2$\times$2 matrix $M$ is
\begin{eqnarray}
M=\frac{1}{\sqrt{2}}\left[v+H+2\ i\ \pi^a\ T^a\right]\ ,\quad\quad  a=1,2,3
\end{eqnarray}
where $\pi^a$ are the Goldstone bosons produced in the chiral symmetry breaking, $v=\mu/\sqrt{\lambda}$ is the corresponding VEV, $H$ is the composite Higgs, and $T^a=\sigma^a/2$, where $\sigma^a$ are the Pauli matrices. The covariant derivative is
\begin{eqnarray}
D_\mu M&=&\partial_\mu M -i\ g\ \widetilde{W}_\mu^a\ T^a M + i\ g^\prime \ M\ \widetilde{B}_\mu\ T^3\ . 
\end{eqnarray}
When $M$ acquires a VEV, the Lagrangian of Eq.~(\ref{eq:boson}) contains mixing matrices for the spin one fields. The mass eigenstates are the ordinary SM bosons, and two triplets of heavy mesons, of which the lighter (heavier) ones are denoted by $R_1^\pm$ ($R_2^\pm$) and $R_1^0$ ($R_2^0$). These heavy mesons are the only new particles, at low energy, relative to the SM.

Now we must couple the SM fermions. The interactions with the Higgs and the spin one mesons are mediated by an unknown ETC sector, and can be parametrized at low energy by Yukawa terms, and mixing terms with the $C_{\rm L}$ and $C_{\rm R}$ fields. Assuming that the ETC interactions preserve parity and do not generate extra flavor violation beyond the SM like Yukawa terms, the most general form for the quark Lagrangian is~\footnote{The lepton sector works out in a similar way, the
only difference being the possible presence of Majorana neutrinos.}
\begin{eqnarray}
{\cal L}_{\rm quark}&=&\bar{q}^i_L\ i \slashed{D} q_{iL} + \bar{q}^i_R\ i \slashed{D} q_{iR} \nonumber \\
&-&\left[\bar{q}^i_L\ (Y_u)_i^j\ M\ \frac{1+\tau^3}{2}\ q_{jR}
+\bar{q}^i_L\ (Y_d)_i^j\ M \ \frac{1-\tau^3}{2}\ q_{jR} + {\rm h.c.}\right] \ ,
\label{eq:quark}
\end{eqnarray}
where $i$ and $j$ are generation indices, $i=1,2,3$, $q_{iL/R}$ are electroweak doublets, $Y_u$ and $Y_d$ are 3$\times$3 complex matrices. The covariant derivatives are the ordinary
electroweak ones,
\begin{eqnarray}
\slashed{D}q_{iL}&=&\left(\slashed{\partial}-i\ g\ \slashed{\widetilde{W}}^a\ T^a
-i\ g^\prime \slashed{\widetilde{B}} Y_{\rm L}\right)q_{iL} \ ,\nonumber \\
\slashed{D}q_{iR}&=&\left(\slashed{\partial}-i\ g^\prime \slashed{\widetilde{B}} Y_{\rm R}\right)q_{iR} \ ,
\end{eqnarray}
where $Y_{\rm L}=1/6$ and $Y_{\rm R}={\rm diag}(2/3,-1/3)$. One can exploit the global symmetries of the kinetic terms to reduce the number of physical parameters in the Yukawa matrices. Thus we can take
\begin{eqnarray}
Y_u={\rm diag}(y_u,y_c,y_t) \ , \quad Y_d= V\ {\rm diag}(y_d,y_s,y_b) \ ,
\end{eqnarray}
and
\begin{eqnarray}
q^i_L=\left(\begin{array}{c} u_{iL} \\ V_i^j d_{jL} \end{array}\right) \ , \quad
q^i_R=\left(\begin{array}{c} u_{iR} \\ d_{iR} \end{array}\right) \ ,
\end{eqnarray}
where $V$ is the CKM matrix.

It is possible to further reduce the number of independent couplings using the WSRs discussed above. For example in NMWT, featuring technifermions with three technicolors transforming according to the two-index 
symmetric representation of the TC gauge group, the {naive} one-loop  $S$ parameter is $S=1/\pi\simeq 0.3$: this is a reasonable input for 
$S$ in Eq.~(\ref{eq:WSR0}).

With $S=0.3$ the  remaining parameters
are $M_A,\ \tilde{g},\ s$ and $M_H$, with $s$ and $M_H$ having a sizable effect in processes involving the composite Higgs \footnote{{
The information on the spectrum
alone is not sufficient to constrain $s$, but it can be measured studying
other physical processes.}}.

\subsection{WW - Scattering in Technicolor and Unitarity}

The simplest argument often used to predict the existence of yet undiscovered particles at the TeV scale comes from unitarity of longitudinal gauge boson scattering amplitudes. If the electroweak symmetry breaking sector is weakly interacting, unitarity implies that new particle states must show up below one TeV, being these spin zero isosinglets (the Higgs boson) or spin one isotriplets (e.g. Kaluza-Klein modes). A strongly interacting electroweak symmetry breaking sector can however change this picture, because of the strong coupling between the pions (eaten by the longitudinal components of the standard model gauge bosons) and the other bound states of the strongly interacting sector. An illuminating example comes from QCD. In \cite{Harada:2003em}  it was shown that for six colors or more, the 770 MeV $\rho$ meson  is enough to delay the onset of unitarity violation of the pion-pion scattering amplitude up to well beyond 1 GeV. Here the 't Hooft large N limit was used, however an even lower number of colors is needed to reach a similar delay of unitarity violation when an alternative large N limit is used \cite{Sannino:2007yp}. Scaling up to the electroweak scale, this translates in a 1.5 TeV technivector being able to delay unitarity violation of longitudinal gauge boson scattering amplitudes up to 4 TeV or more.  As we discussed in the previous sections such a model, however, would not be realistic for other reasons: a large contribution to the $S$ parameter \cite{Peskin:1990zt}, and large FCNC if the ordinary fermions acquire mass via an old fashioned ETC, to mention the most relevant ones. It is therefore interesting to analyze the pion-pion scattering in generic models of Walking TC. We follow the analyses performed in \cite{Foadi:2008xj,Foadi:2008ci}.

In the effective theory for TC the scattering amplitudes for the longitudinal SM gauge bosons approach at large energies the scattering amplitudes for the corresponding eaten pions. We mainly analyze the contribution to the $\pi\pi$ scattering amplitude from a spin zero isosinglet and a spin one isotriplet, and consider the case in which a spin two isosinglet contributes as well.
 
\subsubsection{ Spin zero + spin one}
The isospin invariant amplitude for the pion-pion elastic scattering is \cite{Harada:1995dc}:
\begin{equation}
A(s,t,u)=\left(\frac{1}{F_\pi^2}-\frac{3g_{V\pi\pi}^2}{M_V^2}\right)s
-\frac{h^2}{M_H^2}\frac{s^2}{s-M_H^2}
-g_{V\pi\pi}^2\left[\frac{s-u}{t-M_V^2}+\frac{s-t}{u-M_V^2}\right] \ .
\label{eq:inv_2}
\end{equation}
Note that our normalization for $g_{V\pi\pi}$, which is the heavy vector to two-pions effective coupling, differs by a factor of $\sqrt{2}$ from that of Ref.~\cite{Harada:1995dc}. The scalar $H$ contribution is proportional to the coupling $h$. These couplings are simply related to the ones of the Vanilla TC Lagrangian, but the specific relation is not relevant here.

The amplitude of Eq.~(\ref{eq:inv_2}) has an $s$-channel pole in the Higgs exchange. In the vicinity of this pole the propagator should be modified to include the Higgs width. In order to catch the essential features of the unitarization process we will take the Higgs to be a relatively narrow state, and consider values of $\sqrt{s}$ far away from $M_H$, where the finite width effects can be neglected. If the Higgs or any other state is not sufficiently narrow to be treated at the tree level, it would be relevant to investigate the effects due to unitarity corrections using specific unitarization schemes as done for example in \cite{Black:2000qq}.
\begin{figure}
\begin{center}
\includegraphics[width=0.62\textwidth,height=0.47\textwidth]{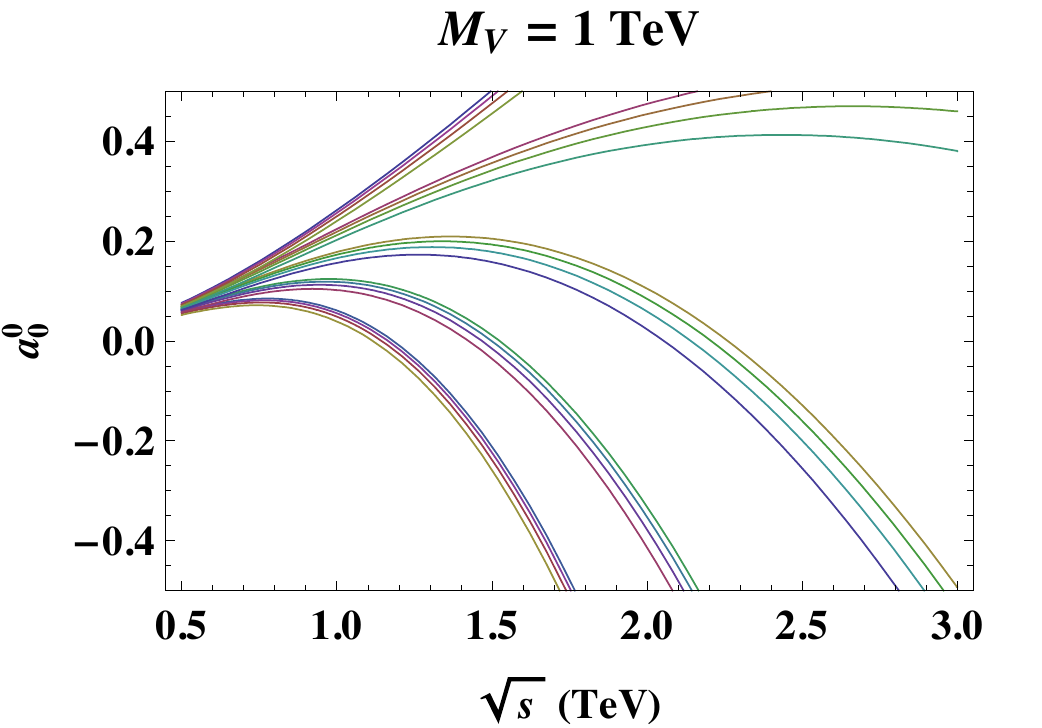}
\caption{$I=0$ $J=0$ partial wave amplitude for the $\pi\pi$ scattering. Here a Higgs with mass $M_H=200$ GeV, and a spin-one vector meson with mass $M_V=1$ TeV contribute to the full amplitude. The different groups of curves correspond, from top to bottom, to $g_{V\pi\pi}=2,2.5,3,3.5,4$. The different curves within each group correspond, from top to bottom, to $h=0,0.1,0.15,0.2$. Nonzero values of $g_{V\pi\pi}$ and $h$ give negative contributions to the linear term in $s$ in the amplitude, and may lead to a delay of unitarity violation.}
\label{fig:a00}
\end{center}
\end{figure}
In order to study unitarity of the $\pi\pi$ scattering the most general amplitude should be expanded in its isospin $I$ and spin $J$ components, $a^I_J$. However the $I=0$ $J=0$ component,
\begin{equation}
a_0^0(s) = \frac{1}{64\pi} \int_{-1}^1 d\cos\theta \left[3A(s,t,u)+A(t,s,u)+A(u,t,s)\right] \ ,
\end{equation}
has the worst high energy behavior, and is therefore sufficient for our analysis. Since we are interested in testing unitarity at few TeV in presence of a light Higgs, we set
$M_H=200$ GeV as a reference value, and study the regions in the $(M_V,g_{V\pi\pi})$ plane in which $a_0^0$ is unitary up to 3 TeV, for different values of $h$. If the Higgs mass is larger than 200 GeV but still smaller than or of the same size of $M_V$, we expect our results to be qualitatively similar, even though finite width effects might be important due to the pole in the $s$-channel. If the Higgs mass is much larger than $M_V$ the theory is Higgsless at low energies. This case was studied in Ref.~\cite{Foadi:2008ci}, and applies also to the light Higgs scenario if $H$ is decoupled from the pions, {\em i.e.} $h=0$.

In order to study the effect of the Higgs exchange on the scattering amplitude, consider the high energy behavior of $A(s,t,u)$,
\begin{equation}
A(s,t,u)\sim\left(\frac{1}{F_\pi^2}-\frac{3g_{V\pi\pi}^2}{M_V^2}-\frac{h^2}{M_H^2}\right)s \ .
\label{eq:leading}
\end{equation}
This shows that the Higgs exchange provides an additional negative contribution at large energies, which, together with the vector meson, contributes to delay unitarity violation to higher energies. In Fig.~\ref{fig:a00} $a_0^0$ is plotted as a function of $\sqrt{s}$ for $M_V=1$ TeV, $M_H=200$ GeV, and different values of $g_{V\pi\pi}$ and $h$. The different groups of curves from top to bottom correspond to $g_{V\pi\pi}=$ 2, 2.5, 3, 3.5, and 4. 
For comparison, the QCD value that follows from $\Gamma(\rho\to \pi\pi)\simeq 150$~MeV would be $g_{V\pi\pi}\simeq 5.6$~\footnote{Fig.~\ref{fig:a00} does not reproduce a scaled up version of QCD $\pi\pi$ scattering. For the latter to occur, the vector resonance should be as large as (246 GeV/93 MeV)$\times$770 MeV $\simeq$ 2 TeV. However in a theory with walking dynamics the resonances are expected to be lighter than in a running setup.}. Within each group, the top curve corresponds to the Higgsless case, $h=0$, while the remaining ones correspond, from top to bottom, to $h=$ 0.1, 0.15, and 0.2. For small values of $g_{V\pi\pi}$ the presence of a light Higgs delays unitarity violation to higher energies: if the partial wave amplitude has a maximum near 0.5 the delay is dramatic.  
 
For a given value of $M_V$, the presence of a light Higgs enlarges the interval of values of $g_{V\pi\pi}$ for which the theory is unitary, provided that $|h|$ is not too large.  

\subsubsection{Spin zero + spin one + spin two}

\begin{figure}
\includegraphics[width=0.52\textwidth,height=0.42\textwidth]{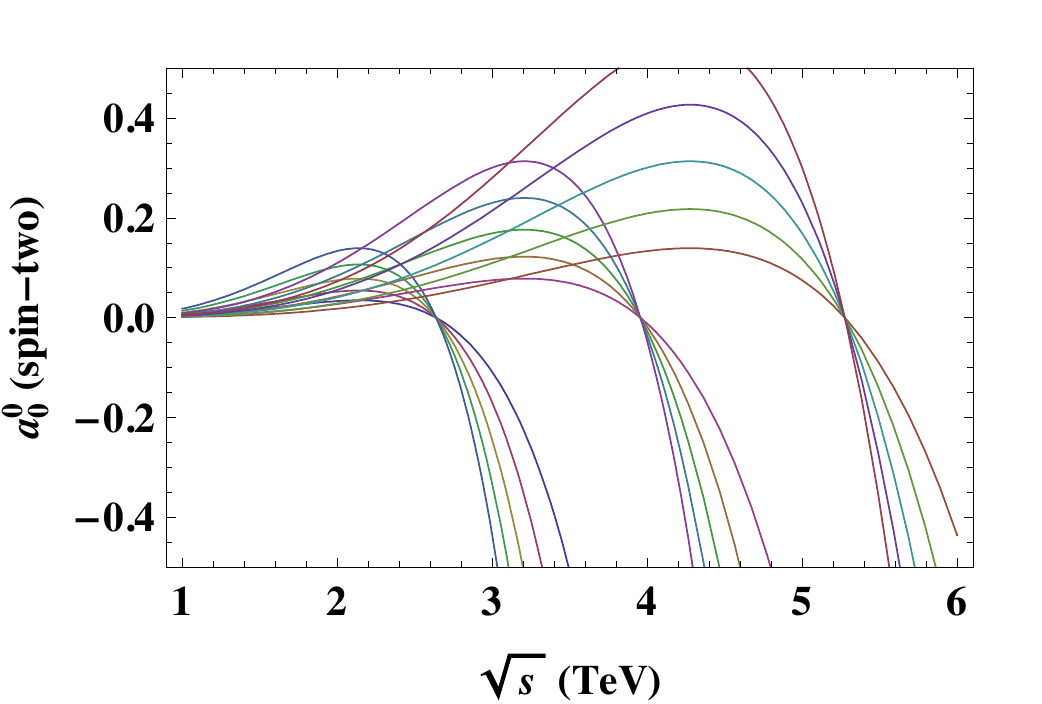}%
\includegraphics[width=0.52\textwidth,height=0.42\textwidth]{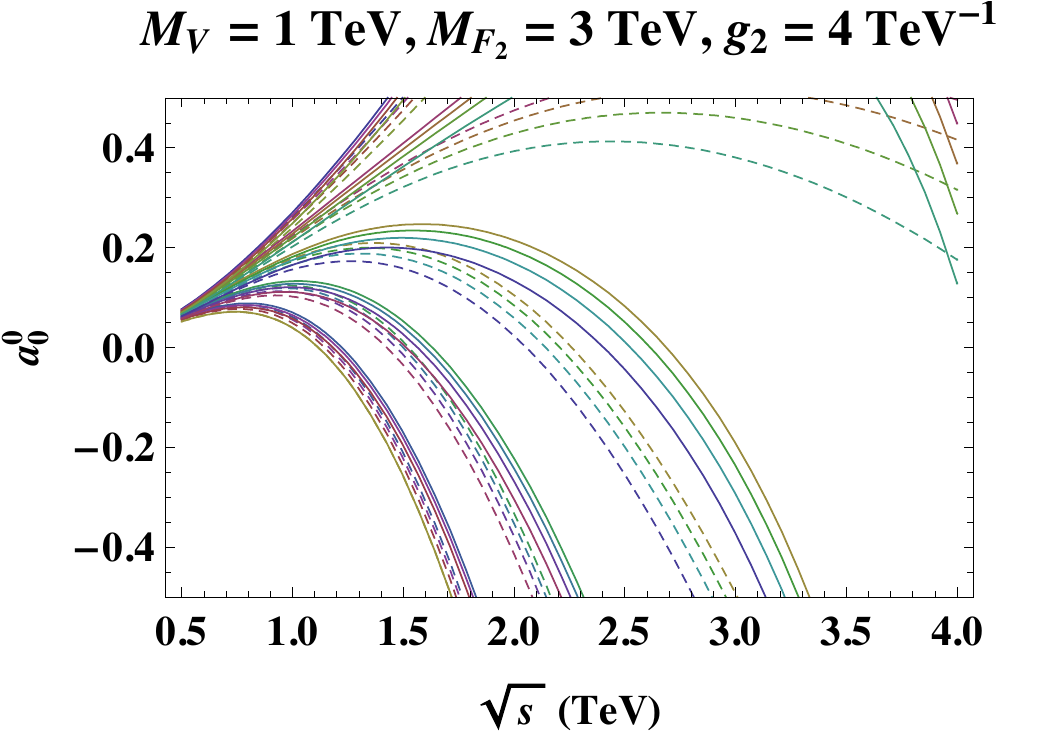}
\caption{Left: contribution from the spin-two exchanges to the $I=0$ $J=0$ partial wave amplitude of the $\pi\pi$ scattering. The different groups of curves correspond, from left to right, to $M_{F_2}=2,3,4$ TeV. Within each group, the different curves correspond, from smaller to wider, to $g_2=2,2.5,3,3.5,4$~TeV$^{-1}$. Right: $I=0$ $J=0$ partial wave amplitude with all channels included (spin-zero, -one, and -two). The dashed curves reproduce Fig.~\ref{fig:a00}, with just the spin-zero and the spin-one channels included. The solid curves contain also the spin-two exchanges, for $M_{F_2}=3$ TeV, and $g_2=4$~TeV$^{-1}$. If unitarity is violated at negative values of $a_0^0$, the spin-two exchanges may lead to a delay of unitarity violation.}
\label{fig:a00spin2}
\end{figure}

In addition to spin-zero and spin-one mesons, the low energy spectrum can contain spin two mesons as well~\cite{Harada:1995dc}. The contribution of a spin-two meson $F_2$ to the invariant amplitude is
\begin{eqnarray}
A_2(s,t,u)=\frac{g_2^2}{2(M_{F_2}^2-s)}\left[-\frac{s^2}{3}+\frac{t^2+u^2}{2}\right]-\frac{g_2^2 s^3}{12 M_{F_2}^4} \ ,
\end{eqnarray}
where $M_{F_2}$ and $g_2$ are mass and coupling with the pions, respectively. A reference value for $g_2$ can be obtained from QCD: $m_{f_2}\simeq 1275$~MeV and $\Gamma(f_2\to \pi\pi) \simeq 160$~MeV give $|g_2| \simeq 13$~GeV$^{-1}$ so that $|g_2|F_\pi \simeq 1.2$. Scaling up to the eletroweak scale results in $|g_2| \simeq 4$~TeV$^{-1}$.
The 
contribution of $F_2$ to the $I=0$ $J=0$ partial wave amplitude is given in Fig.~\ref{fig:a00spin2} (left) for different values of $M_{F_2}$ and $g_2$. Notice that the amplitude is initially positive, and then becomes negative at large values of $\sqrt{s}$. If $M_{F_2}$ is large enough, the positive contribution can balance the negative contribution from the spin-zero and spin-one channels, shown in Fig.~\ref{fig:a00}. This can lead to a further delay of unitarity violation, as shown in Fig.~\ref{fig:a00spin2} (right). Here the curves of Fig.~\ref{fig:a00} are redrawn dashed, while the full contribution from spin-zero, spin-one, and spin-two is shown by the solid lines, for $M_{F_2}=3$~TeV and $g_2=4$~TeV$^{-1}$. If unitarity is violated at negative values of $a_0^0$, then the spin-two contibution delays the violation to higher energies.

 The unitarity analysis presented here is for generic Vanilla TC theories, or any other model, featuring spin zero, one, and two resonances. The specialization to running and walking TC is described in detail in \cite{Foadi:2008ci,Foadi:2008xj}. The bottom line is that it is possible to delay the onset of unitarity violation, at the effective Lagrangian level, for phenomenologically viable values of the couplings and masses of the composite spectrum.

\newpage
\section{Phenomenological benchmarks }

We introduce here a natural classification of the different types of dynamical models which can be constrained at the LHC. We do this by analyzing the constraints that different types of dynamics imposes on the coefficients of the general effective Lagrangian introduced in the earlier sections. We start with counting the number of parameters. The effective Lagrangian \eqref{eq:boson} contains 9 parameters:
\begin{itemize}
\item $g$ and $g'$, the $SU(2)_\text{L} \times U(1)_Y$ gauge couplings;
\item $\mu$ and $\lambda$, the parameters of the scalar potential;
\item $\tilde{g}$, the strength of the spin one resonances interaction;
\item $m^2$, the $SU(2)_\text{L} \times SU(2)_\text{R}$ invariant axial-vector mass squared;
\item $r_2, r_3,s$, the couplings between the Higgs and the vector states;
\end{itemize}
Using the experimental values of $G_F$, $M_Z$ and $\alpha$ we can eliminate 3 of them. The remaining parameters can be rearranged according to the needs of the model builder.

For the Vanilla TC case, we found useful to choose as independent parameters the vector and axial vectors masses $M_V$, $M_A$, since they dictate the scale of the new resonances, as well as the physical mass of the composite Higgs $M_H$, the value of the $S$ parameter of the theory and the two couplings $\tilde{g}$ and $s$. 

The Lagrangian couplings $r_2$, $r_3$, and $m$ can be expressed as functions of $M_A$, $M_V$ and the $S$ parameter in the following way: 
\begin{align}
m^2 &= \left( M_A^2 + M_V^2 - \tilde{g}^2 v^2 s \right) / 2 \ ,
\\
r_2 &= 2 (M_A^2 - M_V^2) / \tilde{g}^2 v^2 \ ,
\\
r_3 &= 4 M_A^2 \left( 1 \pm \sqrt{1 - \tilde{g}^2 S / 8 \pi} \right) / \tilde{g}^2 v^2 \ .
\end{align}

The input parameters for the various models analyzed here as special cases of the Vanilla TC are listed in Table~\ref{tab:Inparam}. 
Under the last column we quote only the constraints that can be used to reduce the number of input parameters. 

\begin{table}[h!]
\begin{center}
\begin{tabular}{>\pnt l | l | l}
\textrm{Model} & \textrm{Input parameters} & Constraints \\
\hline\hline
Vanilla & $M_H$, $\tilde{g}$, $M_A$, $M_V$, $s$, $S$ & \\
Walking & $M_H$, $\tilde{g}$, $M_A$, $s$, $S$ & $1^{\textrm{st}}$ WSR \\
Running & $M_H$, $\tilde{g}$, $M_A$, $s$ & $1^{\textrm{st}}$ and  $2^{\textrm{nd}}$  WSR \\
NMWT & $M_H$, $\tilde{g}$, $M_A$, $s$ & $1^{\textrm{st}}$ WSR, $S = 0.3$ \\
Custodial &  $M_H$, $\tilde{g}$, $M_A$, $s$ & $r_2=r_3=0$ \\
D-BESS &  $M_H$, $\tilde{g}$, $M_A$ & $r_2=r_3=s=0$ \\
\end{tabular}
\caption{\label{tab:Inparam}  List of models  to be used as benchmarks at the LHC. We classified them in  a simple manner for the experiments to be able to easily identify the relevant parameter space. } 
\end{center}
\end{table}

Briefly we comment about the different models:

\begin{itemize}
\item  {\bf Walking TC:} using the first WSR we eliminate $M_V$ from the list of input parameters. In walking dynamics, the second WSR is modified according to \cite{Appelquist:1998xf}. We can use this relation as a constraint on the parameter space requiring that $a>0$ in \eqref{eq:WSR2}.
\item {\bf Running TC:} in this case both the first and the second WSRs can be used to reduce the number of input parameters. We eliminate $M_V$ and $S$.
\item {\bf NMWT:} we apply the same constraints as in the case of the general Walking TC, but now we take $S=0.3$ estimated knowing the underlying TC theory.
\item {\bf Custodial TC:} this scenario requires $r_2=r_3=0$. In this case $S$ is automatically 0 and $M_V = M_A$ \cite{Appelquist:1999dq,Appelquist:1998xf,Duan:2000dy,Foadi:2007se}.
\item {\bf D-BESS model:} in this model \cite{Casalbuoni:1995qt} all couplings involving one or more vector resonances and one or more scalar fields vanish. Effectively it is like the previous case plus the condition $s=0$.
\end{itemize}

As for ordinary QCD besides the simplest spin one and spin zero sectors discussed here the underlying dynamics will produce several composite states of bosonic and fermionic nature. For example, there will be states analog to the familiar baryons of QCD. The physics of these new states is very interesting, it will help selecting the specific underlying gauge dynamics, and needs to be explored in much detail. However, we have chosen to limit ourselves to the most common sectors which have been studied extensively in the literature for an initial analysis. 

We present  basic processes relevant for the collider phenomenology of sensible models of dynamical electroweak symmetry breaking. We have divided the processes according to the spin of the new relevant composite particle which can be potentially discovered at collider experiments.
 
Before presenting the actual predictions we highlight key features of the model which will help understanding the search strategy and the results presented below. When turning off the electroweak interactions some of the interesting decay modes, depending on the parameters of the effective Lagrangian, involving spin one massive vector states and composite scalars (defined in Eqs.(\ref{TM-eigenstates},\ref{TMV-eigenstates}) are: \beq
V \to \Pi\ \Pi \ , \quad A \to H\ \Pi \ , \quad H \to \Pi\ \Pi \ ,
\label{basedecays}
\eeq
with the appropriate charge assignments. We assumed in the equation above that the composite Higgs is lighter than the vector states, however our effective theory allows for a more general spectrum. 
Once the electroweak interactions are turned on the technipions become the longitudinal components of the $W$ and $Z$ bosons. Therefore the processes in \eqref{basedecays} allow to detect the spin one resonances at colliders. The massive spin one states mix with the SM gauge bosons. After diagonalizing the spin one mixing matrices (see \cite{Foadi:2007ue,Belyaev:2008yj}) the lightest and heaviest of the composite spin one triplets are termed $R_{1}^{\pm, 0}$ and $R_{2}^{\pm, 0}$ respectively. In the region of parameter space where $R_{1}$ is mostly an axial-like vector (for a mass less than or about one TeV) and $R_2$ mostly a vector state one observes the following qualitative dependence of the couplings of the SM fields to heavy vector bosons as function of the electroweak gauge coupling $g$ and the heavy vector self interaction coupling $\tilde{g}$:
\beq \label{eq:couplbeh}
g_{R_{1,2} f\bar{f}}\sim \frac{g}{\tilde{g}}  \ , \quad g_{R_2 W W}\sim \tilde{g} \ , \quad g_{R_1 H Z}\sim \tilde{g} \ .
\eeq
Notice that, since the heavy spin one states do not couple directly to SM fermions, the couplings $g_{R_{1,2} f\bar{f}}$ arises solely from the mixing with $W$ and $Z$. This coupling is roughly proportional to $g/\tilde{g}$.

We implemented the effective Lagrangian (\ref{eq:boson}) on event generators such as \href{http://madgraph.phys.ucl.ac.be/}{MadGraph} \cite{MG/ME} and \href{http://theory.sinp.msu.ru/~pukhov/calchep}{CalcHEP} \cite{CalcHEP} in order to compare the predictions of the model for different choices of its parameters with the experimental data that are and will be produced at LHC. The implementation can be freely downloaded from the page \href{http://cp3-origins.dk/research/tc-tools}{http://cp3-origins.dk/research/tc-tools}, together with the calculator needed to change the parameters. The details of the implementation and how to use the different computer packages are provided in Appendix~\ref{TEG}.

The analyses we are going to present are at the parton-level study, not accounting for efficiencies and
for some systematic uncertainties arising from detector effects. 

\subsection{Spin one processes: Decay widths and branching ratios}\label{sec:decay}

The existence of new heavy spin one resonances has been postulated in different extensions of the SM. It is not hard to convince oneself that exploring the entire range of the couplings and masses of the effective Lagrangian presented here one is able to cover virtually the entire spectrum of models presented in the literature. The link to a specific TC model can be obtained using the WSRs or the knowledge of the spectrum deduced via first principle lattice simulations. In this subsection we consider different processes where the spin one resonances appear in the intermediate diagrams in the $s$-channel at the parton level. The event generators are then used to embed the parton process within the proton-proton collision. 
We have also checked that there is no appreciable difference in using MadGraph or CalcHEP, which can be taken as a direct test of the validity of the two implementations.

\begin{figure}
{
\includegraphics[width=0.45\textwidth]{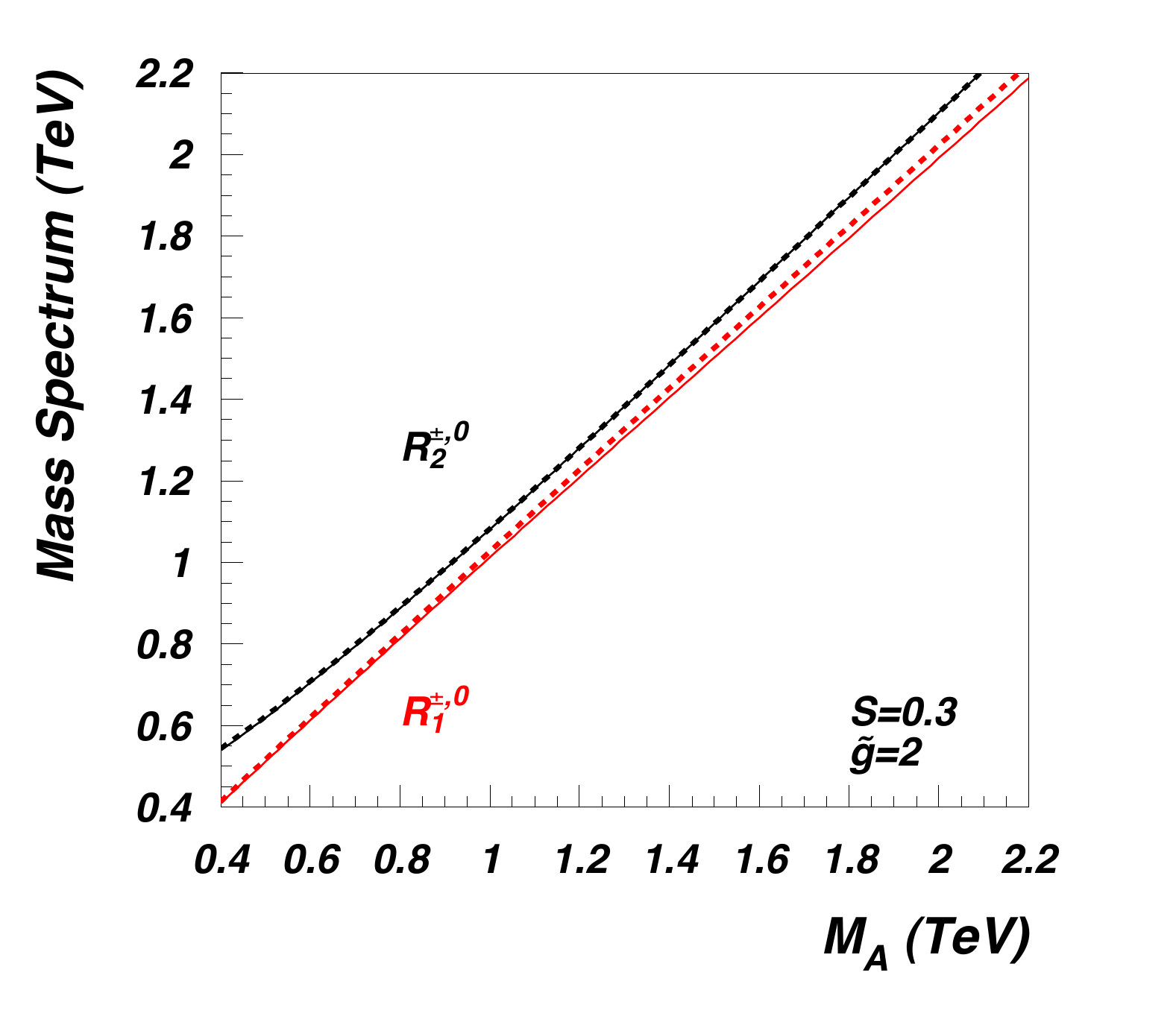}
\includegraphics[width=0.45\textwidth]{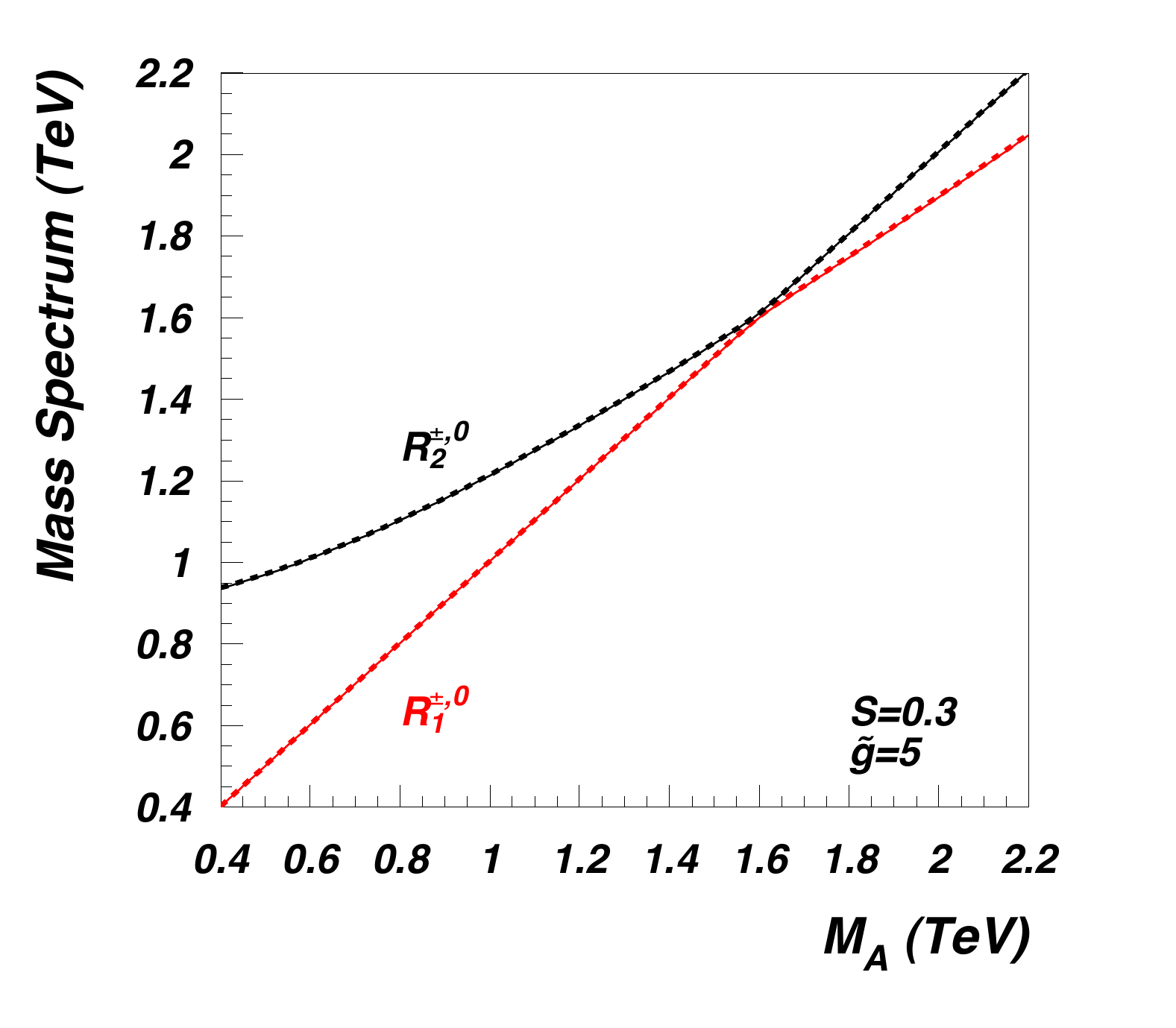}%
} 
 \caption{The mass spectrum of  the $M_{R^{\pm,0}_{1,2}}$
        vector mesons versus $M_A$
	for $\tilde{g}=2$ (left)
	and $\tilde{g}=5$ (right).
	The masses of the charged vector mesons are denoted by solid lines,
	while the masses of the neutral mesons are denoted by dashed lines.
	\label{fig:spectrum}}
\end{figure}

In traditional TC models of running dynamics the enforcement of the first and second WSRs leads to a spin one vector lighter than the axial one. An intriguing possibility  for walking theories \cite{Foadi:2007ue,Appelquist:1998xf} is the
possible mass spectrum inversion of the vector and axial spin one  mesons. This occurs since the second WSR gets modified \cite{Appelquist:1998xf}.
\begin{figure}
{\includegraphics[height=6.5cm,width=7.82cm]{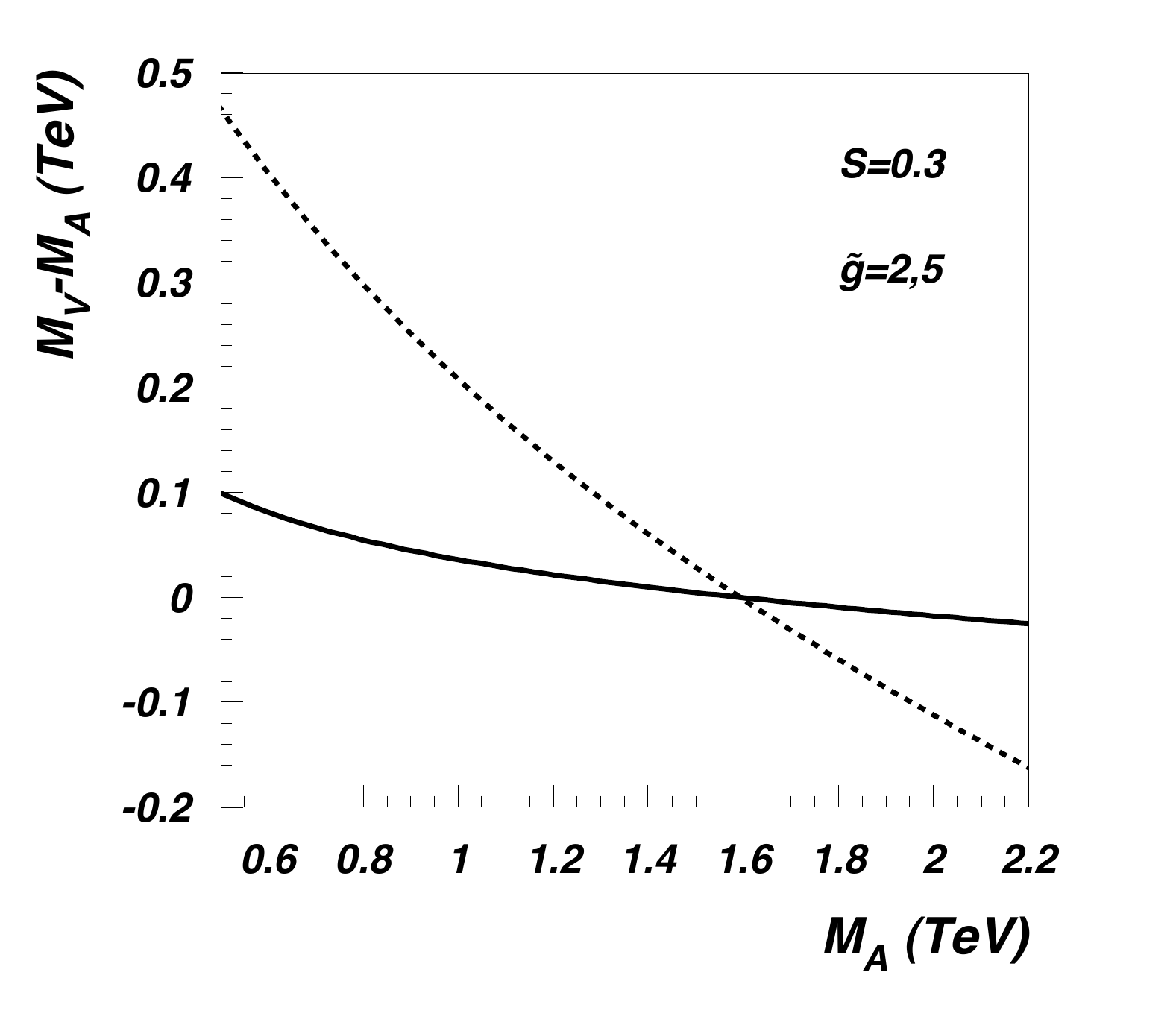}
 \includegraphics[height=6.5cm,width=7.82cm]{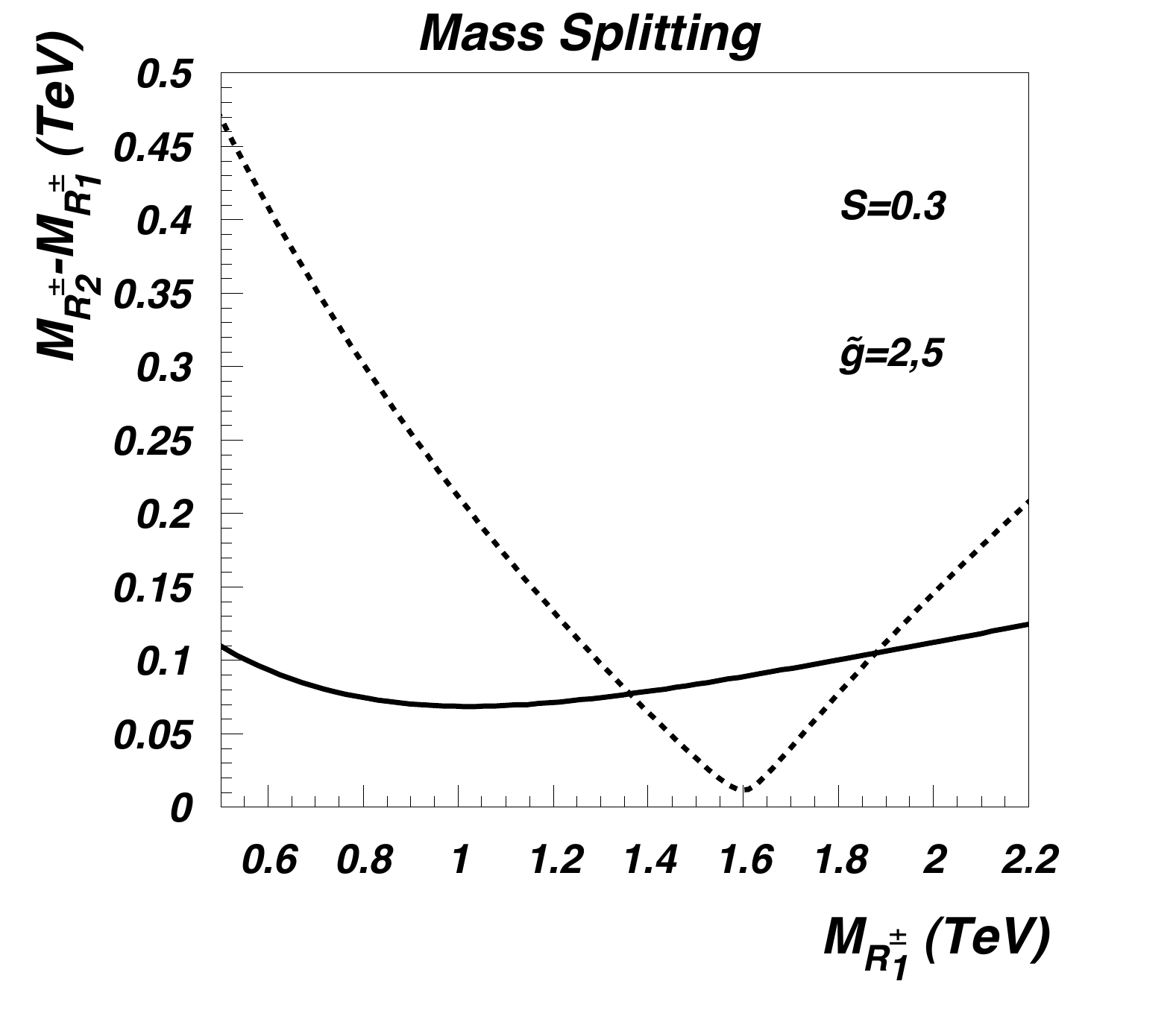}}
\caption{Mass splittings $M_V-M_A$ (left) and $M_{R_2^\pm}-M_{R_1^\pm}$ (right). The dotted lines are for $\tilde{g}=5$ while the solid lines are for $\tilde{g}=2$.}\label{fig:masses}
\end{figure}
In
Fig.~\ref{fig:masses} (left) we plot $M_V-M_A$ as a function
of $M_A$ for two reference values of $\tilde{g}$ and $S=0.3$. For generic values of $S$ the
inversion occurs for
\begin{eqnarray}
{M}^{\rm inv} = \sqrt{\frac{4\pi}{S}}F_\pi \ .
\label{eq:inv}
\end{eqnarray}
This gives ${M}^{\rm inv}\simeq 1.6$ TeV for $S=0.3$, as clearly shown in the plot. 
Fig.~\ref{fig:masses}  (right) shows $M_{R_2^\pm}-M_{R_1^\pm}$ as a function of $M_{R_1^\pm}$, where $R_1^{\pm,0}$ ($R_2^{\pm,0}$) are the lighter (heavier) vector resonances, with tree-level electroweak corrections included. This mass difference is always positive by definition, and the mass inversion becomes a kink in the plot. Away from ${M}^{\rm inv}$ $R_1$ ($R_2$) is 
an axial (vector) meson for $M_A < {M}^{\rm inv}$, and a vector (axial) meson for $M_A > {M}^{\rm inv}$. 
The mass difference in Fig.~\ref{fig:masses} is proportional to $\tilde g^2$, and becomes relatively small for $\tilde g=2$. The effects of the electroweak corrections are larger for 
small $\widetilde{g}$ couplings. For example, the minimum of $M_{R_2^\pm}-M_{R_1^\pm}$ is shifted from  ${M}^{\rm inv}\simeq 1.59$~TeV to about 1~TeV for $\tilde g=2$. To help the reader we plot in Fig.~\ref{fig:spectrum}
 the actual spectrum for the vector boson masses
 versus $M_A$. Preliminary studies on the lattice of MWT and the mass inversion issue appeared in \cite{DelDebbio:2008zf}.
\begin{figure}[tbhp]
 \vskip -0.4cm
 \includegraphics[width=0.45\textwidth,height=0.35\textwidth]{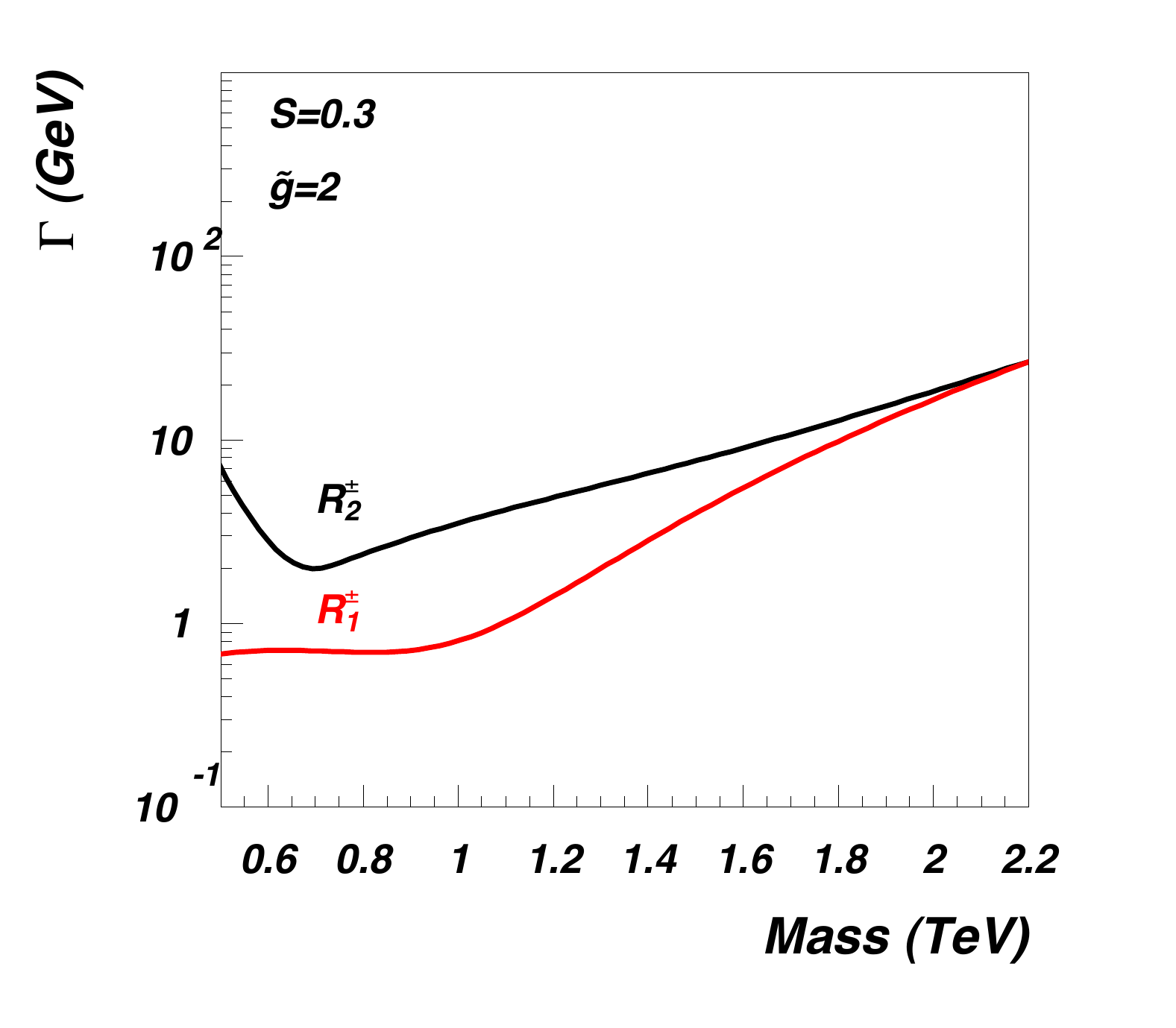}%
 \includegraphics[width=0.45\textwidth,height=0.35\textwidth]{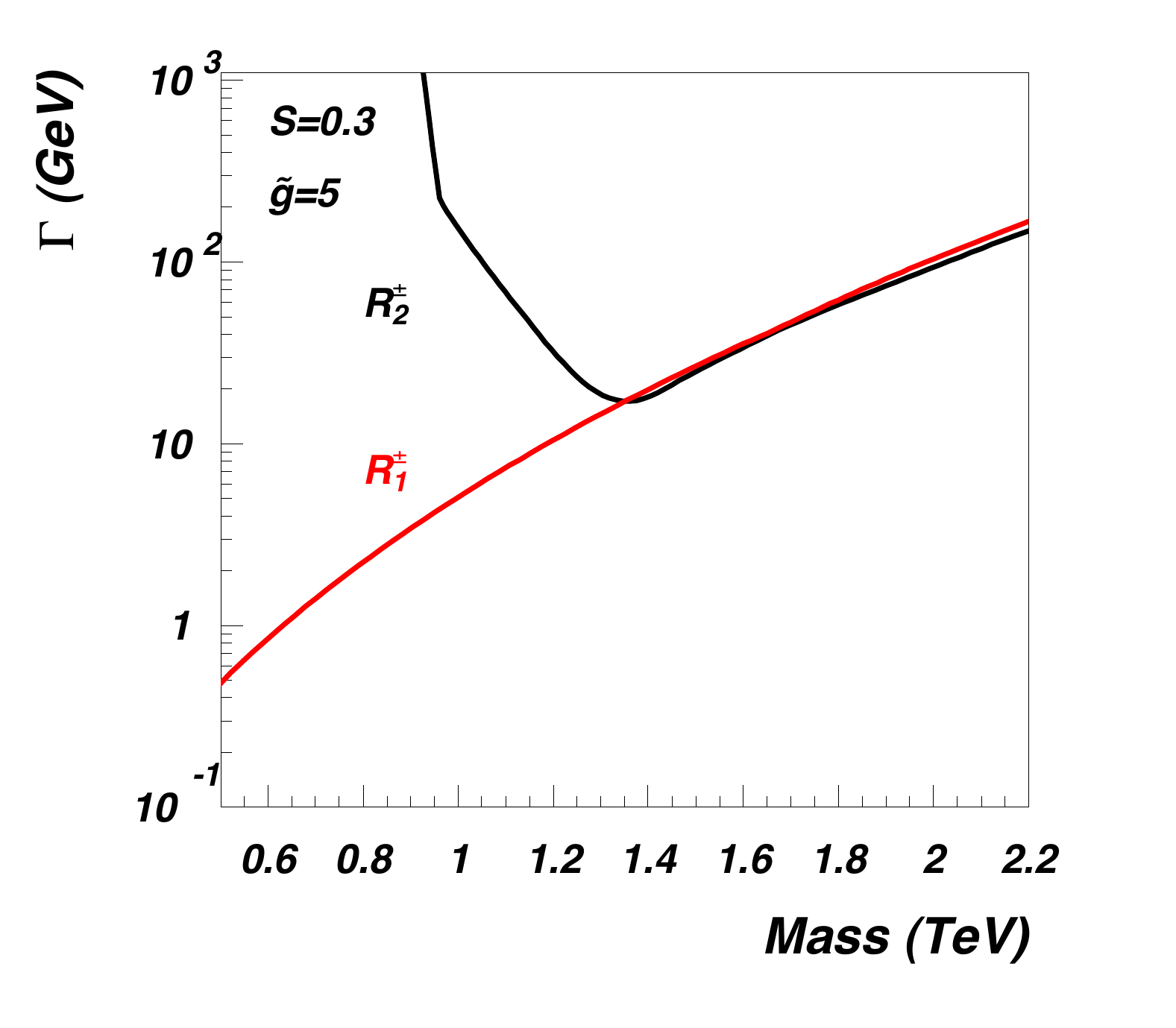}
 \vskip -0.4cm
 \includegraphics[width=0.45\textwidth,height=0.35\textwidth]{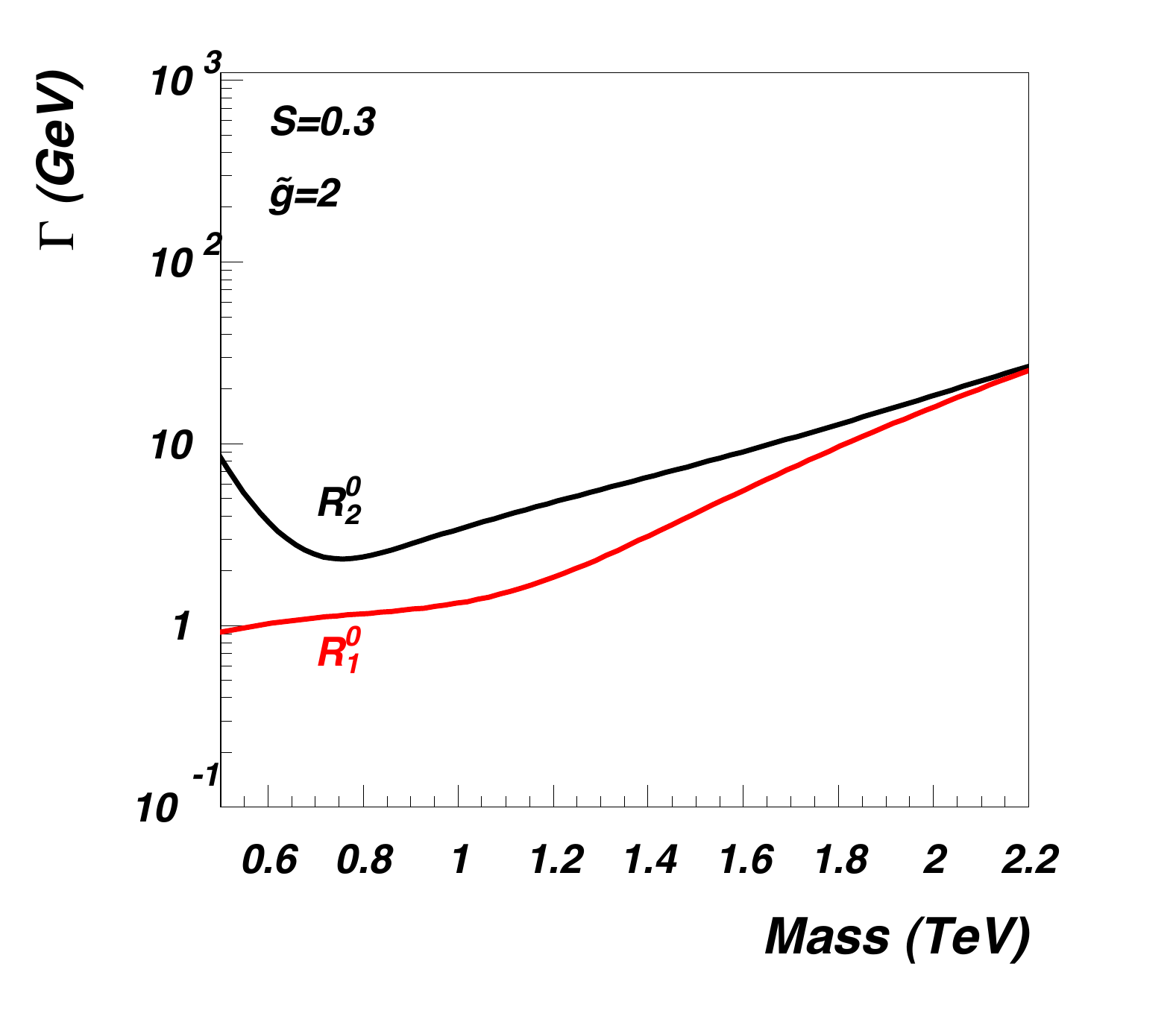}%
 \includegraphics[width=0.45\textwidth,height=0.35\textwidth]{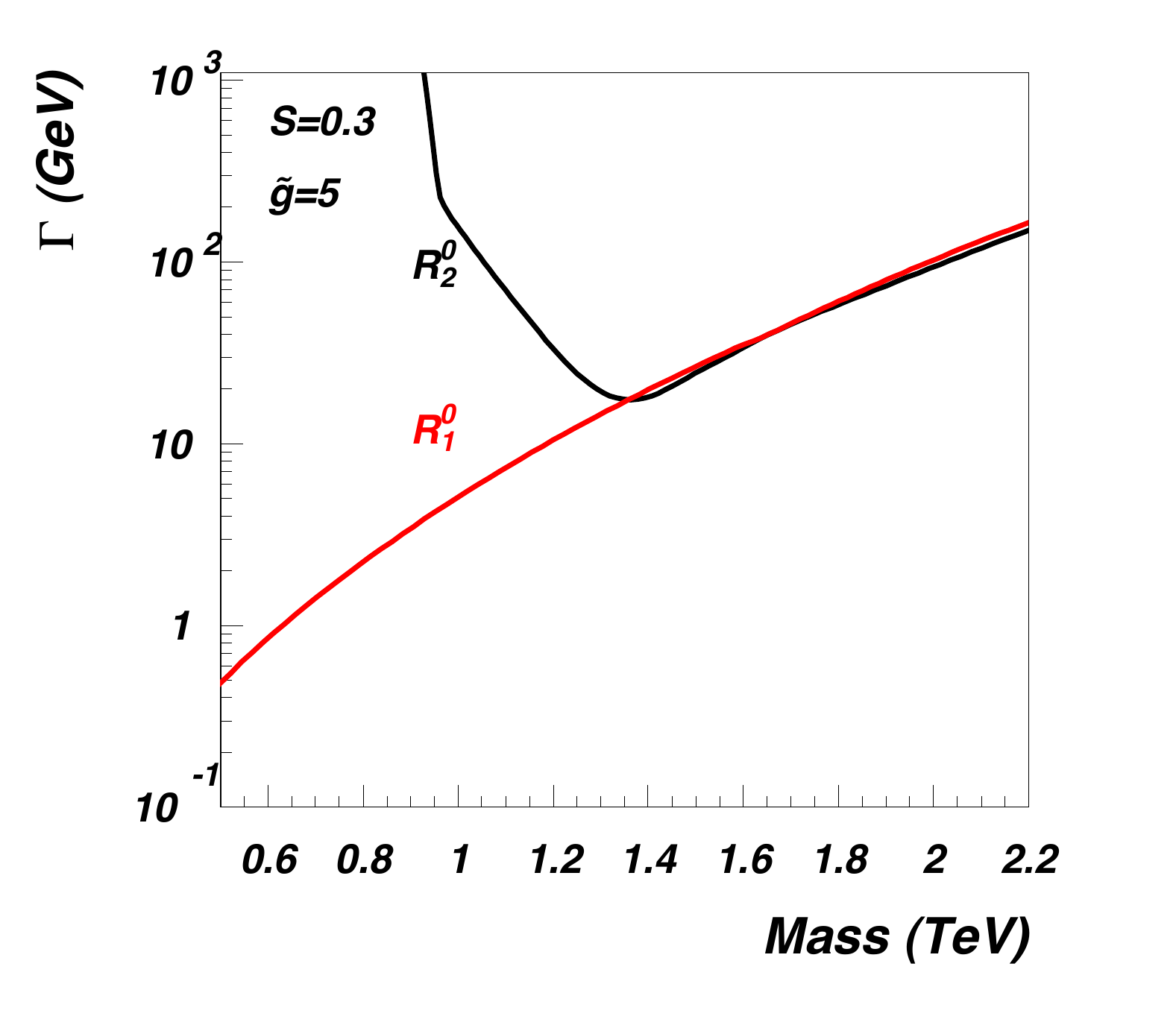}
 \vskip -0.4cm
\caption{Decay width of the charged (first row) and neutral (second row) 
vector mesons for $S=0.3$ and $\tilde{g}=2,5$.  We take $M_H = 0.2 \ \textrm{TeV},\ s=0$.}
\label{fig:decay2}
\end{figure}

 The widths of the heavy vectors are displayed
 in Fig.~\ref{fig:decay2}.  The lighter meson, $R_1$, is very narrow.
 The heavier meson, $R_2$, is very narrow for small values of
 $\tilde{g}$. In fact in this case $M_{R_2} \simeq M_{R_1}$, forbidding decays of $R_2$ to $R_1$ (+anything). For large
 $\tilde{g}$, $R_2$ is very narrow  for large masses, but then becomes
 broader when the $R_2 \to R_1,X$ channels open up, where $X$ is a SM 
 gauge boson.  It becomes very broad when the $R_2 \to 2R_1$ decay channel opens up.
  The former are only important below the inversion point,
 where $R_1$ is not too heavy. The latter is only possible when $R_2$
 is essentially a spin one vector  and $M_{R_2} > 2 M_{R_1}$.

The narrowness of $R_1$ { (and $R_2$, when the $R_2\to R_1,X$ channels are forbidden)} 
is essentially due to the
small value of the $S$ parameter. In fact for $S=0$ the trilinear
couplings of the  vector mesons to two scalar fields of the strongly
interacting sector vanish. This can be understood as follows:  the
trilinear couplings with a vector resonance contain a derivative of
either the Higgs or the technipion,  and this can only come from $r_3$
in Eq.~(\ref{eq:boson}). Since $r_3=0$ implies $S=0$, as
Eqs.~(\ref{eq:S})  and (\ref{eq:chi})  show explicitly, it follows
that the decay width of $R_1$ and $R_2$ to two scalar fields  vanishes
as $S\rightarrow 0$. As a consequence, for $S=0$ the vector meson
decays to the longitudinal SM  bosons are highly suppressed,
because the latter are nothing but the eaten technipions. (The
couplings to the SM  bosons do not vanish exactly because of the mixing with the spin one resonances.)
A known scenario in which the widths of $R_1$ and $R_2$ are  highly
suppressed is provided by the D-BESS model~\cite{Casalbuoni:2000gn},
where the spin one  and the spin zero  resonances do not interact.
Therefore, in D-BESS {\em all} couplings involving one or more vector
resonances and  one or more scalar fields vanish, not just the
trilinear coupling with one vector field. The former scenario requires 
$r_2=r_3=s=0$, the latter only requires $r_3=0$. A somewhat
intermediate scenario is provided by custodial TC, in  which $r_2=r_3=0$ but
$s\neq 0$. Narrow spin one  resonances seem to be a common feature in
various models of dynamical electroweak symmetry breaking. (see for
example \cite{Brooijmans:2008se}). Within our effective Lagrangian~(\ref{eq:boson}) this property is linked to having a
small $S$ parameter.

\bigskip

\begin{figure}[tbhp]
\vskip -0.2cm
\includegraphics[width=0.45\textwidth,height=0.35\textwidth]{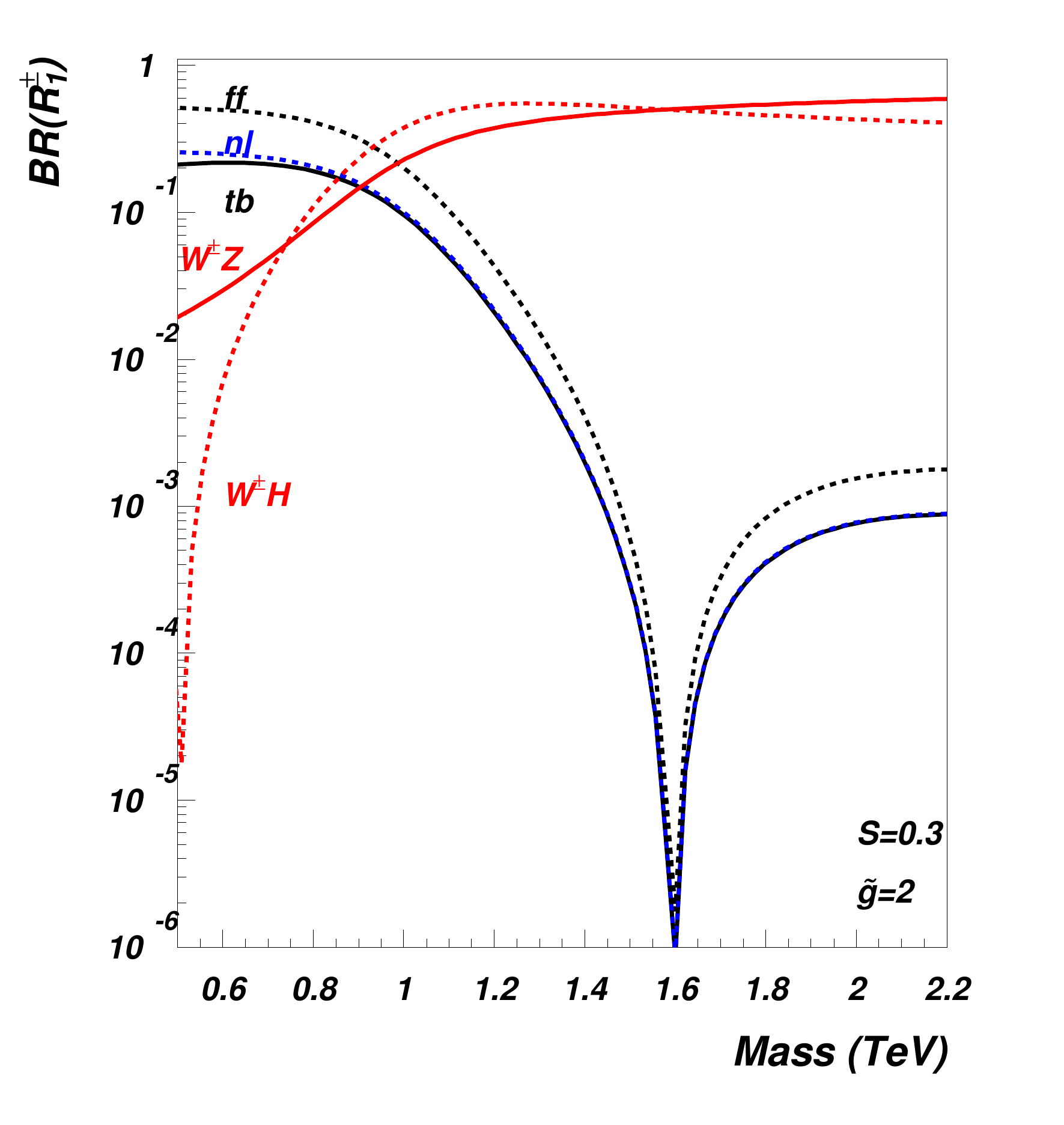}%
\includegraphics[width=0.45\textwidth,height=0.35\textwidth]{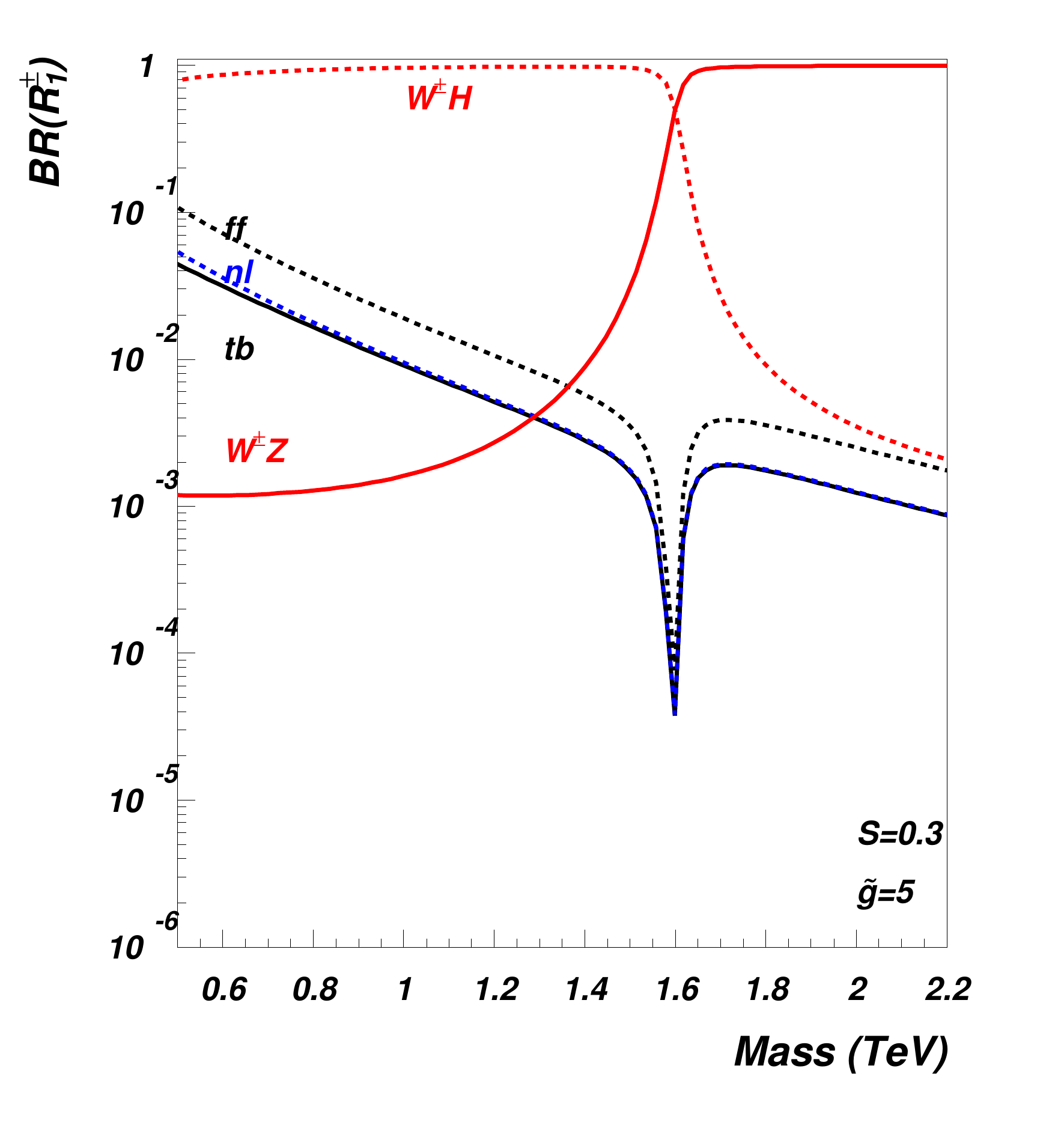}
 \vskip -0.2cm
 \includegraphics[width=0.45\textwidth,height=0.35\textwidth]{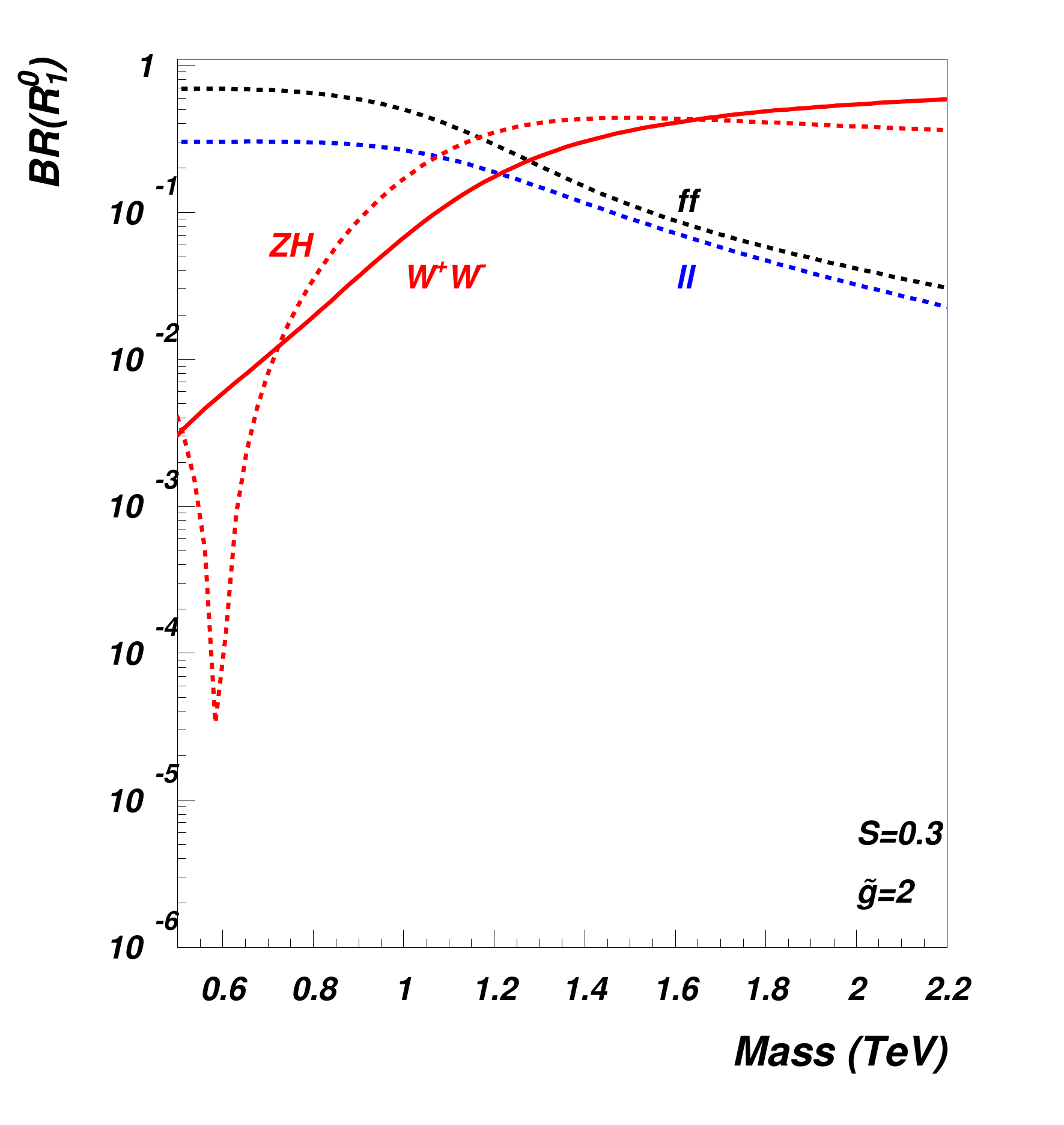}%
 \includegraphics[width=0.45\textwidth,height=0.35\textwidth]{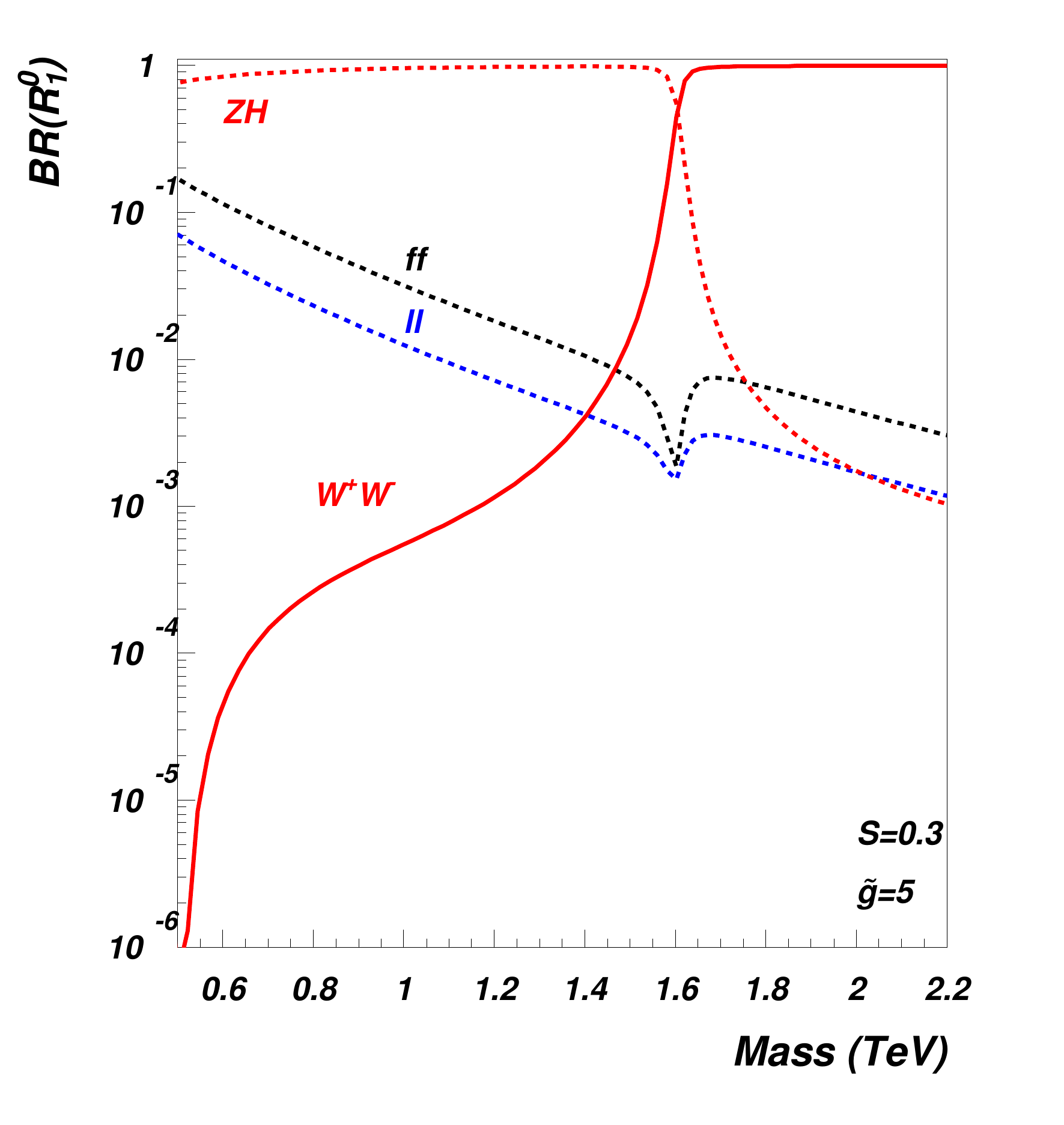}
 \vskip -0.2cm
\caption{Branching ratios of the charged (first row)
and neutral (second row)  $R_1$ resonance for $S=0.3$ and $\tilde{g}=2,5$.  We take $M_H = 0.2 \ \textrm{TeV}, s=0$.}\label{fig:BRR1}
\end{figure}
The  $R_1$ branching ratios are shown in
Fig.~\ref{fig:BRR1}. 
The wild variations observed in the plots around $1.6$~{TeV} reflect the mass inversion discussed earlier. Here the mixing between $R_1^a$ and $\widetilde{W}^a$,  with $a=0,\pm 1$,  vanishes, suppressing the decay to SM fermions. 

The other observed structure for the decays in $ZH$ and $WH$, at low masses, is due to the opposite and competing contribution coming from the TC and electroweak sectors. This is technically possible since the coupling of the massive vectors to the longitudinal component of the gauge bosons and the composite Higgs is suppressed by the small value of S. 

\begin{figure}[tbhp]
 \vskip -0.2cm
 \includegraphics[width=0.45\textwidth,height=0.35\textwidth]{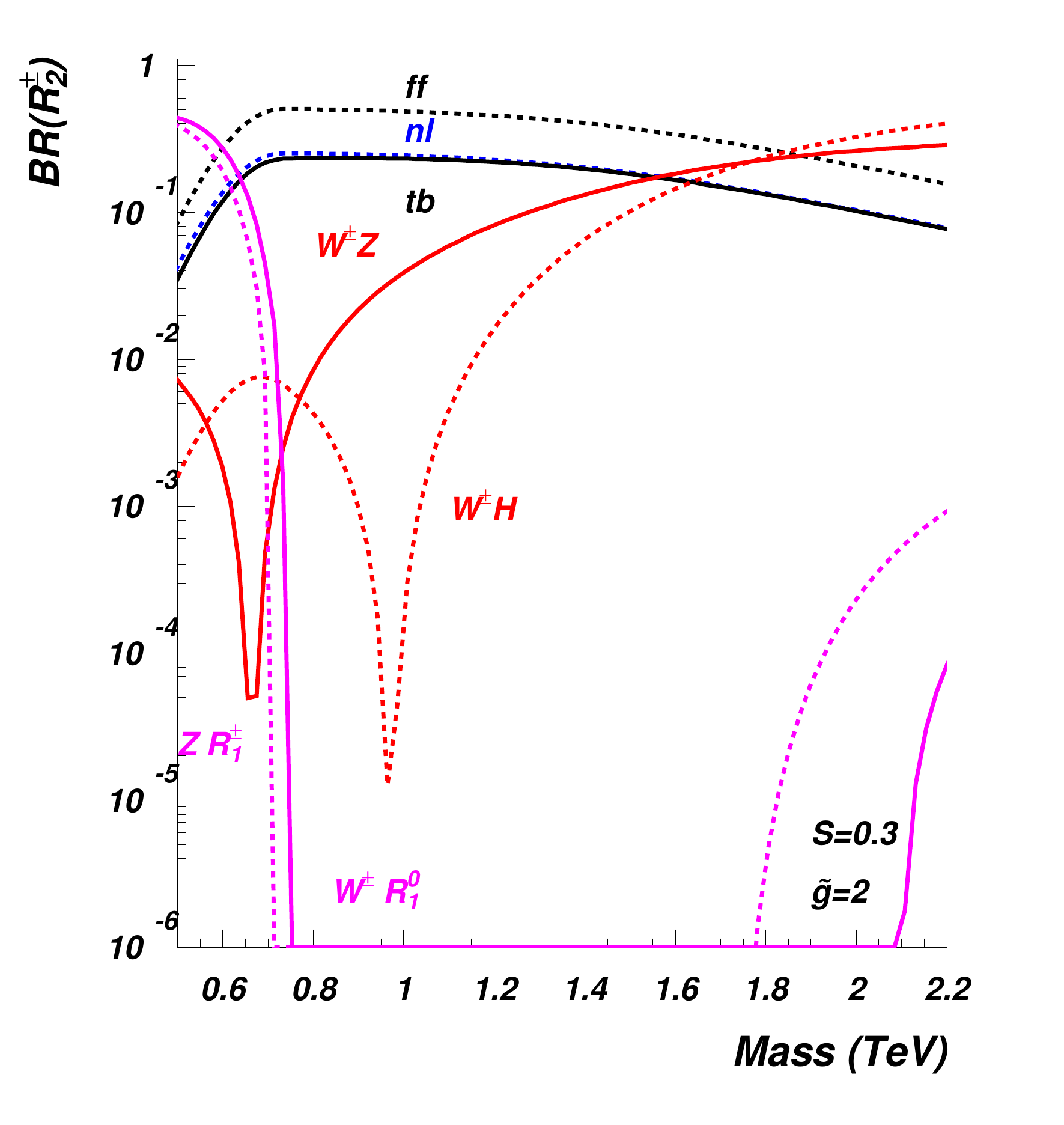}%
 \includegraphics[width=0.45\textwidth,height=0.35\textwidth]{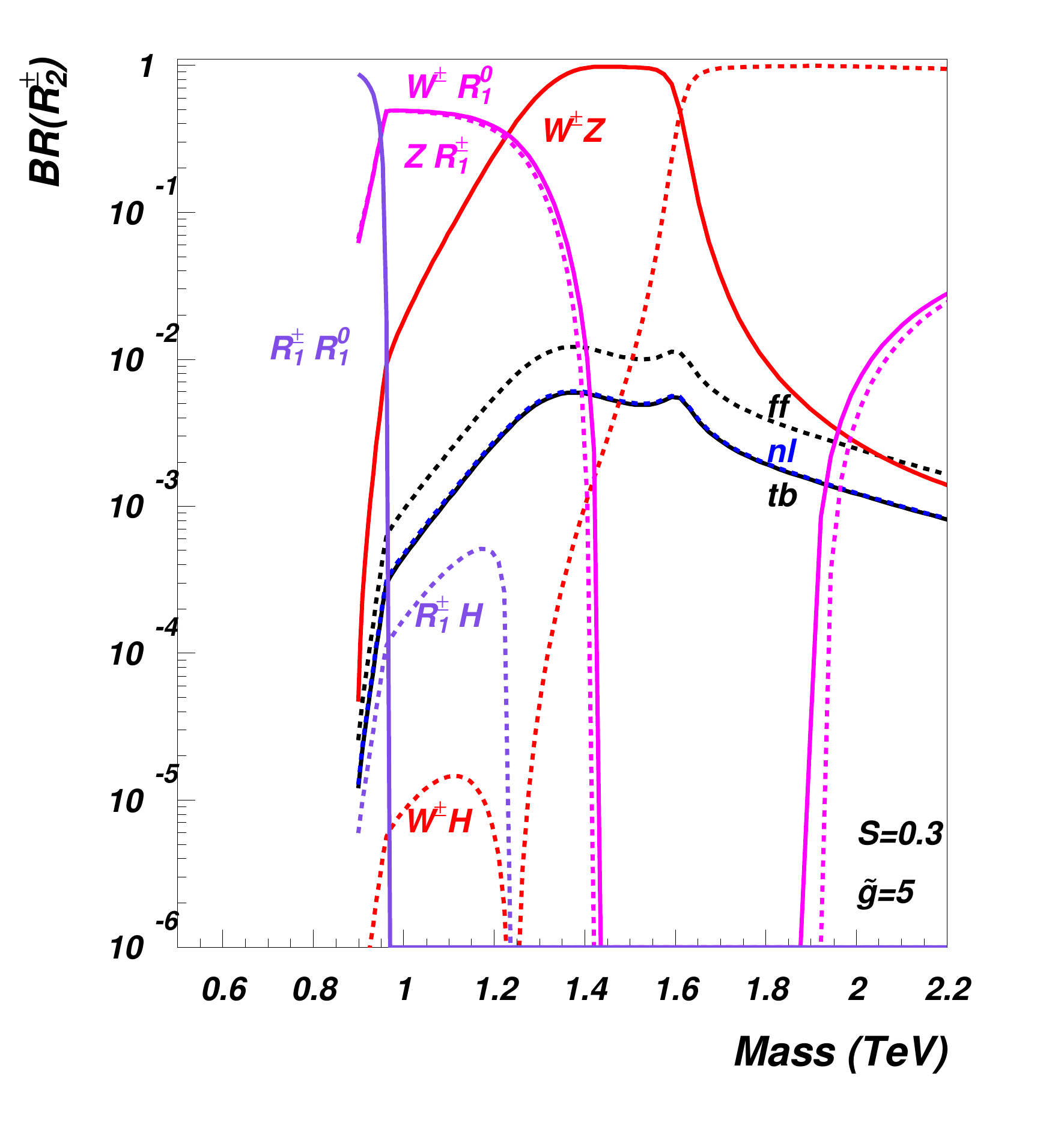}
 \vskip -0.2cm
 \includegraphics[width=0.45\textwidth,height=0.35\textwidth]{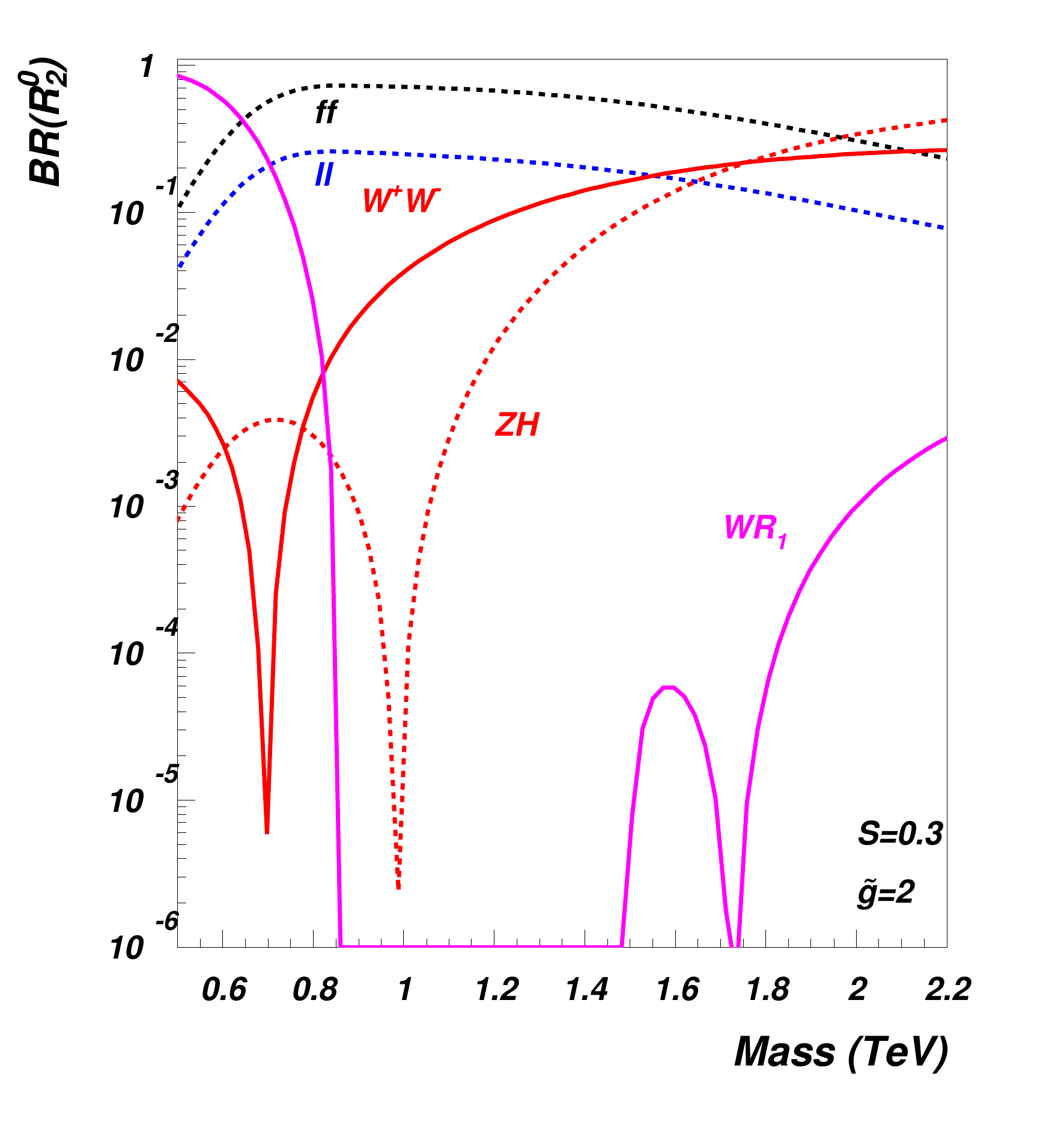}%
 \includegraphics[width=0.45\textwidth,height=0.35\textwidth]{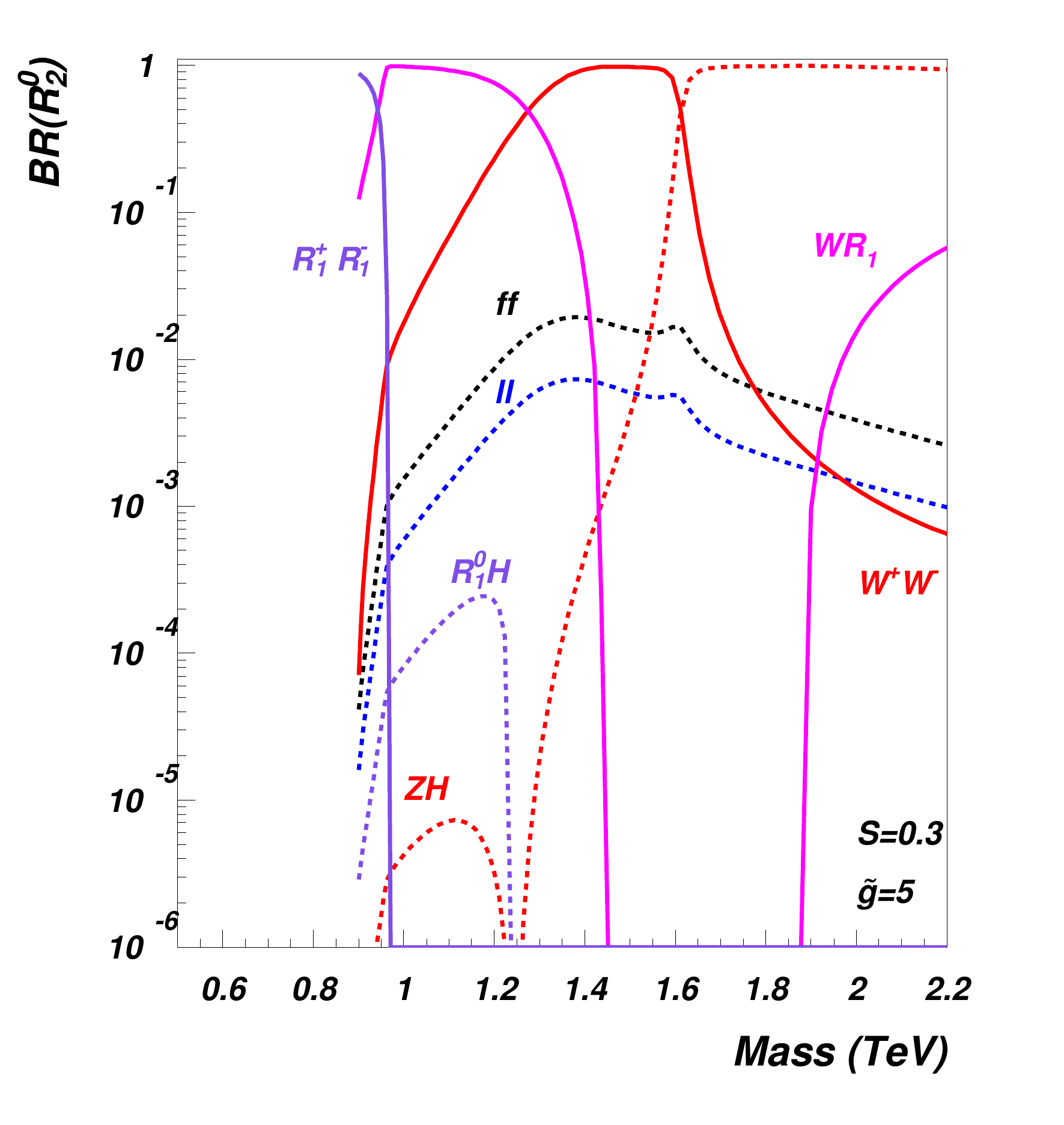}
 \vskip -0.2cm
\caption{Branching ratios of the charged (first row) and neutral (second row) $R_2$ resonance for $S=0.3$ and $\tilde{g}=2,5$ .  
We take $M_H = 0.2 \ \textrm{TeV}, s=0$.}\label{fig:BRR2}
\end{figure} 
 
Now we consider the $R_2$ branching ratios displayed in Fig.~\ref{fig:BRR2}. Being $R_2$ heavier than $R_1$ by definition, new channels like $R_2\rightarrow 2R_1$ and $R_2\rightarrow R_1 X$ show up, where $X$ denotes a SM boson. Notice that there is a qualitative difference in the $R_2$ decay modes for small and large values of $\tilde{g}$. First, for small $\tilde{g}$ the $R_2-R_1$ mass splitting is not large enough to allow the decays $R_2\rightarrow 2R_1$ and $R_2\rightarrow R_1 H$, which are instead present for large $\tilde{g}$. Second, for small $\tilde{g}$ there is a wide range of masses for which the decays to $R_1$ and a SM vector boson are not possible, because of the small mass splitting. 
The branching ratios to fermions do not drop at the inversion point, because {the $R_2-\widetilde{W}$ mixing does not vanish}.

 \subsubsection{$pp \to R \to \ell\ell$}
 
  \begin{figure}[htbp]
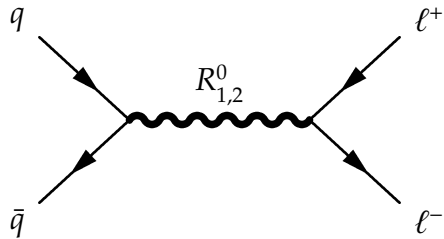

\begin{center}
\parbox{60mm}{\begin{fmfchar*}(60, 22)
\fmflabel{$q$}{qi}
\fmflabel{$\bar q$}{qibar}
\fmflabel{$\ell^-$}{lf}
\fmflabel{$\ell^+$}{lfbar}
\fmfleft{qibar,qi}
\fmfright{lf,lfbar}
\fmf{fermion}{qi,vi,qibar}
\fmf{fermion}{lfbar,vf,lf}
\fmf{photon, label=$R_{1,, 2}^0$, width=1mm, label.side=left}{vi,vf}
\end{fmfchar*}}
\vspace{7mm}

\caption{Feynman diagram of the signal processes for the dilepton production.}
\label{ppll}
\end{center}
\end{figure}
 
\begin{figure}[htbp]
\begin{center}
\includegraphics[width=0.45\textwidth]{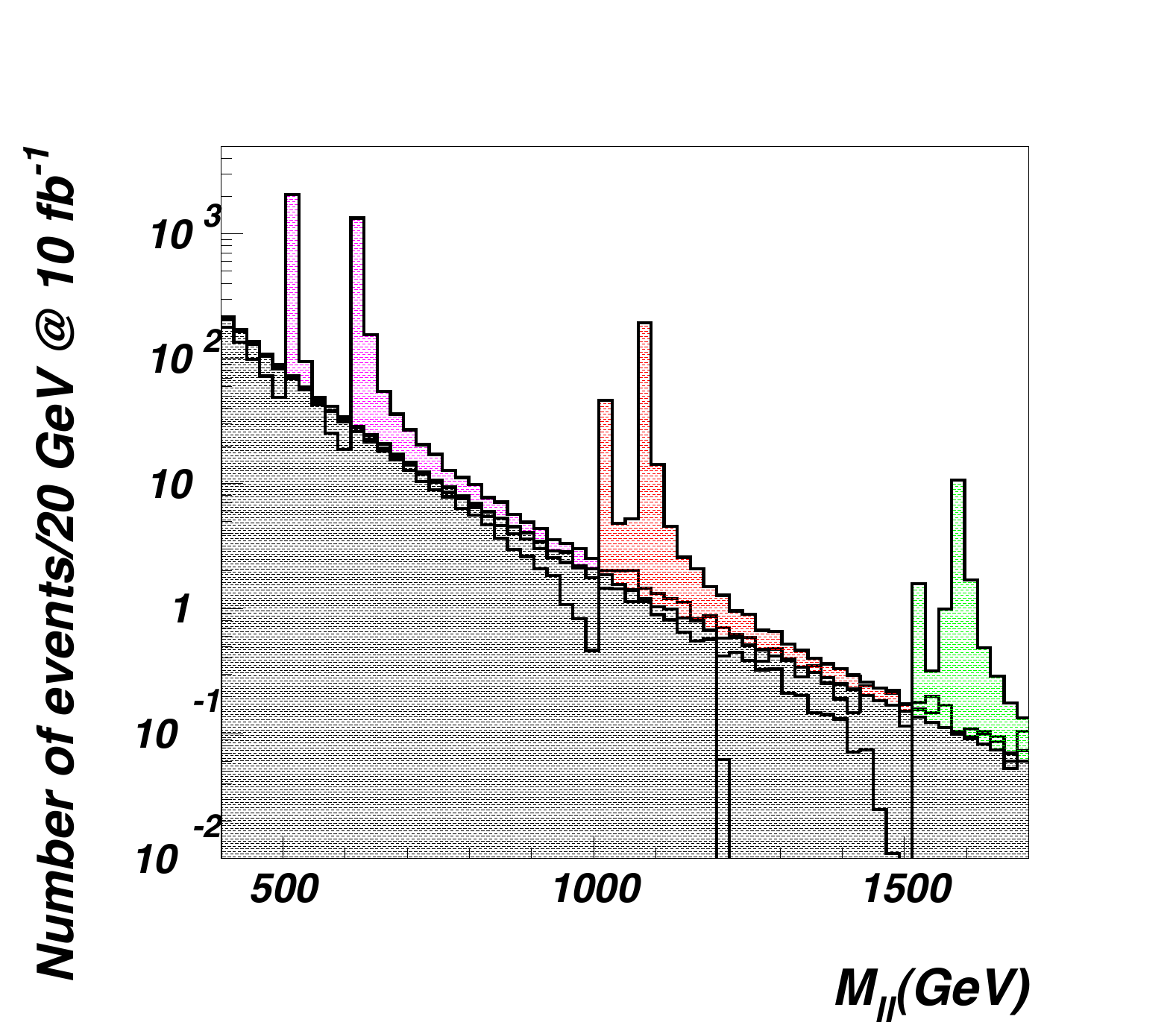}  ~~ \includegraphics[width=0.45\textwidth]{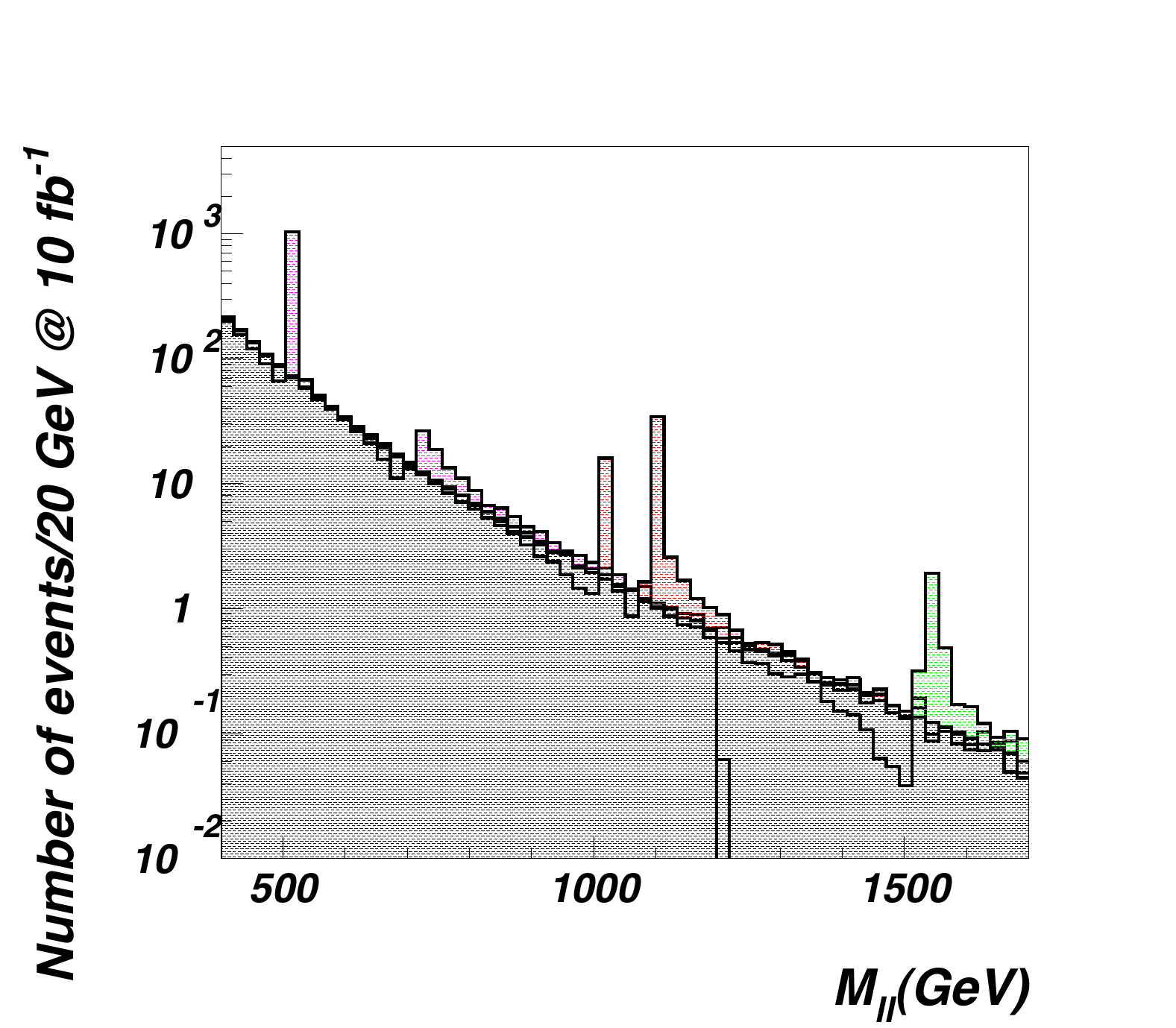} \\~\\
\includegraphics[width=0.45\textwidth]{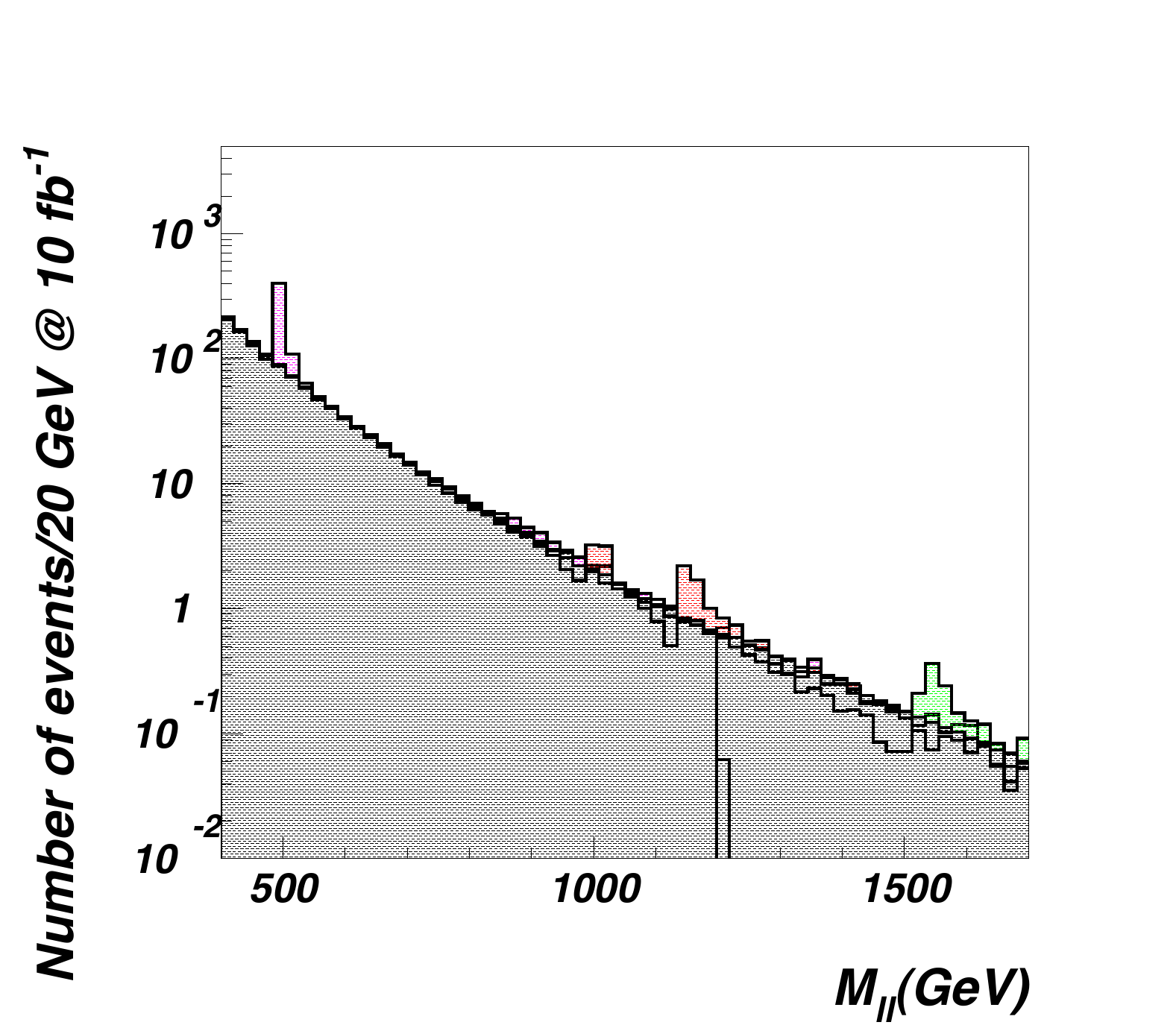}
\caption{Dilepton invariant mass distribution $M_{\ell\ell}$ for $pp \to R_{1,2} \to \ell^{+} \ell^{-}$ signal and background processes. We consider $\tilde{g}=2,3,4$ respectively from left to right and masses $M_{A}=0.5$ TeV (purple), $M_{A}=1$ TeV (red) and $M_{A}=1.5$ TeV (green).}
\label{gt24}
\end{center}
\end{figure}

Dilepton production was discussed also in \cite{Belyaev:2008yj}, with $\sqrt{s}=14$ TeV and 100 fb$^{-1}$ integrated luminosity. The Feynman diagram of this Drell-Yan process is shown in Fig.~\ref{ppll}. We updated that analysis for the near future LHC using the parameters $\sqrt{s}=7$ TeV and 10 fb$^{-1}$. The signal and the background are obviously reduced compared to the earlier studies, but in the optimal region of the parameter space signals are still clearly visible. Increasing the effective TC coupling $\tilde{g}$ quickly flattens out the signal.
In Fig.~\ref{gt24} we plot the number of events with respect to the invariant mass of the lepton pair, using  $\tilde{g}=2,3,4$ and $M_{A}=0.5, 1, 1.5$ TeV, where $M_{A}$ is the mass of axial eigenstate before mixing with the SM gauge bosons. We have applied  cuts of $|\eta^{\ell}| < 2.5$ and $p^{\ell}_{T} > 15$ GeV on the rapidity and transverse momentum of the leptons. The peaks from the $R_{1}$ and $R_{2}$ clearly stand out with signal-to-background ratio $S/B>10$ for several bins over the parameter space under consideration. The background is considered to be the contribution coming from the SM gauge bosons Z and $\gamma$. In Table~\ref{gt2x} the signal and background cross sections are reported, applying the cut

 \begin{table}[htdp]
\caption{$pp\to R_{1,2}\to \ell^{+}\ell^{-}$. Signal and background cross sections for $\tilde{g}=2,3,4$ and estimates for required luminosity for $3\sigma$ and $5\sigma$ signals. $M_{R_{1,2}}$ are the physical masses for the vector resonances in GeV.}
\begin{center}
\begin{tabular}{c|c|c|c|c|c|c}
$\tilde{g}$&$M_{A}$& $M_{R_{1,2}}$&$\sigma_{S}$ (fb) & $\sigma_{B}$ (fb) & $\mathscr{L}({\text{fb}}^{-1})$ for $3\sigma$ & $\mathscr{L}({\text{fb}}^{-1})$ for $5\sigma$ \\
\hline
\hline
  2	&  500 	&$M_{1}=517$	&	194					& 3.43 				&  0.012	& 0.038 \\
  2	& 500  	&$M_{2}=623$	&	118					& 1.34 				&  0.019	 & 0.056\\
  2	& 1000	&$M_{1}=1027$&	4.57					& $9.17 \cdot 10^{-2}$ 	&  0.53	 & 1.8 \\
  2	& 1000	&$M_{2}=1083$&	16.4					& $5.60 \cdot 10^{-2}$ 	&  0.13	 &  0.39\\
  2	& 1500	&$M_{1}=1526$&	0.133				& $5.91 \cdot 10^{-3}$ 	&  26	 &  67 \\
  2	& 1500	&$M_{2}=2103	$&	0.776				& $2.81 \cdot 10^{-3}$ 	& 2.7	 &  8.2 \\
\hline
 3   	& 500	&$M_{1}=507$	&      93.5	 				& 3.71 				&  0.037	& 0.090 \\
 3	& 500	&$M_{2}=715$	&	0.447 				& 0.649 				&  39 	& 81 \\
 3	& 1000	&$M_{1}=1013$&	1.32 					& $8.81 \cdot 10^{-2}$ 	&  2.7	& 7.4 \\
 3	& 1000	&$M_{2}=1097$&	2.94 					& $5.15 \cdot 10^{-2}$ 	&  0.79 	& 2.5 \\
 3	& 1500	&$M_{1}=1514$&	$3.19 \cdot 10^{-3}$ 	& $5.63 \cdot 10^{-3}$ 	&  6300	& 14000  \\
 3	& 1500	&$M_{2}=1541$&	0.120 				& $3.94 \cdot 10^{-3}$ 	&  29 	& 68 \\
\hline
 4	& 500	&$M_{1}=504$	&	34.6	 				& 3.85 				& 0.12	& 0.34 \\
 4	& 500	&$M_{2}=836$	&	0.0 					& 0.649 				& -		& - \\
 4	& 1000	&$M_{1}=1007$&	0.234 				& $8.98 \cdot 10^{-2}$ 	& 30		& 85 \\
 4	& 1000	&$M_{2}=1148$&	0.0 					& $5.15 \cdot 10^{-2}$ 	& -		& - \\
 4	& 1500	&$M_{1}=1509$&	$1.31 \cdot 10^{-3}$		& $3.94 \cdot 10^{-3}$ 	& 25000	& 57000 \\
 4	& 1500	&$M_{2}=1533$&	$1.43 \cdot 10^{-2}$		& $3.94 \cdot 10^{-3}$ 	& 435	& 1200 \\
\end{tabular}
\end{center}
\label{gt2x}
\end{table}%

\begin{equation}
|M_{\ell\ell}-M_{R}| < 5~ \text{GeV}
\end{equation}
separately for the $R_{1}$ and $R_{2}$ peaks. The choice of the value 5 GeV is dictated by the dilepton invariant mass resolution \cite{Bayatian:2006zz}.  The invariant mass resolution drops when the mass of the resonance increases, in any event, we use the same cut for all of the different mass values coming from the  one with worst resolution. Also estimates for the required integrated luminosity for the $3\sigma$ and  $5\sigma$ discoveries are given in the Table. The significance is defined as the number of signal events divided by the square root of the number of background events, when the number of events is large. The Poisson distribution is used for the small event samples.  The dilepton final state should be clearly visible at the LHC in this particular region of the parameter space already with 1 $\text{fb}^{-1}$ integrated luminosity.

\subsubsection{$pp \to R \to WZ \to \ell\ell\ell\nu$} 

  \begin{figure}[htbp]
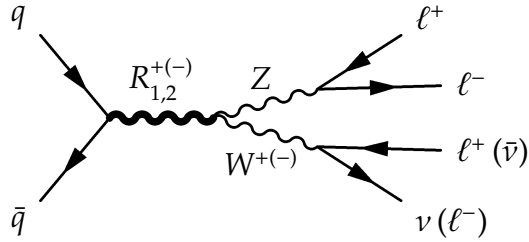

\begin{center}
\parbox{60mm}{\begin{fmfchar*}(60, 22)
\fmflabel{$q$}{q2}
\fmflabel{$\bar{q}$}{q1}
\fmflabel{$\ell^+$}{lp1}
\fmflabel{$\ell^-$}{lm1}
\fmflabel{$\ell^+ \, (\bar \nu)$}{lp2}
\fmflabel{$\nu \, (\ell^-)$}{lm2}
\fmfleft{q1,q2}
\fmfright{lm2,lp2,lm1,lp1}
\fmf{fermion, tension=3}{q2,v1,q1}
\fmf{photon, tension=4, label=$R_{1,, 2}^{+(-)}$, width=1mm}{v1,v2}
\fmf{photon, tension=2, label=$Z$, label.side=right}{v3,v2}
\fmf{photon, tension=2, label=$W^{+ (-)}$, label.side=right}{v2,v4}
\fmf{fermion}{lp1,v3,lm1}
\fmf{fermion}{lp2,v4,lm2}
\end{fmfchar*}}
\vspace{7mm}
\caption{Feynman diagram for the process $pp \to R^{\pm} \to WZ^{\pm} \to \ell\ell\ell\nu$. }
\label{ppwz}
\end{center}
\end{figure}

\begin{figure}[htbp]
\begin{center}
\includegraphics[width=0.5\textwidth]{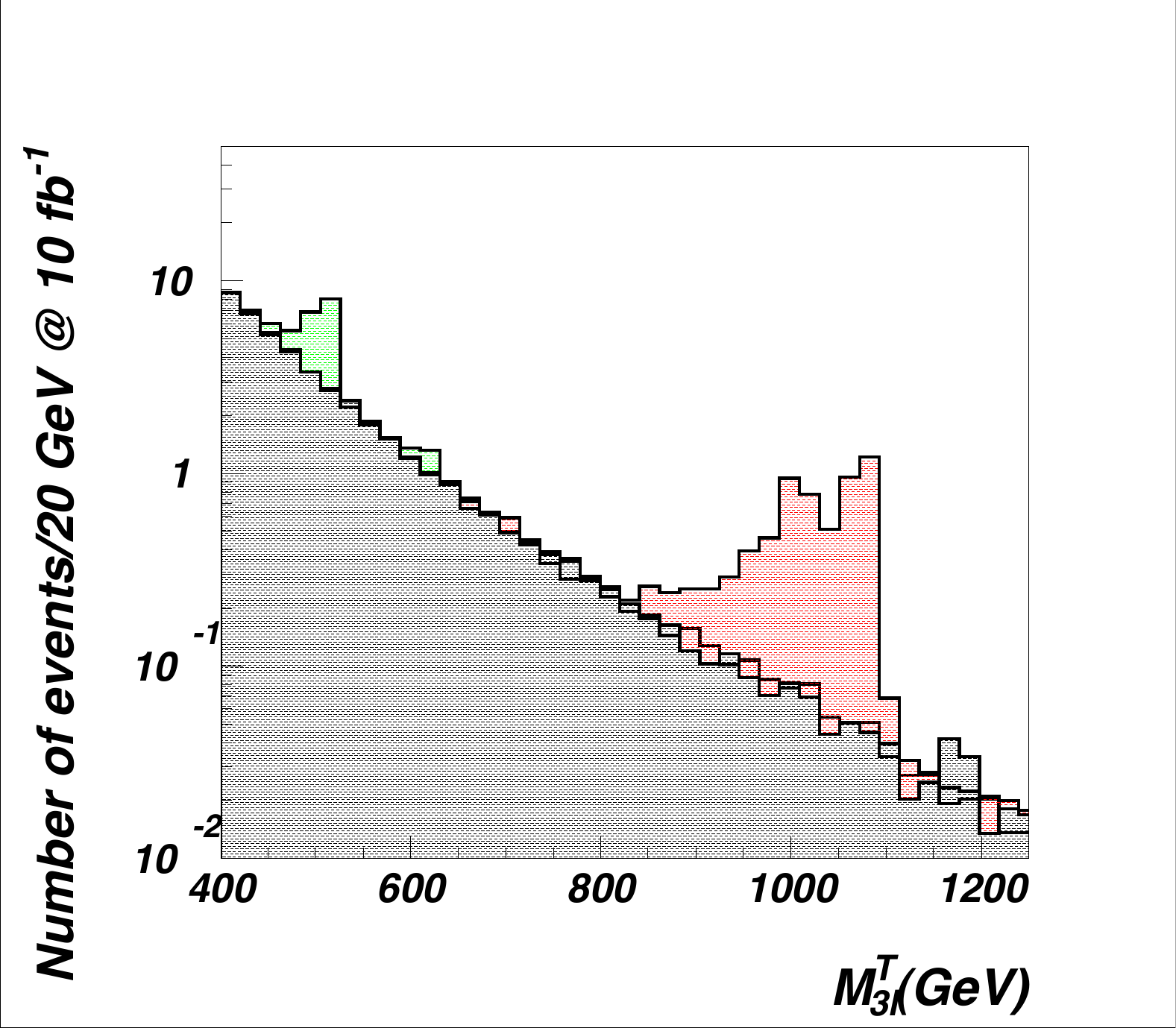}~\includegraphics[width=0.5\textwidth]{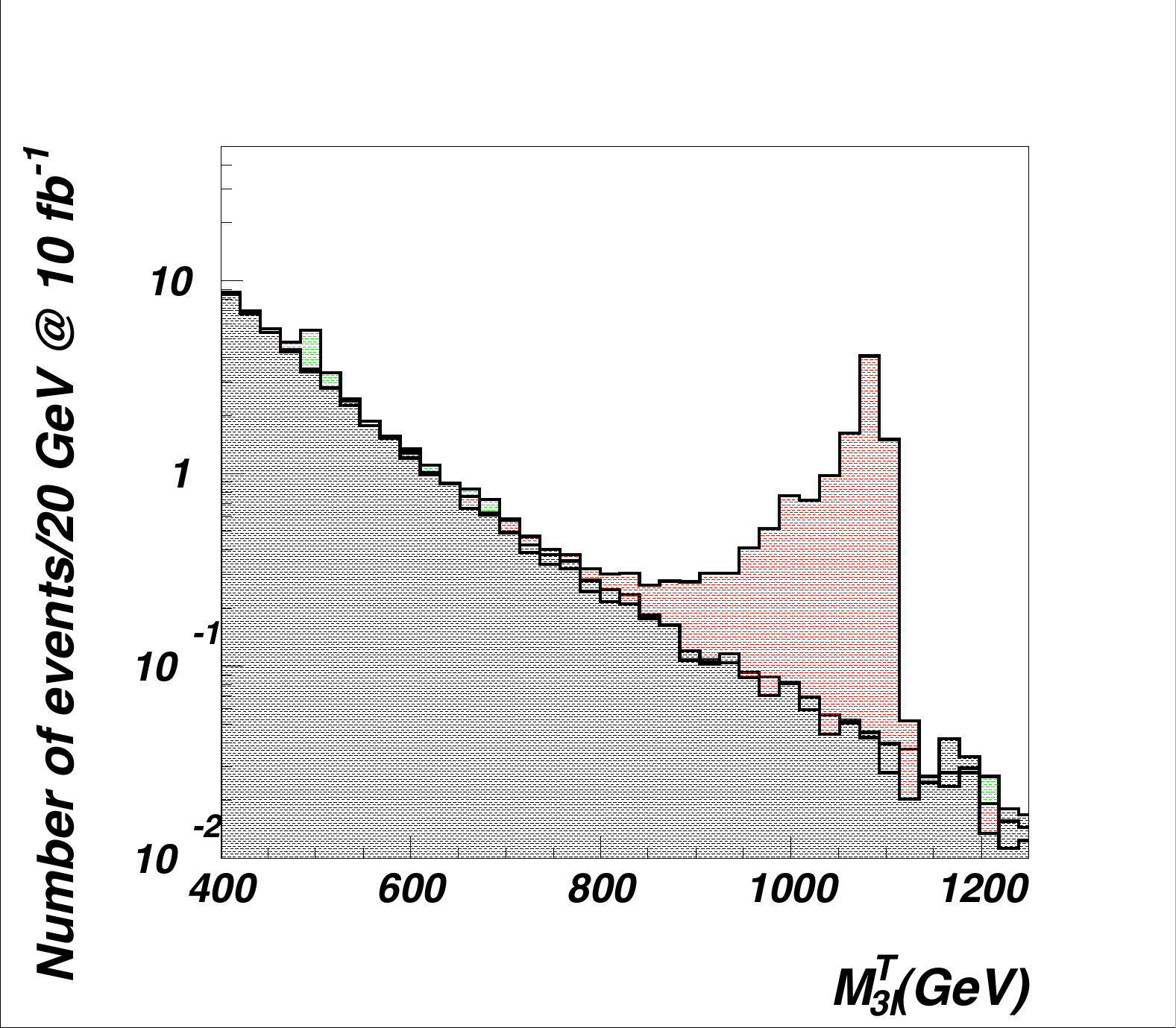} \\~\\
\includegraphics[width=0.5\textwidth]{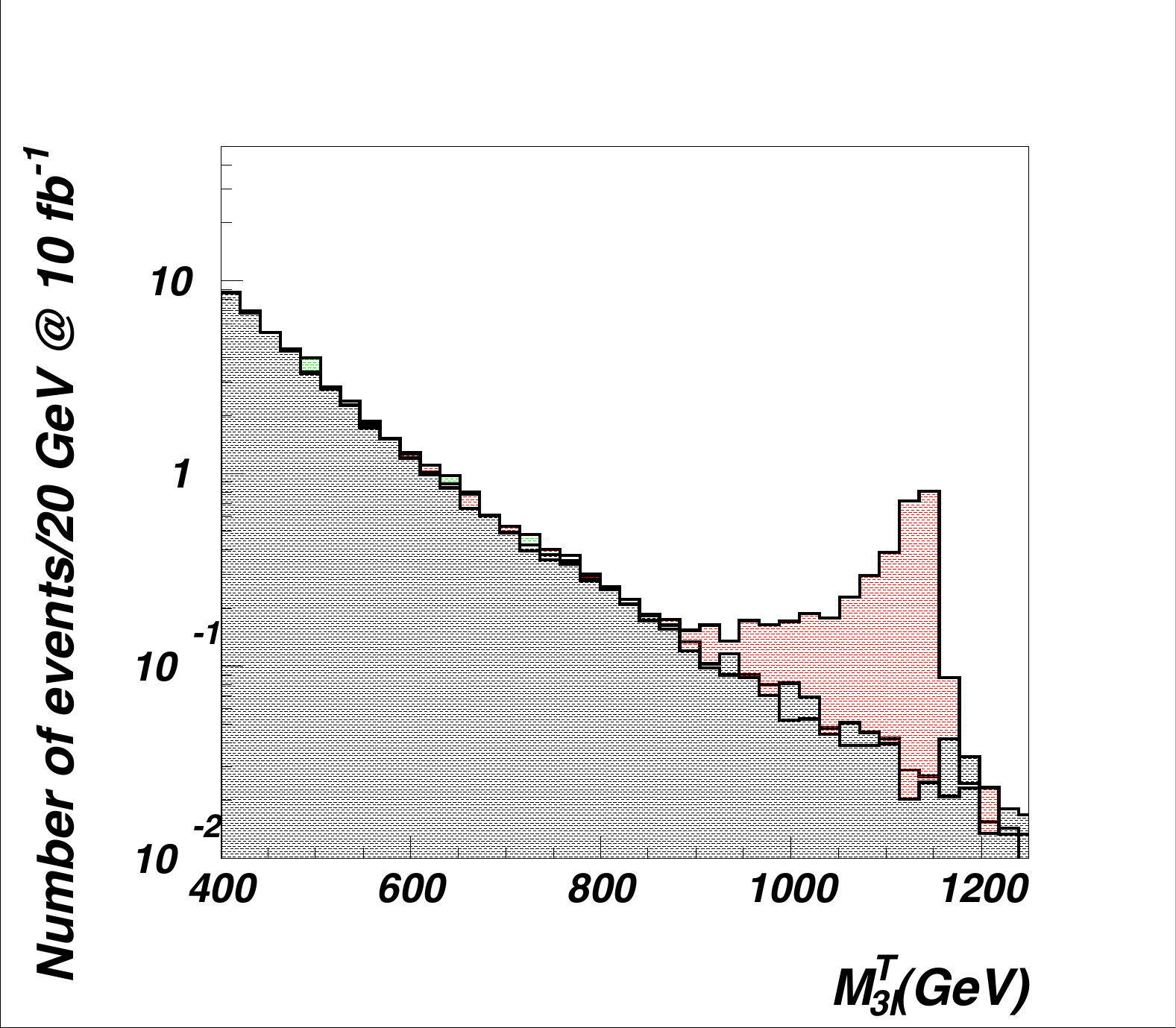}
\caption{Transverse mass distribution $M_{3\ell}^{T}$ for $pp \to R_{1,2}^{\pm} \to ZW^{\pm} \to \ell\ell\ell\nu$ signal and background processes. We consider $\tilde{g}=2,3,4$ respectively from left to right and masses $M_{A}=0.5$ TeV (green), $M_{A}=1$ TeV (red).}
\label{wzgt24}
\end{center}
\end{figure}

The final state signature with three leptons and missing energy arises from the process $pp \to R \to WZ \to \ell\ell\ell\nu$ (see Fig.~\ref{ppwz}), where $\ell$ denotes a muon or an electron and $\nu$ denotes the corresponding neutrino. This was also studied in \cite{Belyaev:2008yj}, with $\sqrt{s}=14$ TeV and 100 fb$^{-1}$, where it was shown to be a promising signature for higher values of $\tilde{g}$ and $M_{A}$. The technivector-fermion couplings are suppressed for large $\tilde{g}$, which makes the dilepton final state  uninteresting in that region of the parameter space. In contrast, the technivector coupling to SM vector bosons is enhanced for large values $\tilde{g}$, balancing the suppression coming from the quark couplings. This can be seen from Fig.~\ref{wzgt24}, where the second peak begins to go down slowly with increasing $\tilde{g}$. Following \cite{Belyaev:2008yj}, we have used the transverse mass variable 

\begin{equation}
(M^{T}_{3\ell})^{2}=[\sqrt{M^{2}(\ell\ell\ell)+p^{2}_{T}(\ell\ell\ell)}+|\slashed{p}_{T}|]^{2}-|\vec{p}_{T}(\ell\ell\ell)+\vec{\slashed{p}}_{T}|^{2},
\end{equation}
where $\slashed{p}_{T}$ denotes the missing transverse momentum. The cuts for the leptons are applied as in the previous subsection and in addition we impose a cut on the missing transverse energy $\slashed{E}_{T} > 15$ GeV. As a background we consider the SM processes with $R_{1,2}^{\pm}$ replaced by the $W^{\pm}$.

In Table~\ref{wzgt4x} we present the signal and background cross sections and quantify the possible observability of the signal. 
 
  \begin{table}[htdp]
\caption{$pp \to R_{1,2}^{\pm} \to ZW^{\pm} \to \ell\ell\ell\nu$. Signal and background cross sections for $\tilde{g}=2,3,4$ and estimates for required luminosity for $3\sigma$ and $5\sigma$ signals. Cuts are applied to take into account only the region of the visible peak. $M_{R_{1,2}}$ are the physical masses (GeV).}
\begin{center}
\begin{tabular}{c|c|c|c|c|c|c|c}
$\tilde{g}$&$M_{A}$&$M_{R_{1}}$ &$M_{R_{2}}$ &$\sigma_{B}$ (fb) & $\sigma_{S}$ (fb) & $\mathscr{L}(\text{fb}^{-1})$ for $3\sigma$ & $\mathscr{L}(\text{fb}^{-1})$ for $5\sigma$ \\
\hline
\hline
2 & 500 	&512	&619	&	1.75	 				& 1.12 				& 16		& 35\\
2 & 1000	&1012	&1081	&	$7.34 \cdot 10^{-2}$ 	& 0.545 				& 8.4		& 24\\
3 & 500	&505	&713	&	1.12	 				& 0.340 				& 89		& 243\\
3 & 1000	&1009	&1094	&	0.117 				& 1.10 				& 3.7		& 11\\
4 & 500	&503	&835	&	0.659 				& $2.15 \cdot 10^{-2}$ 	& 10000	& - \\
4 & 1000	&1005	&1145	&	$8.92 \cdot 10^{-2}$ 	& 0.297			 	& 21	 	& 60 \\
\end{tabular}
\end{center}
\label{wzgt4x}
\end{table}%
 
 \subsubsection{$pp\to R\to \ell\nu$}
 
   \begin{figure}[htbp]
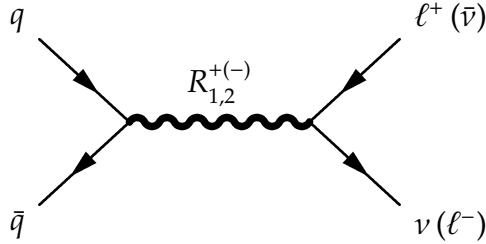

\begin{center}
\parbox{60mm}{\begin{fmfchar*}(60, 22)
\fmflabel{$q$}{qi}
\fmflabel{$\bar q$}{qibar}
\fmflabel{$\nu \, (\ell^-)$}{lf}
\fmflabel{$\ell^+ \, (\bar \nu)$}{lfbar}
\fmfleft{qibar,qi}
\fmfright{lf,lfbar}
\fmf{fermion}{qi,vi,qibar}
\fmf{fermion}{lfbar,vf,lf}
\fmf{photon, label=$R_{1,, 2}^{+ (-)}$, width=1mm, label.side=left}{vi,vf}
\end{fmfchar*}}
\vspace{7mm}
\caption{Feynman diagram for the process $pp\to R^{\pm} \to \ell\nu$.}
\label{pplnu}
\end{center}
\end{figure}
 
\begin{figure}[htbp]
\begin{center}
\includegraphics[width=0.48\textwidth]{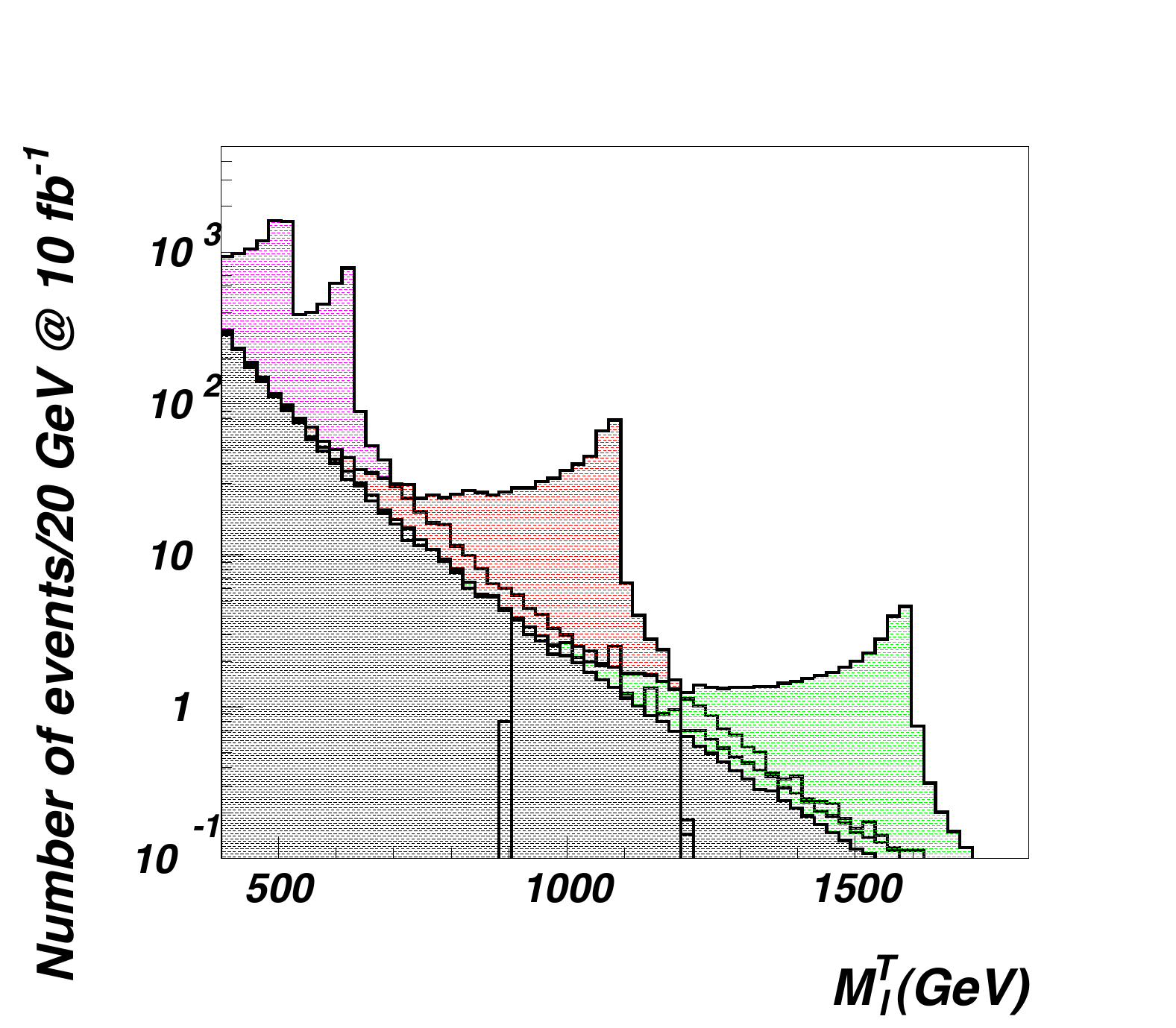} \includegraphics[width=0.48\textwidth]{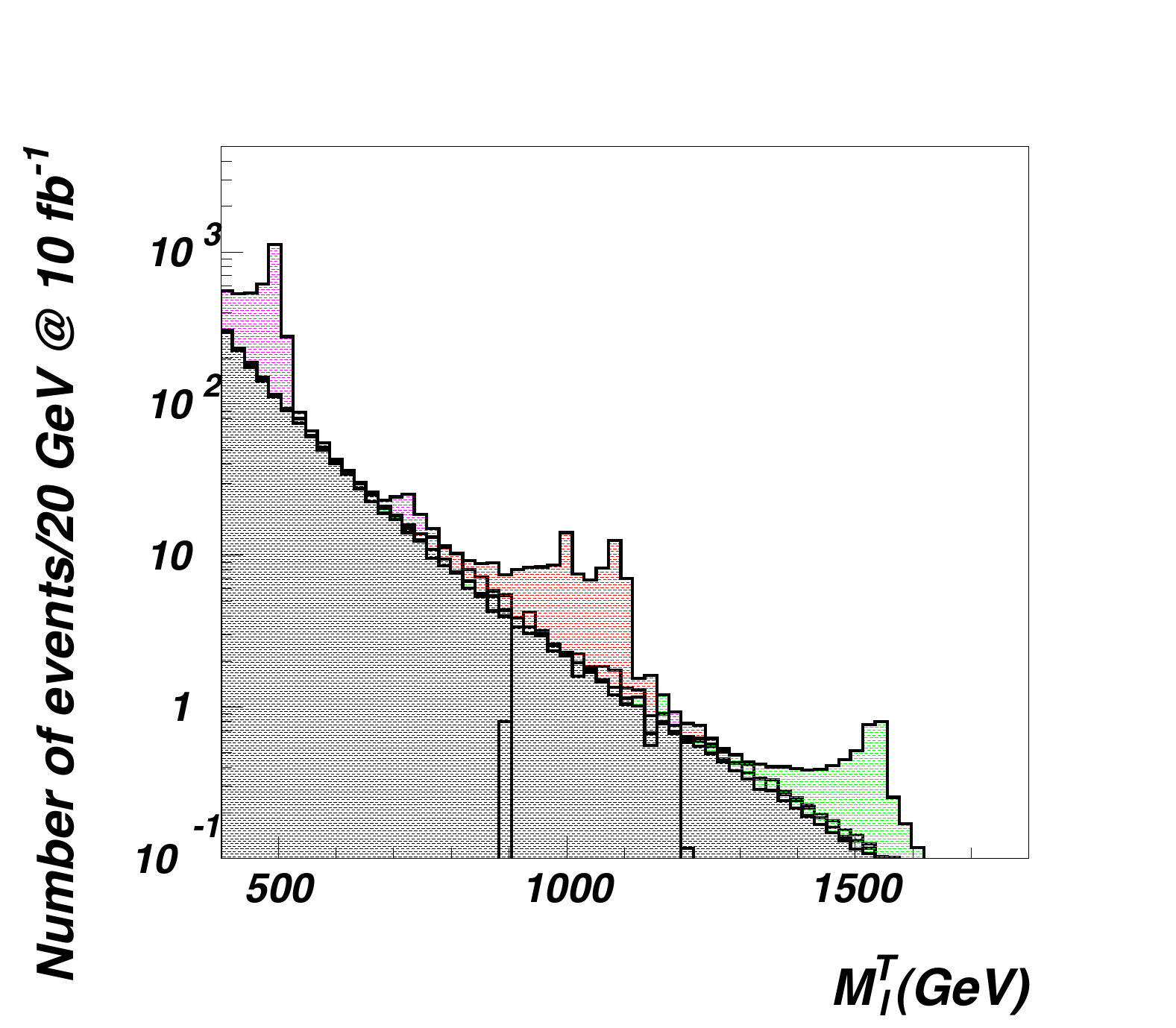} \\
\includegraphics[width=0.48\textwidth]{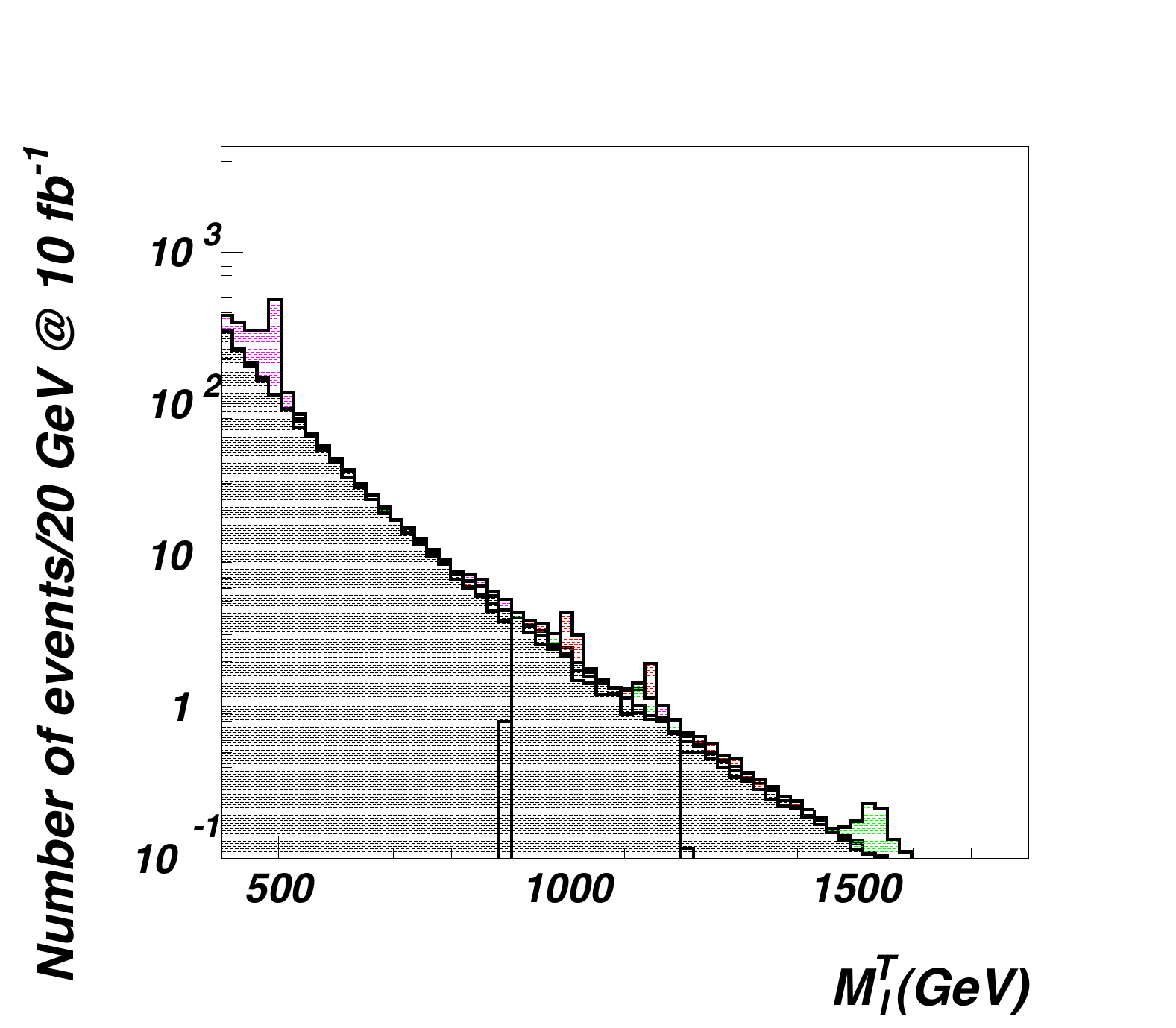}
\caption{Transverse mass distribution $M_{\ell}^{T}$ for $pp \to R_{1,2}^{\pm} \to \ell\nu$ signal and background processes.  We consider $\tilde{g}=2,3,4$ respectively from left to right and masses $M_{A}=0.5$ TeV (purple), $M_{A}=1$ TeV (red) and $M_{A}=1.5$ TeV (green).}
\label{lnugt24}
\end{center}
\end{figure}

 This is the third signature, with leptons in the final state, studied in \cite{Belyaev:2008yj}.  The essential complication compared to the  $\ell^{+} \ell^{-}$ final state is that the longitudinal component of the neutrino momentum cannot be measured. The cuts applied for the charged lepton and on the missing energy are  $|\eta^{\ell}| < 2.5$, $p^{\ell}_{T} > 15$ GeV and $\slashed{E}_{T} > 15$ GeV. In Fig.~\ref{lnugt24}, the number of events is plotted with respect to the transverse mass variable
 \begin{equation}
M^{T}_{\ell}=\sqrt{2 \, \slashed{E}_{T} \, p_{T}(\ell) \, (1-\cos \Delta\phi_{\ell,\nu})} \ ,
\end{equation} 
using the parameter space points  $\tilde{g}=2,3,4$ and $M_{A}=0.5, 1, 1.5$ TeV. This final state signature behaves like the dilepton final state. Due to the suppression of the fermion couplings, the signal is reduced when $\tilde{g}$ is increased.

\subsubsection{$pp\to R \rightarrow jj$}

\begin{figure}[htbp]
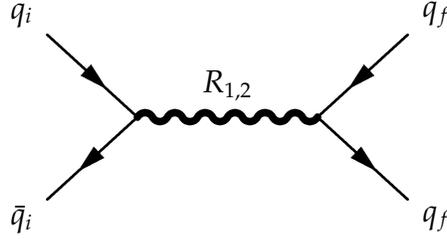

\begin{center}
\parbox{60mm}{\begin{fmfchar*}(60, 22)
\fmflabel{$q_i$}{qi}
\fmflabel{$\bar q_i$}{qibar}
\fmflabel{$q_f$}{qf}
\fmflabel{$\bar q_f$}{qfbar}
\fmfleft{qibar,qi}
\fmfright{qf,qfbar}
\fmf{fermion}{qi,vi,qibar}
\fmf{fermion}{qfbar,vf,qf}
\fmf{photon, label=$R_{1,, 2}$, width=1mm, label.side=left}{vi,vf}
\end{fmfchar*}}
\vspace{7mm}
\caption{Feynman diagram for the Drell-Yan process $pp \rightarrow R \rightarrow jj$.}
\label{Fig:pp>jj diagram}
\end{center}
\end{figure}

The tree-level MWT contributions to this process is the s-channel exchange of a composite vector boson (Drell-Yan process, Fig.~\ref{Fig:pp>jj diagram}).

The quark couplings to the heavy spin one states are generated by the mixing with the SM gauge bosons. Their value can then be expanded perturbatively in powers of $g / \tilde g$, where $g$ is the $SU(2)_\text{L}$ coupling constant, for large $\tilde{g}$. Therefore the cross section for this process is reduced by, at least, a factor of $\tilde{g}^{- 4}$, respect to the SM one, and the signal drowns in the large QCD background for this process.

\subsubsection{$pp\to R\to \gamma V$ and $pp \to R \to ZZ$ }

Up to dimension four operators, and before including operators involving the Lorentz $\epsilon^{\mu\nu\rho\sigma}$ invariant tensor \cite{Sannino:2009za,Duan:2000dy}, the photon coupling $\gamma R_{1}^{+} W^{-} $ is forbidden. This implies that we do not have diagrams with technivectors decaying into photon and vector boson. Higher order operators as well as operators involving  the $\epsilon^{\mu\nu\rho\sigma}$  in a coherent manner will be investigated in the near future. The $pp \to R \to ZZ$ processes are also absent in the present effective description.

\subsection{Ultimate LHC reach for heavy spin resonances}

We conclude this subsection summarizing the results for the ultimate reach of LHC, i.e. $14$~TeV and $100$~$\text{fb}^{-1}$ \cite{Belyaev:2008yj}.  We consider the  representative parameter space points
$\tilde{g}=2,5$ and $M_A=0.5, 1, 1.5, 2$ TeV for our plots and discussion.

\subsubsection{$pp \rightarrow R \rightarrow 2 \ell,  \ell \nu, 3\ell \nu$  at $14$~TeV}
\begin{figure}[tbhp]
\vskip -0.2cm 
\includegraphics[width=0.5\textwidth]{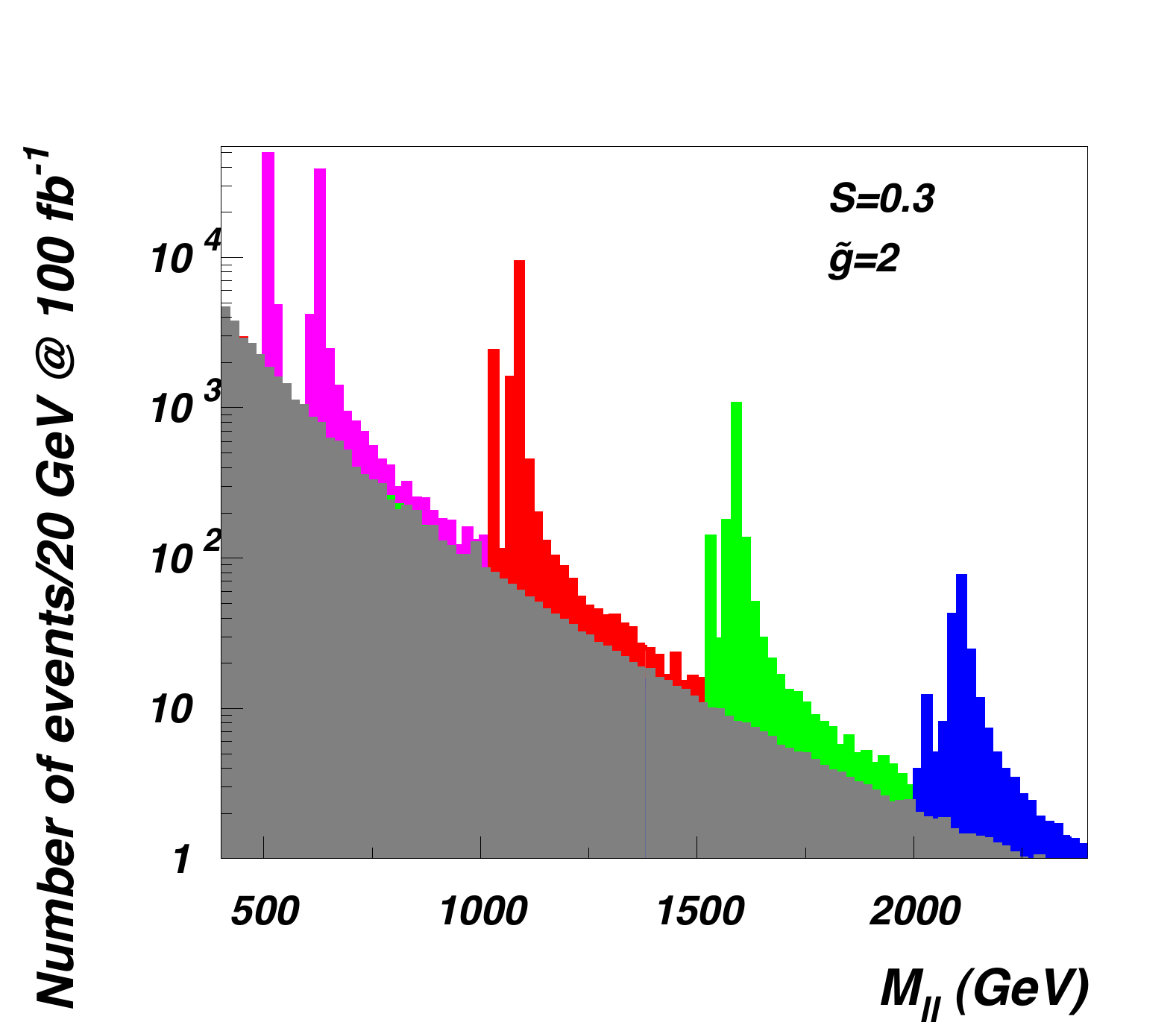}%
\includegraphics[width=0.5\textwidth]{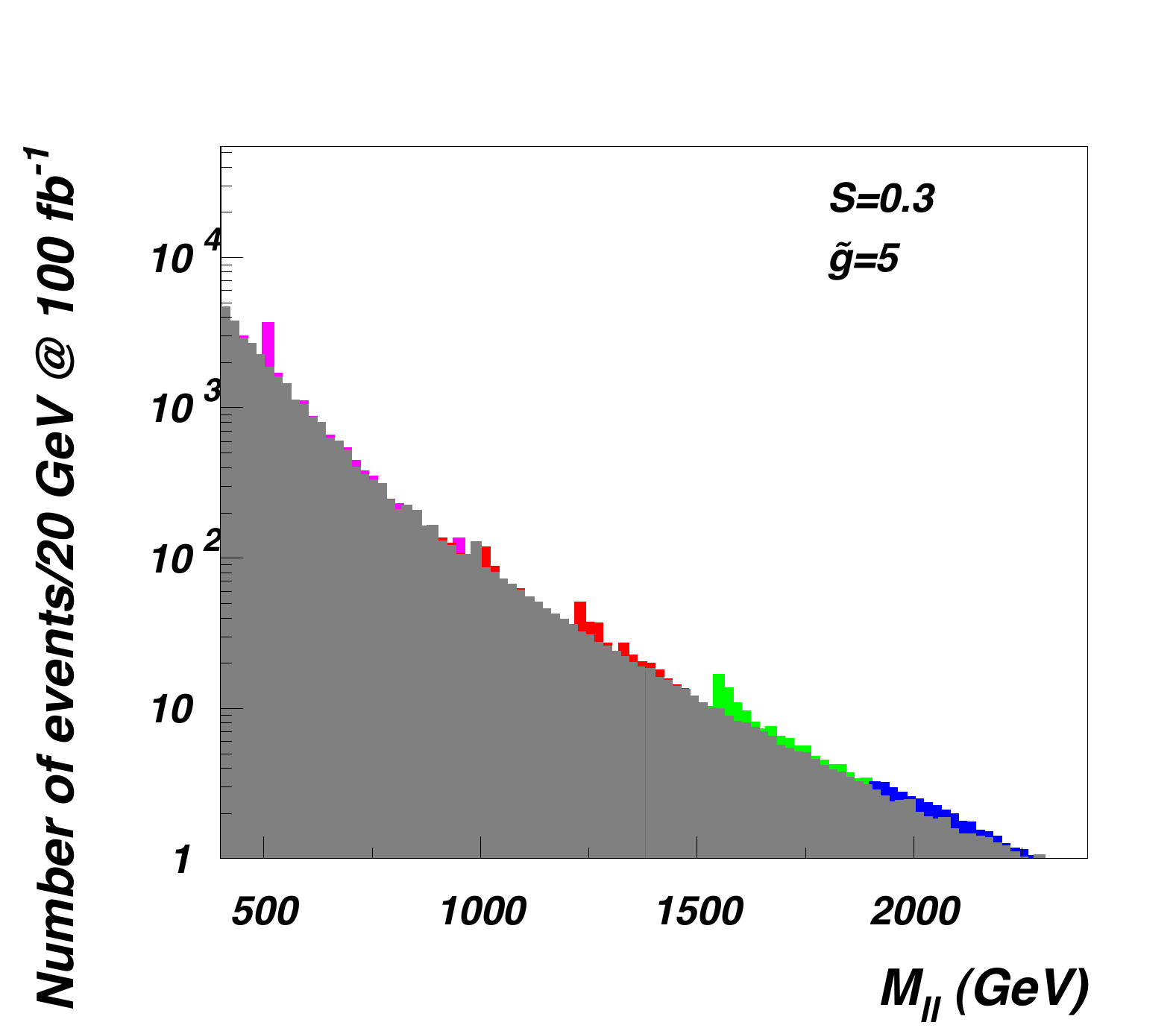}%
\caption{Dilepton invariant mass distribution
$M_{\ell \ell}$ for $pp\to R^{0}_{1,2}\to \ell^+\ell^-$
signal and background processes.
We consider $\tilde{g}=2,5$ respectively
and masses $M_A=0.5$ TeV (purple), $M_A=1$ TeV (red), 
$M_A=1.5$ TeV (green) and $M_A=2$ TeV
(blue). }
\label{fig:sig1}
\end{figure}

\begin{figure}[tbhp]
\vskip -0.2cm
\includegraphics[width=0.5\textwidth]{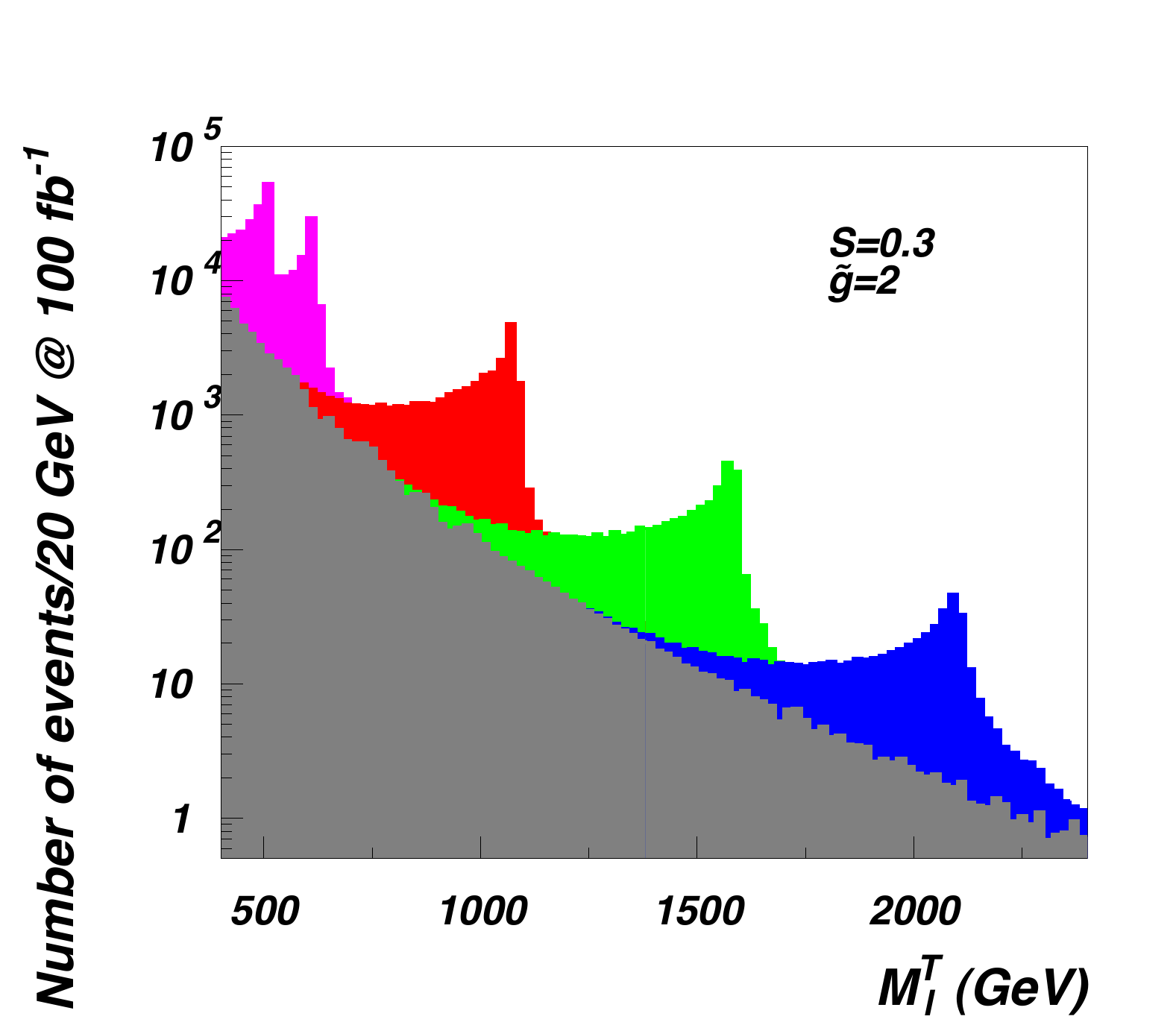}%
\includegraphics[width=0.5\textwidth]{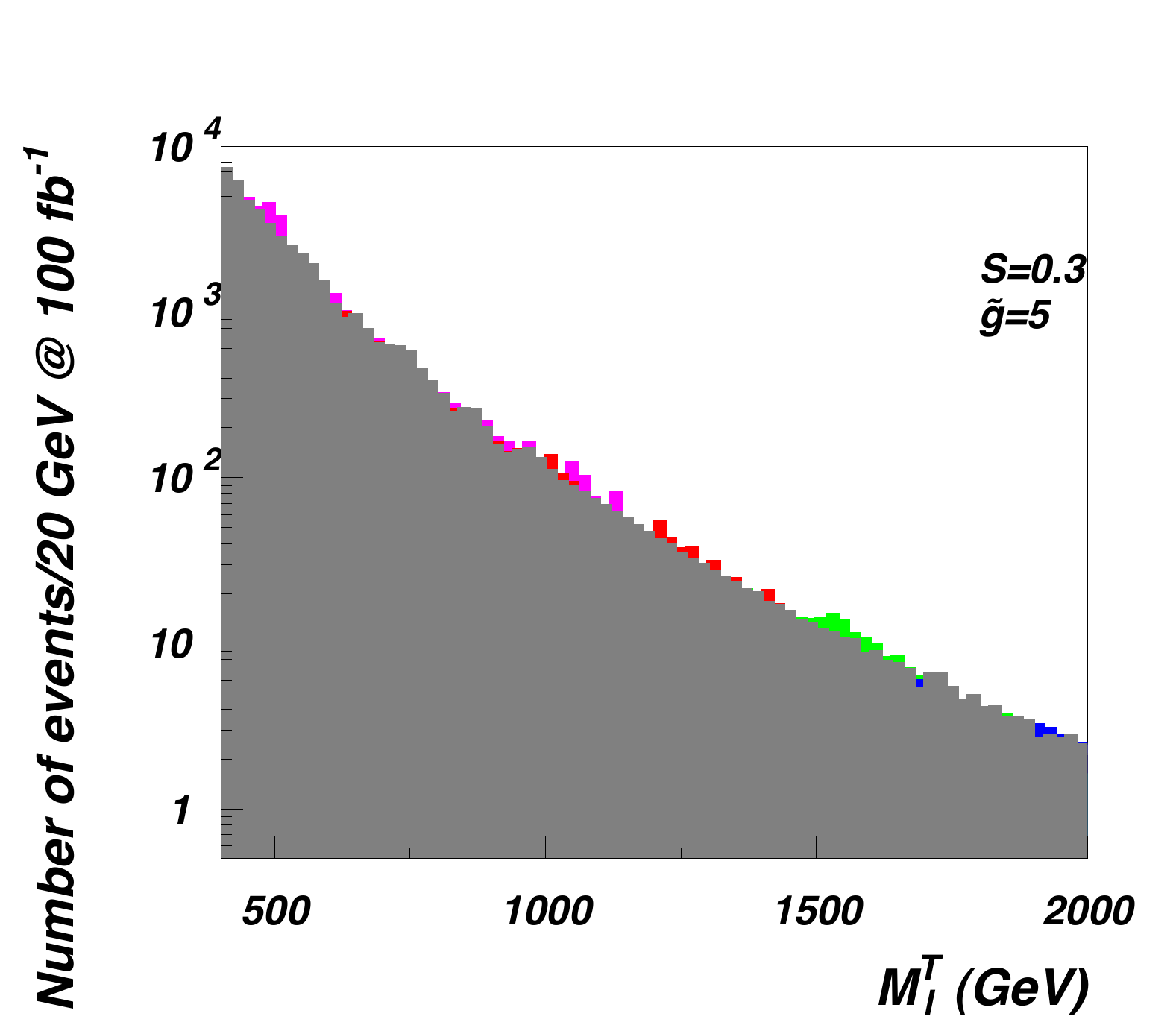}%
\caption{$M^T_\ell$ mass distribution
for $pp\to R^{\pm}_{1,2}\to \ell^\pm\nu$
signal and background processes.
We consider $\tilde{g}=2,5$ respectively
and masses $M_A=0.5$ TeV (purple), $M_A=1$ TeV (red), 
$M_A=1.5$ TeV (green) and $M_A=2$ TeV
(blue). 
\label{fig:sig2}}
\end{figure}

\begin{figure}[tbhp]
\vskip -0.2cm
\includegraphics[width=0.5\textwidth]{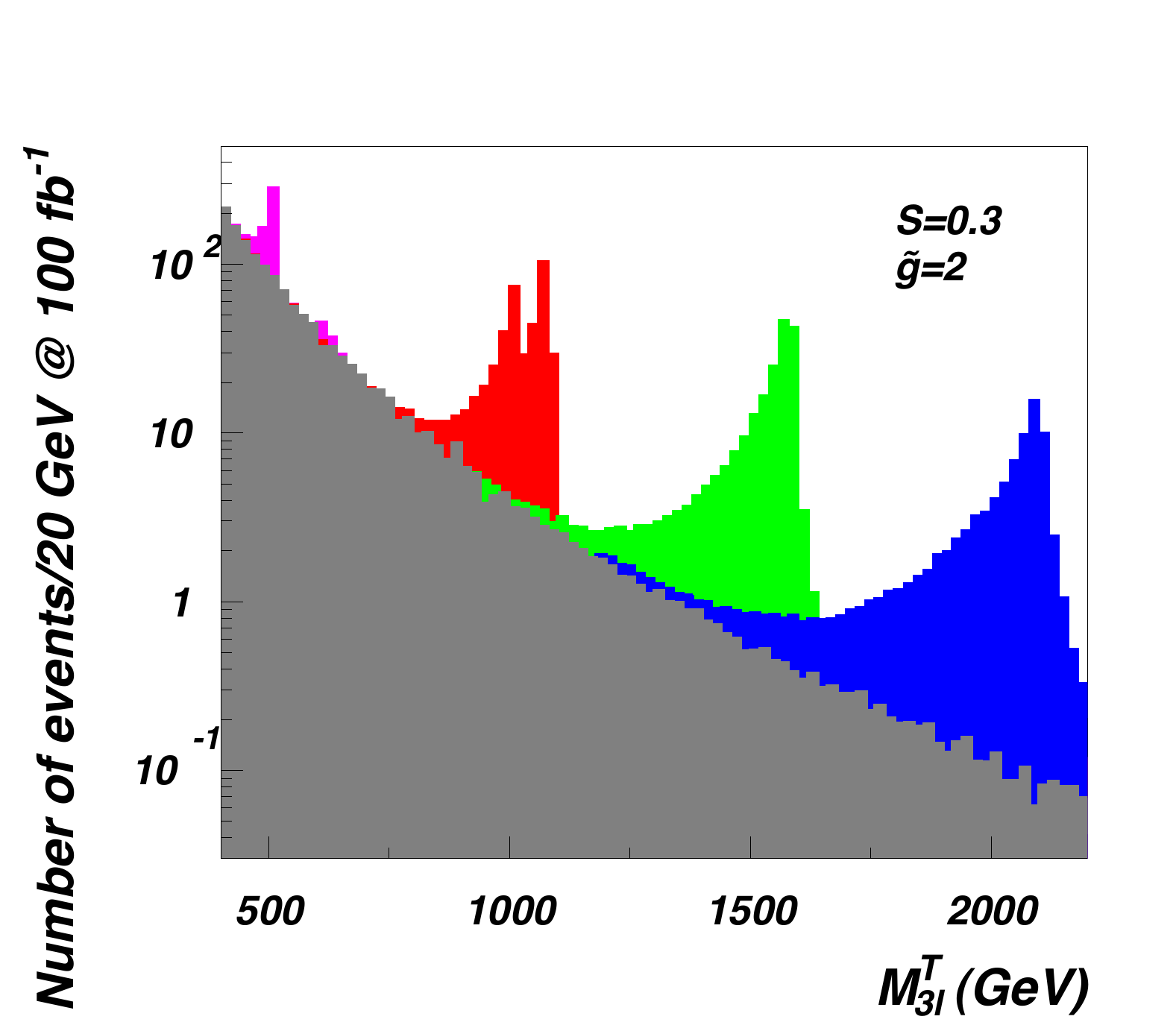}%
\includegraphics[width=0.5\textwidth]{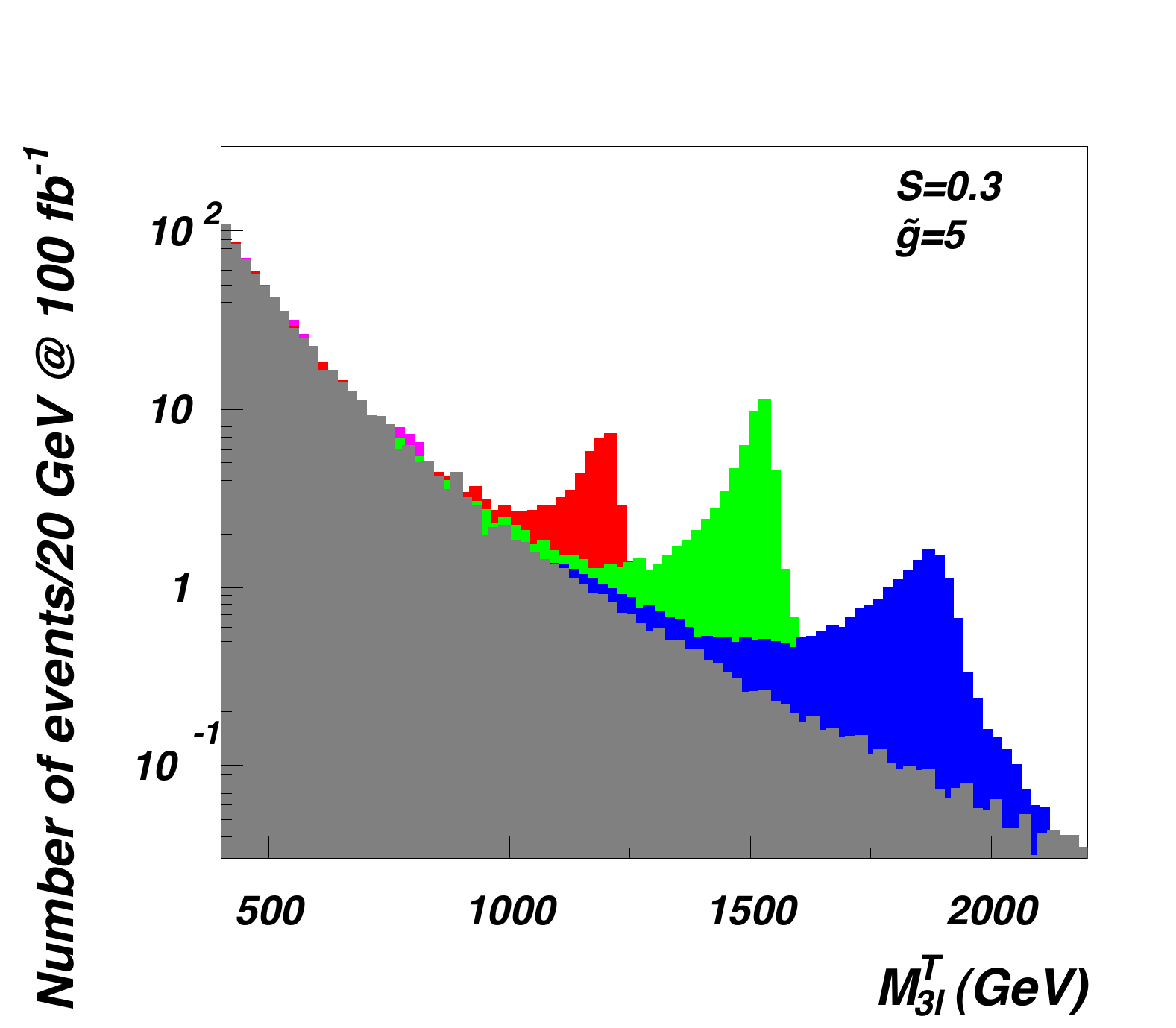}
\caption{$M^T_{3\ell}$ mass distribution
for $pp\to R^{\pm}_{1,2}\to ZW^\pm\to 3\ell\nu$
signal and background processes.
We consider $\tilde{g}=2,5$ respectively
and masses $M_A=0.5$ TeV (purple), $M_A=1$ TeV (red), 
$M_A=1.5$ TeV (green) and $M_A=2$ TeV
(blue). 
\label{fig:sig3}}
\end{figure}

The  invariant mass and transverse 
mass distributions for signatures $pp \rightarrow R \rightarrow 2 \ell,  \ell \nu, 3\ell \nu$ are shown in 
Figs.~\ref{fig:sig1}-\ref{fig:sig3}.
In the left frames of Figs.~\ref{fig:sig1} and ~\ref{fig:sig2},
corresponding to $\tilde g=2$, clear signals from the leptonic decays of 
$R^{0}_{1,2}$ and  $R^{\pm}_{1,2}$ are seen even for 2
TeV resonances. Moreover Fig.~\ref{fig:sig1} demonstrates that
for $\tilde g=2$  both peaks
from  $R^{0}_{1}$ and $R^{0}_{2}$ may be resolved. 
 
Let us now turn to the case of $\tilde g=5$ in the right
frames of Figs.~\ref{fig:sig1} and \ref{fig:sig2}. For large $\tilde g$ the $Rff$ couplings are suppressed, so observing signatures  $pp \rightarrow R \rightarrow 2 \ell$ and  $pp \rightarrow R \rightarrow \ell \nu$ could be problematic. However, for large
$\tilde g$, the triple-vector coupling 
is enhanced,  so one
can observe a clear signal in the $M^{T}_{3\ell}$ distribution presented in Fig.~\ref{fig:sig3}. At low masses 
the decays of the heavy vector mesons to SM gauge bosons are suppressed  and the signal disappears.  This mass range can, however, be covered with signatures   $pp \rightarrow R \rightarrow 2 \ell$ and  $pp \rightarrow R \rightarrow 3\ell \nu$.

\subsubsection{$pp\to Rjj$ at $14$~TeV}

  \begin{figure}[htbp]
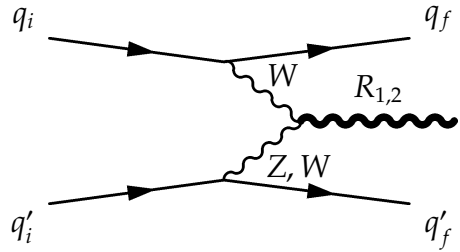

\begin{center}
\parbox{60mm}{\begin{fmfchar*}(60, 22)
\fmflabel{$q_i$}{qi}
\fmflabel{$q_i'$}{qip}
\fmflabel{$q_f$}{qf}
\fmflabel{$q_f'$}{qfp}
\fmfright{qfp,vv,qf}
\fmfleft{qip,qi}
\fmf{fermion, tension=1.5}{qi,v1}
\fmf{fermion}{v1,qf}
\fmf{fermion, tension=1.5}{qip,v2}
\fmf{fermion}{v2,qfp}
\fmf{photon, label=$W$, label.side=left, label.dist=0.8mm}{v1,v}
\fmf{photon, label=$Z,, W$, label.side=right, label.dist=0.8mm}{v2,v}
\fmf{photon, label=$R_{1,, 2}$, width=1mm, label.side=left}{v,vv}
\end{fmfchar*}}
\vspace{7mm}
\caption{Feynman diagram for the vector boson fusion process $pp \rightarrow R jj$.}
\label{Fig:pp>Rjj diagram}
\end{center}
\end{figure}
Vector Boson Fusion (VBF) is, in principle, an interesting channel for the vector meson production (see Fig.~\ref{Fig:pp>Rjj diagram}), especially in theories in which the fermion couplings of the vector resonances are suppressed. Unfortunately, as we shall see, it is not an appealing channel to explore at the LHC due to a small production rate of $R_{1,2}^{\pm,0}$.  Although the coupling between three vector particles increases with $\tilde{g}$ it is hard to produce $R_{1,2}^{\pm,0}$ on-shell. 

Following \cite{Belyaev:2008yj}, to be specific, we consider VBF production of the charged $R_{1}$ and  $R_2$  vectors  impose the following kinematical cuts on 
{ the jet transverse
momentum $p^j_T$, energy $E^j$, and rapidity gap $\Delta \eta^{jj}$, 
as well as rapidity acceptance}
$|\eta^j|$~\cite{He:2007ge,Birkedal:2005yg}:
\begin{eqnarray}
|\eta^j| < 4.5\ , \quad  p^j_T > 30 \mbox{ GeV} \ , \quad   E^j > 300 \mbox{ GeV} \ , \quad  \Delta \eta^{jj} > 4 \ .
\label{eq:vbfcuts}
\end{eqnarray}
\begin{figure}[tbhp]
\begin{center}
\vskip -0.2cm
\includegraphics[width=0.43\textwidth]{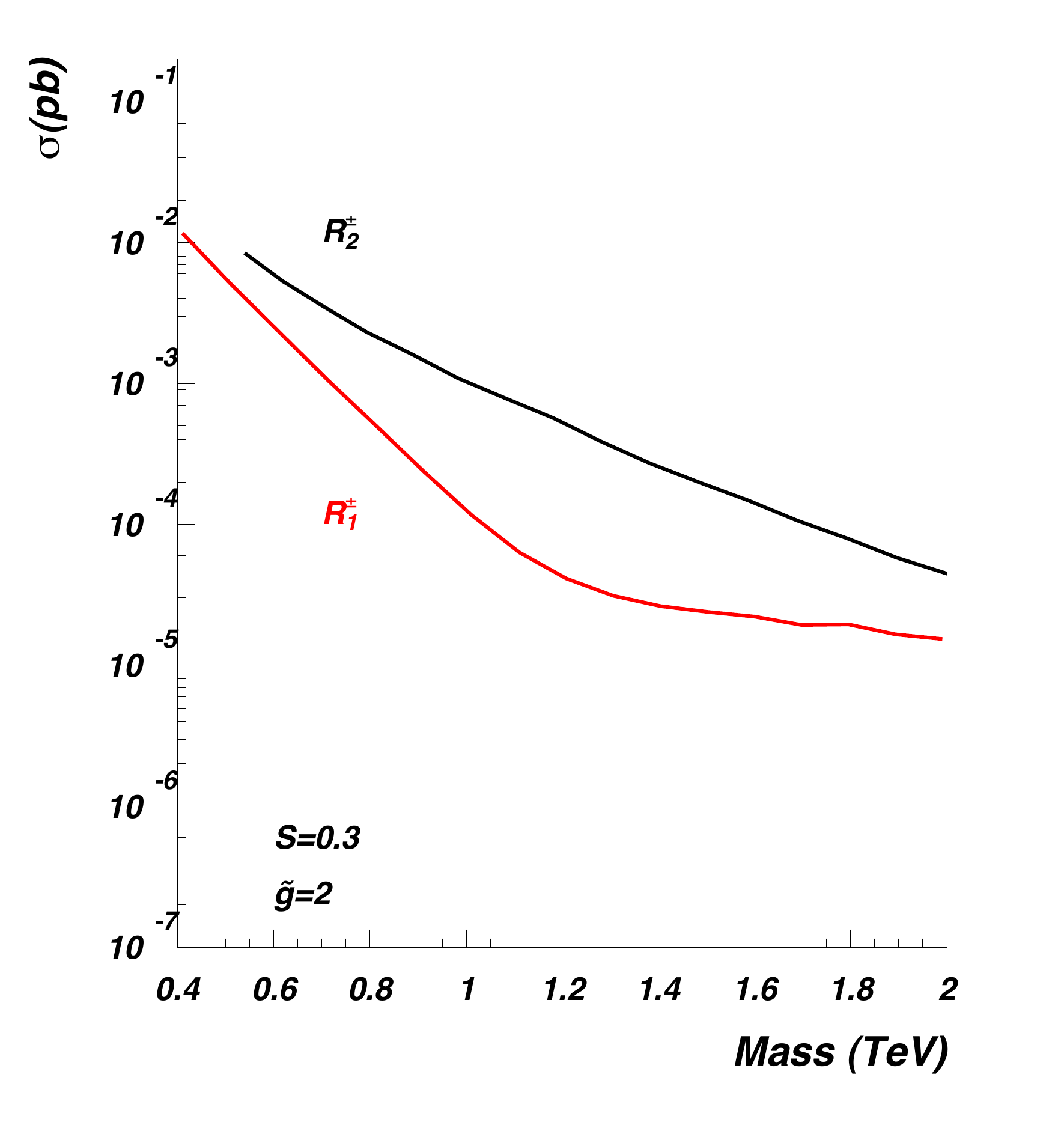}%
\includegraphics[width=0.43\textwidth]{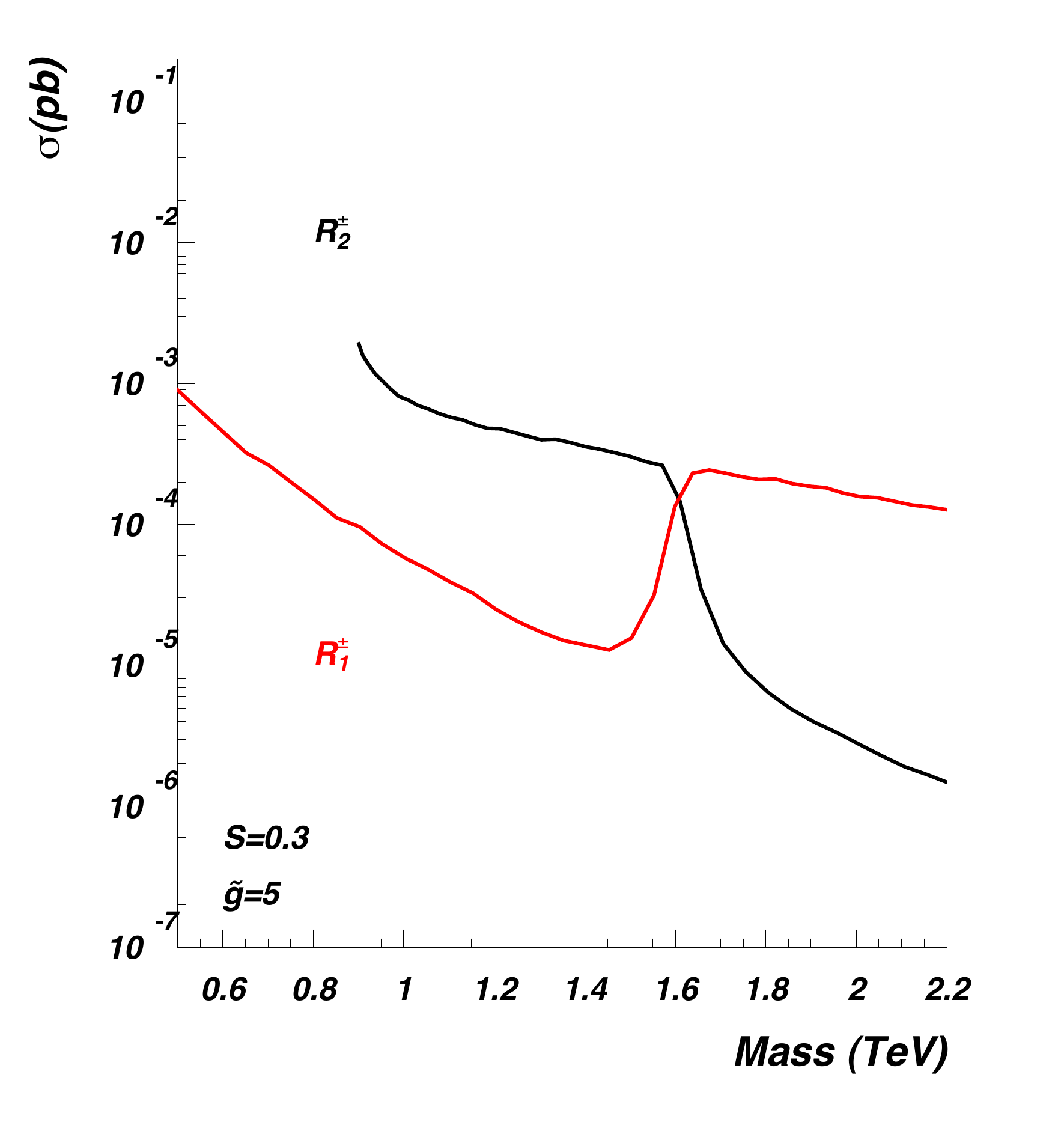}
\vskip -0.2cm
\end{center}
\caption{Vector boson fusion production cross sections for the $R_{1,2}^\pm$ resonances, with $S=0.3$ and $\tilde{g} = 2,5$. The jet cuts are: $|\eta^j| < 4.5$, $p^j_T > 30$ GeV, $E^j > 300$ GeV, $\Delta \eta^{jj} > 4$. See text for details.}
\label{fig:VBF}
\end{figure}
The VBF production cross section for the charged $R_1$ and $R_2$ vector
resonances is shown in Fig.~\ref{fig:VBF} for the set of cuts given
by Eq.~\eqref{eq:vbfcuts}. An interesting feature of the VBF
production is the observed crossover around the mass degeneracy point for $\widetilde{g}=5$. This is a direct consequence of the fact that the $R_{1,2}$ resonances switch their vector/axial nature at the inversion point.  For smal $\tilde{g}$ the crossover does not occur due to the interplay between the electroweak and the TC corrections. 
In D-BESS VBF processes are not very relevant, since there are no direct
interactions between the heavy mesons and the SM vectors. However in fermiophobic
Higgsless models VBF is the main production channel of the heavy resonances.  {Since the production rate of $R^\pm_{1,2}$
is below 1 fb, VBF is not a
promising channel at the LHC.
}

\subsection{Composite Higgs phenomenology}

\begin{figure}
\begin{center}
\includegraphics[width=0.5\textwidth]{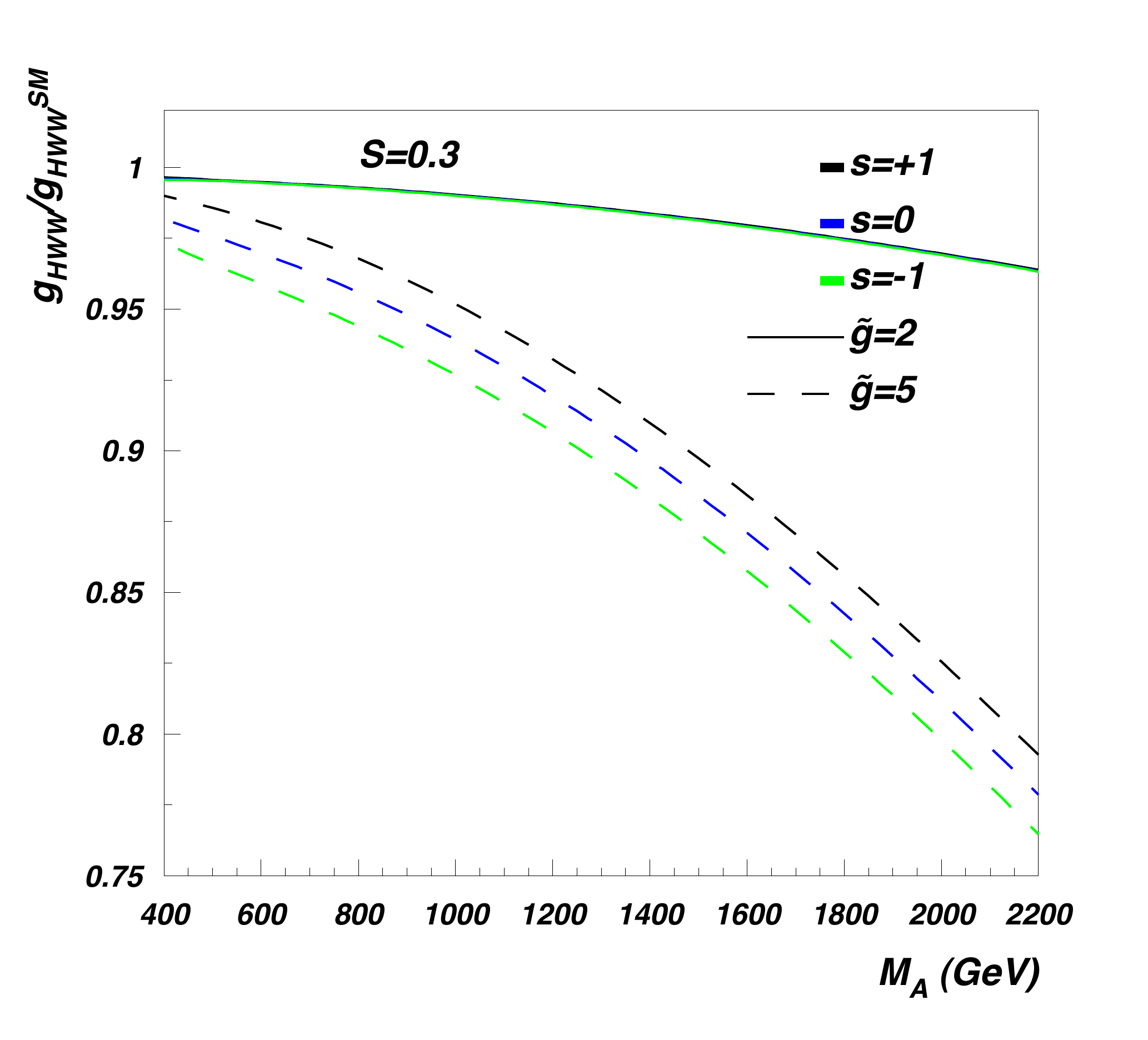}%
\includegraphics[width=0.55\textwidth]{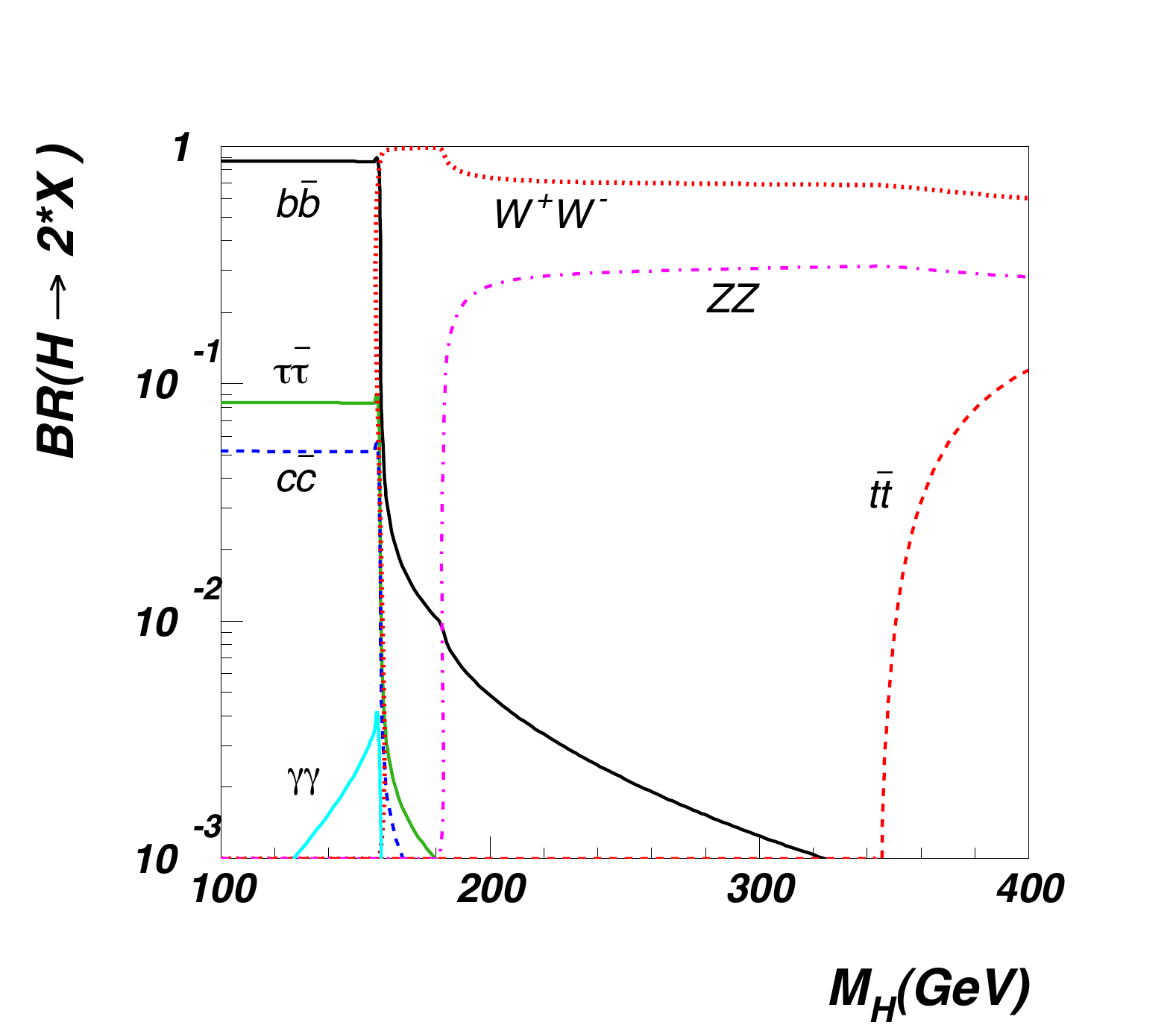}%
\caption{\label{fig:ghww} 
Left: 
$g_{HWW}/g_{HWW}^\text{SM}$
ratio as a function of $M_A$. The behavior of the $g_{HZZ}$ coupling
is identical to the $g_{HWW}$ one.
Results are presented for $S=0.3$,
$\tilde g = 2, 5$ (solid and dashed lines respectively),
 and for $s=(+1,0,-1)$ 
 (black, blue and green colors respectively). 
  Right: branching ratios of the composite Higgs  as function  $M_H$.}
\end{center}
\end{figure}

The composite Higgs phenomenology is interesting due to its interactions with the new massive vector bosons and their mixing with SM gauge bosons. We first analyze the Higgs coupling to the $W$ and $Z$
gauge bosons. In Fig.~\ref{fig:ghww}
we present the $g_{HWW}/g_{HWW}^\text{SM}$
ratio as a function of $M_A$. The behavior of the $g_{HZZ}$ and $g_{HWW}$ 
couplings are identical.  We keep fixed $S=0.3$ 
and consider two values of $\tilde g$,  2 (solid line) and 5 (dashed line). 
We repeat the plots for three choices of the $s$ parameter
$(+1,0,-1)$ depicted in black, blue and green colors respectively.
The deviation of  $g_{HWW}$ from $g_{HWW}^\text{SM}$  increases with $M_A$  due to the
fact that we hold the $S$ parameter fixed. 
One reaches deviations of 20$\%$ from the SM couplings when $M_A\simeq 2$~TeV.
This is reflected in the small deviations of the Higgs branching ratios
when compared with the SM ones. 

\subsubsection{$pp\to WH$ and $pp\to ZH$}
\begin{figure}[htbp]
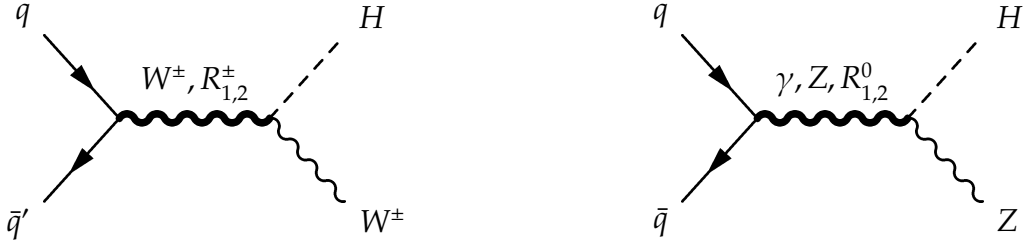

\begin{center}
\vspace{11mm}
\parbox{50mm}{\begin{fmfchar*}(50, 22)
\fmflabel{$q$}{q}
\fmflabel{$\bar{q}'$}{qp}
\fmflabel{$W^\pm$}{vW}
\fmflabel{$H$}{vH}
\fmfleft{qp,q}
\fmfright{vW,vH}
\fmf{fermion}{q,vi,qp}
\fmf{dashes}{vf,vH}
\fmf{photon}{vf,vW}
\fmf{photon, label=$W^\pm,, R_{1,, 2}^\pm$, width=1mm, label.side=left}{vi,vf}
\end{fmfchar*}}
\qquad\qquad\qquad\qquad
\parbox{50mm}{\begin{fmfchar*}(50, 22)
\fmflabel{$q$}{q}
\fmflabel{$\bar{q}$}{qbar}
\fmflabel{$Z$}{vZ}
\fmflabel{$H$}{vH}
\fmfleft{qbar,q}
\fmfright{vZ,vH}
\fmf{fermion}{q,vi,qbar}
\fmf{dashes}{vf,vH}
\fmf{photon}{vf,vZ}
\fmf{photon, label=$\gamma,, Z,, R_{1,, 2}^0$, width=1mm, label.side=left}{vi,vf}
 \end{fmfchar*}}
\vspace{7mm}
 \caption{\label{fig:whdiag}Feynman
 diagrams for the composite Higgs production 
 in association with  SM gauge bosons.}
\end{center}
\end{figure}

\begin{figure}
\includegraphics[width=0.5\textwidth]{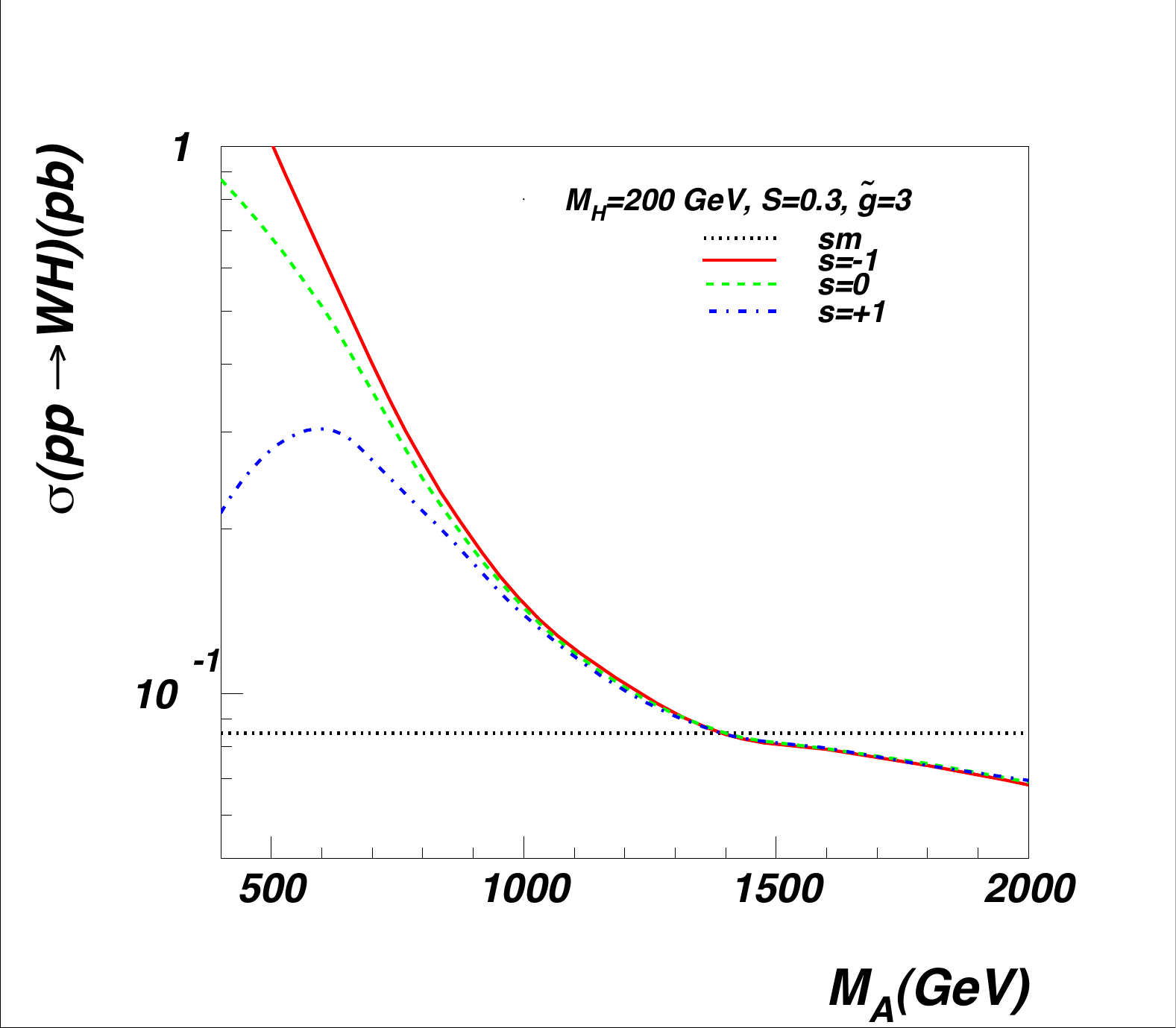}%
\includegraphics[width=0.5\textwidth]{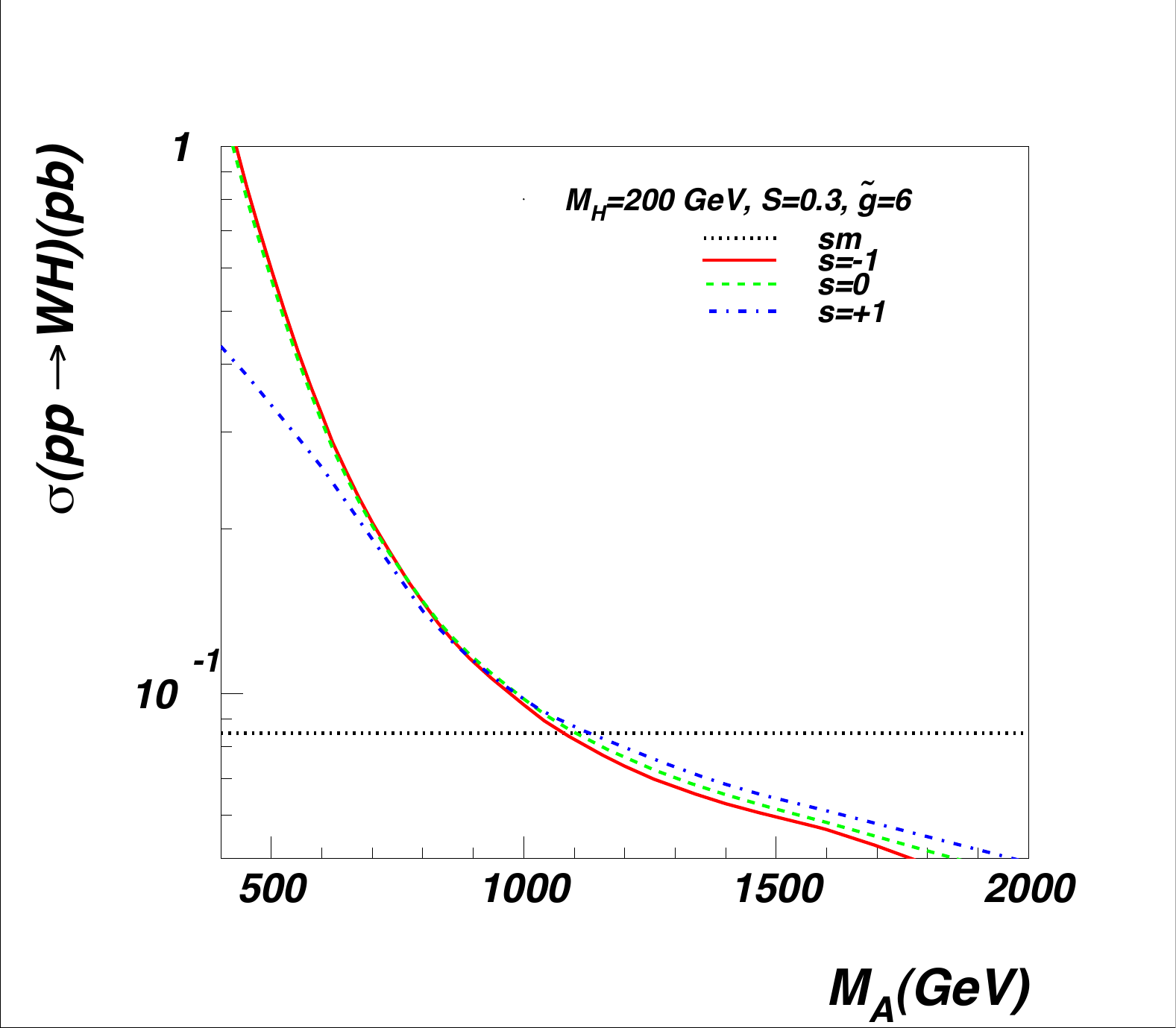}%
\caption{\label{fig:vh7}
The cross section for $pp\to WH$ production at $7$~{TeV} in the center of mass
($W^+H$ and $W^-H$ modes are summed up)
versus $M_A$ for $M_H=200$ GeV, $S=0.3$, $s=(+1,0,1)$
and $\tilde g =3$ (left) and $\tilde g=6$ (right).
The dotted line at the bottom indicates the SM cross section level.}
\end{figure}

The presence of the heavy vectors is prominent in the associate production of the composite Higgs with SM vector bosons, as first pointed out in \cite{Zerwekh:2005wh}. 
Parton level Feynman diagrams for  the $pp\to WH$ and $pp\to ZH$ processes are shown in Fig.~\ref{fig:whdiag} (left)  and
Fig.~\ref{fig:whdiag} (right) respectively.
The resonant production of heavy vectors can enhance $HW$ and
$ZH$ production by a factor 10 as one can see in Fig.~\ref{fig:vh7} (right). This enhancement occurs for low values of the vector meson mass
and large values of $\tilde g$.

\begin{figure} 
\begin{center}
\includegraphics[width=0.5\textwidth,height=0.4\textwidth]{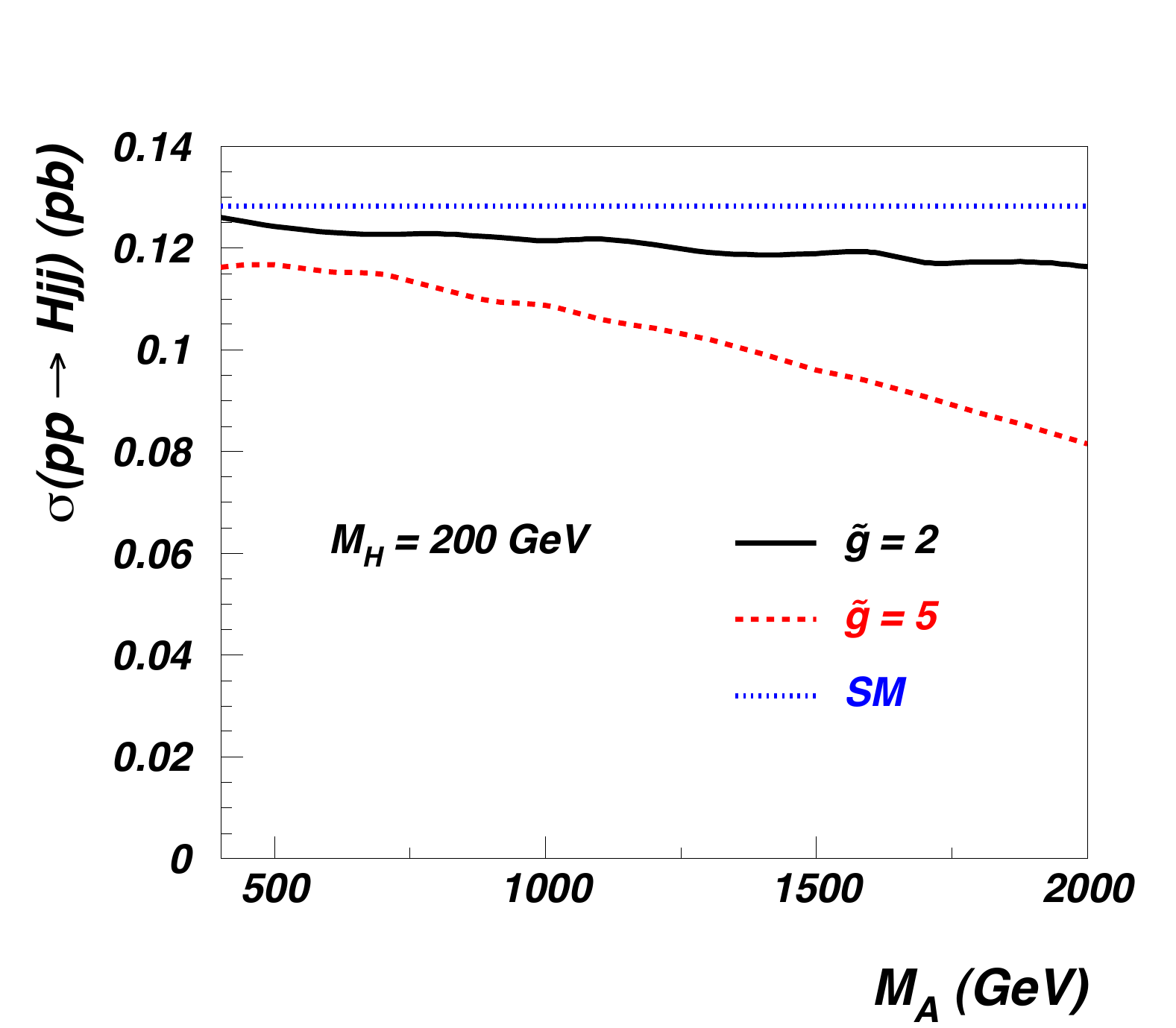}%
\includegraphics[width=0.5\textwidth,height=0.4\textwidth]{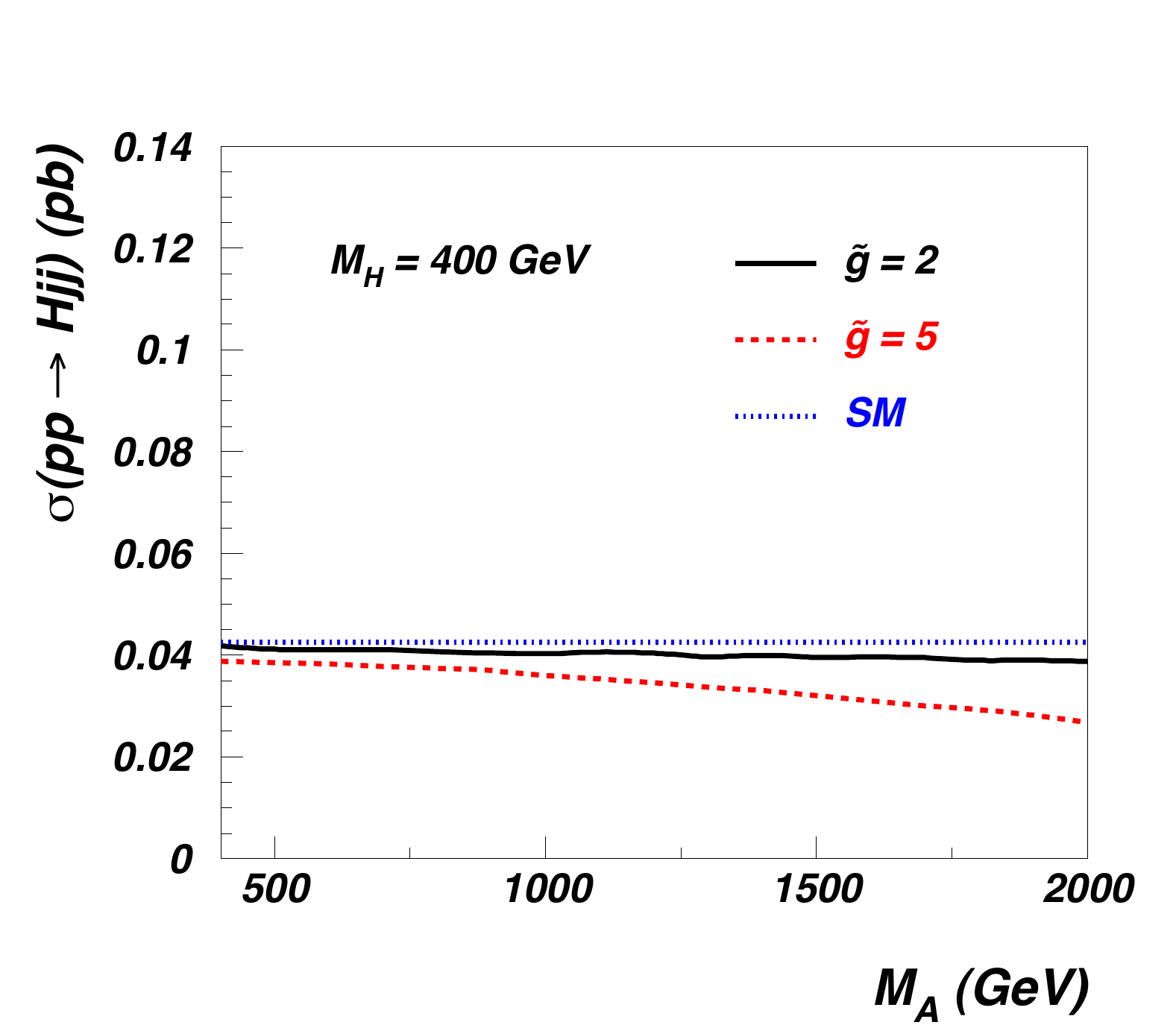}%
\end{center}
\caption{\label{fig:hjj}
Composite Higgs production cross section via the VBF mechanism at $7$~TeV.}
\end{figure}

\subsubsection{Higgs vector boson fusion, $pp\to Hjj$}

We have also analyzed the composite Higgs  production
in VBF processes $pp\to Hjj$. We find that it is not enhanced with respect to the corresponding process in the SM as it is clear from Fig.~\ref{fig:hjj}. The behavior of $\sigma(pp\to Hjj)$ as function of $M_A$  traces the one of the Higgs-gauge bosons coupling shown in Fig.~\ref{fig:ghww}.

We used the same kinematic cuts as for the $pp\rightarrow Rjj$ section except for the transverse momenta and energy of the jets that we take to be $p_T^j>15$~GeV and $E^j>150$~GeV.

\subsubsection{$H  \rightarrow \gamma \gamma$ and $H \rightarrow gg$ }

  \begin{figure}[htbp]
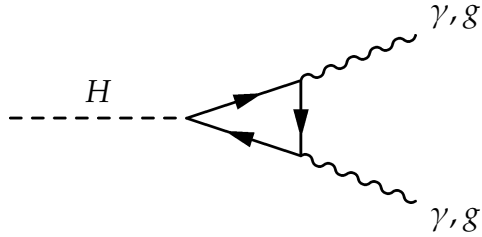

\begin{center}
\parbox{60mm}{\begin{fmfchar*}(60, 22)
\fmflabel{$\gamma, g$}{g1}
\fmflabel{$\gamma, g$}{g2}
\fmfleft{H}
\fmfright{g2,g1}
\fmf{photon}{v1,g1}
\fmf{photon}{v2,g2}
\fmf{fermion}{v2,v,v1}
\fmf{fermion, tension=.1}{v1,v2}
\fmf{dashes, label=$H$, label.side=left, tension=1.3}{H,v}
\end{fmfchar*}}
\vspace{7mm}
\caption{\label{fig:Htogg}Feynman
 diagram for the composite Higgs decay to $\gamma\gamma$ or $gg$ via SM fermion loop.}
\end{center}
\end{figure}

The one loop Higgs decay (see Fig.~\ref{fig:Htogg}) width to two photons for any weakly interacting elementary particle contributing to this process can be neatly summarized via the formula \cite{Gunion:1989we}: 

\begin{equation}
\begin{split}
\Gamma(H\to\gamma\gamma)=\frac{\alpha^2G_{F}M_H^2}{128\sqrt{2}\pi^3}~\left\lvert \sum_i~n_i~Q_i^2~F_{i} \right\rvert^2,
\end{split}
\label{width}
\end{equation}
where i runs over the spins, $n_{i}$ is the multiplicity of each species with electric charge $Q_i$  in units of $e$. The $F_i$ functions are given by
\begin{equation}
F_1=2+3\tau+3\tau(2-\tau)f(\tau)\ ,\quad 
F_{1/2}=-2\tau[1+(1-\tau)f(\tau)]\ ,\quad 
F_0=\tau\left( 1-\tau f(\tau) \right)\ ,
\label{loopfun}
\end{equation}
where $\tau=\frac{4m_i^2}{M_H^2}$ and
\begin{equation}
  f(\tau) = \left\{ 
  \begin{array}{c c}
    \left( \arcsin \sqrt{\frac{1}{\tau}} \right)^{2},  & \quad \text{if $\tau ~\geq~1$}\\
    -\frac{1}{4}\left( \log \left( \frac{1+\sqrt{1-\tau} }{1-\sqrt{1-\tau}} \right) -i\pi \right)^2,  & \quad \text{if $\tau~<~1$}\\
  \end{array} \right. 
\end{equation}
The lower index of the function $F$ indicates the spin of each particle contributing to the process. It is clear from this formula that there can be strong interferences between the different terms contributing. In particular we anticipate that there is an important interplay between the SM gauge bosons and the TC contribution for this process. 

The contribution to the Higgs decay to two-photon process in the SM is due to the $W$s and SM fermion loops contributions (with the leading one given by the top quark). In fact it is an excellent approximation to use just the top in the fermion loop, since the ratio between the corresponding result and the one obtained including all the contributions is around $0.98$, for the reference value of the Higgs mass of $M_H \simeq 120$ GeV. The approximation improves as the Higgs mass increases.  
We consider here the case in which the Higgs is a composite state constituted of some more fundamental matter as in TC. The SM Higgs sector Lagrangian is therefore interpreted simply as a low energy effective theory. We assume the Vanilla TC theory and therefore the composite Higgs ($H$) effective field coincides with the SM  one. It is straightforward to generalize the global symmetries of the TC theories to larger symmetry groups.

The contribution to the sought process is modeled by re-coupling, in a minimal way, the composite Higgs to the techniquarks $Q$ via the following operator: 
\begin{equation} 
\mathcal{L}_{QH} = \sqrt{2}\, \frac{M_Q}{v} \left[ {\overline{Q}_L}^t\cdot H {D_R}_t + {\overline{Q}_L}^t\cdot (i\, \tau_2 \, H^{\ast}) {U_R}_t    \right] + {\rm h.c.} \ ,  
\end{equation}
where $M_{Q}$ is the dynamical mass of the techniquark and $t = 1,\dots, d[r]$ is the technicolor index and $d[r]$ the dimension of the representation under which the techniquarks transform. This type of hybrid models have been employed many times in the QCD literature to extra dynamical information and predictions for different  phenomenologically relevant processes. A close example is  the $\sigma(600)$ decay into two photons \cite{Pennington:2006dg} investigated using, for example, hybrid type models in  \cite{Giacosa:2007bs} and \cite{vanBeveren:2008st}. 

We assumed a single TC doublet with respect to the Weak interactions as predicted by minimal models of TC \cite{Sannino:2004qp}. We added a new lepton family to cure the Witten topological anomaly (with respect to the weak interactions) \cite{Witten:1982fp} when the TC sector features an odd number of doublets charged under the $SU(2)_\text{L}$ weak interactions. This occurs when the dimension of the TC matter representation is an odd number. A further Leptonic interaction term is then added: 
\begin{equation} 
\mathcal{L}_{LH} = \sqrt{2}\, \frac{M_E}{v} \, {\overline{L}_L}\cdot H {E_R}+  \sqrt{2}\, \frac{M_N}{v}{\overline{L}_L}\cdot (i\, \tau_2 \, H^{\ast}) {N_R}    + {\rm h.c.} \ ,
\end{equation}
with $M_E$ and $M_N$ the fermion masses expected to be of the order of the electroweak scale \cite{Frandsen:2009fs}. The detailed spectrum is partially dictated by the precision electroweak constraints.  

The techniquark dynamical mass is intrinsically linked to the technipion decay constant $F_{\pi}$ via the Pagels-Stokar formula  
$M_Q  \approx \frac{2\pi F_{\pi}}{\sqrt{d[r]}}$. 
  To achieve the correct values of the weak gauge bosons  the decay constant is related to the electroweak scale via: 
\begin{equation}
v = \sqrt{N_{f}}F_{\pi}, \qquad \text{with}\qquad v=246~ \text{GeV} \ .
\end{equation}  
 We have checked that  the numerical value of  the contribution of the $M_{E/N}$ states is negligible when the masses are of the order of the electroweak scale. We have used as a reference value for these lepton masses  $500~\text{GeV}$ \footnote{For  $M_{E/N} \gg M_H$ the contribution approaches a constant due to the functional form of $F_{1/2}$.}. 
 
 At this point we determine the contribution of the techniquarks, which do not carry ordinary color charge, and of the new leptons to the diphoton process by evaluating the naive one loop triangle diagrams, in which the coupling of the composite Higgs to two techniquarks, as well as to the new leptons, can be read off from the Lagrangian above. The contribution at the amplitude level for each new particle is identical to the one of a massive SM fermion up to a factor given by the correct electric charge and degeneracy. 

Although we adopted a simple model computation we now argue that the final geometric dependence on the gauge structure, matter representation and number of techniflavors of the composite theory is quite general.

\begin{figure}[htbp]
\begin{center}
 \includegraphics[width=0.6\textwidth]{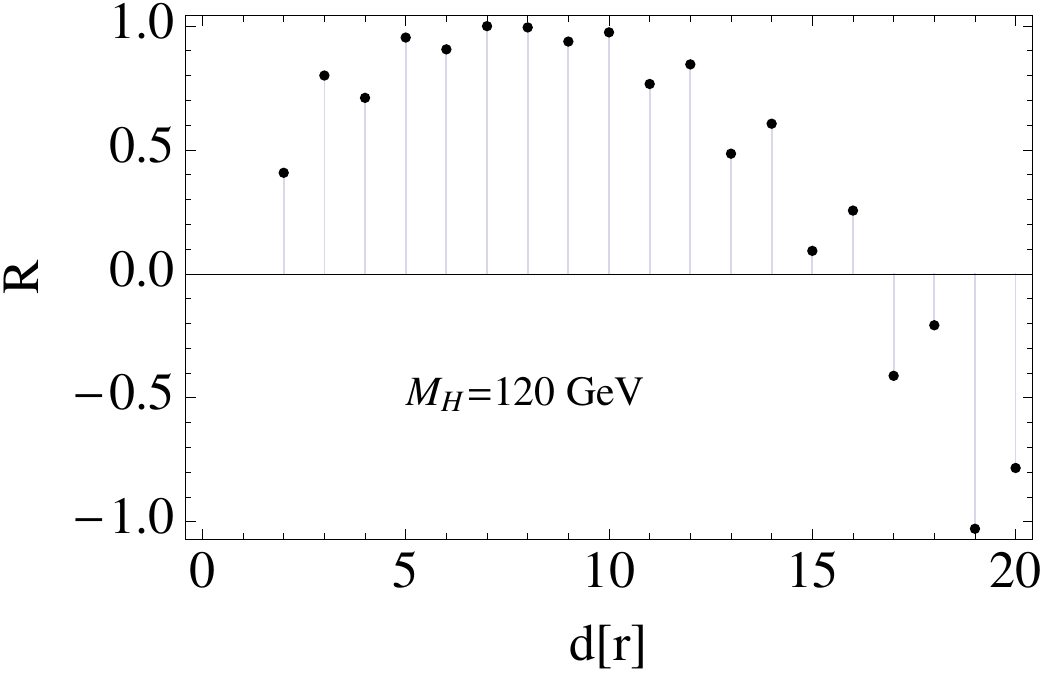}
\caption{Difference of the Walking TC and SM decay widths divided by the SM width plotted with respect to number of colors. $M_H=120$~GeV.}
\label{color}
\end{center}
\end{figure}

\begin{figure}[htbp]
\begin{center}
 \includegraphics[width=0.6\textwidth]{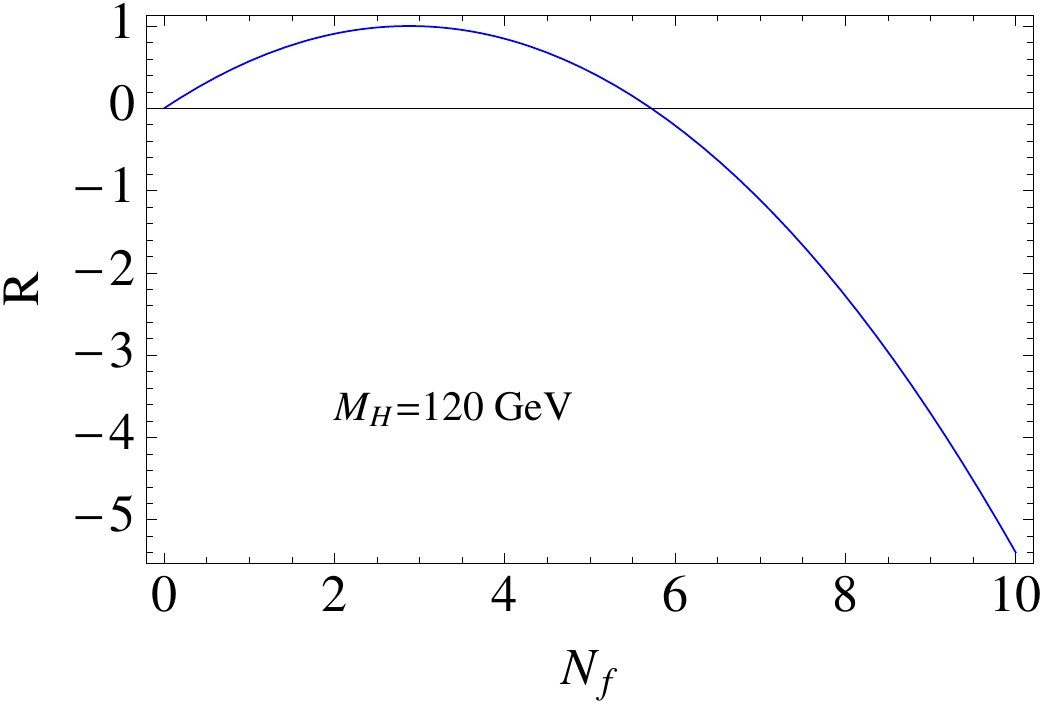}
\caption{Difference of the Walking TC and SM decay widths divided by the SM width plotted with respect to number of flavors. $M_H=120$~GeV.}
\label{flavor}
\end{center}
\end{figure}

\begin{figure}[htbp]
\begin{center}
 \includegraphics[width=0.55\textwidth]{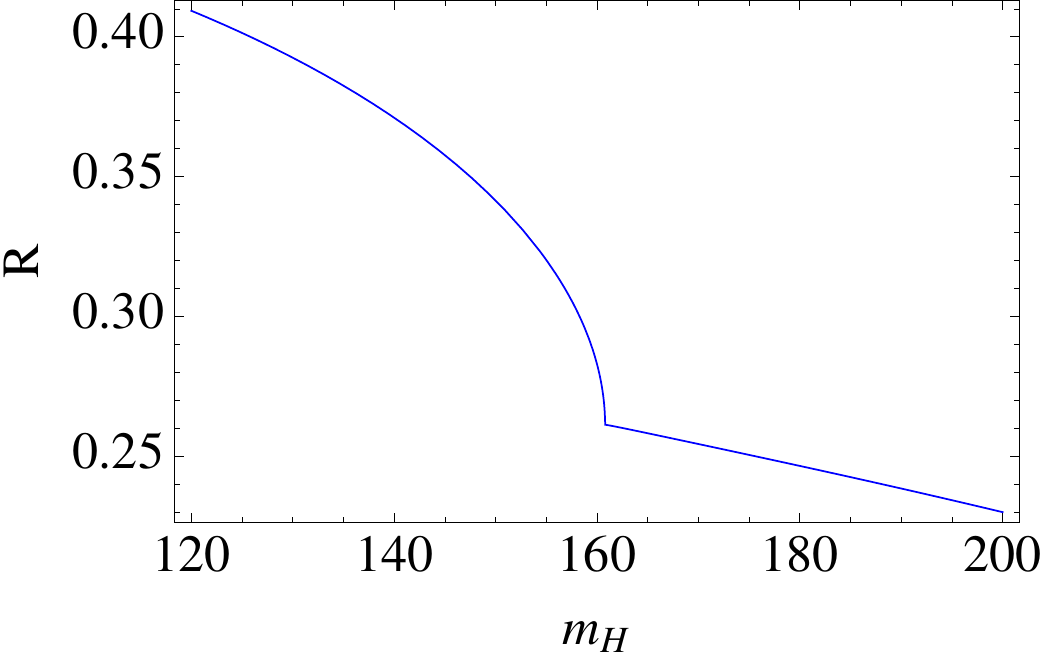}
\caption{ Dependence of $R$ as function of the Higgs mass $M_H$ for a reference value of the techniquark matter representation $d[r] =2$ and two number of Dirac techniflavors.}
\label{Higgs-dependence}
\end{center}
\end{figure}

We plot in Fig.~\ref{color} the intrinsic dependence on the dimension of the TC matter representation $d[r]$ according to the ratio:
\begin{equation}
R=\frac{\Gamma_\text{SM}(H\to \gamma\gamma)-\Gamma_\text{TC}(H\to \gamma\gamma)}{\Gamma_\text{SM}(H\to \gamma\gamma)}.
\label{haa-ratio}
\end{equation}
 For any odd representation we included also the lepton contribution. Interestingly when $d[r]$ is around $7$  the contribution from the techniquarks adds up with that of the SM fermions (i.e. mainly the top) to cancel the contribution of the W,  determining $\Gamma_\text{TC}\approx 0$.  The techniquark matter dominates the process for $d[r]$ larger than around seven.  A similar behavior occurs when we increase the number of flavors as shown in Fig.~\ref{flavor}.   
The dependence on the Higgs mass with $d[r]=2$ and $N_f=2$ is shown in Fig.~\ref{Higgs-dependence}.
The $WW$ threshold is evident in the plot which shows how the ratio decreases with the Higgs mass. 
  
\begin{figure}[htbp]
\begin{center}
 \includegraphics[width=0.6\textwidth]{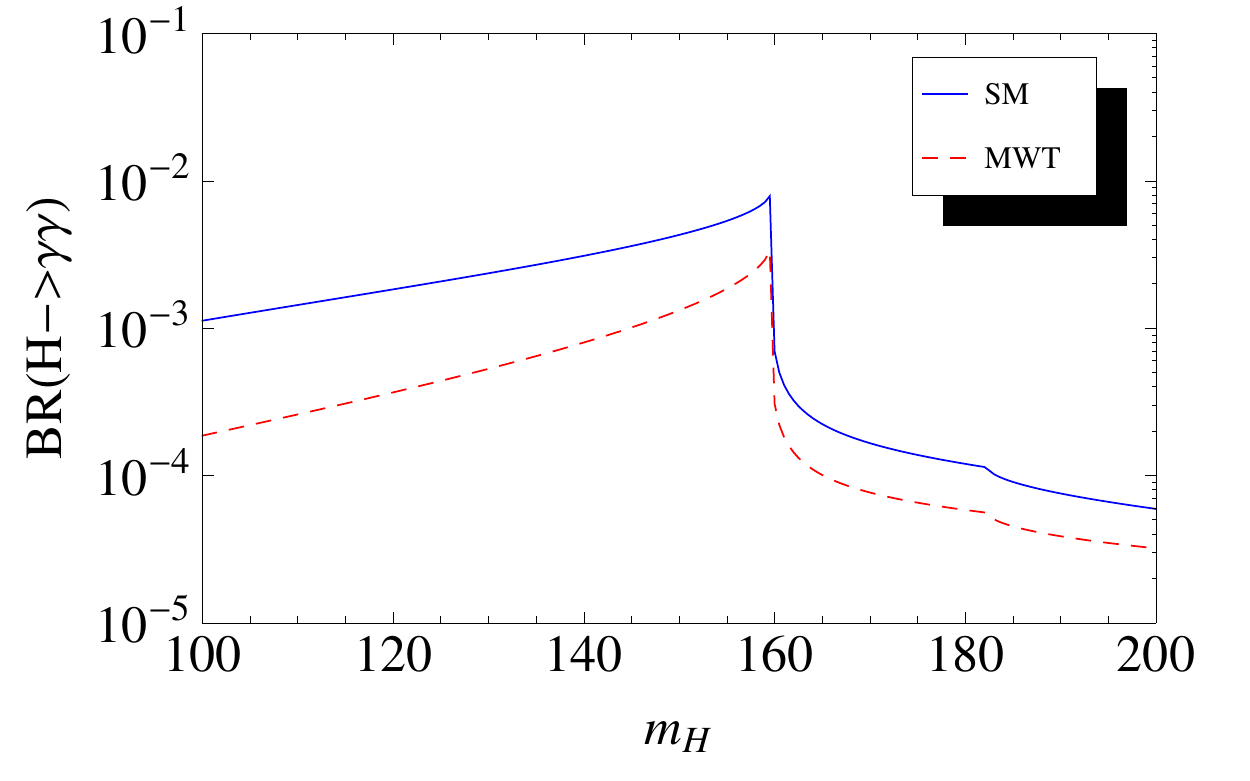}
\caption{Higgs to two photons branching ratio in MWT as function of the composite Higgs mass versus the SM result.}
\label{H2Gbranching}
\end{center}
\end{figure}

 To illustrate our results we plot in Fig.~\ref{H2Gbranching} the Higgs to two-photon branching ratio as function of the Higgs mass for the MWT model and compare it to the SM one.   
It is clear from the plot that  we expect sizable corrections to the SM Higgs to two photons process at the LHC.

 It is interesting to estimate the impact on the digluon process coming from a potentially composite Higgs made by technimatter charged under QCD interactions. This contribution can be estimated, as we have done for the diphoton process, by considering the one loop contribution of the techniquarks arriving at: 
\begin{equation}
\begin{split}
\Gamma(H\to gg)=\frac{\alpha_{s}^2G_{F}M_H^3}{64\sqrt{2}\pi^3}~\left\lvert \sum_i~F_{i}^{\rm SM}+\sum_j~2T[\tilde{r}_{j}]d[r_{j}]F_{j}^{\rm TC} \right\rvert^2,
\end{split}
\end{equation}
where $T[\tilde{r}_{j}]$ is the Dynkin index of the representation of the techniquarks under the SM color group $SU(3)$ and $d[r_{j}]$ is the dimension of  the representation under the TC gauge group.
\begin{figure}[htbp]
\begin{center}
 \includegraphics[width=0.6\textwidth]{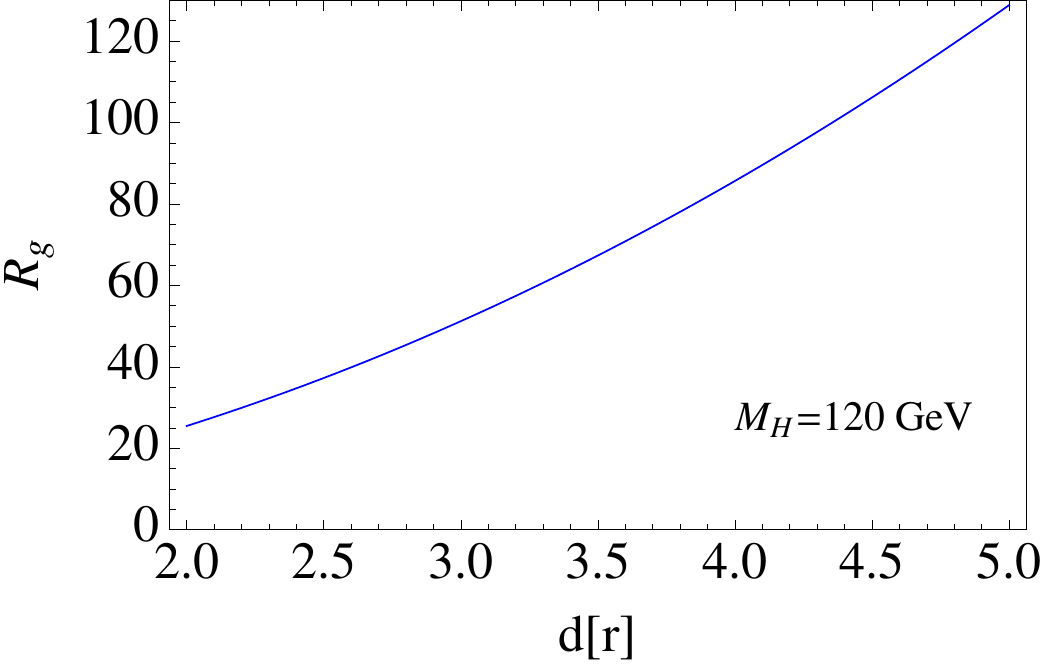}
\caption{Difference of the Walking TC and SM decay widths divided by the SM width plotted with respect to the dimension of the representation. $M_H=120$~GeV.}
\label{ggh-mh}
\end{center}
\end{figure}
For  techniquarks transforming according to the fundamental representation of QCD the two gluon decay width of the Higgs is strongly enhanced. This is presented in Fig.~\ref{ggh-mh} where we have plotted the ratio
 \begin{equation}
R_{g}=\frac{\Gamma_\text{TC}(H\to gg)-\Gamma_\text{SM}(H\to gg)}{\Gamma_\text{SM}(H\to gg)} \ .
\end{equation}
as a function of the dimension of the representation under the TC gauge group. The composite Higgs to two-gluon decay width increases monotonically with respect to the number of technicolors in contrast with the diphoton decay width. The difference resides in the fact that gluons do not couple directly to the other gauge bosons of the SM.

We  now estimate the viability of the models with colored techniquarks by comparing our results with the study performed in \cite{Aaltonen:2010sv}.  Here the authors considered the effects of the fourth fermion SM generation to the process  $\sigma_{gg\to H}\times \text{BR}(H\to W^{+}W^{-})$. The fourth generation quarks enhances the $gg\to H$ production cross section by a factor  of $9$ to $7.5$ in the Higgs mass range $M_H=110-300$ GeV with respect to the SM result. This enhancement leads to a 95\% confidence level exclusion of the Higgs in the mass range $M_H=131-208$ GeV or of the presence of a fourth generation if the Higgs turns out to be within this mass energy range.  Colored techniquakrs are further penalized with respect to the SM fourth generation of quarks since they carry also technicolor leading to a further dramatic enhancement of this process.  \begin{figure}
\begin{center}
 \includegraphics[width=0.5\textwidth]{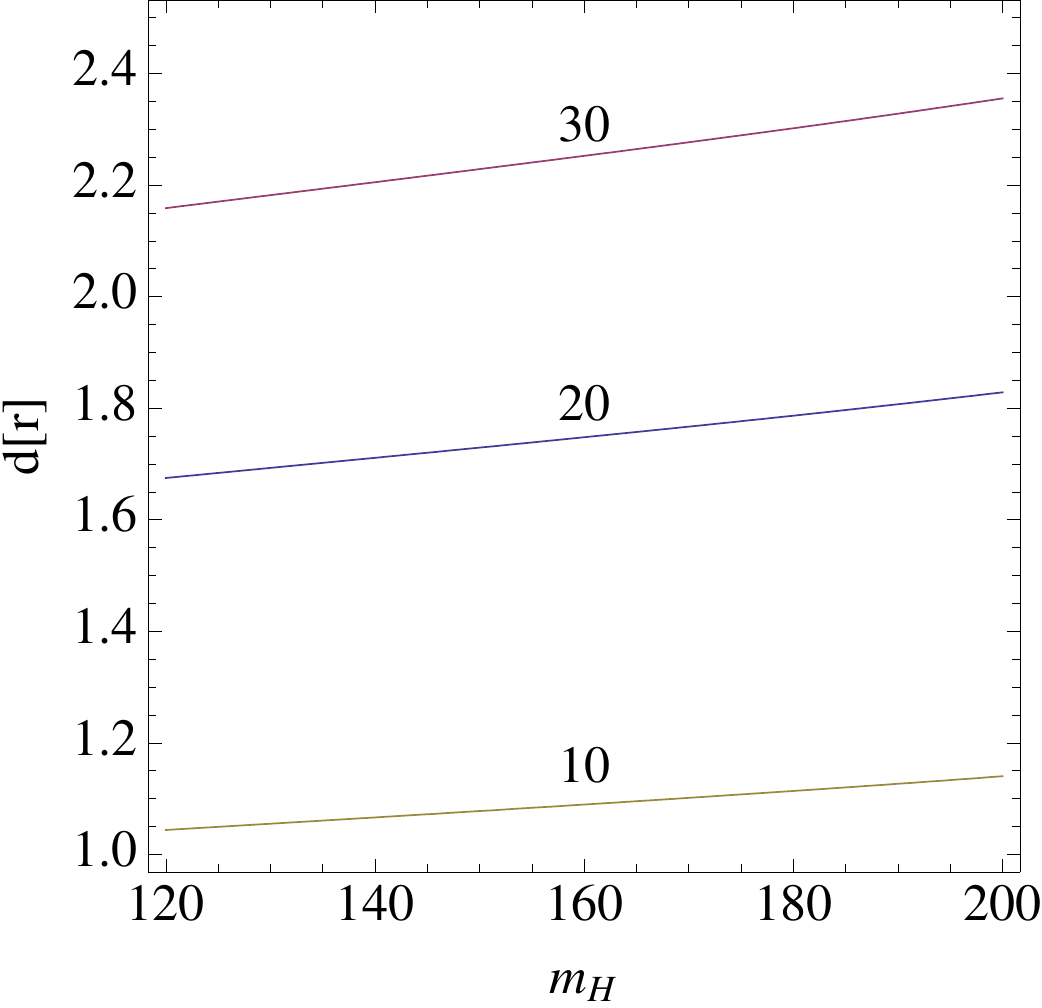}
\caption{Contours in the $d[r]$ versus Higgs mass plane for fixed values of the enhancement of the $gg\to H$ production cross section compared to the SM one. Obviously, in practice,  the dimension of the representation cannot be taken to be continuous and the plot simply shows that already for $d[r]$ around 2 (which is the minimal nontrivial matter group-theoretical dimension) the enhancement factor is around 30.  }
\label{contour}
\end{center}
\end{figure}
In Fig.~\ref{contour} we have plotted contours in the dimension of the techniquark representation versus the Higgs mass when the $gg\to H$ production cross section is enhanced by factors 10, 20 and 30 respectively compared to the SM one. Obviously, in practice,  the dimension of the representation cannot be taken to be continuous and the plot simply shows that already for $d[r]$ around two (which is the minimal nontrivial matter group-theoretical dimension) the enhancement factor is around 30.  Thus we conclude that this process alone strongly indicates that viable TC models should feature {\it colorless} techniquarks as it is the case of MWT or that a light composite Higgs made of colored techniquarks is excluded by the Tevatron experiment. These results are in perfect agreement with the LEP electroweak precision data which require a small $S$ parameter. 
 
\vskip 1cm 

In these two last subsections we introduced a simple framework to estimate the composite Higgs boson coupling to two photons and gluons for generic  TC extensions of the SM.  By comparing the decay rates with the SM ones for the diphoton case we show that the corrections are typically of order one and therefore this process can be efficiently used to disentangle a light composite Higgs from a SM one.  We then turned our attention to the composite Higgs to two gluon process and showed that the Tevatron results for the gluon-gluon fusion production of the Higgs exclude either the techniquarks to carry color charges up to the 95\% confidence level or the existence of a light composite Higgs made of colored techniquarks. 

\subsubsection{Higgs production via gluon fusion}
As we have shown above if the techniquarks do not carry color this process occurs as in the SM, provided that the ETC interactions do not strongly modify the Yukawa sector.

\section{A Natural fourth family of leptons at the TeV scale \label{4thleptons}}
 
If the LHC discovers a new fourth family of leptons but no associated quarks, what would that imply? 
Either the associated quarks are much heavier than the electroweak scale or a new set of fermions are needed 
to account for the 4th family lepton induced gauge anomalies.  The new fermions could address directly the big 
hierarchy problem of the SM if their dynamics leads to a composite Higgs scenario. This is the setup we are 
going to discuss in this chapter, using as a template MWT \cite{Sannino:2004qp,
Frandsen:2009fs}. 

Before investigating the phenomenology of such a theory, we will first review the leptonic sector of the SM
(see {\em e.g.}~\cite{Altarelli:2010fk} for a recent review). 
Secondly, we will consider a general mass structure for the fourth neutrino and both the case of 
mixing and no mixing with the SM neutrinos. We will also summarize the (in)direct phenomenological constraints. 
We then analyze the interplay of the composite Higgs sector with the new lepton family at the LHC.  

We will study the production and decay of the new leptons in proton - proton collisions which is relevant to 
select the LHC signatures for the discovery of these new leptons. We show that the composite Higgs structure 
can affect and differentiate the final signatures with respect to the case in which the Higgs is elementary. 
The bottom line is that one can experimentally determine if the fourth family is associated to a composite 
Higgs sector.

\subsection{The Standard Model leptons: a Mini-review}

In the SM, the three lepton families, $\ell=e,\mu,\tau$, belong to the following representations 
of the gauge group $SU(3)_\text{C} \times SU(2)_\text{L} \times U(1)_Y$:
\beq
L_\ell = ({\nu_\ell}_L ~~ \ell_L)^T \sim (1,2,-1/2) ~~~~~,~~~~~\ell_R \sim (1,1,-1)~~~~~,
\eeq  
where the chirality projectors $P_L=(1-\gamma_5)/2$ and $P_R=(1+\gamma_5)/2$ have been introduced and 
the relation $Q=T^3+Y$ has been adopted in order to define the hypercharge with respect to the electric charge.
The neutral and charged current interactions of the SM leptons are then respectively accounted for by 
the Lagrangian terms: 
\bea
{\cal L}_{NC} &=& 
\frac{g }{\cos \theta_W}  ~\left( \frac{1}{2} \bar \nu_L \gamma^\mu \nu_L - \frac{1}{2} \bar \ell_L \gamma^\mu \ell_L
+ \sin^2 \theta_W ~\bar \ell \gamma^\mu \ell \right) ~Z_\mu ~+ ~e ~\bar \ell \gamma^\mu \ell ~ A_\mu \nn & & \\
{\cal L}_{CC} &=& \frac{g }{\sqrt{2}}  ~\bar \ell_L \gamma^\mu  \nu_L ~W_\mu^-+ h.c.   \no
\eea
where $\ell=\ell_L+\ell_R$ and $\theta_W$ is the Weinberg angle.  

In the SM, due to the absence of gauge singlet fermions with R-chirality, it is not possible to endow neutrinos
with a Dirac-type mass term. An effective Majorana-type mass term can however be identifyed among 
the non-renormalizable operators of dimension $5$. Symbolically: 
\begin{equation}
\frac{ (L H)^T (L H)}{\Lambda}   
\end{equation}
where $\Lambda$ represents the energy scale at which additional interactions and/or fields show up,
which must violate total lepton number (the effective mass operator above violates it by two units). 
The associated effective neutrino masses would thus be suppressed with respect to the electroweak scale, $v$,
by a factor ${\cal{O}}(v/\Lambda)$.  
Experiments on neutrino oscillations, which measure differences of squared masses and mixing
angles, have estabished that at least two of the SM neutrinos have a small mass, 
not larger than the eV scale \cite{Nakamura:2010zzi}. This would indeed suggest a very high scale for
lepton number violation, $\Lambda \sim 10^{11} v$.
 
The neutrino mass matrix is diagonalised in general by a $3 \times 3$ unitary matrix $U$, 
called the PMNS mixing matrix:
\begin{equation}
m_\nu =U^* m_\text{diag} U^\dagger~.
\end{equation} 
Since $U$ and $m_\text{diag}$ are in general complex, neutrino masses can provide three CP violating phases: 
the one in $U$ (which is the analog
of the CP violating phase in the CKM quark mixing matrix) plus the two phase differences in $m_\text{diag}$,
which are denoted as Majorana phases. 
  
Two distinct oscillation frequencies have been first measured in solar and atmospheric neutrino oscillations 
and later confirmed by experiments on earth. Two well separated differences need at least three different 
neutrino mass eigenstates involved in oscillations so that the three known neutrino species can be sufficient. 
Then at least two neutrinos must be massive while, in principle, the third one could still be massless.
The mass eigenstates involved in solar oscillations are $m_1$ and $m_2$ and, by definition, $|m_2| > |m_1|$, 
so that $\Delta m^2_\text{sun} = |m_2|^2 -|m_1|^2 > 0$. The atmospheric neutrino
oscillations involve $m_3$: $\Delta m^2_\text{atm} = |\Delta m^2_{31}|$ with 
$\Delta m^2_{31} = |m_3|^2 -|m_1|^2$ either positive (normal hierarchy) or negative (inverse hierarchy). 
The present data are still compatible with both cases.
The degenerate spectrum occurs when the average absolute value of the masses is much larger than
all mass squared differences. With the standard set of notations and definitions
the present data are summarised by \cite{Nakamura:2010zzi}
\begin{eqnarray}
\Delta m^2_\text{sun} = (7.05-8.34) \times  10^{-5} {\rm eV}^2 ~, ~~
\Delta m^2_\text{atm} = (2.07-2.75) \times  10^{-3} {\rm eV}^2 ~, \nonumber \\
0.25 \le \sin^2\theta_{12} \le 0.37~ ,~~0.36 \le \sin^2\theta_{23} \le 0.67~. ~~~~~~~~~~
\end{eqnarray}
In addition, $\sin^2\theta_{13} < 0.035$ at $90\%$ C.L.
The pattern of neutrino mixing is drastically different from the pattern of quark mixing.
At present no experimental information on the Dirac and Majorana CP violation
phases in the neutrino mixing matrix is available.

Oscillation experiments do not provide information about the absolute neutrino mass scale.
Limits on that are obtained from the endpoint of the tritium beta decay spectrum, from cosmology
and from neutrino-less double beta decay. From tritium we have an absolute upper limit
of $2.2$ eV (at $95\%$ C.L.) on the mass of electron antineutrino, which, combined with the observed
oscillation frequencies under the assumption of three CPT-invariant light neutrinos, also amounts
to an upper bound on the masses of the other active neutrinos.

Oscillation experiments cannot distinguish between Dirac and Majorana neutrinos. The detection
of neutrino-less double beta decay would provide direct evidence of lepton number violation, and
the Majorana nature of neutrinos. It would also offer a way to possibly disentangle the 3 cases of
degenerate, normal or inverse hierachy neutrino spectrum.
The present limit from $0\nu\beta\beta$  (with large ambiguities from nuclear matrix elements) is
about $|m_{ee}| < (0.3 - 0.8)$ eV \cite{Altarelli:2010fk}.

\subsection{Adding a fourth lepton family}

In the following, we will account for the light neutrino masses and mixings by means of an effective Majorana
mass term, namely we add to the SM Lagrangian a dimension-5 non-renormalizable operator. 

Our aim here is to study the phenomenology of an additional heavy lepton family, with masses about the TeV scale
\cite{Frandsen:2009fs}.
Thus, we add 
a 4th-family of leptons - for which we introduce the $\zeta$-flavor - 
composed by a lepton doublet, a charged lepton singlet and a gauge singlet:
\beq
L_\zeta= ({\nu_\zeta}_L ~~ \zeta_L)^T \sim (1,2,-1/2)~~~~,~~~~~{\zeta}_R\sim (1,1,-1)~~~,
~~~~~{\nu_\zeta}_R\sim (1,1,0)~~. \label{newLeptons}
\eeq
The $\zeta$-charged lepton, $\zeta = {\zeta}_L+{\zeta}_R$, will have a Dirac mass term like the other 
three charged leptons of the SM, but large enough to avoid conflict with the experimental limits.  
We work in the basis in which the $4\times4$ charged lepton mass matrix is diagonal.

\subsubsection{Heavy leptons not Mixing with Standard Model neutrinos}

We forbid, in this section, the mixing between the $\zeta$-neutrino and the three light neutrinos of the SM. 
The Lagrangian can be split as ${\cal L} = {\cal L}_\text{SM} + {\cal L}_\zeta$.

The Lagrangian mass terms we take for the $\zeta$-sector reads:
\begin{eqnarray}
{\cal L}_\zeta^{\rm mass} 
= -m_{\zeta} ~ \overline{ {\zeta}}~ {\zeta}  
- \frac{1}{2} \left[ \left( \begin{array}{cc}  \overline{\nu_{\zeta L}} & \overline{(\nu_{\zeta R})^c}  \end{array} \right)
\left( \begin{array}{cc}   0 & m_D \cr  m_D &  m_R \end{array} \right)  
\left( \begin{array}{c} (\nu_{\zeta L})^c \cr \nu_{\zeta R}  \end{array} \right) + h.c.\right] \ ,
\end{eqnarray}
Diagonalizing the neutrino mass matrix above, we obtain two independent Majorana eigenstates, $N_1$ and $N_2$, 
with real and positive masses, $M_1$ and $M_2$ (for convention $M_1 \le M_2$),
\beq
M_1=\frac{m_R}{2} \left( \sqrt{1+4\frac{m_D^2}{m_R^2}} -1 \right)~~~~,~~~~
M_2=\frac{m_R}{2} \left( \sqrt{1+4\frac{m_D^2}{m_R^2}} +1 \right)~~,
\eeq
which are related to the original Dirac and Majorana masses according to
$M_1 M_2 = m^2_D$ and $M_2-M_1 = m_R$.

The $\zeta$-neutrino chiral states will be an admixture of the two Majorana eigenstates $N_1$ and $N_2$:
\begin{equation}
\left( \begin{array}{c}
  \nu_{\zeta L} \cr  (\nu_{\zeta R})^c  \end{array} \right) = 
 \left( \begin{array}{cc}  i \cos\theta  & \sin\theta  \cr  -i \sin\theta & \cos\theta \end{array} \right) 
\left( \begin{array}{c}  P_L N_1 \cr P_L N_2  \end{array} \right) ~~~~,~~~~\tan2\theta = \frac{2m_D}{m_R} ~~.
\end{equation}
In the limit $m_D \ll m_R$ the seesaw mechanism would be at work (leading to $M_1\sim m_D^2 /m_R$, $M_2\sim m_R$, 
${\nu_{\zeta}}_L \sim  i  ~P_L N_1$, $({\nu_{\zeta R}})^c \sim  P_L N_2$). Here however we are more interested 
in the situation $m_D \sim m_R$, in which both Majorana neutrinos have a mass about the TeV scale and hence have
a large $SU(2)$-active component. 

The neutral current interaction of the $\zeta$-leptons in terms 
of the heavy neutrino Majorana mass eigenstates reads:
\beq
{\cal L}^{NC}_\zeta = \frac{g }{\cos\theta_W}~ 
\left(\frac{1}{2} \overline{ {\nu_\zeta}_L} \gamma^\mu {\nu_\zeta}_L 
- \frac{1}{2} \overline{ \zeta_L} \gamma^\mu \zeta_L + \sin^2 \theta_W ~\bar \zeta \gamma^\mu \zeta \right) ~Z_\mu~ 
+ ~e  ~\bar \zeta \gamma^\mu \zeta ~A_\mu~~, 
\eeq
where
\beq
\label{neutral current}
\overline{ {\nu_\zeta}_L } \gamma^\mu {\nu_\zeta}_L= -\frac{\cos^2 \theta}{2} \bar N_1 \gamma^\mu \gamma_5 N_1 
-\frac{\sin^2 \theta}{2} \bar N_2 \gamma^\mu \gamma_5 N_2 + i \cos\theta \sin\theta  \bar N_2 \gamma^\mu  N_1 ~~.
\eeq
The interaction of the $Z$ with a couple of $N_{1}$ or $N_{2}$ is axial, while 
the one with two different $N_i$ is a vector interaction. 
As for the charged current:
\bea
{\cal L}^{CC}_\zeta 
    = \frac{g }{\sqrt{2}} W_\mu^- \bar \zeta_L \gamma^\mu (i \cos\theta ~P_L N_1 + \sin\theta ~P_L N_2)
    + h.c.~~.\no
\label{charged current no mix}
\eea 
The Dirac mass can be written in terms of the Yukawa coupling $y_{\zeta}$ and the Higgs VEV $v$ as $ m_D= y_{\zeta}\,v/\sqrt{2}$. Hence the interaction of the 
new neutrinos with the Higgs field reads:
\bea
{\cal L}^{H}_\zeta 
&=& -\frac{ m_{D}}{v}  
\left( \overline{ {\nu_\zeta}_R}~ {\nu_\zeta}_L  + \overline{ {\nu_\zeta}_L}~ {\nu_\zeta}_R \right) ~H\\
&=& -\frac{ m_{D}}{v}  \left[  \cos\theta \sin\theta \left({\bar{N}_1}N_1 +{\bar{N}_2}N_2  \right) 
- i\, \cos(2\theta)~ {\bar{N}_1}\gamma_5 N_2 \right] ~H~.  \no
\eea

\vskip .5cm

\subsubsection{Promiscuous heavy leptons}
\label{mixing}

In this section we consider the possibility that the new heavy leptons mix with the SM leptons.
For clarity of presentation we assume that the heavy neutrinos mix only with one SM neutrino of flavor 
$\ell$ ($\ell=e,\mu,\tau$). We refer to \cite{Frandsen:2009fs} for the discussion of the general case. 
The entries of the mass matrix are:
\begin{equation}
-{\cal L}= \frac{1}{2} ( \begin{array}{ccc} 
 \overline{\nu_{\ell L}}& \overline{\nu_{\zeta L}} & \overline{(\nu_{\zeta R})^c}  \end{array} )
\left( \begin{array}{ccc}  {\cal O}(\text{eV}) & {\cal O}(\text{eV}) & m' \cr  {\cal O}(\text{eV}) &  {\cal O}(\text{eV}) & m_D \cr 
m' & m_D &  m_R \end{array} \right)  
\left( \begin{array}{c} (\nu_{\ell L})^c \cr (\nu_{\zeta L})^c \cr \nu_{\zeta R}  \end{array} \right) + h.c.~~.
\label{generalmatrix}
\end{equation} 
The measured values of the light 
neutrino masses suggest the entries of the upper 2$\times$2 block  to be of  ${\cal O}(\text{eV})$ while the remaining entries are expected to be at least of the order of the electroweak energy scale. Given such a hierarchical structure and up to small corrections of ${\cal O}(\text{eV} / M_{1,2}) \lesssim 10^{-11}$,
one obtains the following form for the unitary matrix which diagonalises \eqref{generalmatrix}:
\begin{equation}
\left( \begin{array}{c}
\nu_{\ell L} \cr  \nu_{\zeta L} \cr  (\nu_{\zeta R})^c  \end{array} \right) =  V
\left( \begin{array}{c} P_L N_0 \cr P_L N_1 \cr P_L N_2  \end{array} \right) ~~,~~
V=\left( \begin{array}{ccc} \cos \theta' & i \cos\theta \sin\theta' & \sin\theta \sin\theta' 
\cr -\sin\theta' & i \cos\theta \cos\theta' & \sin\theta \cos\theta' \cr 
0 & -i \sin\theta & \cos\theta \end{array} \right).
\label{Eq: mixing matrix}
\end{equation}
$N_{0,1,2}$ are the new (Majorana) mass eigenstates and
\begin{equation}
\tan \theta' = \frac{m'}{m_D} ~~~, ~~~~~\tan 2 \theta= 2 ~\frac{m'_D}{m_R} ~~~,
~~~~~{m'_D}^2\equiv m_D^2+m'^2~~~.
\label{eqtheta}
\end{equation}
The light neutrino $N_0$ has a mass of ${\cal O}(\text{eV})$. Up to corrections of ${\cal O}(\text{eV})$, 
the heavy neutrinos $N_{1,2}$ have masses given by: 
\beq
M_1=\frac{m_R}{2} \left( \sqrt{1+4 \frac{{m'_D}^2}{m_R^2}} -1 \right) ~~,~~~~
M_2=\frac{m_R}{2} \left( \sqrt{1+4 \frac{{m'_D}^2}{m_R^2}} +1 \right) ~.
\label{eq:M1M2}
\eeq
Notice that the smaller is $m_R$, the more the neutrinos $N_1$ and $N_2$ become (the two Weyl components of) 
a Dirac state.

\noindent
Including the neutrino of flavor $\ell$, the neutral fermion current in Eq.~(\ref{neutral current}) 
is replaced with 
\begin{eqnarray}
\bar {\nu_\ell}_L \gamma^\mu {\nu_\ell}_L+
\bar {\nu_\zeta}_L \gamma^\mu {\nu_\zeta}_L \nonumber &=& -\frac{1}{2} \bar N_0 \gamma^\mu \gamma_5 N_0 
-\frac{\cos^2 \theta}{2} \bar N_1 \gamma^\mu \gamma_5 N_1 
-\frac{\sin^2 \theta}{2} \bar N_2 \gamma^\mu \gamma_5 N_2 
\\
&+& i \cos\theta \sin\theta  \bar N_2 \gamma^\mu  N_1 \ ,
\label{eq:NCfull}
\end{eqnarray}
while the charged current terms in the Lagrangian, Eq.~(\ref{charged current no mix}), become 
\beq
{\cal L}^{CC}_\zeta = \frac{g }{\sqrt{2}} W_\mu^- \bar \zeta_L \gamma^\mu 
(-\sin\theta' P_L N_0 + i \cos\theta \cos\theta' ~P_L N_1 + \sin\theta \cos\theta' ~P_L N_2)+ h.c.
\label{eq:CCfull}
\eeq 
and
\beq
{\cal L}^{CC}_\ell = \frac{g }{\sqrt{2}} W_\mu^- \bar \ell_L \gamma^\mu 
(\cos\theta' P_L N_0 + i \cos\theta \sin\theta' ~P_L N_1 + \sin\theta \sin\theta' ~P_L N_2)+ h.c.
\label{eq:CCfull2}
\eeq
Notice that the neutral current remains flavor diagonal at tree-level \footnote{The neutral current is not flavor diagonal in models with TeV scale right handed neutrinos involved in the see-saw mechanism for the light SM neutrino masses, see {\em e.g.}~\cite{del Aguila:2005pf}.}, hence the heavy neutrinos couple to the SM ones only through the charged current interactions at this order.
This is a distinctive feature of our TeV neutrino physics. 
The SM like Yukawa interactions lead to the following terms involving the Higgs: 
\bea
{\cal L}^{H}_\zeta 
  =- \frac{m_D}{v}  (1+\frac{m'^2}{m_D^2}) \cos\theta'
[   \sin\theta \cos\theta  ({\bar{N}_1}N_1 +{\bar{N}_2}N_2) 
  - i  \cos(2\theta)  ~ {\bar{N}_1}\gamma_5 N_2 ]~ H~ .
\label{Eq:Yukawa's}
\eea

\noindent
The low energy effective theory we will use for determining the interesting signals 
for LHC phenomenology \cite{Foadi:2007ue,Appelquist:1999dq}  is the effective Lagrangian for Vanilla Technicolor.

\subsection{LHC phenomenology for the natural heavy lepton family}
In this section we investigate aspects of the phenomenology related to the interplay between 
the new weekly coupled sector, i.e. the heavy leptons with its mixing with the SM fermions, 
and the new strongly coupled sector breaking the electroweak symmetry dynamically \cite{Frandsen:2009fs}. 
We consider only the MWT global symmetries relevant for the electroweak sector, i.e. the subsector  
$SU(2)\times SU(2)$ spontaneously breaking to $SU(2)$. The low energy spectrum contains, besides 
the composite Higgs, two $SU(2)$ triplets of (axial-) vector spin one mesons.  
The effective Lagrangian has been introduced in \cite{Foadi:2007ue,Belyaev:2008yj}. 
The spin one massive eigenstates are indicated with $R_{1}$ and $R_2$ and are linear combinations 
of the composite vector/axial mesons of MWT and the weak gauge boson eigenstates. 
{ We have implemented the $SU(2)\times SU(2)$ TC sector in CalcHEP \cite{Pukhov:2004ca}
using the LanHEP module \cite{Semenov:2008jy} to generate the Feynman rules in \cite{Belyaev:2008yj}. 
We have added the new leptons to this implementation for the present study.}

\subsubsection{Production and decay of the new leptons }

\begin{figure}
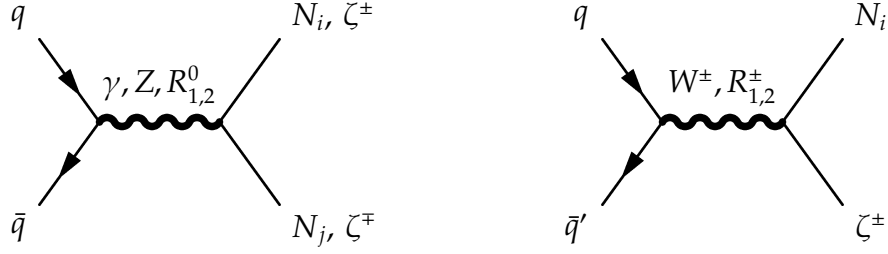

\begin{center}
\parbox{40mm}{\begin{fmfchar*}(40, 22)
\fmflabel{$q$}{qi}
\fmflabel{$\bar q$}{qibar}
\fmflabel{$N_j, \, \zeta^\mp$}{lf}
\fmflabel{$N_i, \, \zeta^\pm$}{lfbar}
\fmfleft{qibar,qi}
\fmfright{lf,lfbar}
\fmf{fermion}{qi,vi,qibar}
\fmf{plain}{lfbar,vf,lf}
\fmf{photon, label=$\gamma,, Z,, R_{1,, 2}^0$, width=1mm, label.side=left}{vi,vf}
\end{fmfchar*}}
\qquad\qquad\qquad\qquad
\parbox{40mm}{\begin{fmfchar*}(40, 22)
\fmflabel{$q$}{qi}
\fmflabel{$q'$}{qibar}
\fmflabel{$\zeta^\pm$}{lf}
\fmflabel{$N_i$}{lfbar}
\fmfleft{qibar,qi}
\fmfright{lf,lfbar}
\fmf{fermion}{qi,vi,qibar}
\fmf{plain}{lfbar,vf}
\fmf{plain}{vf,lf}
\fmf{photon, label=$R_{1,, 2}^\pm$, width=1mm, label.side=left}{vi,vf}
\end{fmfchar*}}
\vspace{7mm}
\caption{Feynman diagrams for direct production of the heavy leptons in MWT with $i,j=1,2.$}
\label{fig:feynman diagrams}
\end{center}
\end{figure}

The heavy leptons may be directly produced through the charged- and neutral current interactions:
\bea
pp &\to& W^\pm/R_{1,2}^\pm \to \zeta^\pm N_i \ ,
\nn
pp &\to& Z/\gamma/R_{1,2}^0 \to \zeta^+ \zeta^- \ , \quad pp \to Z/R_{1,2}^0 \to N_i N_j \ , \quad i,j=1,2 .
\label{Eq:Walking prod}
\eea 
The corresponding Feynman diagrams are given in Fig.~\ref{fig:feynman diagrams}. 
The direct production cross sections for the heavy leptons in MWT are largely independent of the parameters 
associated to the TC sector:
in the direct production of $\zeta^+ \zeta^-$ the only free parameter is the mass of the 
charged lepton $M_\zeta$; the direct production of $N_i N_j$ and $\zeta^\pm N_i$ depends, 
in addition to the masses of the leptons, on the $V$ matrix entries of  Eq.~\eqref{Eq: mixing matrix} 
as follows from Eqs.~\eqref{eq:NCfull} and~\eqref{eq:CCfull}. 
Plots of the LHC cross sections for $pp\rightarrow \zeta^+ \zeta^-, \zeta^\pm N_i, N_i N_i$ 
can be found in \cite{Frandsen:2009fs}.

\begin{figure}[h!]
\begin{center}
{\includegraphics[height=8cm,width=7.5cm]{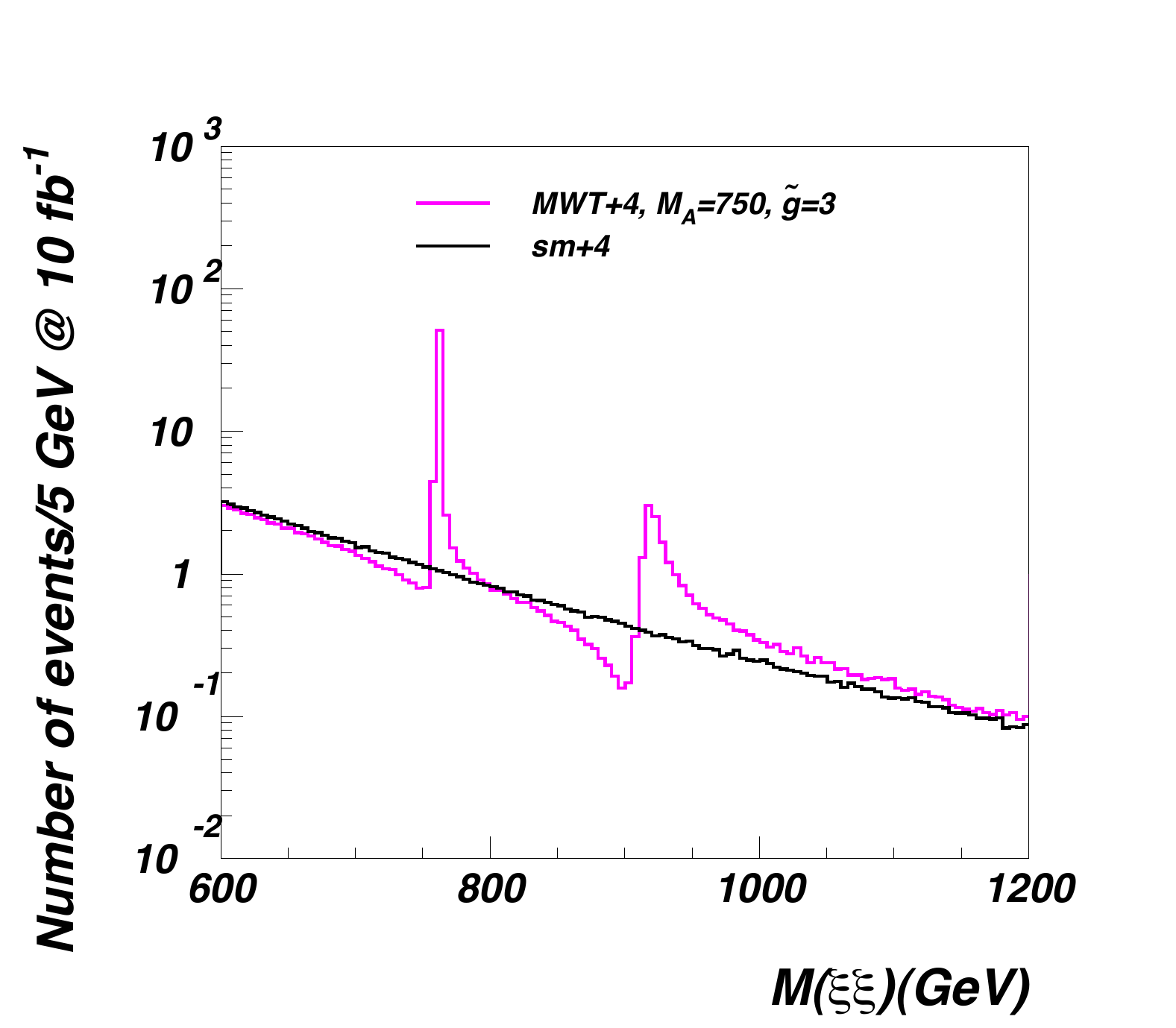}
\hskip .5cm
\includegraphics[height=8cm,width=7.5cm]{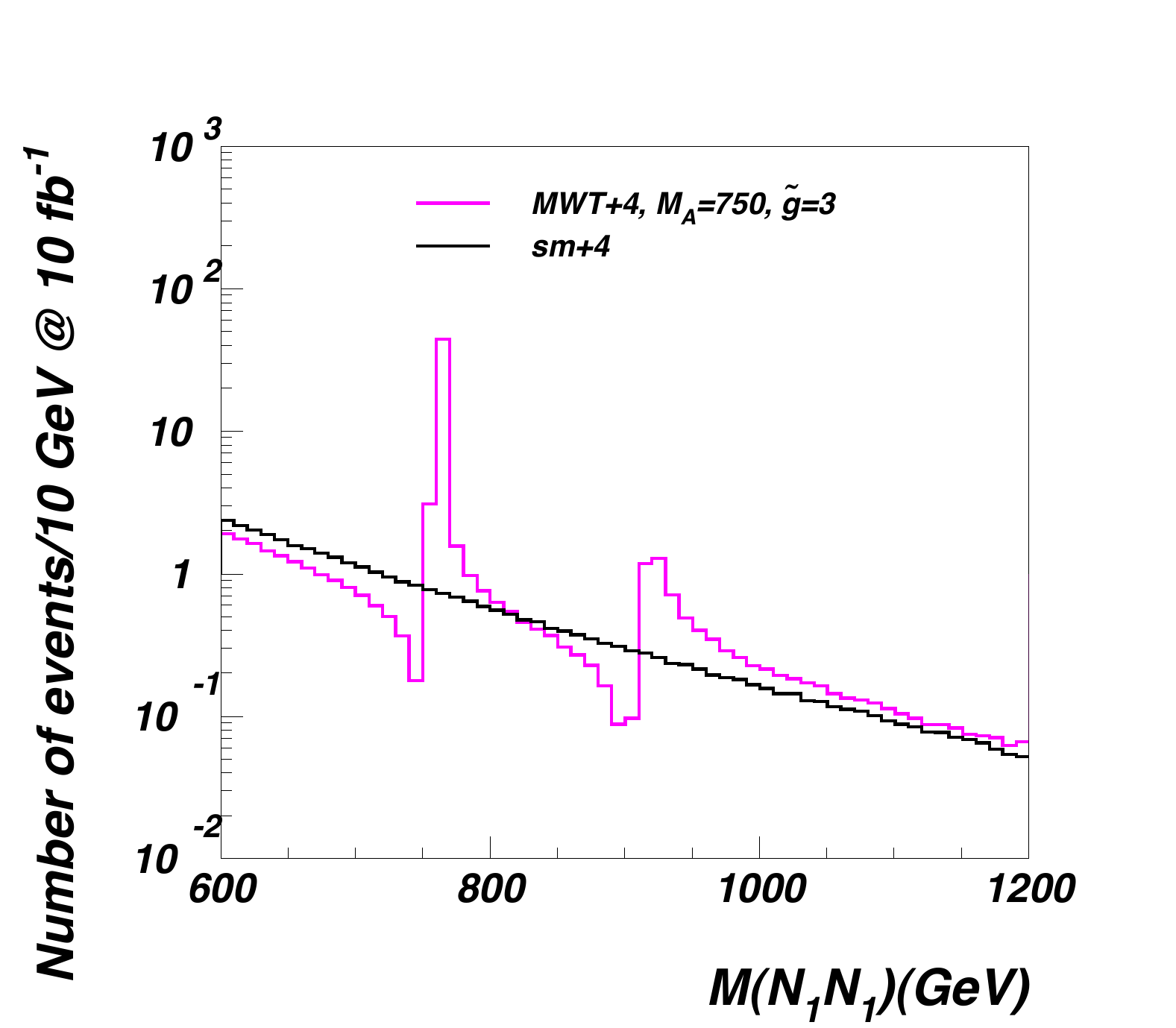}
}
\caption{Invariant mass distributions, $M(\zeta^+ \zeta^-)$ and $M(N_1 N_1)$ in pair production 
of $\zeta^+ \zeta^-$ (left) and $N_1 N_1$ (right) in the MWT (purple/gray) and when the leptons are added to the SM 
(black). The plots are for LHC at $7$~TeV in the centre of mass and $10$ fb$^{-1}$.  For the second plot we assumed unit mixing matrix element. 
In both frames the first peak corresponds to the $R_1$ resonance while the second corresponds to the $R_2$ resonance. 
The neutral current coupling of $N_1 N_1$ is axial and therefore more dominated by the $R_1$ resonance. 
We take the values $\tilde{g}=3, M_A=750, S=0.1$ for the parameters of the TC sector. 
The corresponding masses and widths of $R_{1,2}$ are given in \cite{Belyaev:2008yj}. $sm+4$ and MWT$+4$ in the legend refer to respectively to the SM plus the fourth family and to the MWT with the fourth family implementation on event generators.}
\label{fig:Walking direct prod}
\end{center}
\end{figure}

The final state distributions arising from the direct production of the leptons depend 
on the specific parameters of the TC sector. In particular $R_1$  is a (mostly) axial-resonance 
and $R_2$ is a (mostly) vector-resonance, so $R_1$ mixes mostly with the Z boson while $R_2$ mixes significantly 
with the photon. Consequently, the invariant mass distribution of the heavy neutral leptons $N_i N_i$ 
will be relatively more dominated by the $R_1$ resonance compared to the charged leptons $\zeta^+ \zeta^-$. 
This is demonstrated in Fig.~\ref{fig:Walking direct prod}. 
The masses and widths of $R_{1,2}$ as a function of $M_A, \tilde{g}$ and $S$ are given
in \cite{Belyaev:2008yj}. 

Production of the heavy leptons can also proceed via the Yukawa-type couplings to the composite Higgs 
following from Eq.~\eqref{Eq:Yukawa's}. We are following \cite{Foadi:2007ue} for an effective 
way to give masses to the SM fermions in the MWT setup. The composite Higgs may itself be produced 
through either gluon fusion, vector boson fusion or in association with a SM vector boson:
\bea
gg \to H \ , \quad pp \to q q' H \ , \quad  pp \to H Z / H W     \ .
\eea
The process $gg \to H \to N_i N_i \to  WW\mu\mu $ (within the SM framework) was recently considered 
in \cite{CuhadarDonszelmann:2008jp,Kribs:2007nz,Keung:2011zc} where also a 4th generation of quarks were included that enhance 
the $gg\to H$ cross section \cite{Kribs:2007nz} (for a recent review of the scenario in which the new 
leptons are accompanied by a fourth generation of quarks, see \cite{Holdom:2009rf}).
In the MWT class of models, this channel is not expected to be enhanced compared to the SM 
since the techniquarks are not colored. The same applies for vector boson fusion production 
of the composite Higgs \cite{Belyaev:2008yj}. 

On the other hand, the associate production of the composite Higgs can actually be enhanced in MWT models 
\cite{Belyaev:2008yj,Zerwekh:2005wh}, in particular the $pp \to H Z/ H W$ channel, due to 
the presence of a light axial-vector resonance as shown in \cite{Belyaev:2008yj}. 
We will therefore focus on the associate production of the Higgs.

\bigskip
The relevant expressions for the decay widths of the heavy leptons can be found in \cite{Frandsen:2009fs}. 
Clearly, the decay patterns depend on the mass hierarchy and the mixings between the leptons. 
Notice that, in the regime where $M_\zeta > M_1$ and $M_H > M_1$, 
$N_1$ could decay only via its mixing with SM leptons, $N_1 \to \ell^\pm W^\mp$, whose vertex is proportional 
to $\sin \theta^{\prime}$.
The decay width and lifetime of $N_1$ are displayed in \cite{Frandsen:2009fs}. 
We note that a value of $\sin\theta^{\prime} \sim 10^{-6}$ would yield a decay length of $\sim$ 1 m, 
which is enough for a relativistically boosted particle to escape detection at the LHC and be considered as 
missing energy in the various processes \cite{Basso:2008iv}. 

\subsubsection{Collider signatures of heavy leptons with an exact flavor symmetry}
 
Let us first consider the limit in which the new leptons do not mix appreciably with the SM ones.
If $M_\zeta > M_1$, then $N_1$ constitutes a long lived neutral particle and will give rise to missing momentum 
$\mpt$ and missing energy $\met$ signals. In particular, the decay mode $H\to N_1 N_1$ gives rise to an 
invisible partial width of the composite Higgs.

\begin{figure}[h!]
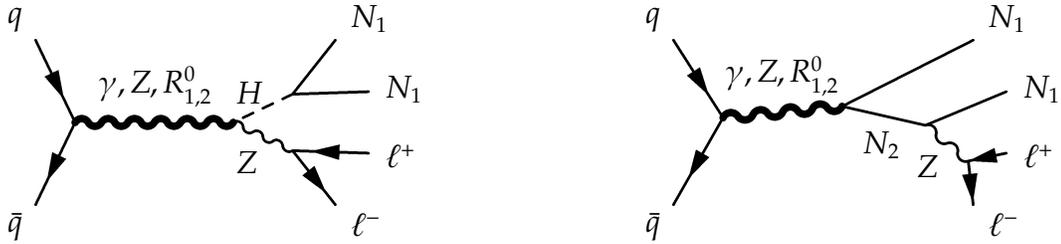

\begin{center}
\vspace{8mm}
\parbox{50mm}{\begin{fmfchar*}(50, 22)
\fmflabel{$q$}{q2}
\fmflabel{$\bar{q}$}{q1}
\fmflabel{$N_1$}{lp1}
\fmflabel{$N_1$}{lm1}
\fmflabel{$\ell^+$}{lp2}
\fmflabel{$\ell^-$}{lm2}
\fmfleft{q1,q2}
\fmfright{lm2,lp2,lm1,lp1}
\fmf{fermion, tension=3}{q2,v1,q1}
\fmf{photon, tension=1.5, label=$\gamma,, Z,, R_{1,, 2}^0$, width=1mm, label.side=left}{v1,v2}
\fmf{dashes, tension=2, label=$H$, label.side=right, label.dist=0.7mm}{v3,v2}
\fmf{photon, tension=2, label=$Z$, label.side=right}{v2,v4}
\fmf{plain}{lp1,v3,lm1}
\fmf{fermion}{lp2,v4,lm2}
\end{fmfchar*}}
\qquad\qquad\qquad\qquad
\parbox{50mm}{\begin{fmfchar*}(50, 22)
\fmflabel{$q$}{q2}
\fmflabel{$\bar{q}$}{q1}
\fmflabel{$N_1$}{lp1}
\fmflabel{$N_1$}{lm1}
\fmflabel{$\ell^+$}{lp2}
\fmflabel{$\ell^-$}{lm2}
\fmfleft{q1,q2}
\fmfright{lm2,lp2,lm1,lp1}
\fmf{fermion, tension=3}{q2,v1,q1}
\fmf{photon, tension=2.5, label=$\gamma,, Z,, R_{1,, 2}^0$, width=1mm, label.side=left}{v1,v2}
\fmf{plain, tension=2, label=$N_2$, label.side=right}{v2,v3}
\fmf{photon, tension=2, label=$Z$, label.side=right}{v3,v4}
\fmf{plain}{lp1,v2}
\fmf{plain}{v3,lm1}
\fmf{fermion, tension=2}{lp2,v4,lm2}
\end{fmfchar*}}
\vspace{7mm}
\caption{Feynman diagrams for the $\ell^+\ell^- +\mpt$ signal due to heavy leptons in the MWT model.}
\label{fig:missEfromH}
\end{center}
\end{figure}

As pointed out in \cite{Belyaev:2008yj,Zerwekh:2005wh}, the cross-section for $ZH$ production can be enhanced 
in MWT models because the axial-vector resonance can be light \cite{Foadi:2007ue,Appelquist:1998xf}.
Here the $\ell^+\ell^- +\mpt$ final state will receive contributions both from $ZH$ and $N_1, N_2$ production, 
as shown in Fig.~\ref{fig:missEfromH}.
We therefore study the proces $pp \to Z N_1 N_1 \to \ell^+ \ell^- + \mpt$. We consider limiting values 
of the parameters such that either the Higgs or the $N_2$ state is too heavy to contribute significantly 
as well as parameters where both contribute in the process. 

The acceptance cuts relevant for LHC are
\begin{eqnarray}
|\eta^{\ell}| < 2.5\ , \quad  p_T^\ell > 10 \mbox{ GeV} \ , \quad   \Delta R(\ell\ell) > 0.4  \ .
\label{eq:cuts1}
\end{eqnarray}
Here $\ell$ is a charged lepton, $\eta^{\ell}$ and $p_T^\ell$ are the pseudo-rapidity and transverse momentum 
of a single lepton while $\Delta R$ measures the separation between two leptons in the detector. 
$\Delta R$ is defined via the difference in azimuthal angle $\Delta\phi$ and rapidity $\Delta\eta$ 
between two leptons as $\Delta R\equiv \sqrt{(\Delta\eta)^2+(\Delta\phi)^2}$. 

The main sources of background come from di-boson production followed by leptonic decays 
\cite{Godbole:2003it,Davoudiasl:2004aj,Meisel:2006}
\beq
ZZ \to \ell^+\ell^- \nu \bar{\nu} \ , \ W^+ W^- \to \ell^+ \nu \ell^-\bar{\nu} \ , \ ZW \to \ell^+\ell^- \ell \nu 
\eeq
where in the last process the lepton from the W decay is missed.
 
We impose the additional cuts
\begin{eqnarray}
 |M_{\ell \ell}-M_Z| < 10 \mbox{ GeV} \ , \quad {\rm and} \quad \Delta \phi(\ell\ell) < 2.5  \ .
\label{eq:cuts2}
\end{eqnarray}
The first is meant to reduce the WW background by requiring the invariant mass of the lepton pair 
to reproduce the Z boson mass. The second cut on the azimuthal angle separation together with taking 
large ~$\mpt$ reduces potential backgrounds such as single Z production + jets 
with fake ~$\mpt$ \cite{Godbole:2003it,Davoudiasl:2004aj}.

\begin{figure}[h!]
{\includegraphics[height=8.5cm,width=9.cm]{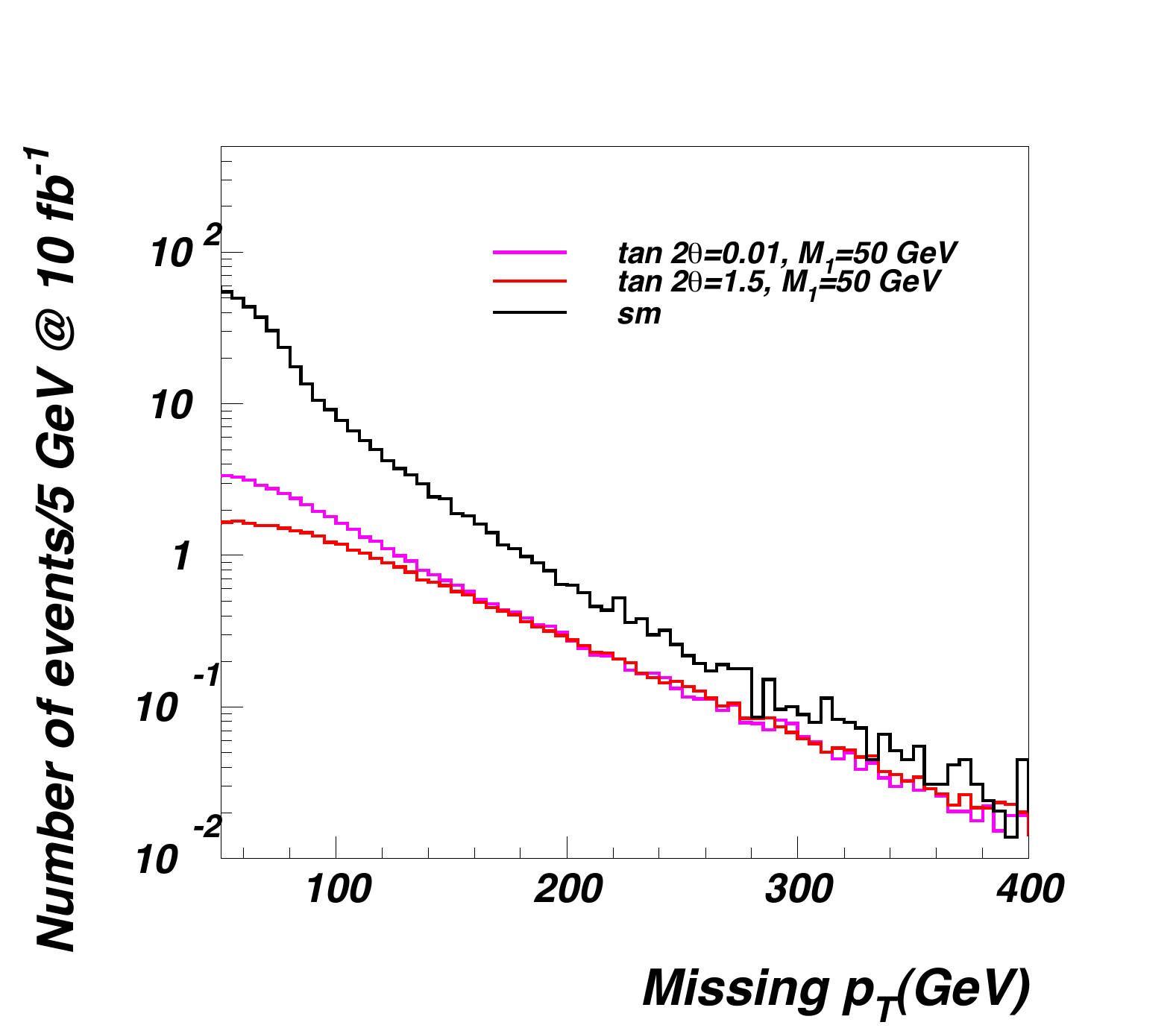}
\hskip .1cm
\includegraphics[height=8.5cm,width=9.cm]{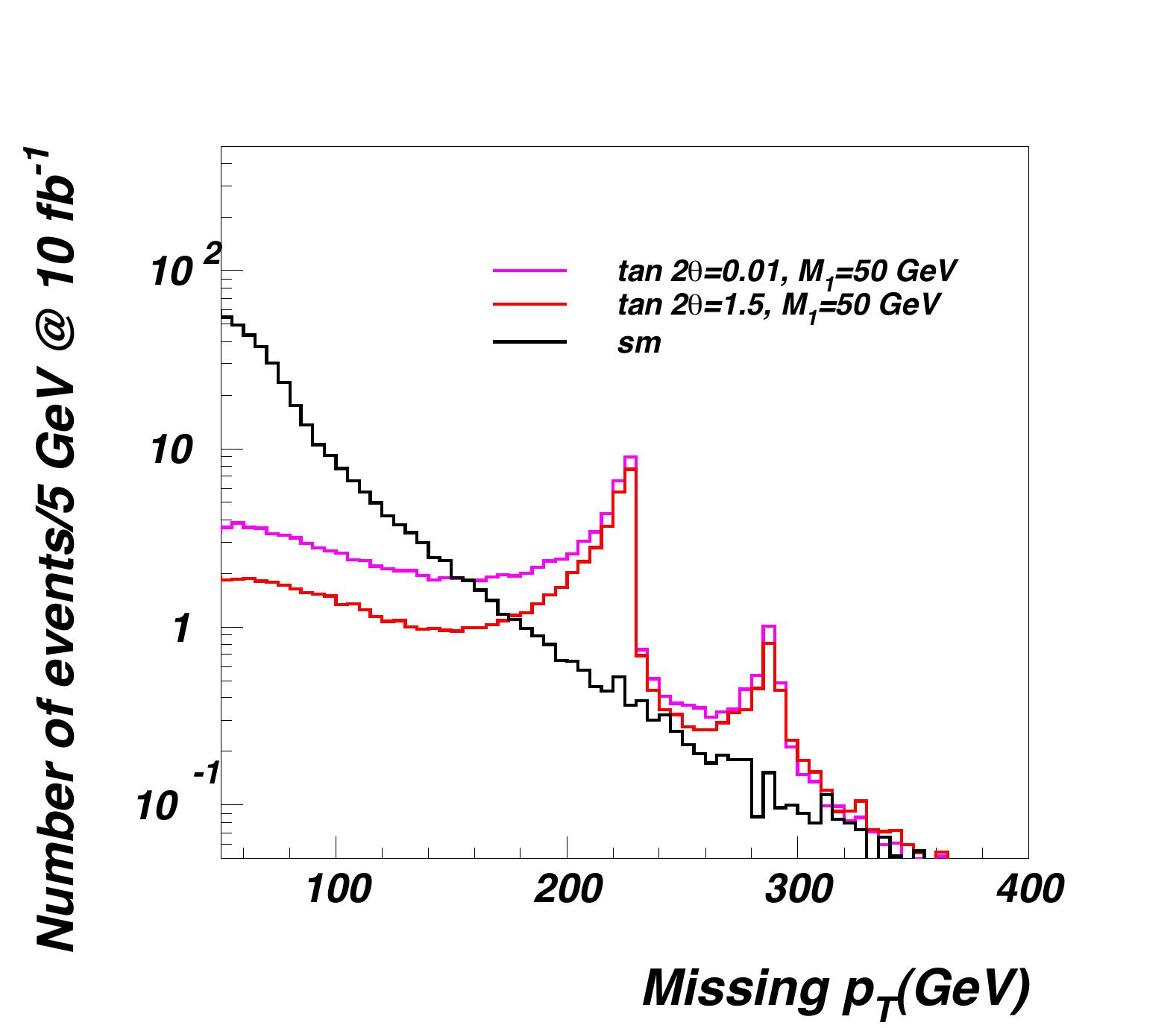}
}
\caption{$\ell^+\ell^- +\mpt$ signal from $pp \to Z N_1 N_1 \to \ell^+ \ell^- N_1 N_1$ { in the SM (left) and in the MWT (right)}. In black the full SM background. The purple/gray line corresponds to $M_H=150$ GeV, $M_1=50$ GeV, $\tan 2\theta$=0.01 (to decouple the second neutrino we set $M_2$ to $2000$ TeV); the red (bottom) line corresponds to $M_H=150$ GeV, $M_1=50$ GeV, $\tan 2\theta$=1.5 ($M_2=175$ GeV).}
\label{fig:Walking invisible higgs}
\end{figure}

\begin{figure}[h!]
{\includegraphics[height=8.4cm,width=9cm]{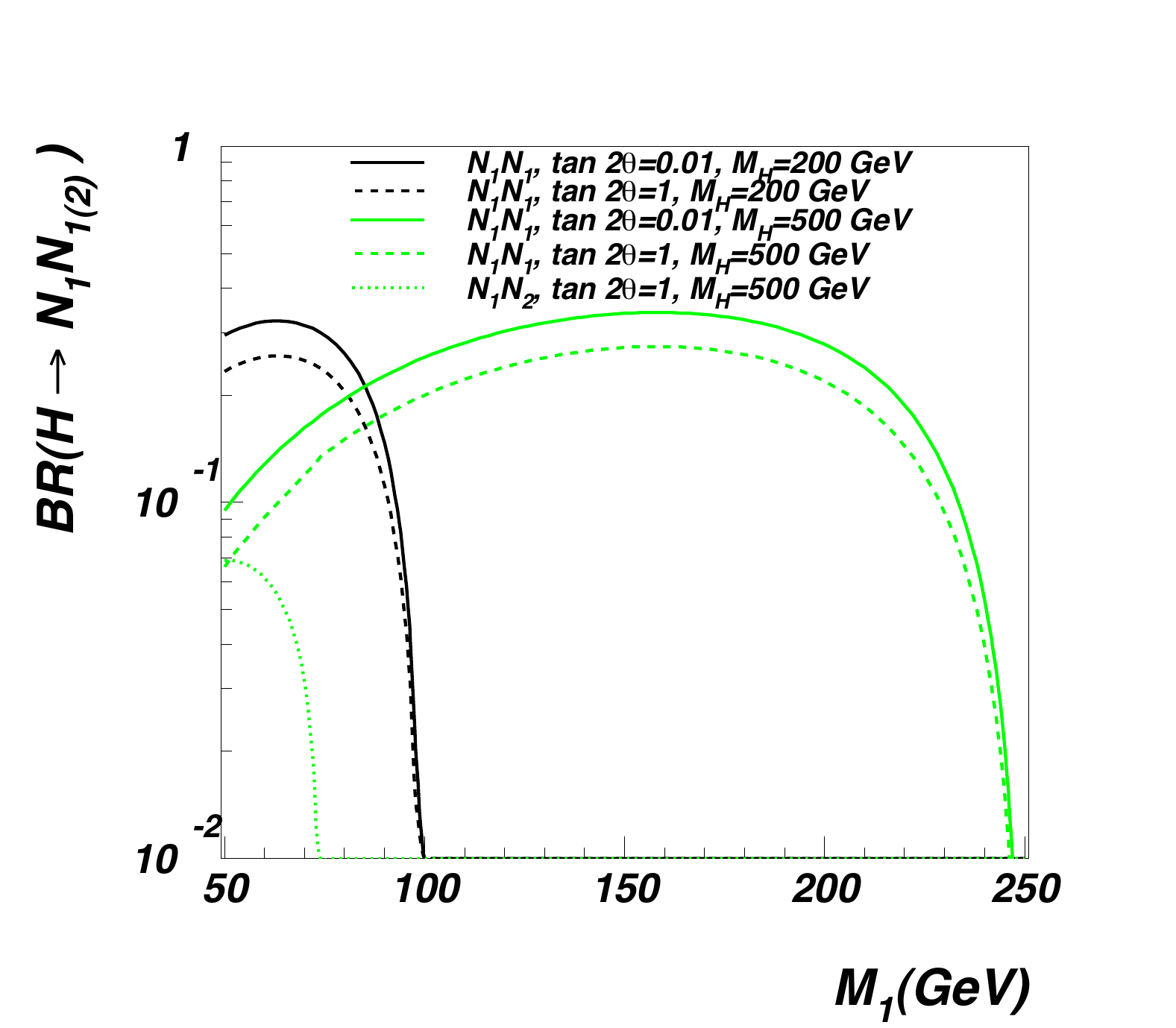}
\includegraphics[height=7.5cm,width=8.5cm]{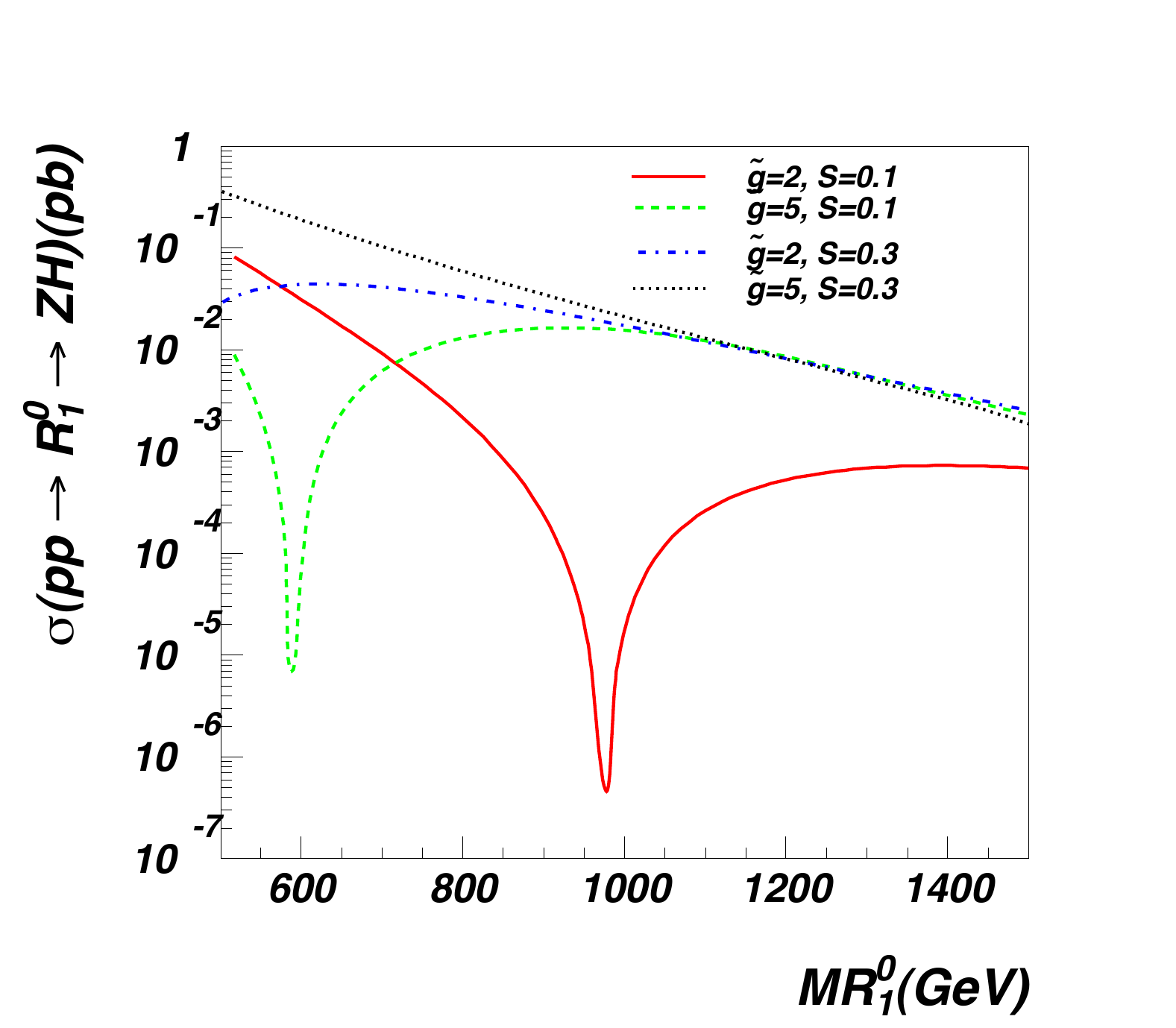}}
\caption{The invisible branching fraction of the composite Higgs to either $N_1 N_1$ or $N_1 N_2$ as a function of $M_1$ (left). The production cross-section times branching ratio $\sigma(pp \to R_1^0 \to ZH)$ as a function of $M_{R_1}$ for $S=0.1$ (red) and $S=0.3$ (black) as well as for $\tilde{g}=2$ (dotted lines) and $\tilde{g}=5$ (solid lines).}
\label{fig:branching}
\end{figure}

The results are given in Fig.~\ref{fig:Walking invisible higgs} assuming a fully invisibly decaying Higgs. 
{ On the left panel we show the signal and background arising from the SM featuring the new heavy leptons.} 
On the right hand panel we show the same signal but in the MWT model. For ~$\mpt > 100$ GeV where the signal 
could potentially be observed, the Higgs production channel dominates and in the MWT model a very 
distinct ~$\mpt$ distribution arises due to the effect of the $R_1$ resonance. 
While invisible decays of a SM model Higgs at most appear as an excess of events compared 
to the background in {\em e.g.}~$\mpt$ distributions, the presence of a light axial-vector resonance 
results in a peaked distribution, different from the shape of the background, making it a more striking signal. 
In general the peak of the $R_1$ resonance will increase for larger values of $\tilde{g}$ and decrease 
with the mass of $R_1$. 
In addition, due to the smallness of the $S$ parameter, the electroweak contributions 
to the $ZH$ production cross section cancel the contribution from the TC sector for small 
$\tilde{g}$ at particular values of $M_{R_1}$, causing a strong depletion of the signal. 
We show this in the left panel in Fig.~\ref{fig:branching}. 
However, for a light axial resonance a relatively large value of $\tilde{g}$ is favored by unitarity 
arguments \cite{Foadi:2008xj,Foadi:2008ci} and electroweak precision observables 
\cite{Belyaev:2008yj,Foadi:2007ue,Foadi:2007se}.

The true invisible branching fraction of the Higgs will depend on the mass of the $N_1$ particle, 
as shown in Fig.~\ref{fig:branching}. 
An invisible partial width of the composite Higgs has been searched for at LEP in the proces $e^+ e^- \to HZ$ 
with $Z$ decaying hadronically \cite{Abbiendi:2006gd}. However, no limits were achieved for $M_H > 114$ GeV. 
The same final state from $HZ/HW$ production in $p p$ collisions has been considered by several authors, 
also recently in the context of TC models (see for instance \cite{Foadi:2008qv} and references therein).
The LHC discovery potential for an invisibly decaying Higgs in this final state at LHC 
has also been investigated at detector level \cite{Meisel:2006,Gagnon:2005}. It was found that the $HW$ 
mode is not promising \cite{Gagnon:2005} while the ZH mode remains challenging. 
With a SM production cross section of $HZ$ a significance of 3.43 $\sigma$ was achieved at $M_H=160$ GeV, 
dropping to a 2 $\sigma$ excess at $M_H=200$ GeV \cite{Meisel:2006}.

{Thanks to the possible increase in the $HZ$ production cross section found in \cite{Belyaev:2008yj} together 
with the resonance like structure in the $\mpt$ distribution displayed in Fig.~\ref{fig:Walking invisible higgs}, 
we believe that this channel could be interesting to investigate the interplay between new long-lived heavy 
neutrinos and composite vector states at LHC. }

\bigskip
\bigskip
If instead $M_\zeta < M_1$, then $\zeta$ can be a long-lived CHAMP (Charged Massive Particle). 
Collider signatures of long lived charged leptons could either be displaced vertices or, if the charged lepton 
decays outside the detector, a muon like signal for which the heavy mass should be reconstructable. 
Such a long-lived CHAMP arises in several scenarios and has been studied in some detail, 
for a review see {\em e.g.}~\cite{Fairbairn:2006gg}.

In \cite{Allanach:2001sd} a Herwig based study of the LHC reach for long-lived leptonic CHAMPs was considered. 
With a discovery criterion of 5 pairs of reconstructed opposite charge heavy leptons a reach of 
$M_\zeta =950$ GeV at $100 \textrm{ fb}^{-1}$ was found, reduced to $800$ GeV without specialized triggers. 
We can expect this reach to be improved in our model by searching also for single $\zeta$ production channel. 
Additionally, it was found that long-lived leptonic and scalar CHAMPS could be distinguished for masses up to 
$580$ GeV. 
      
The discovery potential for long-lived CHAMPS has also been studied at detector level for LHC. 
The CMS and Atlas collaboration has considered various long-lived CHAMPS \cite{Giagu:2008im}. 
From their results we infer that 3 signal events with less than one background event could be observed 
in CMS with 1 $\textrm{fb}^{-1}$ and $M_\zeta \sim 300$ GeV and similarly in Atlas. 
More precisely, in \cite{Giagu:2008im} 3 signal events could be seen in direct pair production of $300$ GeV 
KK taus with a pair production cross-section of $20$ fb, very similar to what we found.    
{Fig.~\ref{fig:Walking direct prod} shows that it is interesting to investigate the invariant 
mass distribution of the leptonic CHAMP.}

\subsection{Collider signatures of promiscuous heavy leptons}

\begin{figure}[h!]
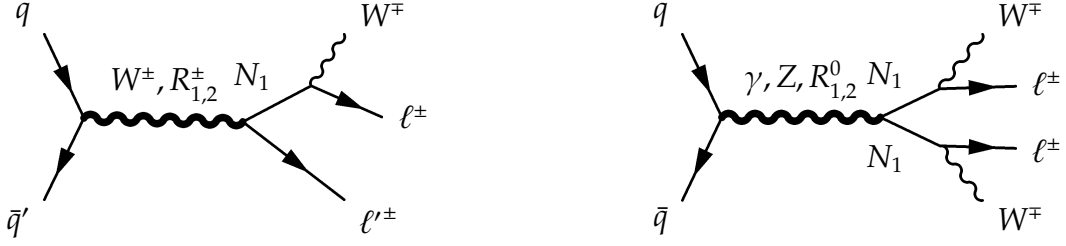

\begin{center}
\vspace{8mm}
\parbox{50mm}{\begin{fmfchar*}(50, 22)
\fmflabel{$q$}{q2}
\fmflabel{$\bar{q}$}{q1}
\fmflabel{$W^\mp$}{W}
\fmflabel{$\ell^\pm$}{lp1}
\fmflabel{${\ell'}^\pm$}{lp2}
\fmfleft{q1,q2}
\fmfright{lp2,lp1,W}
\fmf{fermion, tension=3}{q2,v1,q1}
\fmf{photon, tension=1.5, label=$W^\pm,, R_{1,, 2}^\pm$, width=1mm, label.side=left}{v1,v2}
\fmf{plain, tension=2, label=$N_1$, label.side=right}{v3,v2}
\fmf{photon, tension=2}{v3,W}
\fmf{fermion}{v2,lp2}
\fmf{fermion}{v3,lp1}
\end{fmfchar*}}
\qquad\qquad\qquad\qquad
\parbox{50mm}{\begin{fmfchar*}(50, 22)
\fmflabel{$q$}{q2}
\fmflabel{$\bar{q}$}{q1}
\fmflabel{$W^\mp$}{W1}
\fmflabel{$W^\mp$}{W2}
\fmflabel{$\ell^\pm$}{lp1}
\fmflabel{$\ell^\pm$}{lp2}
\fmfleft{q1,q2}
\fmfright{W2,lp2,lp1,W1}
\fmf{fermion, tension=3}{q2,v1,q1}
\fmf{photon, tension=1.5, label=$\gamma,, Z,, R_{1,, 2}^0$, width=1mm, label.side=left}{v1,v2}
\fmf{plain, tension=2, label=$N_1$, label.side=left}{v2,v3}
\fmf{plain, tension=2, label=$N_1$, label.side=left}{v4,v2}
\fmf{photon}{v3,W1}
\fmf{fermion}{v3,lp1}
\fmf{fermion}{v4,lp2}
\fmf{photon}{v4,W2}
\end{fmfchar*}}
\vspace{7mm}
\end{center}
\caption{Same sign leptons from production of $N_1$. We will consider the case where the W's decay to jets. Such that the final states we consider are $\mu^\pm \mu^\pm jj$ (left) and $\mu^\pm \mu^\pm jjjj$ (right).}
\label{fig:samesign}
\end{figure}

If the heavy leptons mix with the SM leptons, this will give rise to Lepton Number Violating (LNV) 
processes with same sign leptons and jets in the final state, {\em e.g.}
\bea
pp &\to& W^\pm/R_{1,2}^\pm \to \mu^\pm N_i \to  \mu^\pm  \mu^\pm W^\mp \to \mu^\pm \mu^\pm jj
\nn
pp &\to& Z/R_{1,2}^0 \to N_i N_j \to \mu^\pm  \mu^\pm W^\mp  W^\mp \to \mu^\pm \mu^\pm jjjj \ , \quad i=1,2 ,
\label{Eq:mumujjjj}
\eea 
as in Fig.~\ref{fig:samesign}.

The potential for observing the $\mu^\pm \mu^\pm jj$ final state has been extensively studied 
in scenarios with heavy right-handed neutrino singlets, both in the SM with 3 light neutrinos 
\cite{del Aguila:2005pf,del Aguila:2006dx,Han:2006ip,Atre:2009rg} and in the presence of additional new 
physics \cite{Basso:2008iv,delAguila:2007ua}. Same sign lepton final states have been searched for at the 
Tevatron in \cite{Abulencia:2007rd, Aaltonen:2008vt}. 
The $pp \to W^\pm \to \ell^\pm N_i \to \ell^\pm \ell'^\pm j j$ process in the SM with 3 light neutrinos 
was studied in \cite{del Aguila:2007em} at the level of a fast detector simulation. 
While backgrounds for same sign lepton production are smaller than for opposite sign lepton production, 
arising in the Dirac limit, they were found to be significantly larger than previously estimated in parton level 
processes, in particular for $M_i < M_W$.   

Again the production cross-sections are largely unaffected by the presence of heavy vectors. 
However, the shape of the distributions are affected by the presence of the heavy vectors. 

To study these processes we impose jet acceptance cuts in addition to the leptonic acceptance 
cuts given in Eq.~(\ref{eq:cuts1})
\begin{eqnarray}
|\eta^{j}| < 3 \ , \quad  p_T^j > 20 \mbox{ GeV} \ , \quad   \Delta R(\ell j) > 0.5  \ .
\label{eq:cuts3}
\end{eqnarray}

The resulting invariant mass distributions for $\mu^-\mu^- jj$ (left) and $\mu^-\mu^- jjjj$ (right) are 
given below in Fig.~\ref{fig:Walking direct production}. We have taken $\cos\theta=0.7$, $\sin\theta'=0.098$
and $N_2$ decoupled. 
While $\sin\theta'$ determines the mixing between $N_1$ and $\ell$ and therefore is constrained by experiment, 
$\cos\theta$ is not. This means that the production cross section of $N_1 N_1$ potentially is significantly 
larger than the $N_1 \mu$ production cross-section.
If at the same time $N_1$ only decays to $W \mu$, 
we find the result given in the right frame of Fig.~\ref{fig:Walking direct production}.  

\begin{figure}[h!]
{\includegraphics[height=8.5cm,width=8cm]{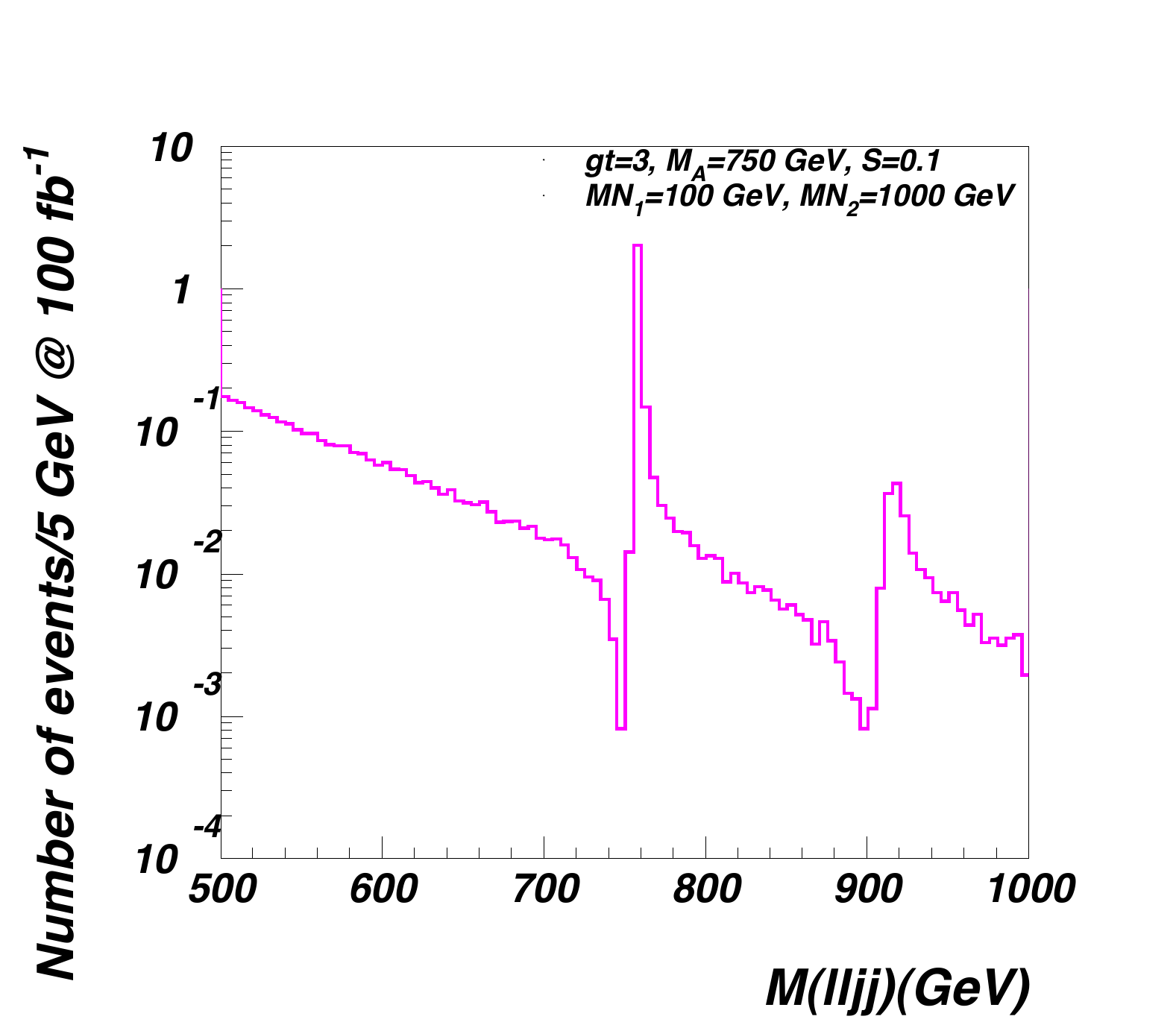}
\hskip .5cm
\includegraphics[height=8.5cm,width=8cm]{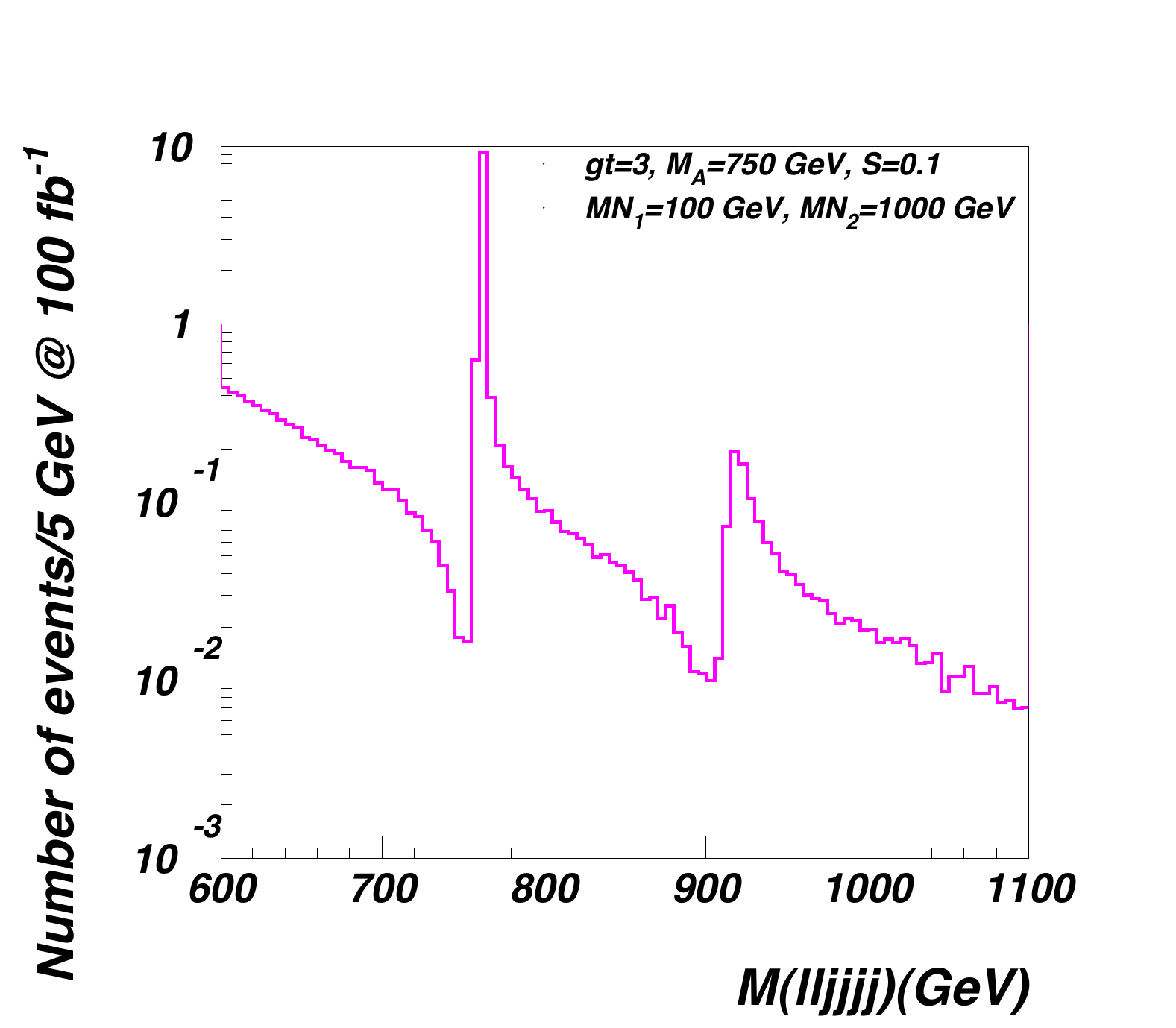}
}
\caption{Invariant mass distributions for LHC at $14$~TeV: $pp \to N_1 \mu^- \to \mu^-\mu^- jj $ (left) and $pp \to N_1 N_1 \to \mu^-\mu^- jjjj$ (right). The parameters in the TC sector are fixed to be $M_A=750$ GeV, $S=0.1, \tilde{g}=3$ while the new lepton sector parameters are $M_1=100 GeV, \cos\theta=0.7, \sin\theta'=0.098$. $N_2$ is decoupled for these parameters.}
\label{fig:Walking direct production}
\end{figure}

The production cross section $\sigma(pp\to R_1)$ scales roughly as $\tilde{g}^{-2}$ and independently of $S$ 
for small $S$. The detailed behavior may be inferred from the branching ratios presented 
in~\cite{Belyaev:2008yj}.
The above shows that the interplay of heavy neutrinos and composite vector mesons can lead to 
striking signatures at the LHC.

\section{Conclusion}

We introduced extensions of the SM in which the Higgs emerges as a composite state. In particular we motivated TC, constructed underlying gauge theories leading to minimal models of TC, compared the different extensions with electroweak precision data and constructed the  low energy effective theory. 

We have then classified the spectrum of the states common to most of the TC models and investigated their decays and associated experimental signals for the LHC. We have  set up the effective description to allow for easy extensions to account for specific features of  given underlying gauge dynamics or for generic models in which the Higgs emerges as a composite state.

We have established important experimental benchmarks for Vanilla, Running, Walking and Custodial Technicolor for the potential discovery of these models at the LHC.

\subsection*{Acknowledgments}
It is a pleasure for us to thank M.~Antola, A.~Belyaev, S.~Catterall, D.D.~Dietrich, R.~Foadi, M.T.~Frandsen,  H.~Fukano-Sakuma, M. Heikinheimo, J.~Giedt, S.B.~Gudnason, C.~Kouvaris, A. Nisati, T.A.~Ryttov, V.~Sanz, J.~Schechter, P.~Schleper, R.~Shrock and K.~Tuominen for having shared part of the work on which this work is based on, long and fruitful collaborations, relevant discussions, criticism and/or careful reading of the manuscript. 

\newpage
\appendix 
 \section{Realization of the generators for MWT and the Standard Model embedding \label{appgen}}

It is convenient to use the following representation of $SU(4)$
\beq S^a = \begin{pmatrix} \bf A & \bf B \\ {\bf B}^\dag & -{\bf A}^T
\end{pmatrix} \ , \qquad X^i = \begin{pmatrix} \bf C & \bf D \\ {\bf
    D}^\dag & {\bf C}^T \end{pmatrix} \ , \eeq
where $A$ is hermitian, $C$ is hermitian and traceless, $B = -B^T$ and
$D = D^T$. The ${S}$ are also a representation of the $SO(4)$
generators, and thus leave the vacuum invariant $S^aE + E {S^a}^T = 0\ $.
Explicitly, the generators read
\beq S^a = \frac{1}{2\sqrt{2}}\begin{pmatrix} \tau^a & \bf 0 \\ \bf 0 &
  -\tau^{aT} \end{pmatrix} \ , \quad a = 1,\ldots,4 \ , \eeq
where $a = 1,2,3$ are the Pauli matrices and $\tau^4 =
\mathbbm{1}$. These are the generators of $SU(2)_\text{V} \times U(1)_\text{V}$.
\beq S^a = \frac{1}{2\sqrt{2}}\begin{pmatrix} \bf 0 & {\bf B}^a \\
{\bf B}^{a\dag} & \bf 0 \end{pmatrix} \ , \quad a = 5,6 \ , \eeq
with
\beq B^5 = \tau^2 \ , \quad B^6 = i\tau^2 \ . \eeq
The rest of the generators which do not leave the vacuum invariant are
\beq X^i = \frac{1}{2\sqrt{2}}\begin{pmatrix} \tau^i & \bf 0 \\
\bf 0 & \tau^{iT} \end{pmatrix} \ , \quad i = 1,2,3 \ , \eeq
and
\beq X^i = \frac{1}{2\sqrt{2}}\begin{pmatrix} \bf 0 & {\bf D}^i \\
{\bf D}^{i\dag} & \bf 0 \end{pmatrix} \ , \quad i = 4,\ldots,9 \ ,
\eeq
with
\beq\begin{array}{r@{\;}c@{\;}lr@{\;}c@{\;}lr@{\;}c@{\;}l}
D^4 &=& \mathbbm{1} \ , & \quad D^6 &=& \tau^3 \ , & \quad D^8 &=& \tau^1 \ , \\
D^5 &=& i\mathbbm{1} \ , & \quad D^7 &=& i\tau^3 \ , & \quad D^9 &=& i\tau^1
\ .
\end{array}\eeq

The generators are normalized as follows
\beq {\rm Tr}\left[S^aS^b\right] =\frac{1}{2}\delta^{ab}\ , \qquad \ , {\rm Tr}\left[X^iX^j\right] =
\frac{1}{2}\delta^{ij} \ , \qquad {\rm Tr}\left[X^iS^a\right] = 0 \ . \eeq

Having set the notation, it is instructive to split the scalar matrix into four two by two blocks as follows:
\begin{equation}  
M=\begin{pmatrix} {\cal X}  & {\cal O}  \\ {\cal O}^T & {\cal Z} \end{pmatrix} \ ,
\end{equation}
with ${\cal X}$ and ${\cal Z}$ two complex symmetric matrices accounting for six independent degrees of freedom each and ${\cal O}$ is a generic complex two by two matrix featuring eight real bosonic fields. ${\cal O}$ accounts for the SM like Higgs doublet and a second copy as well as for the three Goldstones which upon electroweak gauging will become the longitudinal components of the intermediate massive vector bosons. 
The electroweak subgroup can be embedded in $SU(4)$, as explained in detail in \cite{Appelquist:1999dq}. Here $SO(4)$ is the subgroup to which $SU(4)$ is maximally broken. 
The $S^a$ generators, with $a=1, \ldots ,4$, together with the $X^a$ generators, with $a=1,2,3$, generate an $SU(2)_\text{L} \times SU(2)_\text{R} \times U(1)_\text{V}$ algebra. This is easily seen by changing genarator basis from $(S^a,X^a)$ to $(L^a,R^a)$, where
\begin{eqnarray}
L^a \equiv \frac{S^a + X^a}{\sqrt{2}} = \begin{pmatrix}\frac{\tau^a}{2}\ \ \  & 0 \\ 0 & 0\end{pmatrix} \ , \ \
{-R^a}^T \equiv \frac{S^a-X^a}{\sqrt{2}}  = \begin{pmatrix}0 & 0 \\ 0 & -\frac{{\tau^a}^T}{2}\end{pmatrix} \ ,
\end{eqnarray}
with $a=1,2,3$. The electroweak gauge group is then obtained by gauging $SU(2)_\text{L}$, and the $U(1)_Y$ subgroup of $SU(2)_\text{R} \times U(1)_\text{V}$, where
\begin{eqnarray}
Y =  -{R^3}^T + \sqrt{2}\ Y_{\rm V}\ S^4 \ ,
\end{eqnarray}
and $Y_{\rm V}$ is the U(1)$_V$ charge. For example, from Eq.~(\ref{assign1}) and Eq.~(\ref{assign2}) we see that $Y_{\rm V}=y$ for the techniquarks, and $Y_{\rm V}=-3y$ for the new leptons. As $SU(4)$ spontaneously breaks to $SO(4)$, $SU(2)_\text{L} \times SU(2)_\text{R}$ breaks to $SU(2)_\text{V}$. As a consequence, the electroweak symmetry breaks to $U(1)_Q$, where
\begin{eqnarray}
Q = \sqrt{2}\ S^3 + \sqrt{2}\ Y_{\rm V} \ S^4 \ .
\end{eqnarray}
Moreover the $SU(2)_\text{V}$ group, being entirely contained in the unbroken $SO(4)$, acts as a custodial isospin, which insures that the $\rho$ parameter is equal to one at tree-level.

The electroweak covariant derivative for the $M$ matrix is
\begin{eqnarray}
D_{\mu}M =\partial_{\mu}M - i\,g \left[G_{\mu}(y)M + MG_{\mu}^T(y)\right]  \
, \label{covariantderivative}
\end{eqnarray}
where
\begin{eqnarray}
g\ G_{\mu}(Y_{\rm V}) & = & g\ W^a_\mu \ L^a + g^{\prime}\ B_\mu \ Y  \nonumber \\
& = & g\ W^a_\mu \ L^a + g^{\prime}\ B_\mu \left(-{R^3}^T+\sqrt{2}\ Y_{\rm V}\ S^4\right) \ .
\label{gaugefields}
\end{eqnarray}
Notice that in the last equation $G_\mu(Y_{\rm V})$ is written for a general U(1)$_V$ charge $Y_{\rm V}$, while in Eq.~(\ref{covariantderivative}) we have to take the U(1)$_V$ charge of the techniquarks, $Y_{\rm V}=y$, since these are the constituents of the matrix $M$, as explicitly shown in Eq.~(\ref{M-composite}).

\section{Technicolor on Event Generators}

\label{TEG}

We now implement the effective Lagrangian \eqref{eq:boson} on event generators such as \href{http://madgraph.phys.ucl.ac.be/}{MadGraph} \cite{Alwall:2007st,MG/ME} and \href{http://theory.sinp.msu.ru/~pukhov/calchep}{CalcHEP} \cite{Pukhov:1999gg,CalcHEP}.

To implement the models on Monte Carlo event 
generators we used the \href{https://feynrules.phys.ucl.ac.be/wiki}{FeynRules} Mathematica 
package~\cite{Christensen:2008py,Christensen:2009jx,FRweb}\footnote{The 
FeynRules implementation is a complete rewrite of the 
LanHEP~\cite{Semenov:2008jy} implementation that was used 
in~\cite{Belyaev:2008yj,Frandsen:2011hj}.}. 
FeynRules offers interfaces to several Monte Carlo generators. The 
interfaces to   
\href{http://theory.sinp.msu.ru/~pukhov/calchep}{CalcHEP} 
and \href{http://madgraph.phys.ucl.ac.be/}{MadGraph}  have 
been tested for this model. The implementation supports unitary gauge. 
The FeynRules model file and the precompiled model files for MadGraph and CalcHEP can be downloaded here
\begin{center}
\href{https://feynrules.phys.ucl.ac.be/wiki/TechniColor}{https://feynrules.phys.ucl.ac.be/wiki/TechniColor} 
\end{center}
and here 
\begin{center}
\href{http://cp3-origins.dk/research/tc-tools}{http://cp3-origins.dk/research/tc-tools}. 
\end{center}
Unless the user wants to modify the 
model to include some new features, it is advisable to download the 
precompiled input files. These include 
some tweaks with respect to the
output of FeynRules which makes their use easier.

\subsection{Ruling Technicolor with FeynRules}

The model contains the SM fermions and gauge bosons. The 
naming convention of the particles was chosen to follow that of the SM 
implementation in MadGraph. As the implementation is aimed for studying 
the signals from new high energy interactions, the SM Lagrangian was 
taken to be a very simple one. In particular, the electron, the muon, 
and the neutrinos, as well as the up, down and strange quarks, are taken 
to be massless. This increases significantly the speed of Monte Carlo 
computations for some processes. We include only the Cabibbo mixing of 
the quarks. The input parameters (and their corresponding default 
values) of the SM sector are:
\begin{itemize}
 \item $\mathrm{EE}=0.313429$, the electron charge;
 \item $\mathrm{GF}=0.0000116637$, the Fermi coupling constant (in 
units of GeV$^{-2}$);
 \item $\mathrm{MZ}=91.1876$, the mass of the $Z$ boson (in units of GeV);
 \item $\mathrm{aS}=0.118$, the $Z$ pole value of the strong coupling 
constant $\alpha_s$;
 \item $\mathrm{cabi}=0.227736$, the Cabibbo angle;
 \item The masses of the heavy quarks $\mathrm{MC}=1.3$, 
$\mathrm{MB}=4.2$, $\mathrm{MT}=172$, and the tau lepton 
$\mathrm{MTA}=1.777$;
 \item The widths of the weak gauge bosons and the top quark: 
$\mathrm{wZ}=2.4952$,  $\mathrm{wW}=2.141$, and $\mathrm{wT}=1.508$.
\end{itemize}

The TC sector of the model implements the Vanilla Technicolor 
Lagrangian~(\ref{eq:boson}) with the mass of the heavy vector spin one 
state fixed by the first WSR, i.e. the case of generic 
walking TC in Table~\ref{tab:Inparam}. Thus the composite 
states are
\begin{itemize}
 \item H: the composite Higgs $H$;
 \item R1N and R2N: the neutral heavy spin one states $R_1^0$ and $R_2^0$;
 \item R1+, R1-, R2+, and R2-: the charged heavy spin one states 
$R_1^\pm$ and $R_2^\pm$.
\end{itemize}
For these states the effective theories of MWT and NMWT coincide.

The input parameters of the composite TC sector are those 
listed already in Table~\ref{tab:Inparam}:
\begin{itemize}
 \item $\mathrm{MA}$, the mass $M_A$ of the axial spin one states. Its 
allowed range is from about $500$ (GeV) to about $3000$, depending on 
the values of gt and S.
 \item $\mathrm{gt}$ is the coupling strength $\tilde g$ of the 
TC interactions and its allowed value ranges from about 1 to circa 
10, depending slightly on the parameters MA and PS.
 \item $\mathrm{PS}$ indicates the $S$ parameter of the new strongly coupled sector in isolation. This value is expected to be positive and 
small (much less than one). Estimates give $\mathrm{PS}\simeq 0.1$ for 
MWT and $\mathrm{PS} \simeq 0.3$ for NMWT. The full phenomenologically relevant $S$ parameter comes from different contributions such as the one from the new lepton family. 
 \item $\mathrm{MH}$ is the mass $M_H$ of the composite Higgs boson. As 
discussed in the main text, the Higgs is expected to be lighter than the 
vector mesons for walking TC models.
 \item $\mathrm{rs}$ is the parameter $s$ that controls the coupling 
between the composite Higgs and the heavy spin one states which is expected to 
be $\mathcal{O}(1)$.
 \end{itemize}

An important feature of the model is that the widths of the heavy 
vectors and the composite Higgs depend strongly on the parameters MA, 
gt, and MH. This dependence needs to be included in the Monte Carlo 
calculations
in order to obtain reliable results.
The FeynRules implementation does not include the expressions for these 
widths, since they are computed in different ways, for example, in 
CalcHEP and in MadGraph. It is possible to set CalcHEP to calculate and 
update the widths {\it on the fly} automatically whenever the values of the parameters are 
changed (see the CalcHEP manual \cite{Pukhov:2004ca}). In MadGraph, the 
widths can be included in the calculator program which takes care of 
changing the parameters. A calculator for the TC model, which 
includes the widths of the composite particles, can be downloaded from the same webpages of the FeynRules model implementation given above.

\subsection{Madding Technicolor via MadGraph/MadEvent v.4}

\href{http://madgraph.phys.ucl.ac.be/}{MadGraph/MadEvent} (just MadGraph or MG/ME in brief) is a software that allows to generate amplitudes and events in models for particle interactions. The SM as well as some beyond SM models are already included in the MG/ME model files and are ready to be used once installed. The (N)MWT implementation can be downloaded from the links given at the beginning of this Section.

Here the aim is to provide a fast, practical and ready-to-go guide throughout the installation of MadGraph version 4 \footnote{No modifications for MadGraph v. 5 are needed.} and its use with the MWT package; for further informations and a deeper insight refer to the MadGraph manual, downloadable from
\begin{center}
\href{http://cp3wks05.fynu.ucl.ac.be/twiki/bin/view/}{http://cp3wks05.fynu.ucl.ac.be/twiki/bin/view/}.
\end{center}

To download the MadGraph package a registration is needed on the MG/ME website or one of its mirrors. To compute the matrix elements only the MadGraph StandAlone package is needed, while to also generate events, the entire MadGraph Developer's kit is needed.

\begin{figure}
\begin{center}
\includegraphics[width=\textwidth]{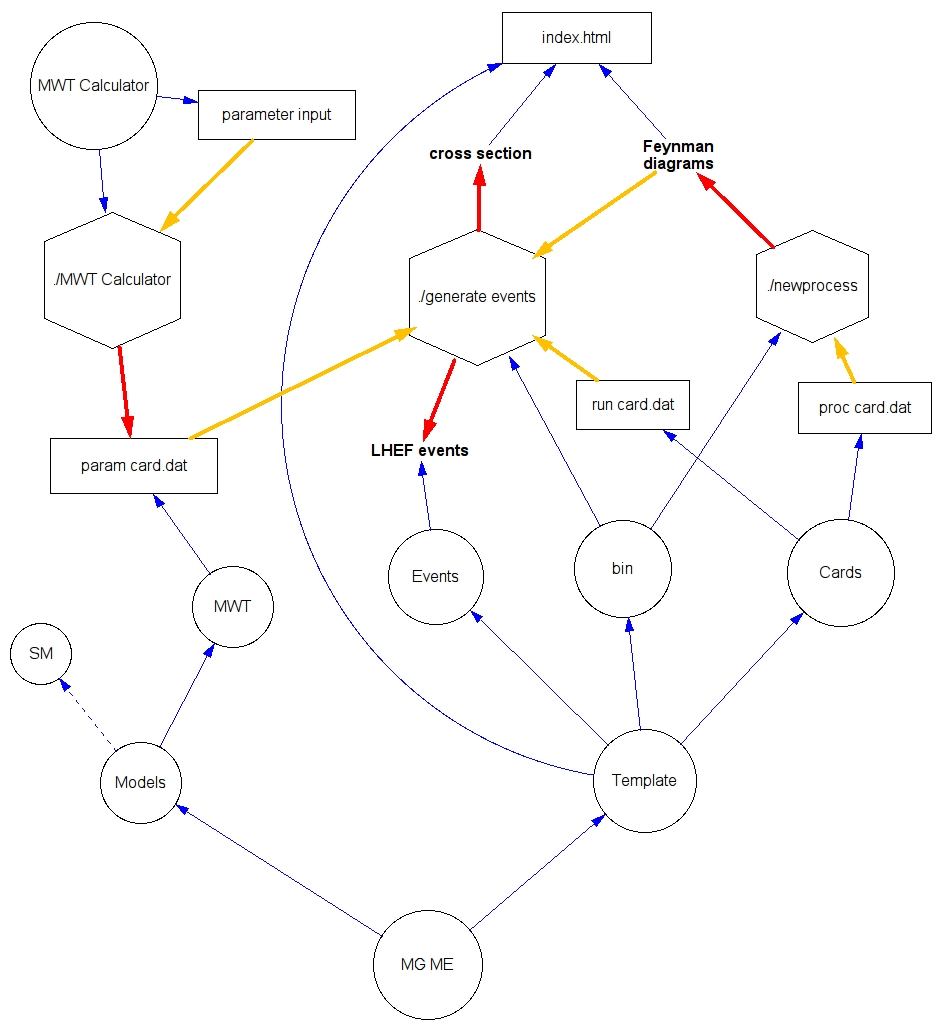}
\caption{\label{Fig:MGscheme}Graph illustrating the usage of the MWT implementation in MG/ME. Circles represent directories, rectangles represent text files while hexagons are executable files. Blue arrows show placement of files and folders into the various directories, thick yellow arrows indicate input for the executable files while thick red ones indicate output.}
\end{center}
\end{figure}

We refer to the MadGraph manual for installation instructions. Inside the MG/ME folder one finds two useful directories: the \textit{Model} directory is the one where the MWT model folder needs to be placed, while the \textit{Template} directory is the one from where the simulations are run. Here the files with the events generated by MadGraph are created.
\\
Fig.~\ref{Fig:MGscheme} shows the position of the directories and files used.

To use the new model the first step is to specify the process or the processes one wants to study. This is done by modifying the \textit{proc\_card.dat} file inside the \textit{Template/Cards} folder. As shown in the default file it is possible to specify more than one process at the same time, provided that the various processes are identified by an integer number following the character \verb+@+. The names of the particles and respective antiparticles are defined in the \textit{particles.dat} file inside the specified model folder, and are case insensitive; in the SM default implementation for instance \verb+e-+, \verb+mu-+, \verb+ta-+ are the charged leptons (with antiparticles \verb+e++, \verb+mu++, \verb+ta++), \verb+ve+, \verb+vm+, \verb+vt+ the neutrinos (with antiparticles denoted by a $\sim$ after the name), the quarks are \verb+u+, \verb+d+, \verb+c+, \verb+s+, \verb+t+, \verb+b+ (again with antiparticles denoted by a $\sim$ after the name), \verb+h+ is the Higgs boson and \verb+g+, \verb+a+, \verb+z+, \verb+w++ and \verb+w-+ are the gauge bosons. In the MWT implementation the SM particles preserve their names, and in addition one has the new neutral vector resonances \verb+r1+, \verb+r2+ plus the charged vector resonances \verb+r1++, \verb+r1-+, \verb+r2++, \verb+r2-+.
 
The process is specified listing initial and final particles separated by the character \verb >.  For example to study $e^+ e^- \to \mu^+ \mu^-$ one should type \verb:e+e->mu+mu-:. At the bottom of this file one can set the definitions for the \virg{multiparticles}, i.e. labels that are given to sets of particles rather than a single one. For instance, one can define the proton as 
\begin{verbatim}
P uu~dd~g
\end{verbatim}
meaning that the proton $p$ is composed by $u$, $\bar u$, $d$, $\bar d$ and the gluon $g$. The process \verb:pp>e+e-: will henceforth consist of all the processes with any combination of two of the following partons $u$, $\bar u$, $d$, $\bar d$ and $g$ in the initial state. Besides modifying the present definitions, the user can provide some new ones. Note that the antiproton is not defined here, and if one wants $p \bar{p}$ in the initial state he should just write \verb+pp+: whether these are protons and/or antiprotons it will be specified later, in the choice of the collider type in the \textit{run\_card.dat} file.

When specifying the process, use the form \verb+xy>z>abc+ to require the particle \verb+z+ to appear in a s-channel intermediate state, and use \verb+xy>abc/z+ to exclude the particle \verb+z+ to appear in any diagram.

Right below the process specification one can assign the maximum number of QED- and QCD-type vertices in each Feynman diagram that will be produced.

The model one wants to use for his simulations shall be specified as the name of the corresponding folder inside the \textit{Model} directory; for example it is \verb+MWT+ for the MWT model (case insensitive).

To generate the Feynman diagrams for the processes specified, one has now to enter the \textit{Template/bin} folder and run the command \verb+./newprocess+ from shell. The result  can be seen by opening the file \textit{Template/index.html} and by clicking on the link \virg{Process Information}; the Feynman diagrams are provided both as \textit{jpg} files (click on \virg{html}) as well as \textit{ps} files (click on \virg{postscript}).

We can now generate events to compute different cross-sections. The result will depend on both the Feynman diagrams that have been generated in the previous step and the value of the couplings and parameters of the model used. To assign different values to the MWT parameters it is necessary to use the calculator, downloadable from the webpages linked at the beginning of this Section; the \textit{readme.txt} file coming along with it contains instructions on how to use it. Once extracted the package one has to compile it typing \verb+make+ from shell; the executable file \textit{MWT\_calculator} is then created. One can assign new values to some or all external parameters editing the \textit{param\_input} file, where all the parameters that can be modified are listed. The calculator will assign a default value to all those parameters that are not requested to be modified in this file. To run the calculator, type
\begin{verbatim}
./MWT_Calculator parameter_input > param_card.dat
\end{verbatim}
from shell, so that the file \textit{parameter\_input} is given as input and a file named \textit{param\_card.dat} is produced. Opening this last file one can see the values of all the external parameters and all the masses of the model; the values of the other parameters will be automatically calculated by MadGraph. Substituting now this file to the existing one in the MWT model folder inside the \textit{Model} directory will update the model with the new parameters values.

The second step before generating events is editing the \textit{Template/Cards/run\_card.dat} file, where one can set the simulation and the collider parameters, choose the parton distribution functions (PDF) and specify some cuts. The simplest entries one can edit are:
\begin{itemize}
\item the tag name for the run, that is just one word providing a label to the run;
\item the number of (unweighted) events one wants;
\item the random seed; it should be always left to zero, unless the user aims to reproduce a specific set of results. If left to zero, it is incremented automatically in each run so that successive runs are statistically independent;
\item the collider type, that means the type of particles in each beam and their PDFs: use $1$ to indicate the proton and $-1$ for the antiproton; setting $0$, the PDFs are switched off. The default setting is $pp$, as in the case of LHC;
\item the beams energy;
\item the PDF choice: to have the list of different possibilities, see the MG/ME manual. Note that the PDF automatically fixes also $\alpha_s$ and its evolution;
\item the cuts.
\end{itemize}

To generate events and get the cross sections of the processes, run now from inside the directory \textit{Template/bin} the command \verb+./generate_events+ from shell; type \verb+0+ to run the simulation on a single machine, and then provide the run name, i.e. a label for the run. Depending on the number of particles and the number of Feynman diagrams involved the run can take from minutes to hours. The result of the simulation can be seen once again opening the file \textit{Template/index.html} and clicking on the link \virg{Results and Event Database}. The files containing the events (in the LHEF format \cite{Alwall:2006yp}) are stored in the \textit{Template/Events} folder.

\medskip
Summarising:
\begin{enumerate}
\item  Install the MadGraph/MadEvent package from the \href{http://madgraph.phys.ucl.ac.be/}{website};
\item download the MWT implementation and the calculator from
\begin{center}
\href{http://cp3-origins.dk/research/tc-tools}{http://cp3-origins.dk/research/tc-tools}, 
\end{center}
and place the \textit{MWT} model folder inside the MadGraph \textit{Models} directory and compile the calculator;
\item specify the processes you want to study and the model in the \textit{Cards/proc\_card.dat} file;
\item run \verb+./newprocess+ from the \textit{Template/bin} directory to produce the Feynman diagrams. You can check the result in the file \textit{Template/index.html};
\item to change the parameters of the MWT model, edit the \textit{parameter\_input} file in the calculator folder and then run
\\
\verb+./MWT_Calculator parameter_input > param_card.dat+
\\
to generate the \textit{param\_card.dat} file, that has to be placed into the MWT model folder in the MadGraph \textit{Models} directory;
\item in order to set the run parameters (collider type, beams energy, PDFs and so on) and impose cuts, edit the \textit{Template/Cards/run\_card.dat};
\item run \verb+./generate_events+ from \textit{Template/bin} to calculate the cross section and generate the events;
\item check the result with the file \textit{Template/index.html}; the event files are stored into the \textit{Template/Events} folder.
\end{enumerate}

\subsection{Calculating Technicolor with CalcHEP}

\href{http://theory.sinp.msu.ru/~pukhov/calchep}{CalcHEP} is a package that allows the user to go from the Lagrangian to the physical observables calculated at the lowest order in perturbation theory. The package can be downloaded from the CalcHEP website, where also the manual and information about new features are available. In addition users need {\it Fortran} and {\it C} compilers\footnote{For example {\it g77} and {\it gcc}.} and {\it X Window System} ({\it X11})\footnote{In some operating systems additional {\it X11} development libraries are required.} to install the program. 

The current CalcHEP version 2.5.7 includes by default the SM and large extra dimensional model implementations. Other models can be also downloaded from the CalcHEP web page. The (N)MWT implementation can be downloaded from the links given at the beginning of this Section.

First of all one has to download the tar file that contains the source code from the CalcHEP web page, and extract the package. One has then to open  the file {\it getFlags} in the new CalcHEP folder and make sure that the names of the {\it Fortran} and {\it C} compilers correspond to those which are installed in the computer (and in case they do not, one has to change the names to the corresponding ones). To install the program one should move to the CalcHEP folder and run the command
\verb+ make+. If all works properly the message {\it CalcHEP is installed successfully and can be started} should appear. It is convenient to create a working directory placed outside the CalcHEP folder by using the command\verb+ ./mkUsrDir ../WorkDir+. At this point the program is ready to use and a Graphical User Interface (GUI) session can be launched with the command
\verb+ ./calchep+, to be entered preferably from the working directory. One can navigate from menu to menu by using the arrow and esc keys. Instructions and help can be found pressing the F1 and F2 keys.
  
To install a new model in the current CalcHEP installation one should download the model files and copy them to the models folder inside the CalcHEP main folder. Among those, the file prtcls.mdl contains description of the particles, vars.mdl defines the model parameters, func.mdl contains additional parameters which depend on those defined in the previous file, and lgrng.mdl determines the interactions. To install the new model one has to launch CalcHEP, choose from the second menu {\it Import models}, and type \verb+CALCHEP/models+: in the resulting menu one can scroll the available models and choose the one to import.
 
 \begin{figure}[h]
\includegraphics[scale=0.65]{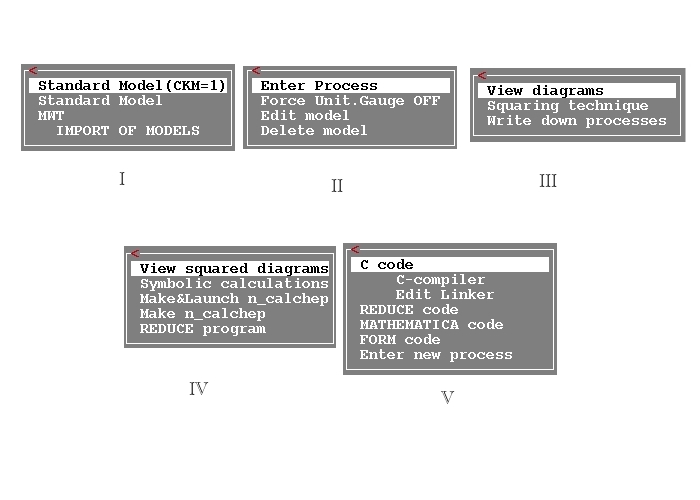}
\caption{Menus of the symbolic part.}
\label{symb}
\end{figure}

\begin{figure}[]
\hskip -2.3cm
\includegraphics[scale=0.58]{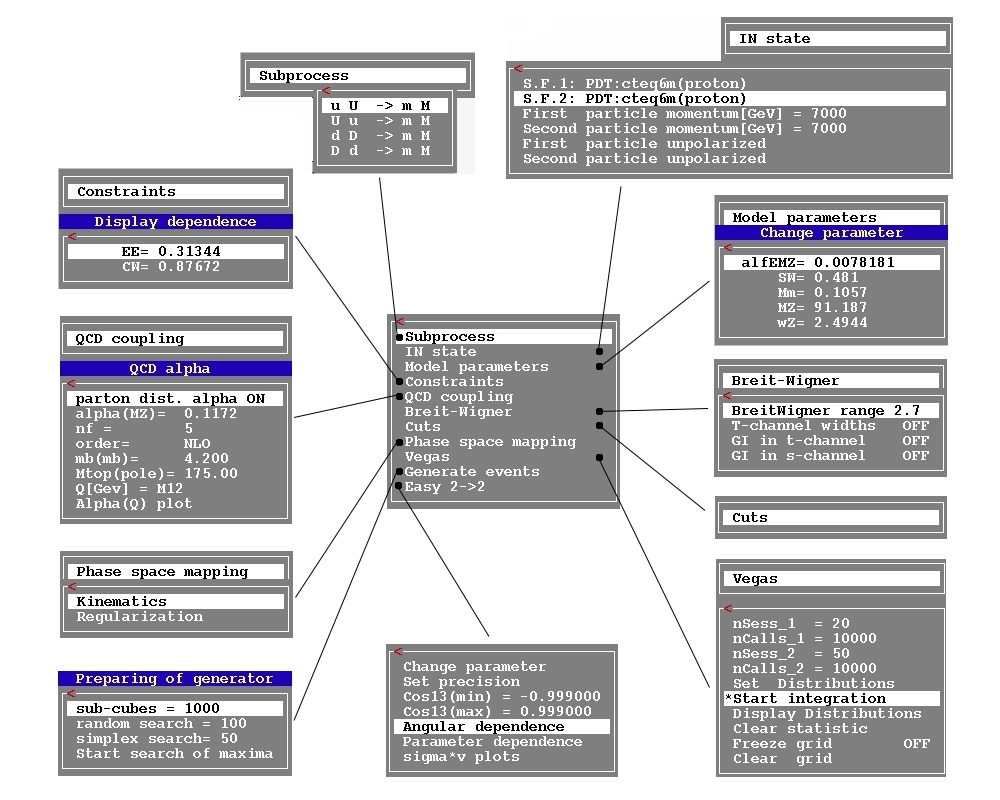}
\caption{Menus of the numerical part.}
\label{num}
\end{figure}

CalcHEP contains two parts: a symbolic and a numerical one. Menus of the symbolic part are presented in Fig.~\ref{symb}. In the second menu in Fig.~\ref{symb} the user can choose the gauge (either Feynman-'t Hooft or unitary), or edit the model. Choosing {\it Enter Process} the user can define the process. The number of initial state particles have to be either one or two. While the number of the final state particles is not limited in principle, the increasing amount of time required by the simulation sets a practical limit between four and six on that number, for a typical machine and process. It is also possible to exclude some particles from the Feynman diagrams used to calculate the process amplitude.

In the third menu the user can view the created diagrams and also save them in a separate file. Choosing {\it Squaring techniques} CalcHEP squares the diagrams. The numerical session can now be started directly by choosing {\it Make\&Launch n\_calchep} or the user can ask CalcHEP to calculate directly the squared matrix elements by choosing {\it Symbolic calculations}. The resulting analytic expressions for the squared matrix elements can be saved in different formats. One can then move to to the numerical session by choosing C-compiler from the fifth menu in Fig.~\ref{symb}.

In numerical sessions the user can define the momentum of the initial state particles and choose if and which parton distribution functions (PDF's) to use. Several cuts can be imposed and regularization of the propagators can be done before the numerical integration. Also the model parameters can be changed at this point. CalcHEP automatically re-calculates values of the related parameters if the model parameters are changed. The structure of the menus in the numerical part is presented in Fig.~\ref{num}.

CalcHEP uses Vegas to perform the numerical integrals: the user has to choose the number of Vegas calls and iteration steps. The actual number depends on how much time the user allows and how well the integrals converge. Some convergence problems can be fixed by the a wise choice of the regularization scheme and appropriate cuts. The program can also generate plots of the differential cross section associated with a given process in function of one of the independent variables. The important part of the program is the event generator which allows to make predictions for collider phenomenology. The method for producing the events is explained in the CalcHEP manual. 

Though intuitive,  the graphical interface is not the only way to use CalcHEP: the user can write at once all the information relative to the process to be simulated in the batch file and then execute it in the working directory. For example, if the name of the batch file is ee-mumu-batch one has to use the command \verb+ ./calchep_batch ee-mumu-batch+ to run the batch. 

For more information on CalcHEP usage, visit the \href{http://theory.sinp.msu.ru/~pukhov/calchep}{CalcHEP} homepage where the official manual can be downloaded.


\includepdf[noautoscale=true, scale=1.06, offset=35 -11]{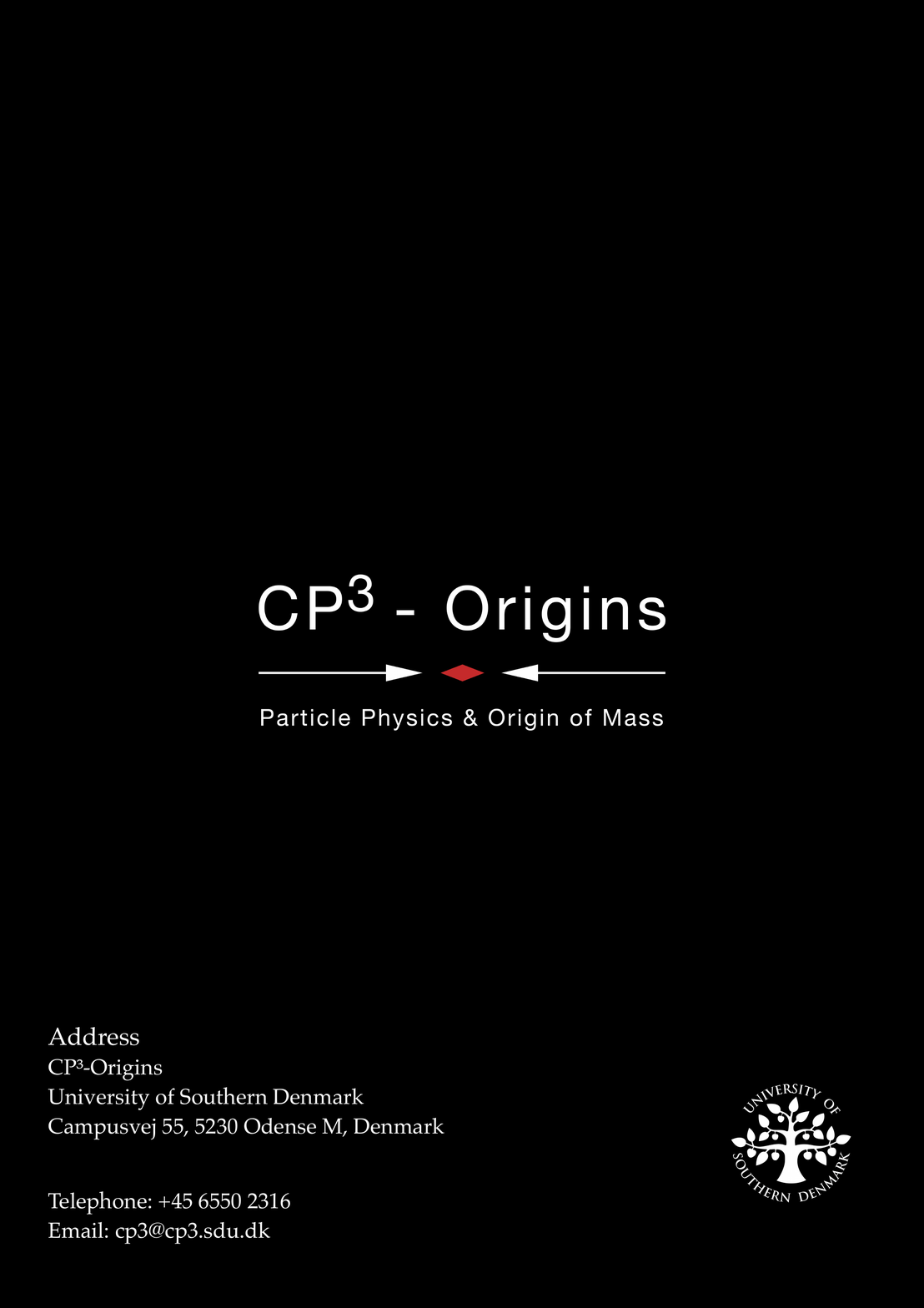}

\end{fmffile}
\end{document}